\documentclass[%
 article,
nofootinbib,
 amsmath,amssymb,
 aps,
prd,
]{revtex4-2}

\usepackage{dcolumn}
\usepackage{bm}

\allowdisplaybreaks

\usepackage[T1]{fontenc}
\usepackage[english]{babel}
\usepackage{amssymb,amsmath}
\usepackage[top=0.6in, bottom=0.6in, left=0.6in, right=0.6in]{geometry}
\usepackage{float}
\usepackage{graphicx}
\usepackage{parskip}
\usepackage{tikz}
\usepackage[compat=1.1.0]{tikz-feynman}
\tikzfeynmanset{warn luatex=false}
\tikzfeynmanset{/tikzfeynman/warn luatex = false}
\usepackage{hyperref}
\usepackage[font=scriptsize]{subcaption}
\usepackage[font=small]{caption,subcaption}
\usepackage{pgf}
\usepackage{physics}
\usetikzlibrary{arrows.meta}
\usepackage{mathtools}
\usepackage{sidecap}
\usepackage{cleveref}
\usepackage{xfrac}
\usepackage[normalem]{ulem}
\usepackage{svg}
\newcommand\redout{\bgroup\markoverwith
{\textcolor{red}{\rule[1ex]{2pt}{0.8pt}}}\ULon}

\usepackage{empheq}
\allowdisplaybreaks

\usepackage{calrsfs}
\newcommand{\La}{\mathcal{L}}

\tikzset{
    ->-/.style={decoration={
  markings,
  mark=at position .6 with {\arrow{>}}},postaction={decorate}},
    -<-/.style={decoration={
  markings,
  mark=at position .5 with {\arrow{<}}},postaction={decorate}},
}

\usepackage{cleveref}

\usetikzlibrary{positioning}
\tikzset{main node/.style={circle,fill=blue!20,draw,pattern=crosshatch dots,pattern color=black!20!white,minimum size=1.2cm,inner sep=0pt, line width=0.4mm},
            }
            
\tikzset{main node2/.style={circle,fill=blue!40,draw,pattern=crosshatch dots,pattern color=black!20!white,minimum size=3cm,inner sep=0pt, line width=0.4mm},
            }
            
\tikzset{main node3/.style={circle,fill=blue!40,draw,pattern=crosshatch dots,pattern color=black!20!white,minimum size=2.3cm,inner sep=0pt, line width=0.4mm},
            }

\tikzfeynmanset{
  fermion1/.style={
    /tikz/postaction={
      /tikz/decoration={
        markings,
        mark=at position 0.5 with {
          \node[
            transform shape,
            xshift=-0.5mm,
            fill,
            inner sep=0.5pt,
            draw=none,
            dart
          ] { };
        },
      },
      /tikz/decorate=true,
    },
  },
  fermion2/.style={
    /tikz/postaction={
      /tikz/decoration={
        markings,
        mark=at position 0.5 with {
          \arrow{[length=4pt, width=3pt]};
        },
      },
      /tikz/decorate=true,
    },
  },
}

\usetikzlibrary{automata,positioning,arrows,matrix,backgrounds,calc}
\usetikzlibrary{decorations.text}
\usetikzlibrary{decorations.pathmorphing}
\usetikzlibrary{shapes.geometric}

\usepackage{enumitem}

\expandafter\def\expandafter\normalsize\expandafter{%
    \normalsize
    \setlength\abovedisplayskip{10pt}
    \setlength\belowdisplayskip{10pt}
    \setlength\abovedisplayshortskip{10pt}
    \setlength\belowdisplayshortskip{10pt}
}

\date{}
\usepackage{slashed}
\usepackage{dsfont}

\usepackage{mmacells}

\mmaDefineMathReplacement[≤]{<=}{\leq}
\mmaDefineMathReplacement[≥]{>=}{\geq}
\mmaDefineMathReplacement[≠]{!=}{\neq}
\mmaDefineMathReplacement[→]{->}{\to}[2]
\mmaDefineMathReplacement[⧴]{:>}{:\hspace{-.2em}\to}[2]
\mmaDefineMathReplacement{∉}{\notin}
\mmaDefineMathReplacement{∞}{\infty}
\mmaDefineMathReplacement{times}{\times}

\mmaSet{
  morefv={gobble=2},
  linklocaluri=mma/symbol/definition:#1,
  morecellgraphics={yoffset=1.9ex}
}
\hypersetup{colorlinks=false, citecolor=red, urlcolor=blue, linkcolor=cyan}

\begin{document}


\begin{center}


\end{center}


\title{Classical worldlines from scattering amplitudes}

\date{\today}

\author{Zeno Capatti}
 \email{zeno.capatti@unibe.ch}
\affiliation{%
 Institute for Theoretical Physics, \\
 University of Bern, \\
Bern, Switzerland
}%

\author{Mao Zeng}
 \email{mao.zeng@ed.ac.uk}
\affiliation{%
 Higgs Centre for Theoretical Physics, \\
 University of Edinburgh, \\
 Edinburgh, United Kingdom
}%

\begin{abstract}
  We present a systematic diagrammatic investigation of the classical
  limit of observables computed from scattering amplitudes in quantum
  field theory through the Kosower-Maybee-O'Connell (KMOC) formalism,
  motivated by the study of gravitational waves from black hole
  binaries. We achieve the manifest cancellation of divergences in the
  $\hbar \to 0$ limit at the integrand level beyond one loop by
  employing the Schwinger parametrisation to rewrite both cut and
  uncut propagators in a worldline-like representation before they are
  combined. The resulting finite classical integrand takes the same
  form as the counterpart in the worldline formalisms such as
  post-Minkowskian effective field theory (PMEFT) and worldline
  quantum field theory (WQFT), and in fact exactly coincides with the
  latter in various examples, showing explicitly the equivalence
  between scattering amplitude and worldline formalisms. The classical
  causality flow, as expressed by the retarded propagator
  prescription, appears as an emergent feature. Examples are presented
  for impulse observables in electrodynamics and a scalar model at two
  loops, as well as certain subclasses of diagrams to higher orders
  and all orders.
\end{abstract}

\maketitle

\begin{center}
\rule{14cm}{0.5pt}
\end{center}

\linespread{0.9}\selectfont
\tableofcontents
\linespread{1.3}\selectfont

\section{Introduction}

Since the historic detection of gravitational waves in 2015
\cite{LIGOScientific:2016aoc}, gravitational wave astronomy has
progressed rapidly, with a total of more than 200 detected events
expected by the end of the LIGO-Virgo-KAGRA O4 run in early
2025. Planned future detectors ~\cite{Punturo_2010, LISA:2017pwj,
  Reitze:2019iox} will significantly increase the signal-to-noise
ratio and require vastly improved theoretical predictions for
waveforms to confront the detection data \cite{Purrer:2019jcp,
  Buonanno:2022pgc}. In the inspiral regime of the gravitational wave
signals, the compact objects (black holes or neutron stars) have a
separation that is much larger than their Schwarzschild radii, and
perturbative descriptions of conservative and dissipative orbital
dynamics is needed, as numerical relativity simulations are generally
too expensive to cover a large number of orbit cycles over a large
parameter space.

Among a variety of perturbative expansion schemes, this paper concerns
the post-Minkowskian (PM) expansion \cite{Bertotti:1956pxu,
  Kerr:1959zlt, Bertotti:1960wuq, Westpfahl:1979gu, Portilla:1980uz,
  Bel:1981be}, which is an expansion in the gravitational coupling
constant only, while treating the velocities and mass ratios of
two-body systems exactly. Perturbative general relativity, being a
non-linear theory of spin-2 fields, is inherently challenging.  In
recent years, quantum field theory (QFT) methods have greatly advanced
the state of the art in computing the post-Minkowskian expansion.
Compact objects are mapped to massive point-like particles, either as
massive fields in the scattering amplitude approach or as
semi-classical worldlines in the worldline approach. Both approaches
have produced cutting-edge results. For example, for conservative and
time-integrated dissipative dynamics of spinless binaries, see
Refs.~\cite{Cheung:2018wkq, Kosower:2018adc, Bern:2019nnu,
  Bern:2019crd, Cristofoli:2019neg, Bjerrum-Bohr:2019kec,
  Brandhuber:2021eyq, Bern:2021dqo, Bern:2021yeh, Damgaard:2023ttc} in
the scattering amplitude approach and Refs.~\cite{Kalin:2020mvi,
  Kalin:2020fhe, Kalin:2020lmz, Mogull:2020sak, Jakobsen:2021smu,
  Dlapa:2021npj, Dlapa:2021vgp, Mougiakakos:2021ckm, Jakobsen:2022psy,
  Riva:2021vnj, Kalin:2022hph, Dlapa:2022lmu, Dlapa:2023hsl,
  Driesse:2024feo} in the worldline approach. Both approaches have
their advantages. For example, the scattering amplitudes approach
benefits from modern advances in scattering amplitudes including
generalised unitarity \cite{Bern:1994zx, Bern:1994cg, Bern:1995db, Bern:1997sc,
  Britto:2004nc} and the double copy \cite{Bern:2008qj, Bern:2010ue},
while the worldline approach makes the classical limit more manifest
and bypasses unphysical terms that diverge in the $\hbar \to 0$ limit,
which we refer to as ``super-leading divergences'' or ``super-leading
terms'' (also known as ``superclassical terms'' or ``classically
divergent terms'' in the literature). Such terms are present in
intermediate calculation steps in the scattering amplitude approach
but cancel in the final result.

A main goal of this paper is comparing the scattering amplitude
approach and the worldline approach and demonstrate they are
equivalent at a very strong level, more precisely, at the level of
loop integrands. Such an equivalence is not entirely surprising, as
both approaches involve Feynman integrals with the same type of
propagators, i.e.\ (usual) quadratic propagators for the massless
graviton lines and linearized matter propagators. In the scattering
amplitudes, linearized matter propagators arise from the classical
(soft) expansion \cite{Kosower:2018adc, Parra-Martinez:2020dzs} using
the method of regions \cite{Beneke:1997zp}, while the same linearized
propagators emerge in the worldline approach from a perturbative
expansion of the worldline trajectory. In the end, however, the
\emph{same type} of Feynman integral are produced. This paper takes it
further and demonstrate that the integrands in the two approaches
exactly coincide after simple manipulations. To be concrete, we will
compare the Kosower-Maybee-O'Connell (KMOC) formalism
\cite{Kosower:2018adc}, which establishes that classical observables can be obtained as a soft limit of quantum observables computed in the IN-IN scattering formalism~\cite{Keldysh:1964ud,Schwinger:1960qe}, with 
worldline quantum field theory (WQFT) \cite{Mogull:2020sak,
  Jakobsen:2022psy} in the worldline approach. An earlier take on the
worldline approach tailored for scattering observables is the
post-Minkowskian effect field theory (PMEFT) proposed by K\"alin and
Porto \cite{Kalin:2020mvi}. The WQFT formalism is equivalent and
recasts iterations by equations of motion into a diagrammatic
representation, which is convenient as a target for diagrammatic
comparisons in our study. Both variants of the worldline approach
ultimately descend from Goldberger and Rothstein's non-relativistic
general relativity (NRGR) formalism \cite{Goldberger:2004jt} targeting
the post-Newtonian expansion, while making several technical
refinements and adaptations.

As a byproduct, we give new prescriptions for making the KMOC
formalism manifestly free of super-leading divergences, using a
systematic procedure which turns quantum currents into worldline-like
expressions via the Schwinger parametrisation. For brevity, we refer
to such expressions as ``quantum worldlines'', which are the central
objects in our formalism. The diagrams in the linear-in-amplitude
(uncut) and quadratic-in-amplitude\footnote{In the KMOC formalism, classical
observables generally receive two types of contributions, namely
\emph{linear-in-amplitude} contributions and
\emph{quadratic-in-amplitude} contributions. The latter type of
contributions involves additional cuts, i.e.\ on-shell phase space
constraints, and will be sometimes referred to as the \emph{cut}
contributions. Furthermore, mirroring prevalent terminology in the
perturbative QCD literature, we also refer to the two types of
contributions as \emph{virtual} and \emph{real} contributions,
respectively. We will often use the superscript $V$ and $R$ to denote
virtual and real contributions, respectively.} contributions in the KMOC formalism
are then combined in a natural way, as worldline-like integrals with
different vertex orderings. For our purpose, we find this to be more
convenient than applying distributional identities to combine linear-
and quadratic-in-amplitude contributions, e.g.\ as in the original KMOC
study \cite{Kosower:2018adc} and studies of an alternative
exponential representation of the S matrix \cite{Damgaard:2021ipf, Damgaard:2023ttc}.

To focus on the essential physics rather than the algebraic
complexities of gravity amplitudes, we use model theories in this
paper, including a scalar model and electrodynamics. The scalar model,
in the QFT setting, involves cubic vertices between two massive matter
lines and a massless scalar line that mediates the long-range
interaction. The scalar model is similar to the one considered in
Refs.~\cite{Barack:2022pde, Barack:2023oqp} but without a coupling to
gravity. Electrodynamics is another useful model that demonstrates that
our methods work in the presence of nontrivial numerators in the
Feynman diagrams. Electrodynamics has often been used as a playground for
gaining insights into post-Minkowskian gravity in the literature,
e.g.\ in Refs.~\cite{Kosower:2018adc, Saketh:2021sri, Bern:2021xze,
  Bern:2023ccb, Jakobsen:2023tvm, Akpinar:2024meg}.

We briefly comment on related work. A previous diagrammatic comparison
between the KMOC formalism and the worldline EFT of
ref.~\cite{Kalin:2020mvi} was published in
ref.~\cite{Damgaard:2023vnx}, where one-loop diagrams are compared
after loop integration. We will instead compare at the integrand level
and go beyond one loop. Another important result was obtained in
ref.~\cite{Comberiati:2022ldk}, which showed the equivalence between
tree-level off-shell currents in (second-quantized) field theory and
worldline QFT while ignoring the $i \epsilon$ prescriptions for matter
propagators. In this work, we will write the currents in a worldline
form while fully preserving the $i \epsilon$ prescription, i.e.\
without neglecting what happens when the matter propagators go
on-shell. Furthermore, exponentiation properties of amplitudes is
linked to WQFT in the recent works of Refs.~\cite{Du:2024rkf,
  Ajith:2024fna}, which also studied several other issues investigated
in this paper, especially the cancellation of divergences in the
classical limit, from a different viewpoint. (See also
ref.~\cite{Kim:2024svw}.)

Some of the techniques at the core of this paper originate from analogous ones studied in the context of three-dimensional representations of Feynman diagrams such as the Loop-Tree Duality~\cite{Aguilera_Verdugo_2021,Catani_2008,Bierenbaum_2010,Runkel_2019,Sborlini_2021,Capatti:2019ypt,Capatti:2022mly,Capatti:2023shz}, the Cross-Free Family representation~\cite{Capatti:2022mly,Capatti:2023shz} as well as Time-Ordered Perturbation Theory~\cite{Sterman:1993hfp,tasi_sterman,Sterman:2023xdj}. Much like Feynman diagrams can be expressed in terms of tree diagrams using the Loop-Tree Duality, classical quantities are also known to be expressible in terms of tree diagrams. In fact, it is plausible that the results presented in this paper could similarly be obtained by considering the classical expansion of trees within the Loop-Tree Duality framework. The emergence of a causality flow is also manifestly linked to the acyclicity property observed in the context of three-dimensional representations~\cite{Capatti:2023shz,Sborlini_2021,Borinsky:2022msp,Capatti:2022mly}. Additionally, the cancellation of super-leading divergences and the appearance of a worldline description depend on the democratic combination of virtual and real contributions that is enabled by the use of a Schwinger/Fourier representation of propagators and Dirac delta functions, reminiscent of the way in which interference diagrams are combined within the Local Unitarity method~\cite{Capatti:2020xjc,Capatti:2021bsm,Capatti:2022tit} (see~\cite{Capatti:2023omc} in particular for a proof of the KLN cancellation mechanism based on the Fourier representation of Dirac delta functions appearing in phase-space integrals). Throughout this paper, however, we strive to maintain covariance at all steps so as to facilitate the applicability of analytic techniques to our results, as well as to ease the comparison with classical worldline formalism.

The outline of the paper is as follows: in
sect.\ref{sec:scalar_worldline} we work with a scalar toy model and
show that the off-shell currents, essentially off-shell Compton
amplitudes, can be recast into a worldline-like form, i.e. the
``quantum worldline''. In sect.~\ref{sec:one_loop_impulse} we discuss
the next-to-leading order impulse in the scattering of spinless
charged bodies in electrodynamics, by sewing the quantum worldlines
for two massive particles in scalar QED. These sections allow the
reader to immediately get familiar with the main ideas of the
paper. In sect.~\ref{sec:formalism}, we review the KMOC formalism,
including the precise way the classical limit has to be taken and the
existence of cancellations of super-leading terms. In
sect.~\ref{sec:quantum_worldline}, we define the fundamental object of
our formalism, the quantum worldline, which is a worldline-like
representation of off-shell currents, and study its classical
expansion. We use the quantum worldline to detail, in
sect.~\ref{sec:method}, the general method we apply to impulse
observables in order to realise \emph{locally}, i.e.\ point by point
in loop momentum space, the cancellation of super-leading
contributions and cast the integrand in the worldline form. Finally,
in sect.~\ref{sec:applications_examples}, we unfold the procedure for
a variety of examples: we work out the classical limit for the
two-loop impulse in the scalar model and in electrodynamics, as well
as the retarded propagator prescription of multi-loop ``non-iteration'' diagrams. We discuss the local
comparison between the integrand found from KMOC and that found in the
worldline approach. We further test local cancellation of
super-leading contributions for the three and four-loop ladders in
scalar QED.

\section{Warmup: Quantum worldline in scalar field theory}
\label{sec:scalar_worldline}
We start by introducing the simplest object that one can study in the
classical limit. We consider a spinless massive body coupled to a
massless scalar field that mediates long-range interactions. In the
scattering amplitudes approach, the massive body is mapped to a
massive scalar field $\phi$, and the massless scalar field is
$\Phi$. The Lagrangian is
\begin{equation}
  -\phi(\Box+m^2)\phi-\Phi\Box\Phi-g \phi^2\Phi,
  \label{eq:scalarModelLagrangian}
\end{equation}
We consider the tree-level time-ordered expectation value of two
massless scalar currents with one incoming and one outgoing on-shell
massive scalar lines. The incoming scalar has momentum $p$. The two currents couple to vertices $v_1$ and
$v_2$, and we sum over the permutations of the two vertices to
facilitate making connections with the worldline approach,
\begin{equation}
\mathcal{L}=\raisebox{-0.6cm}{\input{2att_1_scalar}} +
  \raisebox{-1cm}{\scalebox{0.9}{\input{2att_2_scalar}}}
  =
 \left[\frac{i}{(q_1+p)^2-m^2+i\varepsilon}+\frac{i}{(q_2+p)^2-m^2+i\varepsilon}\right] \tilde{\delta}((q_{12} + p)^2-m^2),
  \label{eq:scalarWorldline}
\end{equation}
having stripped the numerator of the coupling factor $(-2ig)^2$, and having introduced the notations
\begin{align}
  \tilde{\delta}(x) &= 2\pi\delta(x) \, , \label{eq:notationNormalizedDelta} \\
  q_{ij \dots} &= q_i + q_j \dots \, , \quad \text{e.g.,\quad } q_{12} = q_1 + q_2 \, .
\end{align}
Throughout this paper, we
implicitly keep the incoming massive line on-shell and omit the
on-shell constraint $\tilde{\delta}(p^2-m^2)$.

We consider the classical limit of the massive particle, with its
Compton wavelength being much smaller than the wavelength of the
massless particles it interacts with,
\begin{equation}
  \frac {\hbar} {m} \ll \frac{\hbar}{|q_i|} \, ,
\end{equation}
i.e.
\begin{equation}
  |q_i| \ll m \, . \label{eq:classicalLimit}
\end{equation}
Therefore the classical expansion is the soft expansion summarised by
the rescaling $q_i=\lambda \bar{q}_i$ with $\lambda\rightarrow 0$.
We will look at the zeroth and first orders of the expansion.
In order to do so, we introduce the Schwinger parametrisation of the
propagators and the Fourier representation of the Dirac delta function in eq.~\eqref{eq:scalarWorldline},
\begin{align}
&\frac{i\tilde{\delta}((q_{12}+p)^2-m^2)}{(q_1+p)^2-m^2+i\varepsilon}
  = \int_{-\infty}^{\infty} \mathrm{d}\Delta \tau_{12} e^{- i\Delta
  \tau_{12}[(q_1+p)^2-m^2]} \Theta(-\Delta \tau_{12})\int_{-\infty}^{\infty} \mathrm{d}\tau_2 e^{-i\tau_2[(q_{12}+p)^2-m^2]}, \label{eq:scalar_worldline_ordering1} \\
&\frac{i\tilde{\delta}((q_{12}+p)^2-m^2)}{(q_2+p)^2-m^2+i\varepsilon}
  = \int_{-\infty}^{\infty} \mathrm{d}\Delta \tau_{21} e^{-i\Delta \tau_{21}[(q_2+p)^2-m^2]}\Theta(-\Delta \tau_{21})\int_{-\infty}^{\infty} \mathrm{d}\tau_1 e^{-i\tau_1[(q_{12}+p)^2-m^2]}, \label{eq:scalar_worldline_ordering2}
\end{align}
i.e.,
\begin{align}
\begin{split}
  \mathcal{L} &= \int_{-\infty}^{\infty} \mathrm{d} \Delta \tau_{12}\mathrm{d}\tau_2 e^{-
                i[(q_1+p)^2-m^2]\Delta\tau_{12} - i[(q_{12}+p)^2-m^2]\tau_2}\Theta(-\Delta\tau_{12}) \\
              &+\int_{-\infty}^{\infty} \mathrm{d}\Delta\tau_{21}\mathrm{d}\tau_1
                e^{- i[(q_2+p)^2-m^2]\Delta\tau_{21} - i[(q_{12} + p)^2-m^2]\tau_1} \Theta(-\Delta\tau_{21})
                \ 
\end{split},            
\end{align}
where propagators have been rewritten as one-dimensional integrals
from $0$ to $\infty$ through the Heaviside theta function, and the Dirac delta function has been rewritten
as one-dimensional integrals from $-\infty$ to $\infty$. In the following, we will drop the bounds of integration for these time integrals as it is assumed the domain is the entire real line. So far, this
is just a rewriting of mathematical expressions using the Schwinger
parametrisation, but now we give a physical interpretation. Each of
the two diagrams in eq.~\eqref{eq:scalarWorldline} is associated with
an ordering of vertices $v_1$ and $v_2$. To make a connection with the
worldline picture, the two vertices are assigned proper times $\tau_1$
and $\tau_2$, respectively; then, the propagator with momentum $q_1+p$
is associated to the proper time difference
$\Delta\tau_{12} = \tau_1 - \tau_2$, while that with momentum $q_2+p$
is associated to the proper time difference
$\Delta\tau_{21} = \tau_2 - \tau_1$. Finally, the leftover delta
function is parametrised using the proper time $\tau_i$ corresponding
to the last vertex in each case.

We now change variables in the second integral eq.~\eqref{eq:scalar_worldline_ordering2}, so that it matches that of the first integral eq.~\eqref{eq:scalar_worldline_ordering1}, by writing $\Delta\tau_{12}=-\Delta\tau_{21}$ and $\tau_1=\Delta\tau_{12}+\tau_2$. As a consequence:
\begin{align}
  \frac{i}{(q_2+p)^2-m^2+i\varepsilon}&\tilde{\delta}((q_{12}+p)^2-m^2) = \int \mathrm{d}\Delta \tau_{12} e^{-i\Delta \tau_{12}[q_1^2 + 2q_1 \cdot p + 2q_1 \cdot q_2]}\Theta(\Delta \tau_{12})\int \mathrm{d}\tau_2 e^{-i\tau_2[(q_{12}+p)^2-m^2]} \nonumber \\
  &= \int \mathrm{d}\Delta \tau_{12} e^{-i\Delta \tau_{12}[(q_1 + p)^2 - m^2 + 2q_1 \cdot q_2]}\Theta(\Delta \tau_{12})\int \mathrm{d}\tau_2 e^{-i\tau_2[(q_{12}+p)^2-m^2]} \, .
  \label{eq:scalar_worldline_ordering2_alt}
\end{align}
The sum of the two integrals become
\begin{align}
\mathcal{L}& = \int \mathrm{d}\Delta\tau_{12}\mathrm{d}\tau_2
             e^{-i[(q_1+p)^2-m^2]\Delta\tau_{12} -
             i[(q_1+q_2+p)^2-m^2]\tau_2}(\Theta(-\Delta\tau_{12}) + e^{-2i q_1\cdot q_2 \Delta\tau_{12}}\Theta(\Delta\tau_{12})), \label{eq:scalar_worldline_sum_alt}
\end{align}
where we have factorised common exponential factors. It is now time to
expand in $q_i=\lambda \bar{q}_i$. We first observe that proper times
are a conjugate variables and thus scale like
$\Delta\tau_{12},\tau_2\sim \lambda^{-1}$. Furthermore, instead of
expanding the full integrand, we will only expand the quantity
contained in round brackets, leaving the exponential exact:
\begin{align}
  \mathcal{L} &= \int \mathrm{d}\Delta\tau_{12}\mathrm{d}\tau_2
                e^{-i[(q_1+p)^2-m^2]\Delta\tau_{12} -i[(q_1+q_2+p)^2-m^2]\tau_2}(1
                - 2i(q_1\cdot
                q_2)\Delta\tau_{12}\Theta( \Delta\tau_{12}))+\mathcal{O}(\lambda^0) \,
  \label{eq:scalarWorldlineTauSpace}
\end{align}
where we have kept expansion terms of $\mathcal O(1/\lambda^2)$ and
$\mathcal O(1/\lambda)$, keeping in mind that the integration measure
scales as $1/\lambda^2$. Note that we used the identity
\begin{equation}
  \Theta(\Delta \tau_{12}) + \Theta(-\Delta \tau_{12}) = 1 \label{eq:theta_2att}
\end{equation}
to eliminate one of the
Heaviside functions. Furthermore, \emph{we have kept the exponentials that factor both graphs un-expanded}. When we integrate back, this results in the fact that the result is still expressed in terms of quadratic propagators:
\begin{align}
\mathcal{L} = \tilde{\delta}((q_1+p)^2-m^2)
  \tilde{\delta}((q_1+q_2+p)^2-m^2)) + 2i (q_1\cdot q_2) \frac{\tilde{\delta}((q_1+q_2+p)^2-m^2))}{((q_1+p)^2-m^2-i\varepsilon)^2}+\mathcal{O}(\lambda^0).
\label{eq:scalarWorldlineMomSpace}
\end{align}
Readers familiar with WQFT \cite{Mogull:2020sak, Jakobsen:2022psy} may
already notice that eq.~\eqref{eq:scalarWorldlineTauSpace} contains
the expression $\Delta\tau_{12}\Theta( \Delta\tau_{12})$ which is
proportional to the retarded propagator of the worldline deflection
field.\footnote{The PMEFT formalism, a variant of the worldline
  approach, also incorporates retarded propagators to incorporate
  classical dissipation \cite{Kalin:2022hph}, following earlier work
  in NRGR \cite{Galley:2015kus}.}  The result can be written
diagrammatically as
\begin{align}
  \raisebox{-0.6cm}{\input{2att_1_scalar}} +
  \raisebox{-1cm}{\scalebox{0.9}{\input{2att_2_scalar}}} = \raisebox{-0.6cm}{\input{2att_wqft_0}}+2(q_1\cdot
  q_2) \raisebox{-0.6cm}{\input{2att_wqft_1}} +\ldots
  \, ,
\end{align}
where the red vertical line indicates an additional \emph{cut}, i.e.\
the new delta function $\tilde{\delta}((q_1+p)^2-m^2)$, and the
additional black dot indicates a squared power of the propagator
denominator.
Observe that all propagators and arguments of Dirac
delta functions are quadratic, and the leading term of $\mathcal{L}$
involving a cut still contains subleading terms in the soft
expansion. However, this way of organising subleading terms in lower
orders of the expansion makes the result much more compact and will
turn out to be extremely fruitful in studying the classical expansion
of quantum observables. In the next section, we will extend the
quantum worldline formalism to scalar QED which follows similar steps
except for handling additional numerators which do not appear in
scalar field theory. Then we will multiply the quantum worldlines for
two distinct matter fields to study observables in classical two-body
interactions in the KMOC formalism.

\section{Further Warm up: Next-to-Leading Order impulse in scalar QED}
\label{sec:one_loop_impulse}

\subsection{Quantum worldline with two attachments in scalar QED}
\label{sec:worldline_QED}

We use scalar QED to study the classical interaction between classical
massive bodies that are charged and spinless. This model was already
studied in the original KMOC paper \cite{Kosower:2018adc} which
demonstrated the cancellation of one-loop super-leading divergences
locally, i.e.\ at the integrand level. We will re-derive the
cancellation and demonstrate the equivalence of the resulting finite
integrand with the integrand from classical worldline formalisms, using our quantum worldline
formalism that allows a systematic treatment beyond one loop in later
sections of the paper.

In scalar QED, we define the quantum worldline with two attachments as
the time-ordered expectation value of two photon currents with an
incoming and an outgoing on-shell massive scalar line.
\begin{align}
  \mathcal{L}^{\mu_1\mu_2}(\{v_1,v_2\},\{\},p)=\raisebox{-0.5cm}{\input{2att_1_QED}}+\raisebox{-0.5cm}{\input{2att_2_QED}} \, ,
  \label{eq:qed_two_attachment_diags}
\end{align}
where the first pair of curly brackets encloses the set of vertices on the left of the cut, while the second pair of curly bracket encloses the set of vertices on the right of the cut (empty for the linear-in-amplitude contributions to the impulse in the KMOC formalism, but non-empty for quadratic-in-amplitude contributions). In the following, when the cut lies at the utmost right of a graph, we will drop it.
More precisely, in terms of Feynman rules in the Lorentz gauge,
leaving out the overall factor $(-i e)^2$ from the two vertices,
\begin{align}
\begin{split}
\mathcal{L}^{\mu_1\mu_2}(\{v_1,v_2\},\{\},p)= i&\Bigg[\frac{(q_1+2p)^{\mu_1}(q_2+2(q_1+p))^{\mu_2}}{(q_1+p)^2-m^2+i\varepsilon}\\ 
&+\frac{(q_2+2p)^{\mu_2}(q_1+2(q_2+p))^{\mu_1}}{(q_2+p)^2-m^2+i\varepsilon}\Bigg]\tilde{\delta}\left((q_1+q_2+p)^2-m^2\right).
\end{split}
\label{eq:qed_two_attachment_expression}
\end{align}
Again, we have implicitly imposed the on-shellness of the incoming
momentum $p$ and only written out the on-shell constraint for the
outgoing momentum $q_1 + q_2 + p$ after the scalar particles receive
momentum kicks from the photon attachments.
We write
\begin{align}
\begin{split}
&V_1^{\mu_1}=(q_1+2p)^{\mu_1}, \quad V_2^{\mu_2}=(q_2+2(q_1+p))^{\mu_2}, \\
&(q_2+2p)^{\mu_2}=V_2^{\mu_2}-2q_1^{\mu_2}, \quad (q_1+2(q_2+p))^{\mu_1}=V_1^{\mu_1}+2 q_2^{\mu_1}.
\end{split}
\end{align}
This rewriting amounts to a canonicalisation of the vertex factors with respect to the reference ordering of the vertices, $(1,2)$, i.e.\ the first diagram on the RHS of eq.~\eqref{eq:qed_two_attachment_diags}.
Rewriting the propagator denominators and delta function using the Schwinger parametrisation, as we did in eqs.~\eqref{eq:scalar_worldline_ordering1} and \eqref{eq:scalar_worldline_ordering2_alt}, we end up with an expression analogous to eq.~\eqref{eq:scalar_worldline_sum_alt} for scalar field theory but with additional numerators from QED,
\begin{align}
\label{eq:2att_QED_s}
\begin{split}
\mathcal{L}^{\mu_1\mu_2}(\{v_1,v_2\},\{\},p) &= \int \mathrm{d}\Pi_2(p)\big(\Theta(- \Delta\tau_{12}) V_1^{\mu_1}V_2^{\mu_2} \\
&+ e^{-2i\Delta\tau_{12}(q_1\cdot q_2)} \Theta(\Delta\tau_{12})(V_1+2q_2)^{\mu_1}(V_2-2q_1)^{\mu_2}\big).
\end{split}
\end{align}
with
\begin{equation}
\mathrm{d}\Pi_2(p) = \mathrm{d}\Delta \tau_{12}\mathrm{d}\tau_2 e^{-i\Delta \tau_{12}[(q_1+p)^2-m^2]}e^{-i \tau_{2}[(q_{12}+p)^2-m^2]}.
\end{equation}
We now expand \emph{everything but the integration measure $\mathrm{d}\Pi_2(p)$ and the vertex factors $V_i^{\mu_i}$} in the classical limit $q_i=\lambda\bar{q}_i$, $\lambda\rightarrow 0$. To establish the correct power-counting in the classical limit eq.~\eqref{eq:classicalLimit}, proper time needs to be rescaled as $\tau_2=\lambda^{-1}\bar{\tau}_2$, $\Delta\tau_{12}=\lambda^{-1}\overline{\Delta\tau_{12}}$. We obtain, keeping the first two terms in the expansion,
\begin{align}
  \mathcal{L}^{\mu_1\mu_2}(\{v_1,v_2\},\{\},p) &= \int \mathrm{d}\Pi_2(p) \Big[ V_1^{\mu_1}V_2^{\mu_2}+(-2i\Delta\tau_{12}(q_1\cdot q_2) V_1^{\mu_1}V_2^{\mu_2} + 2q_2^{\mu_1}V_2^{\mu_2}-2q_1^{\mu_2}V_1^{\mu_1})\Theta(\Delta\tau_{12}) \Big] \nonumber \\
  & \qquad + \mathcal{O}(\lambda^0) \, , \label{eq:scalar_worldline_sum_expanded}
\end{align}
where we combined $\Theta(\Delta\tau_{12}) + \Theta(-\Delta\tau_{12}) = 1$ for the $V_1^{\mu_1}V_2^{\mu_2}$ term. The expansion terms can be written diagrammatically in terms of forests, which we introduce here as they will be central in the general treatment. Any term of eq.~\eqref{eq:scalar_worldline_sum_expanded} multiplying $\Theta(\Delta\tau_{12})$ is drawn in terms of a directed edge connecting $v_1$ and $v_2$:
\begin{equation}
V_1^{\mu_1}V_2^{\mu_2} \rightarrow \raisebox{-0.35cm}{\begin{tikzpicture}
    \node[inner sep=0pt] (V1) {};
    \node[inner sep=0pt] (V2) [right=1.5cm of V1] {};
    \node[inner sep=0pt] (L1) [below=0.2cm of V1] {$v_1$};
    \node[inner sep=0pt] (L2) [below=0.2cm of V2] {$v_2$};
    \path[draw=black, fill=black] (V1) circle[radius=0.05];
 \path[draw=black, fill=black] (V2) circle[radius=0.05];
\end{tikzpicture}} \hspace{2cm} (...)\Theta(\Delta\tau_{12})\rightarrow \raisebox{-0.35cm}{\begin{tikzpicture}
    \node[inner sep=0pt] (V1) {};
    \node[inner sep=0pt] (V2) [right=1.5cm of V1] {};
    \node[inner sep=0pt] (L1) [below=0.2cm of V1] {$v_1$};
    \node[inner sep=0pt] (L2) [below=0.2cm of V2] {$v_2$};
    \draw[->-, thick, line width=0.4mm] (V1) to (V2);
    \path[draw=black, fill=black] (V1) circle[radius=0.05];
 \path[draw=black, fill=black] (V2) circle[radius=0.05];
\end{tikzpicture}}
\end{equation}
In other words, we can think of expansion terms in terms of relative orderings of the vertices, an idea that is extremely fruitful in organising the expansion itself. Furthermore, this way of drawing expansion terms enjoys a direct relationship with the diagrammatic expansion in WQFT.

If we were to integrate eq.~\eqref{eq:scalar_worldline_sum_expanded} back from Schwinger parameter space, we would obtain: 
\begin{align}
\label{eq:sQED_2att}
\mathcal{L}^{\mu_1\mu_2}(\{v_1,v_2\},\{\},p)&=\Bigg[V_1^{\mu_1}V_2^{\mu_2}\tilde{\delta}((q_1+p)^2-m^2) + \frac{2 i (q_1\cdot q_2) V_1^{\mu_1}V_2^{\mu_2}}{((q_1+p)^2-m^2-i\varepsilon)^2} \nonumber \\
&-\frac{2  i (q_2^{\mu_1}V_2^{\mu_2} - q_1^{\mu_2}V_1^{\mu_1})}{(q_1+p)^2-m^2-i\varepsilon}\Bigg]\tilde{\delta}((q_1+q_2+p)^2-m^2)+\mathcal{O}(\lambda^0),
\end{align}

We may also introduce the cut versions of the quantum worldline, which are basically the sub-diagrams for a single matter line in the quadratic-in-amplitude terms in the KMOC formalism, with the coupling factor $(-ie)(+ie)$ factored out,
\begin{align}
\raisebox{-0.6cm}{\scalebox{0.75}{\input{2att_1_QED_cut}}}=\mathcal{L}^{\mu_1\mu_2}(\{v_1\},\{v_2\},p) &= V_1^{\mu_1}V_2^{\mu_2}\tilde{\delta}((q_1+p)^2-m^2)\tilde{\delta}((q_{12}+p)^2-m^2) \nonumber \\
&= \int \mathrm{d}\Pi_2(p) V_1^{\mu_1}V_2^{\mu_2}, \label{eq:qedCutWL1} \\
\raisebox{-0.6cm}{\scalebox{0.75}{\input{2att_2_QED_cut}}}=\mathcal{L}^{\mu_1\mu_2}(\{v_2\},\{v_1\},p) &= (V_1+2q_2)^{\mu_1}(V_2-2q_1)^{\mu_2}\tilde{\delta}((q_1+p)^2-m^2)\tilde{\delta}((q_{12}+p)^2-m^2) \nonumber \\
&= \int \mathrm{d}\Pi_2(p) e^{-2i\Delta\tau_{12}(q_1\cdot q_2)}(V_1+2q_2)^{\mu_1}(V_2-2q_1)^{\mu_2} \, .  \label{eq:qedCutWL2}
\end{align}
As a reminder of our notation, the above two equations contain non-empty sets of vertices enclosed in the second curly bracket, which means that there are nontrivial expressions on the right of the cut for the complex-conjugated amplitude, indicating that we are looking at the quadratic-in-amplitude terms in the KMOC formalism. We may now proceed to the computation of the impulse observable in two-body scattering, by multiplying together the two quantum worldline for the two massive bodies.

\subsection{Classical limit of Next-To-Leading order impulse in scalar QED}

We now set ourselves the task of computing the Next-to-Leading order impulse in electrodynamics in the scattering between particle 1, with mass $m_1$ and charge $e_1$, and particle 2, with mass $m_2$ and charge $e_2$. We always draw particle 1 legs and propagators as a horizontal line on the top of a diagram, and particle 2 on the bottom. We divide the computation in equivalence classes of diagrams: diagrams in each equivalence class differ by a permutation of the vertices on the upper or lower matter line. Each equivalence class scales classically in the limit of all massless particles being soft, i.e.\ exhibits cancellation of super-leading divergences. We retrieve the worldline integrand by performing purely local manipulations.

In the KMOC formalism, soft power-counting establishes that the
Next-to-Leading order impulse for particle 1, i.e.\ the change in its
momentum during scattering, receives contributions from the following diagrams
\begin{align}
&\langle \mathds{P}^\mu\rangle^{(1)}=\int\mathrm{d}^4q_1\mathrm{d}^4q_2 \Bigg( e^{i(q_1+q_2)\cdot b} \Bigg[(q_1+q_2)^\mu\raisebox{-0.75cm}{\scalebox{0.55}{\input{box}}}+(q_1+q_2)^\mu\raisebox{-0.75cm}{\scalebox{0.55}{\input{cross-box}}} \nonumber \\
&-q_1^\mu\raisebox{-0.8cm}{\scalebox{0.55}{\input{box_cut}}} 
+(q_1+q_2)^\mu\raisebox{-0.75cm}{\scalebox{0.55}{\input{triangle}}} 
+(q_1+q_2)^\mu\raisebox{-0.75cm}{\scalebox{0.55}{\input{triangle_ud}}}\Bigg] \times \nonumber \\
&\times\tilde{\delta}((p_1+q_{12})^2-m_1^2)\tilde{\delta}((p_2-q_{12})^2-m_2^2) \nonumber \\
&+e^{iq_1\cdot b}\Bigg[q_1^\mu\raisebox{-0.75cm}{\scalebox{0.55}{\input{mushroom1}}}+q_1^\mu\raisebox{-1.2cm}{\scalebox{0.55}{\input{mushroom1_ud}}}\Bigg]\tilde{\delta}((p_1+q_{1})^2-m_1^2)\tilde{\delta}((p_2-q_{1})^2-m_2^2)\Bigg), \label{eq:qedNLOIntegrand}
\end{align}
where we chose the momenta of the photons to be the integration variables. 
The impulse relates to the change in momentum of the upper particle, and each diagram corresponds to an integrand obtained by applying the Feynman rules of scalar QED. We expect to retrieve the Next-to-Leading Order correction to the classical impulse by taking the leading term of this expression in the soft expansion $|q_1|,|q_2| \sim \lambda\rightarrow 0$. Before taking into account the scaling of the coupling constant, the observable contribution from the box, cross-box and cut box integrand scale like $\lambda^{1}$, the triangle contributions scale like $\lambda^{2}$ and the mushroom integrands scale like $\lambda^{1}$. However, we will see that the entire integrand scales like $\lambda^{2}$, which is the classical order (an overall observable scaling of $\lambda^0$ is obtained if, following the KMOC convention, the coupling is rescaled as $e=\bar{e}/\sqrt{\lambda}$). The mushroom diagrams turn out to give zero contributions, contrary to the naive power counting argument.

\textbf{Box equivalence class} We start by looking at the virtual box contributions to the integrand in eq.~\eqref{eq:qedNLOIntegrand}:
\begin{equation}
F_\Box^V=(q_1+q_2)^\mu\left[\raisebox{-1cm}{\scalebox{0.75}{\input{box}}}+\raisebox{-1cm}{\scalebox{0.75}{\input{cross-box}}}\right]\tilde{\delta}((p_1+q_{12})^2-m_1^2)\tilde{\delta}((p_2-q_{12})^2-m_2^2).
\end{equation}
The photon edge stretching from $w_1$ to $v_1$ has momentum $q_1$,
while that stretching from $w_2$ to $v_2$ has momentum $q_2$. As it is written now, the expression is not symmetric under permutation of the vertices on the upper line. We symmetrise and obtain a new integrand
\begin{align}
F_{\Box}^V&=\frac{(q_1+q_2)^\mu}{2}\Bigg[\raisebox{-1cm}{\scalebox{0.75}{\input{box}}}+\raisebox{-1cm}{\scalebox{0.75}{\input{cross-box}}}+\raisebox{-1cm}{\scalebox{0.75}{\input{box_sym}}} \nonumber \\
&+\raisebox{-1cm}{\scalebox{0.75}{\input{cross-box-sym}}}\Bigg]\tilde{\delta}((p_1+q_{12})^2-m_1^2)\tilde{\delta}((p_2-q_{12})^2-m_2^2).
\end{align}
This quantity can be written as the product of the ``upper'' and ``lower'' quantum worldlines along with the photon propagators in between,
\begin{equation}
F_{\Box}^V = (-ie_1)^2 (-ie_2)^2 \frac{(q_1+q_2)^\mu}{2}\frac{-ig^{\mu_1\nu_1}}{q_1^2+i\varepsilon}\frac{-ig^{\mu_2\nu_2}}{q_2^2+i\varepsilon}\mathcal{L}^{\mu_1 \mu_2}(\{v_1,v_2\},\{\},p_1)\mathcal{L}^{\nu_1 \nu_2}(\{w_1,w_2\},\{\},-p_2)
\end{equation}
where the factor $(-ie_1)^2 (-ie_2)^2$ comes from the Feynman rules of the vertices. $F_\Box^V$ scales like $\lambda^{-7}$ in the classical limit, implying that it has super-leading classical terms. For what concerns the cut contribution, we may perform an analogous symmetrisation and define
\begin{equation}
F_{\Box}^R=\frac{1}{2}\left[q_1^\mu\raisebox{-1cm}{\scalebox{0.75}{\input{box_cut}}}+q_2^\mu\raisebox{-1cm}{\scalebox{0.75}{\input{box_cut_sym}}}\right]\tilde{\delta}((p_1+q_{12})^2-m_1^2)\tilde{\delta}((p_2-q_{12})^2-m_2^2).
\end{equation}
We write $F_{\Box}^R$ in terms of the cut versions of the quantum worldline $\mathcal{L}^{\mu_1 \mu_2}$ given in eqs.~\eqref{eq:qedCutWL1} and \eqref{eq:qedCutWL2},
\begin{align}
F_{\Box}^R &= \frac{(-ie_1)(-i e_2) (ie_1)(i e_2)}{2} \Bigg[q_1^\mu\frac{-ig^{\mu_1\nu_1}}{q_1^2+i\varepsilon}\frac{ig^{\mu_2\nu_2}}{q_2^2-i\varepsilon}\mathcal{L}^{\mu_1 \mu_2}(\{v_1\},\{v_2\};p_1)\mathcal{L}^{\nu_1 \nu_2}(\{w_1\},\{w_2\};-p_2) \nonumber \\
&+q_2^\mu\frac{ig^{\mu_1\nu_1}}{q_1^2-i\varepsilon}\frac{-ig^{\mu_2\nu_2}}{q_2^2+i\varepsilon}\mathcal{L}_2^{\mu_1 \mu_2}(\{v_2\},\{v_1\};p_1)\mathcal{L}_2^{\nu_1 \nu_2}(\{w_2\},\{w_1\};-p_2)\Bigg] \label{eq:nloQEDBoxReal}
\end{align}
Note that the photon propagators on the right of the cut has complex-conjugated $i \varepsilon$ prescription. Analogously, the vertex factors amount to $(-ie_1)(-ie_2)$ for the two vertices on the left of the cut and $(ie_1)(ie_2)$ for the two vertices on the right of the cut.
In order to combine virtual and cut contributions, we rewrite the quantum worldlines in Schwinger parameter space, following eq.~\eqref{eq:2att_QED_s}. Specifically, we rewrite
\begin{align}
F_{\Box}^V &=-e_1^2 e_2^2\frac{g^{\mu_1\nu_1}}{q_1^2+i\varepsilon}\frac{g^{\mu_2\nu_2}}{q_2^2+i\varepsilon}\frac{(q_1+q_2)^\mu}{2}\int \mathrm{d}\Pi_2(p_1)\mathrm{d}\Pi_2(p_2)' \nonumber \\
&\quad \times \left[V_1^{\mu_1}V_2^{\mu_2} \, \Theta(\Delta\tau_{12})+(V_1^{\mu_1} +2q_2^{\mu_1})(V_2^{\mu_2}-2q_1^{\mu_2})e^{-2i(q_1\cdot q_2)\Delta\tau_{12}} \, \Theta(\Delta\tau_{12})\right] \label{eq:line1_exp} \\
&\quad \times \left[W_1^{\nu_1}W_2^{\nu_2} \, \Theta(\Delta\tau_{12}')+(W_1^{\nu_1} -2q_2^{\nu_1})(W_2^{\nu_2}+2q_1^{\nu_2})e^{-2i(q_1\cdot q_2)\Delta\tau_{12}'} \, \Theta(\Delta\tau_{12}')\right]\label{eq:line2_exp},
\end{align}
where
\begin{align}
  & \mathrm{d}\Pi_2(p_1) = \mathrm{d}\Delta\tau_{12}\mathrm{d}\tau_2 e^{-i[(q_1+p_1)^2-m_1^2]\Delta\tau_{12}-i[(q_1+q_2+p_1)^2-m_1^2]\tau_2} \, ,  \\
  & \mathrm{d}\Pi_2(p_2)' = \mathrm{d}\Delta\tau_{12}'\mathrm{d}\tau_2' e^{-i[(q_1-p_2)^2-m_2^2]\Delta\tau_{12}'-i[(q_1+q_2-p_2)^2-m_2^2]\tau_2'} \, , \\
  & V_1 = q_1+2p_1, \quad V_2=q_2+2(q_1+p_2), \quad W_1=-q_1+2p_2, \quad W_2=-q_2+2(-q_1+p_2) \, .
\end{align}
The numerator has already been expressed in canonicalised form, namely
in terms of vertex numerators $V_1^{\mu_1}$ and $V_2^{\mu_2}$
appearing in the ordering $v_1,v_2$ of the vertices in the upper line
(similarly for the bottom line). We now expand in
$q_i=\lambda \bar{q}_i$. We take care in expanding
lines~\eqref{eq:line1_exp} and~\eqref{eq:line2_exp} only, taking
advantage of the expansion result for an individual worldline in
eq.~\eqref{eq:scalar_worldline_sum_expanded}. \emph{We keep the
  measure un-expanded until the very end, as the cancellation of
  super-leading contributions does not rely on it being expanded}:
\begin{align}
&F_{\Box}^V=-\frac{e_1^2 e_2^2}{2} \frac{g^{\mu_1\nu_1}}{q_1^2+i\varepsilon}\frac{g^{\mu_2\nu_2}}{q_2^2+i\varepsilon} (q_1+q_2)^\mu \int \mathrm{d}\Pi_2(p_1)\mathrm{d}\Pi_2(p_2)'  \nonumber \\
&\Bigg[V_1^{\mu_1}V_2^{\mu_2}W_1^{\nu_1}W_2^{\nu_2}(1 - 2i(q_1\cdot q_2)\Delta\tau_{12}\Theta(\Delta\tau_{12}) - 2i(q_1\cdot q_2)\Delta\tau_{12}'\Theta(\Delta\tau_{12}')) \nonumber \\
&+(2q_2^{\mu_1}V_2^{\mu_2}-2q_1^{\mu_2}V_1^{\mu_1}) W_1^{\nu_1}W_2^{\nu_2} \Theta(\Delta\tau_{12})+(2q_1^{\nu_2}W_1^{\nu_1}-2q_2^{\nu_1}W_2^{\nu_2}) V_1^{\mu_1}V_2^{\mu_2} \Theta(\Delta\tau_{12}') \Bigg]+\ldots \label{eq:nloQEDVirt}
\end{align}
where the ellipsis include higher orders in the expansion. Let us repeat the same procedure for the real contributions. The term of the integral that reads $V_1^{\mu_1}V_2^{\mu_2}W_1^{\nu_1}W_2^{\nu_2}$ is super-leading. Substituting eqs.~\eqref{eq:qedCutWL1} and \eqref{eq:qedCutWL2} into eq.~\eqref{eq:nloQEDBoxReal}, we have 
\begin{align}
&F_{\Box}^R=\frac{e_1^2 e_2^2}{2}\int \mathrm{d}\Pi_2(p_1)\mathrm{d}\Pi_2(p_2)'  \nonumber \\
&\Bigg[V_1^{\mu_1}V_2^{\mu_2}W_1^{\nu_1}W_2^{\nu_2}\left(q_1^\mu \frac{g^{\mu_1\nu_1}}{q_1^2+i\varepsilon}\frac{g^{\mu_2\nu_2}}{q_2^2-i\varepsilon}+q_2^\mu \frac{g^{\mu_1\nu_1}}{q_1^2-i\varepsilon}\frac{g^{\mu_2\nu_2}}{q_2^2+i\varepsilon}\right)  \nonumber \\
&+q_2^\mu(-2i (q_1\cdot q_2)\Delta\tau_{12} - 2i(q_1\cdot q_2)\Delta\tau_{12}')\frac{g^{\mu_1\nu_1}}{q_1^2+i\varepsilon}\frac{g^{\mu_2\nu_2}}{q_2^2-i\varepsilon} \nonumber \\
&+q_2^\mu[(2q_2^{\mu_1}V_2^{\mu_2}-2q_1^{\mu_2}V_1^{\mu_1}) W_1^{\nu_1}W_2^{\nu_2}+q_2^\mu(2q_1^{\nu_2}W_1^{\nu_1}-2q_2^{\nu_1}W_2^{\nu_2}) V_1^{\mu_1}V_2^{\mu_2}]\frac{g^{\mu_1\nu_1}}{q_1^2+i\varepsilon}\frac{g^{\mu_2\nu_2}}{q_2^2-i\varepsilon}  \Bigg]+\ldots.
\end{align}
Before combining real and virtual contributions together, let us observe that at the present loop order, the $i \epsilon$ prescription of photon propagators can be essentially ignored. Physically speaking, this is due to the fact that the virtual photon propagators can never go on-shell, as radiative effects only enters through two-loop amplitudes in the context of classical observables. To see this, by writing the complex-conjugated Feynman propagator as the sum of the original Feynman propagator and a delta function term, any difference caused by the two different $i \epsilon$ terms is captured by terms involving a cut propagator, which causes the integral to vanish as a zero-measured cut (see sect.~\ref{sec:zero_measured} for more details). An example of zero-measured cut, in the context of this calculation, would be the following
\begin{equation}
\raisebox{-1cm}{\input{zero_measured_box}}
\end{equation}
At the vertex $v_1$, a kinematically degenerate process occurs: a massive particle splitting into a particle with equal mass plus a massless particle. These diagrams are set to zero in our procedure. In other words, we can immediately replace the conjugate propagators by ordinary Feynman propagators,
\begin{align}
  F_{\Box}^R &= F_{\Box}^R \big |_{1/(q_i^2 - m_i^2 - i \varepsilon) \rightarrow 1/(q_i^2 - m_i^2 + i \varepsilon)} \nonumber \\
  &= \frac{e_1^2 e_2^2}{2} \frac{g^{\mu_1\nu_1}}{q_1^2+i\varepsilon}\frac{g^{\mu_2\nu_2}}{q_2^2+i\varepsilon} \int \mathrm{d}\Pi_2(p_1)\mathrm{d}\Pi_2(p_2)'  \nonumber \\
&\Bigg[V_1^{\mu_1}V_2^{\mu_2}W_1^{\nu_1}W_2^{\nu_2} \bigg( q_1^\mu + q_2^\mu -q_2^\mu(2i (q_1\cdot q_2)\Delta\tau_{12}+2i(q_1\cdot q_2)\Delta\tau_{12}') \bigg) \nonumber \\
&+q_2^\mu[(2q_2^{\mu_1}V_2^{\mu_2}-2q_1^{\mu_2}V_1^{\mu_1}) W_1^{\nu_1}W_2^{\nu_2}+q_2^\mu(2q_1^{\nu_2}W_1^{\nu_1}-2q_2^{\nu_1}W_2^{\nu_2}) V_1^{\mu_1}V_2^{\mu_2}] \Bigg] + \ldots \label{eq:nloQEDRealChangedIEps}
\end{align}
The real contribution eq.~\eqref{eq:nloQEDRealChangedIEps} is now in a form which can be easily combined with the virtual contribution eq.~\eqref{eq:nloQEDVirt}. The expression we obtain, after a purely \emph{algebraic} simplification of the integrand considered of as a polynomial in $\Delta\tau_{12}$, $\Delta\tau_{12}'$, $q_1^{\mu_2}$, $q_2^{\mu_2}$, $V_i^{\mu_i}$, $W_i^{\mu_i}$ and the photon propagator is
\begin{align}
&F_{\Box}^V+F_{\Box}^R=-\frac{e_1^2 e_2^2}{2}\int \mathrm{d}\Pi_2(p_1)\mathrm{d}\Pi_2(p_2)' \frac{g^{\mu_1\nu_1}}{q_1^2+i\varepsilon}\frac{g^{\mu_2\nu_2}}{q_2^2+i\varepsilon} \nonumber \\
\Bigg[ & V_1^{\mu_1} V_2^{\mu_2}W_1^{\mu_1} W_2^{\mu_2}\Big(-2 i q_1^\mu (q_1\cdot q_2)\Delta\tau_{12}\Theta(\Delta\tau_{12}) + 2 i q_2^\mu (q_1\cdot q_2)\Delta\tau_{12}\Theta(-\Delta\tau_{12}) \nonumber \\
-&2 i q_1^\mu (q_1\cdot q_2)\Delta\tau_{12}'\Theta(\Delta\tau_{12}') + 2 i q_2^\mu (q_1\cdot q_2)\Delta\tau_{12}'\Theta(-\Delta\tau_{12}')\Big) \nonumber \\
+&q_1^\mu[(2q_2^{\mu_1}V_2^{\mu_2}-2q_1^{\mu_2}V_1^{\mu_1}) W_1^{\nu_1}W_2^{\nu_2} \Theta(\Delta\tau_{12})+(2q_1^{\nu_2}W_1^{\nu_1}-2q_2^{\nu_1}W_2^{\nu_2}) V_1^{\mu_1}V_2^{\mu_2} \Theta(\Delta\tau_{12}')] \nonumber \\
-&q_2^\mu[(2q_2^{\mu_1}V_2^{\mu_2}-2q_1^{\mu_2}V_1^{\mu_1}) W_1^{\nu_1}W_2^{\nu_2} \Theta(-\Delta\tau_{12})+(2q_1^{\nu_2}W_1^{\nu_1}-2q_2^{\nu_1}W_2^{\nu_2}) V_1^{\mu_1}V_2^{\mu_2} \Theta(-\Delta\tau_{12}')]
\Bigg]+\ldots. \label{eq:nloQEDSum}
\end{align}
All the terms of this expression proportional to $q_1^\mu$ are also present in exactly the same form in the virtual contribution $F_{\Box}^V$ in eq.~\eqref{eq:nloQEDVirt}. (The only term of the real contribution proportional to $q_1^\mu$ cancels exactly with the equal term in the virtual contribution.) Conversely, the terms proportional to $q_2^\mu$ arise from combining terms from real and virtual contributions using the simple identity $1-\Theta(-\Delta\tau)=\Theta(\Delta\tau)$.
The result is free of super-leading contributions and is furthermore cast in the worldline form. In particular, we may integrate it back to obtain
\begin{align}
F_{\Box}^V&+F_{\Box}^R=-\frac{e_1^2 e_2^2}{2}\frac{g^{\mu_1\nu_1}}{q_1^2+i\varepsilon}\frac{g^{\mu_2\nu_2}}{q_2^2+i\varepsilon}\tilde{\delta}((q_1+q_2+p_1)^2-m_1^2)\tilde{\delta}((q_1+q_2-p_1)^2-m_2^2) \nonumber \\
  \times \Bigg[ & \Big( 2(q_1\cdot q_2)V_1^{\mu_1} V_2^{\mu_2} - ((q_1+p_1)^2-m_1^2)(2q_2^{\mu_1}V_2^{\mu_2}-2q_1^{\mu_2}V_1^{\mu_1}) \bigg) W_1^{\nu_1}W_2^{\nu_2} \nonumber \\
  &\quad \times
    \left(\frac{iq_1^\mu}{((q_1+p_1)^2-m_1^2-i\varepsilon)^2} +
    \frac{iq_2^\mu}{((q_1+p_1)^2-m_1^2+i\varepsilon)^2} \right)  \tilde{\delta}((q_1-p_2)^2-m_2^2) \nonumber \\
  & + V_1^{\mu_1}V_2^{\mu_2} \bigg( 2(q_1\cdot q_2)
    W_1^{\mu_1}W_2^{\mu_2} - ((q_1-p_2)^2-m_2^2)(2q_1^{\nu_2}W_1^{\nu_1}-2q_2^{\nu_1}W_2^{\nu_2}) \Big ) \nonumber \\
  &\quad \times \left(
    \frac{iq_1^\mu}{((q_1-p_2)^2-m_2^2-i\varepsilon)^2} +
    \frac{iq_2^\mu}{((q_1-p_2)^2-m_2^2+i\varepsilon)^2} \right) \tilde{\delta}((q_1+p_1)^2-m_1^2)
 \Bigg]+\ldots
\end{align}
Note that the terms multiplying $q_2^\mu$ are almost the same as the terms multiplying the terms multiplying $q_1^\mu$ except for a flipped $i \epsilon$ prescription, which originates from the sign flip between these two types of terms in eq.~\eqref{eq:nloQEDSum}.
This expression still contains subleading quantum terms that arise from the fact that we did not expand the integration measure in Schwinger-parameter space, but no super-leading terms. This allows us to straight-forwardly expand the result again and simply drop any subleading correction, and substitute $V_i^{\mu_i}\rightarrow 2p_1^{\mu_i}$, $W_i^{\mu_i}\rightarrow -2p_2^{\mu_i}$, $(q_{I}\pm p_i)^2-m_i^2\rightarrow \pm 2 q_I\cdot p_i$,
\begin{align}
F_{\Box}^V+F_{\Box}^R=&-\frac{e_1^2 e_2^2}{2}\frac{g^{\mu_1\nu_1}}{q_1^2+i\varepsilon}\frac{g^{\mu_2\nu_2}}{q_2^2+i\varepsilon}\tilde{\delta}((q_1+q_2)\cdot p_1)\tilde{\delta}((q_1+q_2)\cdot p_2) \nonumber \\
  \times \Bigg[ & \Big((q_1\cdot q_2)p_1^{\mu_1} p_1^{\mu_2} - (q_1\cdot p_1)(q_2^{\mu_1}p_1^{\mu_2}-q_1^{\mu_2}p_1^{\mu_1}) \Big) p_2^{\nu_1}p_2^{\nu_2} \nonumber \\
  &\quad \times \left( \frac{iq_1^\mu} {(q_1\cdot p_1-i\varepsilon)^2} + \frac{iq_2^\mu} {(q_1\cdot p_1+i\varepsilon)^2} \right) \tilde{\delta}(q_1\cdot p_2) \nonumber \\ 
  & + p_1^{\mu_1}p_1^{\mu_2}\Big( (q_1\cdot q_2)p_2^{\mu_1}p_2^{\mu_2}
    - (- q_1\cdot p_2)(q_1^{\nu_2}p_2^{\nu_1}-q_2^{\nu_1}p_2^{\nu_2}) \Big ) \nonumber \\
  &\quad \times \left( \frac{iq_1^\mu} {(-q_1\cdot p_2 - i\varepsilon)^2} + \frac{iq_2^\mu} {(-q_1\cdot p_2 + i\varepsilon)^2} \right) \tilde{\delta}(q_1\cdot p_1)
 \Bigg] + \ldots
 \label{eq:VplusRdropQuantum}   
\end{align}
We have cancelled various factors of 2 to reach this expression. Performing the contractions and imposing the constraints from the Dirac delta functions, we obtain 
\begin{align}
F_{\Box}^V+F_{\Box}^R=& -\frac{i e_1^2 e_2^2}{2} \frac{1}{(q_1^2 +i\varepsilon) (q_2^2 +i\varepsilon)}\tilde{\delta}((q_1+q_2)\cdot p_1)\tilde{\delta}((q_1+q_2)\cdot p_2)(p_1\cdot p_2)^2 \nonumber \\
\times \Bigg[ & q_1^\mu\frac{(q_1\cdot q_2)}{(q_1\cdot p_1-i\varepsilon)^2}\tilde{\delta}(q_1\cdot p_2) +q_2^\mu\frac{(q_1\cdot q_2)}{(q_1\cdot p_1+i\varepsilon)^2}\tilde{\delta}(q_1\cdot p_2) \nonumber \\
&  +q_1^\mu\frac{(q_1\cdot q_2)}{(-q_1\cdot p_2-i\varepsilon)^2}\tilde{\delta}(q_1\cdot p_1) +q_2^\mu\frac{(q_1\cdot q_2)}{(-q_1\cdot p_2+i\varepsilon)^2}\tilde{\delta}(q_1\cdot p_1)\Bigg] + \ldots
\end{align}
Note that the delta functions have killed the term proportional to
$(q_1 \cdot p_1)$ on the 2nd line and the term proportional to $(-q_1
\cdot p_2)$ on the 4th line of eq.~\eqref{eq:VplusRdropQuantum}.
Finally, we can undo the symmetrisation and simply consider the integrand
\begin{align}
F_{\Box}^V+F_{\Box}^R\quad \rightarrow \quad &-\frac{ie_1^2 e_2^2}{(q_1^2 +i\varepsilon) (q_2^2 +i\varepsilon)}\tilde{\delta}((q_1+q_2)\cdot p_1)\tilde{\delta}((q_1+q_2)\cdot p_2)(p_1\cdot p_2)^2 \nonumber \\
\times \Bigg[ & q_2^\mu\frac{(q_1\cdot q_2)}{(q_1\cdot p_1+i\varepsilon)^2}\tilde{\delta}(q_1\cdot p_2)  +q_2^\mu\frac{(q_1\cdot q_2)}{(-q_1\cdot p_2+i\varepsilon)^2}\tilde{\delta}(q_1\cdot p_1)\Bigg]+\ldots
\end{align}

\textbf{Triangle equivalence class} The second relevant class in the
classical limit is that featuring the triangle topology. In this case
the real contribution $F^R_\triangle=0$, since an intermediate
two-particle cut does not exist for the triangle diagram. Thus
\begin{equation}
F^V_{\triangle}+F^R_{\triangle}=\frac{(q_1+q_2)^\mu}{2}\left[\raisebox{-1cm}{\scalebox{0.75}{\input{triangle}}}+\raisebox{-1cm}{\scalebox{0.75}{\input{triangle_sym}}}\right].
\end{equation}
For this class of diagrams, the classical limit is straight-forward to obtain, as both diagrams shown above start at the classical order, and there is no need to cancellation of super-leading divergences. Upon expansion, the two matter propagators of the two symmetrised contributions add up to a Dirac delta function, corresponding to applying the identity $\Theta(\tau_{12}) + \Theta(\tau_{21}) = 1$ in Schwinger parameter space,
\begin{align}
  &\quad \frac i {(p_1+q_1)^2 - m_1^2 + i\varepsilon} + \frac i {(p_1+q_2)^2 - m_1^2 + i\varepsilon} \approx \frac i {2 q_1 \cdot p_1 + i\varepsilon} + \frac i {-2 q_1 \cdot p_1 + i\varepsilon} = \tilde \delta (2 q_1 \cdot p_1) \, ,
\end{align}
where the ``$\approx$'' symbol denotes the leading order in the soft expansion.
The classical integrand thus reads, again keeping only the leading order in the soft expansion
\begin{align}
  F^V_{\triangle}+F^R_{\triangle} &= \frac{e_1^2 e_2^2}{2}(q_1+q_2)^\mu \frac{-i g^{\mu_1\nu_1}}{q_1^2+i\varepsilon}\frac{-i g^{\mu_2\nu_2}}{q_2^2+i\varepsilon}\tilde{\delta}(2 q_1\cdot p_1)\tilde{\delta}(2 (q_1+q_2)\cdot p_1)\tilde{\delta}(2 (q_1+q_2)\cdot p_2) \nonumber \\
  &\quad \times (-2 i p_1^{\mu_1}) (-2 i p_1^{\mu_2}) (2 i g_{\nu_1\nu_2})  \\
& = \frac{ie_1^2 e_2^2}{2}(q_1+q_2)^\mu \frac{p_1^2}{q_1^2 q_2^2}\tilde{\delta}(q_1\cdot p_1)\tilde{\delta}((q_1+q_2)\cdot p_1)\tilde{\delta}((q_1+q_2)\cdot p_2) \\
&\rightarrow ie_1^2 e_2^2 q_2^\mu \frac{p_1^2}{q_1^2 q_2^2}\tilde{\delta}(q_2\cdot p_1)\tilde{\delta}((q_1+q_2)\cdot p_1)\tilde{\delta}((q_1+q_2)\cdot p_2) \, .
\end{align}
Similarly, the ``inverted triangle'' diagrams with matter propagators on the $m_2$ line instead of the $m_1$ line give
\begin{equation}
  F^V_{\triangledown} + F^R_{\triangledown} \rightarrow ie_1^2 e_2^2 q_2^\mu \frac{p_2^2}{q_1^2 q_2^2}\tilde{\delta}(q_2\cdot p_2)\tilde{\delta}((q_1+q_2)\cdot p_1)\tilde{\delta}((q_1+q_2)\cdot p_2) \, .
\end{equation}

\textbf{Comparison with worldline QFT} Summing the results for the box and triangle equivalence classes, including the vertically flipped version for the triangle, and expressing everything in terms of velocities through the rescaling $p_i=m_i u_i$, we obtain the combined integrand:
\begin{align}
& - \frac{ie_1^2 e_2^2}{q_1^2 q_2^2}\tilde{\delta}((q_1+q_2)\cdot u_1)\tilde{\delta}((q_1+q_2)\cdot u_2) \times \nonumber \\
\times \Bigg[ & q_2^\mu\frac{ (u_1\cdot u_2)^2(q_1\cdot q_2)- (q_1\cdot u_1)^2}{m_1(q_1\cdot u_1+i\varepsilon)^2}\tilde{\delta}(q_1\cdot u_2)  +q_2^\mu\frac{(u_1\cdot u_2)^2(q_1\cdot q_2)-(q_1\cdot u_2)^2}{m_2(-q_1\cdot u_2+i\varepsilon)^2}\tilde{\delta}(q_1\cdot u_1)\Bigg]
\end{align}
The above expression agrees with the integrand in the worldline QFT formalism from the two Feynman diagrams eqs.~\eqref{eq:oneLoopQEDIntegrand1} and \eqref{eq:oneLoopQEDIntegrand2} reviewed in Appendix \ref{sec:wqft}, which we report here for completeness:
\begin{equation}
\raisebox{-1cm}{\input{box_wqft1}} + \raisebox{-1cm}{\input{box_wqft_2}}
\end{equation}
Importantly, we have reproduced the correct retarded propagator prescriptions for the matter lines corresponding to the classical causality flow in classical worldline formalisms \cite{Jakobsen:2022psy,Kalin:2020mvi}. This is reflected in Schwinger parameter space by the correct Heaviside theta functions for the ordering of worldline proper time parameters in eq.~\eqref{eq:nloQEDSum}.

\textbf{Mushroom equivalence class} The last power-counting-wise relevant contribution corresponds to diagrams in the mushroom equivalence class:
\begin{equation}\label{eq:NLO_mushrooms}
\scalebox{0.75}{
\input{mushroom1}}\scalebox{0.75}{
\input{mushroom2}}\scalebox{0.75}{
\input{mushroom3}}
\end{equation}
The last two contributions are meant to account for the LSZ-induced field renormalisation factors for the external legs. We can show that each individual mushroom diagram appearing at this order vanish at all orders in the soft region. Indeed, in the soft region, all three diagrams of eq.~\eqref{eq:NLO_mushrooms} schematically read
\begin{align}
\sum_{n_1,n_2,n_3=0}^\infty\frac{c_{n_1,n_2,n_3}(q_1,q_2,p_1,p_2)}{q_1^2 q_2^2 (q_2\cdot p_1)^{n_1}}\delta^{n_2}(q_1\cdot p_1)\delta^{n_3}(q_1\cdot p_2),
\end{align}
with $c_{n_1,n_2,n_3}(q_1,q_2,p_1,p_2)$ being polynomial in its entries and diagram-dependent and $\delta^{n}$ being the $n$-th derivative of the Dirac delta function. As a consequence, all three diagrams of eq.~\eqref{eq:NLO_mushrooms} can be written as sums of scaleless integrals, which vanish in dimensional regularisation.

\section{Formalism}
\label{sec:formalism}

We review quantum observables in the IN-IN scattering formalism and their associated classical limit. The method for deriving classical observables from their quantum counter-parts in the IN-IN scattering formalism has been originally developed in~\cite{Kosower:2018adc}. We present it here highlighting specifically the generality of the power-counting procedure as well as the graph-theory underlying the cancellation of super-leading contributions.

\subsection{Perturbative in-in quantum observables from in-out scattering amplitudes}

In the in-in scattering formalism, quantum observables take the form:
\begin{equation}
\langle \mathds{O} \rangle=\langle \psi |\hat{S}^\dagger \hat{\mathds{O}} \hat{S}|\psi \rangle-\langle \psi | \hat{\mathds{O}} |\psi \rangle.
\end{equation}
The observable is equivalently expressed in terms of the transition matrix $i\hat{T}=(\hat{S}-\hat{1})$ and its hermitian conjugate. Using the unitarity relation $\hat{S}^\dagger \hat{S}=\hat{1}$, or equivalently $i\hat{T}-i\hat{T}^\dagger=- \, \hat{T}^\dagger \hat{T}$, it reads
\begin{equation}
\label{eq:observable}
\langle \mathds{O} \rangle=i\langle \psi | [\hat{\mathds{O}}, \hat{T}]|\psi \rangle+\langle \psi |\hat{T}^\dagger [\hat{\mathds{O}},\hat{T}]|\psi \rangle.
\end{equation}
Let us work in a relativistic field theory with two uncoupled massive fields $\phi_1$ and $\phi_2$ and a massless field $\Phi$ that couples to both $\phi_1$ and $\phi_2$. Such fields may carry a non-trivial representation of the Lorentz group. However, we will require that every vertex of the theory is either quadratic in $\phi_1$ or it is quadratic in $\phi_2$, or it does not depend on them. Examples of theories satisfying these constraint are scalar QED with two massive fields, ordinary QED with two massive spinors, QCD with two massive quarks, gravity with two massive scalars, scalar theory with two massive and one massless scalars interacting in Yukawa fashion, et cetera. 

The IN-states, denoted as $|\psi\rangle$, are superpositions of the tensor product of two asymptotic free particle states, 
\begin{equation}
\label{eq:state}
|\psi \rangle=\int \frac{\mathrm{d}^4 p_1}{(2\pi)^4} \frac{\mathrm{d}^4 p_2}{(2\pi)^4} \tilde{\delta}^+(p_1^2-m_1^2)\tilde{\delta}^+(p_2^2-m_2^2)\psi(p_1,p_2) |\phi_1(p_1) \phi_2(p_2) \rangle,
\end{equation}
with
\begin{align}
\tilde{\delta}^+(p^2-m^2)=2\pi \delta(p^2-m^2) \Theta(p^0), 
\end{align}
Let us stress that $\phi_1$ and $\phi_2$ are two distinct fields, even if they have the same mass. This differentiation is needed in order to define the impulse observable $\hat{\mathds{O}}=\hat{\mathds{P}}_{1}^\mu$, which measures the deflection of one of the two scalars, $\phi_1$:
\begin{equation}
\hat{\mathds{P}}_{1}^\mu |X_1(q_1),...,X_n(q_n) \rangle=\sum_{\substack{j\in\{1,...,n\}\\ \text{with }X_j=\phi_1}} q_j^\mu |X_1(q_1),...,X_n(q_n) \rangle
\end{equation}
where $X_i\in\{\phi_1,\phi_2,\Phi\}$. Assuming a perturbative expansion in the coupling(s) $g$ is allowed, the impulse reads
\begin{equation}
\langle \mathds{P}_1^\mu \rangle=\sum_{L=0}^\infty g^{2(L+1)}\langle \mathds{P}_1^\mu \rangle^{(L)}.
\end{equation}
Expressing the $L$-loop corrections in terms of Feynman diagrams and
massaging the expression, we obtain the following integral:
\begin{align}
&\langle \mathds{P}_1^\mu \rangle^{(L)}=\int  \mathrm{d}\Pi(p_1,p_2,q)\psi(p_1,p_2)\psi(p_1+q,p_2-q)^\star\times \nonumber \\
&\times \int \left[\prod_{i=1}^L \frac{\mathrm{d}^4k_i}{(2\pi)^4} \right]   \sum_{G\in\Gamma^L} \frac{1}{\text{Sym}(G)}F_G(\{k_j\}_{j=1}^L,p_1,p_2,p_1+q,p_2-q),
\end{align}
with $\text{Sym}(G)$ being the symmetry factor of the graph, and 
\begin{align}
 \mathrm{d}\Pi(p_1,p_2,q)=&\frac{\mathrm{d}^4 p_1}{(2\pi)^4} \frac{\mathrm{d}^4 p_2}{(2\pi)^4} 
 \frac{\mathrm{d}^4 q}{(2\pi)^4}\tilde{\delta}^+(p_1^2-m_1^2)\tilde{\delta}^+(p_2^2-m_2^2).
\end{align}
The correction is expressed in terms of the function $F_G$, which receives contributions from a graph $G=(V,E)$ and its cuts. $F_G$ implicitly depends on a Lorentz index $\mu$, which we keep implicit for simplicity of notation. More, precisely, $\Gamma^L$ is the set of all connected graphs that have $L$-loops, two external lines of type $\phi_1$ and two external lines of type $\phi_2$. In principle, $\Gamma^L$ may feature diagrams that have closed massless loops, but if we are only interested in the classical limit we may exclude them. 

The graphs in $\Gamma^L$ can be classified in terms of a connectedness criterion; for example, consider the following graph:
\begin{center}
\input{connected_components}
\end{center}
where the blobs are connected graphs, denoted by $G_{\mathrm{c}}$, with massless particles as the only possible external particles:
\begin{equation}
\raisebox{-0.3cm}{\resizebox{1cm}{!}{%
\begin{tikzpicture}

    \node[main node]  (C1) {\scalebox{1.5}{$G_{\mathrm{c}}$}};
    
\end{tikzpicture}
}}\in\Bigg \{ \raisebox{-0.6cm}{
\begin{tikzpicture}

    \node[]  (1) {};
    \node[] (F1) [below = 1cm of 1] {};

    \begin{feynman}
    \draw[dashed, thick, black, line width=0.2mm] (1) to (F1);
    \end{feynman}
    
\end{tikzpicture}
},\raisebox{-0.4cm}{
\begin{tikzpicture}

    \node[inner sep=0pt, minimum size=0.1mm]  (1) {};
    \node[inner sep=0pt, minimum size=0.1mm] (F1) [below = 0.5cm of 1] {};
    \node[inner sep=0pt, minimum size=0.1mm] (F2) [below right = 0.5cm and 0.3cm of F1] {};
    \node[inner sep=0pt, minimum size=0.1mm] (F3) [below left = 0.5cm and 0.3cm of F1] {};

    \begin{feynman}
    \draw[dashed, thick, black, line width=0.2mm] (1) to (F1);
    \draw[dashed, thick, black, line width=0.2mm] (F1) to (F2);
    \draw[dashed, thick, black, line width=0.2mm] (F1) to (F3);
    \end{feynman}
    
\end{tikzpicture}
},\raisebox{-0.4cm}{
\begin{tikzpicture}

    \node[inner sep=0pt, minimum size=0.1mm]  (1) {};
    \node[inner sep=0pt, minimum size=0.1mm] (F1) [below = 1.01cm of 1] {};
    \node[inner sep=0pt, minimum size=0.1mm] (F2) [right = 0.3cm of 1] {};
    \node[inner sep=0pt, minimum size=0.1mm] (F3) [left = 0.3cm of 1] {};

    \begin{feynman}
    \draw[dashed, thick, black, line width=0.2mm] (F1) to (F2);
    \draw[dashed, thick, black, line width=0.2mm] (F1) to (F3);
    \end{feynman}

    \path[draw=black, fill=black] (F1) circle[radius=0.05];
    
\end{tikzpicture}
}, \raisebox{-0.4cm}{
\begin{tikzpicture}

    \node[inner sep=0pt, minimum size=0.1mm]  (1) {};
    \node[inner sep=0pt, minimum size=0.1mm] (F1) [right = 0.5cm of 1] {};
    \node[inner sep=0pt, minimum size=0.1mm] (F2) [below  = 0.5cm of  1] {};
    \node[inner sep=0pt, minimum size=0.1mm] (F3) [below  = 0.5cm of F1] {};
    \node[inner sep=0pt, minimum size=0.1mm] (F4) [below  = 0.5cm of  F2] {};
    \node[inner sep=0pt, minimum size=0.1mm] (F5) [below  = 0.5cm of F3] {};

    \begin{feynman}
    \draw[dashed, thick, black, line width=0.2mm] (1) to (F2);
    \draw[dashed, thick, black, line width=0.2mm] (F2) to (F4);
    \draw[dashed, thick, black, line width=0.2mm] (F2) to (F3);
    \draw[dashed, thick, black, line width=0.2mm] (F1) to (F3);
    \draw[dashed, thick, black, line width=0.2mm] (F3) to (F5);
    \end{feynman}
    
\end{tikzpicture}
} \, , \raisebox{-0.4cm}{
\begin{tikzpicture}

    \node[inner sep=0pt, minimum size=0.1mm]  (1) {};
    \node[inner sep=0pt, minimum size=0.1mm] (F1) [right = 0.5cm of 1] {};
    \node[inner sep=0pt, minimum size=0.1mm] (F2) [below  = 0.5cm of  1] {};
    \node[inner sep=0pt, minimum size=0.1mm] (F3) [below  = 0.5cm of F1] {};
    \node[inner sep=0pt, minimum size=0.1mm] (F4) [below  = 0.5cm of  F2] {};
    \node[inner sep=0pt, minimum size=0.1mm] (F5) [below  = 0.5cm of F3] {};

    \begin{feynman}
    \draw[dashed, thick, black, line width=0.2mm] (1) to (F3);
    \draw[dashed, thick, black, line width=0.2mm] (F2) to (F4);
    \draw[dashed, thick, black, line width=0.2mm] (F2) to (F3);
    \draw[dashed, thick, black, line width=0.2mm] (F1) to (F2);
    \draw[dashed, thick, black, line width=0.2mm] (F3) to (F5);
    \end{feynman}
    
\end{tikzpicture}
} \, , \ldots
\Bigg\}
\end{equation}
We say that the example graph has \emph{four massless blobs}, or $N_G=4$. A massless blob is rigorously defined as a collection of massless edges of the graph that is connected, and that is maximal with respect to this property. The allowed types of connected massless blobs are specified by the field theory under scrutiny.

Once $\Gamma^L$ is determined, the next step is to write down $F_G$ as given by direct application of Feynman rules. $F_G$ contains the integrand contribution arising from the Feynman diagram $G=(V,E)$ as well as the cut contributions that may arise from $G$. We label the vertices of the graph $G$ that lie on the upper massive line as $V_{\text{up}}=\{v_1,...,v_n\}$ and those that lie on the lower massive line as $V_{\text{bot}}=\{w_1,...,w_m\}$. For any vertex $v$, let $E_\text{m}(v)$ be the set of \emph{massless} edges that connect to that vertex. By extension, for any subset of the vertices $V'\subseteq V$, let $E_\text{m}(V')$ be the set of \emph{massless} edges connecting to any of the vertices in $V'$. Then, we can write (as usual, using superscripts $V$ and $R$ to denote virtual and real contributions, respectively)
\begin{align}
F_G=&F_G^V+F_G^R \, , \\
F_G^V=& \left(\sum_{v\in V_{\text{up}}}\sum_{e\in E_\text{m}(v)}q_{e}^\mu\right) D_G \,  \tilde{\delta}^+((p_1+q)^2-m_1^2)\tilde{\delta}^+((p_2-q)^2-m_2^2) \, , \\
F_G^R=& \sum_{ c\in\mathcal{C}_G } \left(\sum_{v\in V_{\text{up}}\cap V_L^c}\sum_{e\in E_\text{m}(v)}q_{e}^\mu \right) D_{G,c} \, \tilde{\delta}^+((p_1+q)^2-m_1^2)\tilde{\delta}^+((p_2-q)^2-m_2^2),
\end{align}
where $\mathcal{C}_G$ is the set of all Cutkosky cuts (described by collections of cut edges) that divide the graph in two connected components, $G_L^c=(V_L^c,E_L^c)$ and $G_R^c=(V_R^c,E_R^c)$. $V_L^c$ is the set of vertices belonging to the connected component $G_L^c$ \emph{on the left} (meaning: containing the incoming particles). $D_G$ is the Feynman integrand that depends on the external momenta $p_1,p_2,p_1'=p_1+q,p_2'=p_2-q$ and on the loop-momenta $\{k_i\}_{i=1}^L$,
\begin{equation}
D_G=\mathcal{N}_G\prod_{e\in E}G_F(q_e,m_e)=\mathcal{N}_G\prod_{e\in E}\frac{i}{q_e^2-m_e^2+i\varepsilon}
\end{equation}
and $\mathcal{N}_G$ is the numerator obtained by application of Feynman rules. $D_{G,c}$ is the cut contribution obtained from $D_G$ by applying the cut Feynman propagator rule and the appropriate complex conjugation:
\begin{equation}
\label{eq:cut_diag}
D_{G,c}=\left[\prod_{e\in c} \tilde{\delta}^+(q_e^2-m_e^2)\right] D_{G_L^c} D_{G_R^c}^\star ,
\end{equation}
where we implicitly assumed a specific orientation of the cut momenta in order to express $D_{G,c}$ in terms of $\tilde{\delta}^+$ only, so that no $\tilde{\delta}^-$ appears. A few comments are in order: 
\begin{itemize}
\item[a)] If we insist, as we do, that the numerator $\mathcal{N}_G$ for the cut contributions is the same as that for the non-cut ones, we will be forced, in gauge theories, to allow for cuts that cross ghost propagators.\footnote{In other words, in line with usual unitarity considerations, we allow for the polarisation sum over external vector states to equal the form of the numerator for the virtual propagator, at the expense of having to consider ghosts as external states (see the analogous discussion relating to the KLN cancellation of infrared singularities in ref.~\cite{Capatti:2022tit}). In this paper, we will work in the Feynman gauge. Furthermore, it is important to note that the external ghosts are only needed to argue that certain classes of diagrams do not contribute in the classical limit (specifically, those having a loop made entirely by massless gauge particles), and is superfluous for what concerns the construction of the classical integrand.} Equivalently, if one instead starts from an integrand for the virtual amplitude constructed from generalised unitarity, the same diagram numerators are simply reused to obtain the integrand for the cut contributions.
\item[b)] We require that the Cutkosky cuts belonging to $\mathcal{C}_G$ leave two connected components, a notion that we take here to be strict, in the sense that no spectator state is allowed on the left or right of the graph. For example, the couplet of the following graph and Cutkosky cut 
\begin{center}
\scalebox{0.8}{
\input{example_bad_cut}} 
\end{center}
are forbidden. In simpler terms, the Cutkosky cut must cut at least one \emph{internal} line of type $\phi_1$ and one \emph{internal} line of type $\phi_2$. The types of cuts that do not satisfy this property are ill-defined as we will see in~\ref{sec:zero_measured}.
\end{itemize}
We now introduce the notion of classical limit.

\subsection{Classical limit}

KMOC formulated the classical limit of perturbative quantum observables in the IN-IN scattering formalism in terms of a soft region expansion of the integrand~\cite{Kosower:2018adc}. For each graph, choose a loop momentum basis that does not contain any edge of the upper or lower massive lines and consider the re-scaled quantity under the classical limit eq.~\eqref{eq:classicalLimit}:
\begin{align}
\langle \mathds{P}^\mu_1 \rangle^{(L)}(\lambda)=&\lambda^{-D_{\text{soft}}} \int  \mathrm{d}\Pi(p_1,p_2,\lambda \bar{q})\psi(p_1,p_2)\psi(p_1+\lambda \bar{q},p_2-\lambda \bar{q})^\star \times \nonumber \\  &\times\int \left[\prod_{i=1}^L \frac{\mathrm{d}^4(\lambda \bar{k}_i)}{(2\pi)^4} \right]   \sum_{G\in\Gamma^L} \frac{F_G(\{\lambda \bar{k}_j\}_{j=1}^L,p_1,p_2,p_1+\lambda \bar{q},p_2-\lambda \bar{q})}{\text{Sym}(G)},
\end{align}
where $D_{\text{soft}}$ is the only integer such that $\lim_{\lambda\rightarrow 0}\langle \mathds{P}^\mu_1 \rangle^{(L)}(\lambda)$ exists and is finite. We have, of course $\langle \mathds{P}^\mu_1 \rangle^{(L)}(1)=\langle \mathds{P}_1^\mu \rangle^{(L)}$. The \emph{classical expansion} of $\langle \mathds{P}^\mu_1 \rangle^{(L)}$ reads: 
\begin{align}
  \langle \mathds{P}^\mu_1 \rangle^{(L)}=\langle \mathds{P}^\mu_1 \rangle^{(L)}(\lambda)\big|_{\lambda=1}
  =\sum_{n=0}^\infty \frac{1}{n!} \frac{\mathrm{d}^n}{\mathrm{d}\lambda^n}\langle \mathds{P}^\mu_1 \rangle^{(L)}(\lambda)\Bigg|_{\lambda=0}.
\end{align}
where we expanded $\langle \mathds{P}^\mu_1 \rangle^{(L)}(\lambda)$ about $\lambda=0$ and evaluated the resulting Taylor expansion at $\lambda=1$. In scalar QED or gravity, the main statement inherent to the KMOC formalism reads: 
\begin{center}
\begin{minipage}{14cm}
\centering
$D_{\text{soft}}=L+1$\emph{ and $\langle \mathds{P}^\mu_1 \rangle^{(L)}(0)$ is the $L$-th order correction to the classical impulse.}
\end{minipage}
\end{center}
If we include a rescaling of the coupling $g=\tilde{g}/\sqrt{\lambda}$ to the power-counting, this results in the overall impulse scaling like $\lambda^0$. The hypothesis above, though compact, hides important subtleties. To get an idea of these issues, we may define a rescaled quantity for the individual contribution of each diagram to the observable, namely 
\begin{align}
\langle \mathds{P}^\mu_1 \rangle^{(L)}_G(\lambda)=&\lambda^{4L-D_{\text{soft}}^G} \int  \mathrm{d}\Pi(p_1,p_2,\lambda \bar{q})\psi(p_1,p_2)\psi(p_1+\lambda \bar{q},p_2-\lambda \bar{q})^\star \times\nonumber \\
&\times\int \left[\prod_{i=1}^L \frac{\mathrm{d}^4\bar{k}_i}{(2\pi)^4} \right]  F_G(\{\lambda \bar{k}_j\}_{j=1}^L,p_1,p_2,p_1+\lambda \bar{q},p_2-\lambda \bar{q}),
\end{align}
where $D_{\text{soft}}^G$ is again the only integer such that the $\lambda\rightarrow 0$ limit of $\langle \mathds{P}^\mu_1 \rangle^{(L)}_G(\lambda)$ exists and is finite, and the factor of $\lambda^{4L}$ arises from the loop integration measure. In general there will be diagrams such that
\begin{equation}
D_{\text{soft}}^G<D_{\text{soft}}
\end{equation}
In other words, the scaling behaviour of the full observable is better than the scaling behaviour of each single diagram, and cancellations across different diagrams must take place in order for $\langle \mathds{P}^\mu_1 \rangle^{(L)}(\lambda)$ to possess the expected scaling of $D_\mathrm{s}=L+1$. If that is the case, we say that the observable undergoes the \emph{cancellation of super-leading contributions}. Conversely, there will be diagrams such that
\begin{equation}
D_{\text{soft}}^G>D_{\text{soft}},
\end{equation}
implying that they do not contribute to the classical observable. A sharper statement comes from simple power-counting arguments. \\
\textbf{Power-counting} Let us consider three different theories: scalar QED, gravity (coupled to a scalar massive field) and a scalar theory with two massive fields and one massless field. Let $D_{\text{soft}}^G$ be the soft degree of divergence for a graph with $N_G$ massless blobs. A massless blob may itself be a loop diagram. We say that a graph $G$ has a 
quantum loop if, by stripping away all the edges on the upper or lower massive line, we are left with a graph that has at least one loop.

In general, the power-counting reads
\begin{align}
D_{\text{soft}}^G&=\text{rk}_{E_\text{m}(V)}(\mathcal{N}_G)+4 (L+1)-2|E_\text{m}(V)|-|V_{\text{up}}|-|V_{\text{bot}}|,
\end{align}
where $\text{rk}_{E_\text{m}(V)}(\mathcal{N}_G)$ is the rank (i.e.\ the number of powers) of the numerator in the momenta of the massless particles. The quantity $D_{\text{soft}}^G$ allows for bounds expressed entirely in terms of $L$, $N_G$, and the presence of quantum loops.  

\begin{itemize}
\item \textbf{Scalar field theory without massless self-interaction} For a scalar field theory with potential $\phi_1^2 \Phi+\phi_2^2 \Phi$, the power-counting reads
\begin{equation}
D_{\text{soft}}^G=L-N_G+2=1.
\end{equation}
for all the graphs without quantum loops, i.e. for all ladder diagrams and their crossed versions. We expect these diagrams to undergo $L=N_G-1$ orders of cancellations. If the diagrams have quantum loops, then
\begin{equation}
D_{\text{soft}}^G>L-N_G+2.
\end{equation}
If $N_G=1$, this lower bound implies that the graph will not contribute classically. If $N_G>1$ and the diagram also undergoes $N_G-1$ orders of cancellations, then it will not contribute classically, accounting for purely quantum effects.

\item \textbf{Scalar QED} 
If $N_G=1$, for scalar QED, we have
\begin{equation}
D_{\text{soft}}^G=L+1,
\end{equation}
if the diagram has no quantum loops, i.e., if the diagram is a zig-zag diagram such as
\begin{equation}
\scalebox{0.65}{\input{VV}}
\end{equation}
and, if the diagram has quantum loops, we have 
\begin{equation}
D_{\text{soft}}^G > L+1,
\end{equation}
if the diagram has a quantum loop; in particular, self-energy corrections aside, there is only one diagram with a quantum loop and $N_G=1$, namely:
\begin{equation}
\scalebox{0.75}{
\input{almond_diagram}}
\end{equation}
This implies that for $N_G=1$ only diagrams with no quantum loops will contribute to the leading order in the classical expansion, and will do so with the leading term in their expansion. 

If we consider a graph with $N_G$ connected blobs and no quantum loops, the power-counting will change to
\begin{equation}
D_{\text{soft}}^G=L-N_G+2,
\end{equation}
implying that these diagrams lead to super-leading contributions that need to cancel in the sum over all diagrams in order to retrieve the observable scaling $D_{\text{soft}}=L+1$. In particular, $N_G-1$ orders in the soft expansion of these diagrams should cancel. Again, if the graph has a quantum loops, a strict lower bound holds instead:
\begin{equation}
D_{\text{soft}}^G> L-N_G+2.
\end{equation}
Again, if it is shown that $N_G-1$ orders of cancellations also hold for these diagrams, then diagrams with quantum loops will not contribute to the classical limit.

\item \textbf{Gravity} Graviton vertices scale like two powers of the momentum. Although this scaling makes gravity notoriously non-renormalisable, it gives a well-defined classical limit. For $N_G=1$, it holds that
\begin{equation}
D_{\text{soft}}^G \ge L+1,
\end{equation}
with the equality corresponding to the case of $G$ not having quantum loops. This implies that any diagram with one-connected blob that is also a loop diagram will not contribute to the classical limit. When the graph has $N_G$ massless blobs, the power-counting reads
\begin{equation}
D_{\text{soft}}^G\ge L-N_G+2.
\end{equation}
Again, equality holds when the graph $G$ has no quantum loops. In other words, we find that in gravity the bounds are specular to those found in scalar QED.

\item \textbf{Scalar field theory with massless self-interaction}  However, if we also add a super-renormalisable self-interaction term of the type $\Phi^3$ to the scalar theory we studied first, we obtain
\begin{equation}
D_{\text{soft}}^G\le L+1.
\end{equation}
Every diagram will thus give super-leading contributions, whether or not the diagram has quantum loops. In this case, the notion of a classical limit breaks down. If, in place of the super-renormalisable self-interaction $\Phi^3$ we were to introduce a renormalisable self-interaction $\Phi^4$, we would then obtain the power-counting 
\begin{equation}
D_{\text{soft}}^G=1,
\end{equation}
for any diagram, regardless of the fact that they have quantum loops. This fact is counter-intuitive with respect to the idea that only massless trees should be allowed in the connected blobs.
\end{itemize}
This simple power-counting procedure highlights the following two facts: 
\begin{itemize}
\item\emph{If we are invested in the idea that only diagrams with no quantum loops should contribute to the classical limit, then we can only consider non-renormalisable self-interaction terms for the massless field.\footnote{In non-Abelian gauge theories that have renormalisable self-interaction terms for the massless field, a notion of a well-defined classical limit nevertheless exists, since colour charges also have nontrivial scaling in the classical limit \cite{delaCruz:2020bbn, delaCruz:2021gjp}.}}
\item In order for $D_{\text{soft}}=L+1$, for theories with non-renormalisable self-interactions of the massless field, the sum over all diagrams with $N_G$ connected components needs to undergo $N_G-1$ orders of cancellations in the soft expansion.
\end{itemize}
The second statement is referred to as the cancellation of super-leading contributions.
 
\textbf{Cancellation of super-leading contributions.} The cancellation property discussed above can be formulated in an independent way than that of the classical limit. Given two graphs $G$ and $G'$ in $\Gamma^L$ contributing to $\langle \mathds{O} \rangle^{(L)}_G$, we say that they are equivalent if they differ by a permutation of the vertices in $V_{\text{up}}$ or $V_{\text{bot}}$. Let $\Gamma^L/\sim$ be the quotient of the space of graphs with respect to the equivalence relation established above. Let $[G]\in\Gamma^L/\sim$ be an equivalence class. The equivalence class collects graphs that all have the same massless blobs. Furthermore, all graphs in an equivalence class have the same degree of divergence $D_\text{s}^G$. 

The statement of the cancellation of super-leading contribution is 
\begin{equation}
\label{eq:sym}
\langle \mathds{P}_1^\mu \rangle^{(L)}_{[G]}(\lambda)=\sum_{G'\in[G]}\frac{1}{\text{Sym}(G')}\langle \mathds{P}_1^\mu \rangle^{(L)}_{G'}(\lambda)=\mathcal{O}\left(\lambda^{L+1-D_{\text{soft}}^G}\right)=\mathcal{O}\left(\lambda^{N_G-1}\right) \, ,
\end{equation}
i.e. $N_G-1$ orders of $\lambda$ beyond the leading soft limit. Since $[G]$ is not a gauge-invariant subset, this statement, in gauge theories, holds at a local level for internal propagators in the Feynman gauge and external states that include longitudinal gauge bosons as well as ghosts (see the discussion below eq.~\eqref{eq:cut_diag}). In other words, while each individual graph in $[G]$ in scalar QED scales at worst with $L+2-N_G$, the sum over all the graphs in the equivalence class scales at worst like $L+1$, i.e. super-leading divergences cancel within each equivalence class of graphs.

\textbf{Local cancellation of super-leading contributions.} One of the aims of this paper concerns the problem of realising cancellation of super-leading contributions at the integrand level. A local formulation, as we will see, requires a symmetrisation procedure. 

Let us consider the equivalence class $[G]$. Graphs contributing to it will have the same number of vertices in the upper massive line and the same number of vertices in the lower massive line. The representative graph $G$ is associated with the order $V_{\text{up}}^G=(v_1,...,v_n)$ and $V_{\text{bot}}^G=(w_1,...,w_m)$ of the vertices, ``from left to right''. Another graph in the equivalence class will vary by a permutation of exactly these vertices, so it will be associated to an order $V_{\text{up}}^{G'}=(v_{\pi(1)},...,v_{\pi(n)})$, $V_{\text{bot}}^{G'}=(w_{\pi'(1)},...,w_{\pi'(m)})$, and may be denoted as $G_{\pi,\pi'}$. There is not a one-to-one relation between graphs in $[G]$ and permutations. Two distinct permutations of the same graph may yield isomorphic graphs. 

Symmetrising over all the permutations of vertices in the upper or lower massive line leads to a new representation for $\langle \mathds{O} \rangle^{(L)}_{[G]}$. Let $G_1,...,G_{N_G}$ be the massless blobs of the representative graph $G$ (which are shared by all the graphs in the equivalence class). Then
\begin{equation}
\langle \mathds{O} \rangle^{(L)}_{[G]}(\lambda)=\frac{1}{|V_{\text{up}}|!|V_{\text{bot}}|!\prod_{j=1}^{N_G}\text{Sym}(G_j)}\sum_{\substack{\pi\in\mathfrak{S}(V_{\text{up}})\\ \pi'\in\mathfrak{S}(V_{\text{bot}})}}\langle \mathds{O} \rangle^{(L)}_{G_{\pi,\pi'}}(\lambda).
\end{equation}
Note that upon symmetrising, there is an overall symmetry factor, as opposed to what was the case in eq.~\eqref{eq:sym}.

To write an integrand for $\langle \mathds{O} \rangle^{(L)}_{[G]}$ we need to choose a convenient loop-momentum basis for each of the $\langle \mathds{O} \rangle^{(L)}_{G_{\pi,\pi'}}$ contributions. Let $e$ be an edge of the representative graph $G$. The permutations of the vertices in the upper and lower line map the edge $e$ in the graph $G$ to the edge $\varphi_{\pi,\pi'}(e)$ in the graph $G_{\pi,\pi'}$. We require the routing for the graph $G_{\pi,\pi'}$ to be such that $q_{\varphi_{\pi,\pi'}(e)}=q_e$ for any edge $e$ that is not in the upper or lower massive line. In other words, the vertex permutation naturally induces a corresponding change in loop momentum routing. Finding one example of such routing is simple: for the graph $G$ that is the representative of $[G]$, we choose a loop momentum basis $B\subset E$ that does not contain edges in the upper of lower massive line and instead only contains massless edges. The induced loop momentum basis for $G_{\pi,\pi'}$ is then $\{\varphi_{\pi,\pi'}(e)\}_{e\in B}$. 

In such a loop momentum routing there exists an integrand \emph{for the equivalence class }$[G]$, defined by
\begin{equation}
\langle \mathds{O} \rangle^{(L)}_{[G]}(\lambda)=\frac{\lambda^{4(L+1)-D_{\text{soft}}^{[G]}}}{|V_{\text{up}}|!|V_{\text{bot}}|!\prod_{j=1}^{N_G}\text{Sym}(G_j)}  \int  \mathrm{d}\Pi(p_1,p_2,\bar{q})\psi(p_1,p_2)\psi(p_1+\lambda\bar{q},p_2-\lambda\bar{q})^\star  \int \left[\prod_{i=1}^L \frac{\mathrm{d}^4\bar{k}_i}{(2\pi)^4} \right] F_{[G]}
\end{equation}
with
\begin{equation}
F_{[G]}=F_{[G]}^V+F_{[G]}^R=\sum_{\substack{\pi\in\mathfrak{S}(V_{\text{up}})\\ \pi'\in\mathfrak{S}(V_{\text{bot}})}} F_{G_{\pi,\pi'}}(\{\lambda \bar{k}_j\}_{j=1}^L,p_1,p_2,p_1+\lambda \bar{q},p_2-\lambda \bar{q}),
\label{eq:integrandEquivClass}
\end{equation}
and $F_{[G]}^V$, $F_{[G]}^R$ denoting the sums of all virtual and real contributions arising from any of the graphs in the equivalence class. We may now require \emph{the integrand} to have the scaling that reproduces the cancellation of super-leading contributions:
\begin{equation}
F_{[G]}\sim d \lambda^{-3(L+1)}, \quad \lambda\rightarrow 0 \, ,
\end{equation}
which multiplies the scaling of the integration measure, $\lambda^{4(L+1)}$, to produce $\lambda^{L+1}$ matching the degree of divergence $D_s^{[G]} \leq L+1$.
In this paper, we will take important steps towards showing this statement at arbitrary orders. The general presentation will be supported by several multi-loop examples.

\section{The quantum worldline}
\label{sec:quantum_worldline}

\begin{figure}
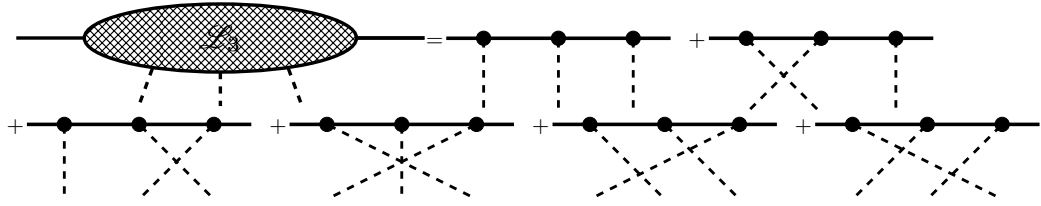

\centering
\scalebox{0.9}{
\begin{minipage}{14cm}
\raisebox{-0.9cm}{\begin{tikzpicture}
\draw[fill=black!25!white, line width=0.5mm, pattern=crosshatch, fill opacity=0.1] (0,0) ellipse (2cm and 0.5cm);
\draw[line width=0.5mm] (-3,0) -- (-2,0);
\draw[line width=0.5mm] (2,0) -- (3,0);
\draw[line width=0.5mm] (2,0) -- (3,0);
\draw[line width=0.5mm,dashed] (1.,-0.45) -- (1.2,-1);
\draw[line width=0.5mm,dashed] (0,-0.5) -- (0,-1);
\draw[line width=0.5mm,dashed] (-1.,-0.45) -- (-1.2,-1);
\node[] at (0, 0)  (c)  {\scalebox{1.5}{$\mathcal{L}_3$}};
\end{tikzpicture}}$=$\raisebox{-1cm}{\input{3att_1}}$+$\raisebox{-1cm}{\input{3att_2}} 
\\
$+$\raisebox{-1cm}{\input{3att_3}}$+$\raisebox{-1cm}{\input{3att_4}}$+$\raisebox{-1cm}{\input{3att_5}}$+$\raisebox{-1cm}{\input{3att_6}}
\end{minipage}
}
\caption{Diagrammatics of the quantum worldline with three attachments}
\label{fig:qWorld}
\end{figure}
We now introduce the quantum worldline, an object closely related to the time-ordered tree-level correlator with two external massive legs
and $n$ off-shell massless legs. The Feynman diagram
representation consists of a symmetrised version of a massive line
with $n$ vertices for attachments to (amputated) off-shell massless lines. For $3$ attachments, for example, we define the
quantum worldline as in fig.~\ref{fig:qWorld}. The symmetrised virtual
integrand for the impulse observable in the KMOC formalism will
contain terms which, expressed in terms of the quantum worldline, read
\begin{align}\label{eq:virtual_worldline}
\raisebox{-1.8cm}{\scalebox{0.75}{
\begin{tikzpicture}
\draw[fill=black!05!white, line width=0.5mm, pattern=crosshatch, fill opacity=0.1] (0,-2) ellipse (1cm and 1cm);
\draw[fill=black!25!white, line width=0.5mm, pattern=crosshatch, fill opacity=0.1] (0,0) ellipse (2cm and 0.5cm);
\draw[line width=0.5mm] (-3,0) -- (-2,0);
\draw[line width=0.5mm] (2,0) -- (3,0);
\draw[line width=0.5mm,dashed] (1.,-0.45) -- (0.8,-1.4);
\draw[line width=0.5mm,dashed] (0,-0.5) -- (0,-1);
\draw[line width=0.5mm,dashed] (-1.,-0.45) -- (-0.8,-1.4);
\draw[fill=black!25!white, line width=0.5mm, pattern=crosshatch, fill opacity=0.1] (0,-4) ellipse (2cm and 0.5cm);
\draw[line width=0.5mm] (-3,-4) -- (-2,-4);
\draw[line width=0.5mm] (2,-4) -- (3,-4);
\draw[line width=0.5mm,dashed] (1.,-3.55) -- (0.8,-2.6);
\draw[line width=0.5mm,dashed] (-1.,-3.55) -- (-0.8,-2.6);
\node[] at (0, 0)  (c)  {\scalebox{1.5}{$\mathcal{L}_3$}};
\node[] at (0, -4)  (c)  {\scalebox{1.5}{$\mathcal{L}_2$}};
\node[] at (0, -2)  (c)  {\scalebox{1.4}{$G_{\text{m-less}}$}};
\end{tikzpicture}}}\subset F_{[G]}^V,
\end{align}
where $G_{\text{m-less}}$ is a product of massless blobs. In the
following, we will study the classical expansion of the quantum
worldline. 


\subsection{The quantum worldline and its cuts}

We start by rigorously introducing the quantum worldline, writing a
Schwinger parametrisation for it and \emph{canonicalising} it,
i.e. writing it in a unique representation. The idea that Schwinger parameter space is particularly apt for studying the soft expansion of this object is inspired by the derivation Steven Weinberg proposes in chapter 13 of ref.~\cite{Weinberg:1995mt} for the leading-order term of the quantum worldline ($\lambda^{-n}$ in the soft expansion), which he used to extract the Coulomb potential from the classical limit of symmetrised ladder diagrams. The first sub-leading correction to the quantum worldline in the soft parameter for a massless probe particle moving in the gravitational background of a massive object (e.g.\ bending of light) has been computed in~\cite{Akhoury:2013yua} ($\lambda^{-n+1}$ in the soft expansion). Here we derive all leading and sub-leading soft corrections up to $\lambda^0$ for the scattering of two objects of arbitrary masses, show their connection with graph-theoretical constructions such as forests, and express them in a compact resummed form. This derivation and the extension presented here also share common tools with diagrammatic derivations of Time-Ordered Perturbation Theory~\cite{tasi_sterman,Sterman:1993hfp}, Flow-Oriented Perturbation Theory~\cite{Borinsky:2022msp}, as well as the Cross-Free Family representation and its variants~\cite{Capatti:2022mly,Capatti:2023shz,Sterman:2023xdj}.
The general quantum worldline with $n$ vertices $V=\{1,...,n\}$ reads:
\begin{equation}
\mathcal{L}(V,\{\},p)=\sum_{\pi\in\mathfrak{S}_n} \bar{\mathcal{N}}_{\pi}^V(\{q_i\}_{i=1}^n)\left[\prod_{i=1}^{n-1} \frac{i}{(p+\sum_{j=1}^i q_{\pi(j)})^2-m^2+i\varepsilon}\right]\tilde{\delta}\left(\Bigg(p+\sum_{j=1}^n q_j\Bigg)^2-m^2\right),
\end{equation}
where $\{\}$ indicates an empty set of vertices on the right side of
the cut, as we are looking at the uncut worldline, and $\bar{\mathcal{N}}_{\pi}^V$ is the numerator associated with the quantum worldline (specifically vertices on the quantum worldline), stripped of $i$ vertex factors. This expression also includes the case in which a vertex couples the matter lines to multiple massless particles, instead of one only. Indeed, $\mathcal{L}$ implicitly depends on the number of cubic and quartic vertices contained in the ground set $V$. In this case, one simply has to identify the momentum $q_j$ injecting at the $j$-th vertex with the sum of the momenta of the massless particles injecting at that vertex.

We rewrite $\mathcal{L}(V,\{\},p)$ by expressing propagators in their Schwinger parametrisation 
\begin{equation}
\frac{\pm i}{p^2-m^2\pm i\varepsilon}=\int \mathrm{d}\Delta\tau e^{-ip^2\Delta\tau}\Theta(\mp \Delta\tau).
\end{equation}

More specifically, for the permutation $\pi$ corresponding to the ordering $\pi(1)<\pi(2)<...<\pi(n)$, the propagator with momentum $q_{\pi(1)}+p$ will be labelled by the proper time difference $\Delta\tau_{\pi(1)\pi(2)}$, while the propagator with momentum $q_{\pi(1)}+...+q_{\pi(j)}+p$ will be labelled by the proper time difference $\Delta\tau_{\pi(j)\pi(j+1)}$. Finally, for the summand labelled by $\pi$ we will use the Fourier representation for the Dirac delta function integrated over the proper time $\tau_{\pi(n)}$:
\begin{equation}
\tilde{\delta}\left(\Bigg(p+\sum_{j=1}^n q_j\Bigg)^2-m^2\right)=\int \mathrm{d}\tau_{\pi(n)} e^{-i\tau_{\pi(n)}((p+\sum_{j=1}^n q_j)^2-m^2)}.
\end{equation}
In summary, we obtain:
\begin{align}
\mathcal{L}(V,\{\},p)=\sum_{\pi\in\mathfrak{S}_n} \int \left[\prod_{j=1}^{n-1} \mathrm{d}\Delta \tau_{\pi(j)\pi(j+1)}\Theta\big(-\Delta \tau_{\pi(j)\pi(j+1)}\big)e^{-i ((p+\sum_{l=1}^j q_{\pi(l)})^2-m^2)\Delta \tau_{\pi(j)\pi(j+1)} }\right]\times\nonumber \\
\times \int \mathrm{d}\tau_{\pi(n)} e^{-i\tau_{\pi(n)}((p+\sum_{j=1}^n q_j)^2-m^2)} \bar{\mathcal{N}}_{\pi}^V(\{q_i\}_{i=1}^n),
\end{align}
Let us now perform a change of variable. For the product of integrals corresponding to a fixed permutation $\pi$, we change variables to 
\begin{equation}
\label{eq:ch_of_vars}
\Delta \tau_{\pi(j)\pi(j+1)}=\begin{cases}\displaystyle
\sum_{i=\pi(j)}^{\pi(j+1)-1}\Delta \tau_{i,i+1}\quad  &\text{if } \pi(j+1)>\pi(j) \\ \displaystyle
-\sum_{i=\pi(j+1)}^{\pi(j)-1}\Delta \tau_{i,i+1}\quad  &\text{if } \pi(j)>\pi(j+1)
\end{cases}, \quad \tau_{\pi(i)}=\tau_n+\sum_{j=\pi(i)}^{n-1}\Delta\tau_{j,j+1}.
\end{equation}
 When $n=2$, we saw in our warm-up example that this change of variables maps the $\Delta\tau_{12},\tau_2$ variables of the identity permutation to themselves, while it maps the $\Delta\tau_{21},\tau_1$ variables of the non-trivial permutation to $\Delta\tau_{21}=-\Delta\tau_{12}$, $\tau_1=\Delta\tau_{12}+\tau_2$. Under this change of variables, the integration sign and summation sign can be exchanged, giving 
\begin{align}
\mathcal{L}(V,\{\},p)=& \int \left[\prod_{j=1}^{n-1} \mathrm{d}\Delta \tau_{jj+1}e^{-i ((p+\sum_{l=1}^j q_{l})^2-m^2)\Delta \tau_{j j+1} }\right]\Bigg[\mathrm{d}\tau_{n} e^{-i\tau_{n}((p+\sum_{j=1}^n q_j)^2-m^2)} \Bigg]\times\nonumber \\
\times&\sum_{\pi\in\mathfrak{S}_n}\left[\prod_{j=1}^{n-1}\Theta\big(-\Delta \tau_{\pi(j)\pi(j+1)}\big)\right]\exp\left\{-2 i \sum_{\substack{l,l'=1 \\ l<l'}}^{n} \Delta \tau_{l,l'} (q_l\cdot q_{l'}) \Theta(\pi(l)-\pi(l'))  \right\}\bar{\mathcal{N}}_{\pi}^V(\{q_i\}_{i=1}^n), \label{eq:quantumWorldlineExact}
\end{align}
where $\tau_{\pi(j)\pi(j+1)}$ should now be thought of as a linear function of the integration variables, as given in eq.~\eqref{eq:ch_of_vars}.
The advantage of this representation is that it \emph{canonicalises} the sum over permutations of the attachments to the matter line.

As a simple example, for the scalar model eq.~\eqref{eq:scalarModelLagrangian} in which numerators are trivial, eq.~\eqref{eq:quantumWorldlineExact} coincides with eq.~\eqref{eq:scalar_worldline_sum_alt}. Specifically, the identity permutation in eq.~\eqref{eq:quantumWorldlineExact} contributes to the term proportional to $\Theta(-\Delta \tau_{12})$ in eq.~\eqref{eq:scalar_worldline_sum_alt}, while the other permutation, with $\pi(1)=2, \pi(2)=1$, contributes to the term proportional to $\Theta(\Delta \tau_{12})$ in eq.~\eqref{eq:scalar_worldline_sum_alt}. Only the latter term is accompanied by an extra exponential $e^{-2i q_1\cdot q_2 \Delta\tau_{12}}$ in eq.~\eqref{eq:scalar_worldline_sum_alt}, since such terms only arise in the presence of reversed ordering of vertices according to eq.~\eqref{eq:quantumWorldlineExact}.

For future use, let us use the notation
\begin{align}
\mathrm{d}\Pi_n(p)&=\left[\prod_{j=1}^{n-1} \mathrm{d}\Delta \tau_{j,j+1}e^{-i ((p+\sum_{l=1}^j q_{l})^2-m^2)\Delta \tau_{j,j+1} }\right]\Bigg[\mathrm{d}\tau_{n} e^{-i\tau_{n}((p+\sum_{j=1}^n q_j)^2-m^2)}\Bigg], \\
\ell(V,\{\},p)&=\sum_{\pi\in\mathfrak{S}_n}\left[\prod_{j=1}^{n-1}\Theta\big(-\Delta \tau_{\pi(j)\pi(j+1)}\big)\right]\exp\left\{-2 i \sum_{\substack{l,l'=1 \\ l<l'}}^{n} \Delta \tau_{l,l'} (q_l\cdot q_{l'}) \Theta(\pi(l)-\pi(l'))  \right\}\bar{\mathcal{N}}_{\pi}^V(\{q_i\}_{i=1}^n),
\end{align}
so that
\begin{equation}
\mathcal{L}(V,\{\},p)= \int \mathrm{d}\Pi_n(p)\ell(V,\{\},p).
\end{equation}
The advantage of these re-definitions is that, in order to derive the cancellation of super-leading contributions, only the expansion of $\ell(V,\{\},p)$ needs to be studied. In particular, the inverse quadratic propagators in the exponents of $\mathrm{d}\Pi_n(p)$ will not need to be expanded. Once super-leading contributions cancel in the sum of $\ell(V,\{\},p)$ for different permutations $\pi$, all terms induced by the expansion of the shared factor $\mathrm{d}\Pi_n(p)$ will eventually only introduce quantum corrections, simplifying considerably the combinatorics.

For the purposes of studying the classical limit of cut diagrams in the KMOC formalism, we also have to introduce the cut version of the quantum worldline. Consider a partition of the set $\{1,...,n\}$, $S_1\cup S_2=\{1,...,n\}$, where $S_1$ is the set of vertices on the left side of the cut, and $S_2$ is the set of vertices on the right side of the cut. Then consider all the permutations that leave the partition invariant 
\begin{equation}
\mathfrak{S}(S_1,S_2)=\{\pi\in\mathfrak{S}_n \, | \, \pi(\{1,...,|S_1|\})=S_1, \, \pi(\{|S_1|+1,...,|S_1|+|S_2|\})=S_2\}.
\end{equation}
Each of these permutations also corresponds to a total ordering of $\{1,...,n\}$, and the collection $\mathfrak{S}(S_1,S_2)$ identifies all orderings such that any element of $S_1$ is lower than any element of $S_2$. The cut quantum worldline reads
\begin{align}
&\mathcal{L}(S_1,S_2,p)=\sum_{\pi\in\mathfrak{S}(S_1,S_2)}\bar{\mathcal{N}}_{\pi}^V(\{q_i\}_{i=1}^n)\left[\prod_{i=1}^{|S_1|-1}\frac{i}{(p+\sum_{j=1}^{i}q_{\pi(j)})^2-m^2+i\varepsilon}\right]\times \nonumber \\
&\times \tilde{\delta}\left(\Bigg(p+\sum_{i=1}^{|S_1|}q_{\pi(i)}\Bigg)^2-m^2\right) \left[\prod_{i=|S_1|+1}^{|S_1|+|S_2|}\frac{-i}{(p+\sum_{j=1}^{i}q_{\pi(j)})^2-m^2-i\varepsilon}\right] \tilde{\delta}\left(\Bigg(p+\sum_{j=1}^n q_j\Bigg)^2-m^2\right) \,
\end{align}
where the propagators on the right side of the cut are complex-conjugated, and there is an extra Dirac delta function for the cut matter line.
In this case, the proper-time parametrisation reads 
\begin{align}
&\mathcal{L}(S_1,S_2,p)=\sum_{\pi\in\mathfrak{S}(S_1,S_2)} \bar{\mathcal{N}}_{\pi}^V(\{q_i\}_{i=1}^n)\int \left[\prod_{j=1}^{n-1} \mathrm{d}\Delta \tau_{\pi(j)\pi(j+1)}e^{-i ((p+\sum_{l=1}^j q_{\pi(l)})^2-m^2)\Delta \tau_{\pi(j)\pi(j+1)} }\right] \nonumber \\
&\times\Bigg[\mathrm{d}\tau_{\pi(n)} e^{-i\tau_{\pi(n)}((p+\sum_{j=1}^n q_j)^2-m^2)} \Bigg]\left[\prod_{j=1}^{|S_1|-1}\Theta\big(-\Delta \tau_{\pi(j)\pi(j+1)}\big)\right] \left[\prod_{j=|S_1|+1}^{|S_1|+|S_2|-1}\Theta\big(\Delta \tau_{\pi(j)\pi(j+1)}\big)\right].
\end{align}
We perform the change of variables given in eq.~\eqref{eq:ch_of_vars}, exchange the sum over permutations and the integration sign. Eventually, we obtain
\begin{align}
\mathcal{L}&(S_1,S_2,p)= \int \mathrm{d}\Pi(p)\ell(S_1,S_2;p),
\end{align}
with
\begin{align}
\ell(S_1,S_2;p)&=\sum_{\pi\in\mathfrak{S}_n}\bar{\mathcal{N}}_{\pi}^V(\{q_i\}_{i=1}^n)\exp\left\{-2 i \sum_{\substack{l,l'=1\\ l<l'}}^{n} \Delta \tau_{\substack{l,l'}} (q_l\cdot q_{l'}) \Theta(\pi(l)-\pi(l'))  \right\} \times \nonumber \\
&\times
 \left[\prod_{j=1}^{|S_1|-1}\Theta\big(-\Delta \tau_{\pi(j),\pi(j+1)}\big)\right]  \left[\prod_{j=|S_1|+1}^{|S_1|+|S_2|-1}\Theta\big(\Delta \tau_{\pi(j),\pi(j+1)}\big)\right].
\end{align}
This provides the canonicalised expression for the cut quantum worldline. The integration measure of the cut contribution is completely aligned with that of the virtual contribution, analogously to the principle underlying the Local Unitarity method~\cite{Capatti:2020xjc,Capatti:2021bsm}.

\subsection{Classical expansion of the quantum worldline}
\label{sec:classical_expansion_quantum_worldline}

\begin{figure}
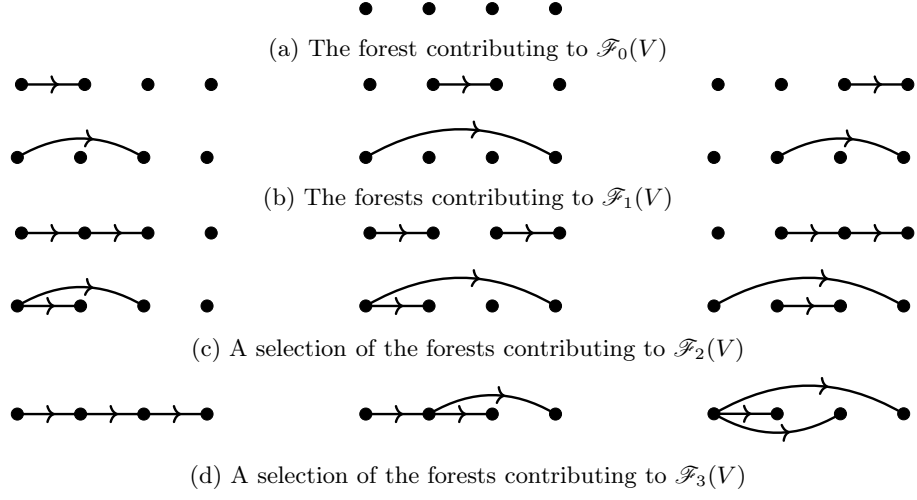

\centering
\begin{subfigure}{\textwidth}
\centering
\input{struct1}
\subcaption{The forest contributing to $\mathcal{F}_0(V )$}
\end{subfigure}
\begin{subfigure}{\textwidth}
\centering
\input{struct2}\hspace{1cm}\input{struct3}\hspace{1cm}\input{struct4} \\[0.3cm]
\input{struct5}\hspace{1cm}\input{struct6}\hspace{1cm}\input{struct7}
\subcaption{The forests contributing to $\mathcal{F}_1(V )$}
\end{subfigure}
\begin{subfigure}{\textwidth}
\centering
\input{struct8}\hspace{1cm}\input{struct9}\hspace{1cm}\input{struct10} \\[0.3cm]
\input{struct11}\hspace{1cm}\input{struct12}\hspace{1cm}\input{struct13}
\subcaption{A selection of the forests contributing to $\mathcal{F}_2(V )$}
\end{subfigure}
\begin{subfigure}{\textwidth}
\centering
\input{struct14}\hspace{1cm}\input{struct15}\hspace{1cm}\raisebox{-0.25cm}{\input{struct16}}
\subcaption{A selection of the forests contributing to $\mathcal{F}_3(V )$}
\end{subfigure}
\caption{Examples of forests contributing to $\mathcal{F}_k(V)$ for $|V|=4$}
\label{fig:forests}
\end{figure}

Studying the expansion of $F_{[G]}$ is equivalent to studying the expansion of $\ell(V,\{\},p)$ and $\ell(S_1,S_2;p)$. Due to the power-counting argument, the order at which $\ell(V,\{\},p)$ and $\ell(S_1,S_2;p)$ have to be expanded upon is at most $n-1$ relative to the leading order.

\textbf{Triangulation identities} While a naive expansion of $\ell$ features an exploding number of terms, many of them undergo significant simplifications from identities of the same type as that in eq.~\eqref{eq:theta_2att}. Another example of such identities is
\begin{equation}
\sum_{\pi\in \mathfrak{S}_n}\prod_{i=1}^{n-1}\Theta(\tau_{\pi(i+1)}-\tau_{\pi(i)})=1.
\end{equation}
The identities are effectively based on the following triangulation: let $F$ be a set on the ordered ground set $(1,...,n)$, that is a collection of couplets $(j,j')$, with orientation fixed such that $j<j'$, and with no loops. $F$ corresponds to a partial ordering of the ground set $\{1,...,n\}$, or equivalently to an acyclic subgraph of the complete directed acyclic graph on the ground set $(1,...,n)$ (a directed acyclic graph is actually a richer object than a partial ordering, see the notion of transitive closure). These partial orderings can be organised according to their cardinality. In fig.~\ref{fig:forests}, we provide examples of partial orderings written in terms of forests: two vertices are ordered, one being located at a higher time than another, if a directed edge departs from the first to end on the other. For example, in (a), the vertices are not ordered at all, while in (b), only two vertices have an ordering, but their ordering relative to any other vertex is not fixed. Accordingly, consider the set of all directed acyclic graphs with $l$ edges that can be constructed on the \emph{ordered} ground set $V=(1,...,n)$:
\begin{equation}
\mathcal{P}_l(V)=\left\{F\subset (V\times V) \, \big| \,  |F|=l, \ \forall e=(v_i,v_j)\in F, \ i<j  \right\}. \label{eq:partialOderingLEdges}
\end{equation}
We also define the restriction of this set under the request that its elements have no loops
\begin{equation}
\mathcal{F}_l(V)=\left\{F\subset (V\times V) \, \big| \, F \text{ is a directed forest (arborescence)}, \ |F|=l, \ \forall e=(v_i,v_j)\in F, \ i<j  \right\}.
\end{equation}
$\mathcal{F}_l(V)$ describes the subset of elements of $\mathcal{P}_l(V)$ whose contribution to the observable survives the classical limit in the KMOC formalism. 
Now, let $\mathfrak{S}_n (F)$ be the set of all permutations that preserve the partial ordering implied by a specific element $F\in \mathcal{P}_l(V)$. If the reference ordering $(1,...,n)$ gives a total ordering of vertices, $F$ only gives partial ordering relations between the vertices. Then 
\begin{align}
\sum_{\pi\in\mathfrak{S}_n(F)} \prod_{j=1}^{n-1}\Theta(\tau_{\pi(j)}-\tau_{\pi(j+1)})=\prod_{a=(v',v)\in F} \Theta(\tau_{v}-\tau_{v'}).
\end{align}
See refs.~\cite{Weinberg:1995mt,tasi_sterman,Capatti:2022mly,Sterman:2023xdj,Capatti:2023shz} for uses of sub-classes of this set of identities, and in particular ref.~\cite{Sterman:2023xdj} for what concerns partial orderings and ref.~\cite{Capatti:2022mly,Capatti:2023shz} for what concerns acyclic graphs. We proceed to the expansion of $\ell(V,\{\},p)$ in the scalar case, namely when $\bar{\mathcal{N}}_\pi^V=1$. We will later discuss the case of scalar QED. We consider the soft expansion 
\begin{equation}
\ell(V,\{\},p)=\sum_{m=0}^\infty \lambda^{m}\ell^{(m)}(V,\{\},p)=\sum_{m=0}^\infty \lambda^{m} \left[\frac{\mathrm{d}^m}{\mathrm{d}\lambda^m}\ell(V,\{\},p)\bigg|_{q_j=\lambda \bar{q}_j, \tau_i=\lambda^{-1}\bar{\tau}_i}\right]_{\lambda=0},
\end{equation}
we have, of course, that
\begin{align}
\ell^{(0)}(V,\{\},p)&=\sum_{\pi\in\mathfrak{S}_n} \prod_{j=1}^{n-1} \Theta(-\Delta \tau_{\pi(j),\pi(j+1)})=1.
\end{align}

This term, once integrated, gives the analogous of eq.~(13.6.7) of ref.~\cite{Weinberg:1995mt} with \emph{quadratic propagators} instead of eikonal ones:
\begin{align}
\int \mathrm{d}\Pi(p)\ell^{(0)}(V,\{\},p)&=\tilde{\delta}\left(\left(p+ q_{1n}\right)^2-m^2\right)\prod_{j=1}^n\tilde{\delta}\left(\left(p+ q_{1j}\right)^2-m^2\right).
\end{align}
In the context of this section, we only integrate back from Schwinger parameters for pedagogical reasons. As explained before, integration should really follow the step at which real and virtual contributions are combined. The first order coefficient in the expansion reads 
\begin{align}
\ell^{(1)}(V,\{\},p)&=-\sum_{\pi\in\mathfrak{S}_n}\left[\prod_{j=0}^{n}\Theta\big(-\Delta \tau_{\pi(j),\pi(j+1)}\big)\right] \sum_{\substack{l,l'=1 \\ l<l'}}^{n} 2i\Delta \tau_{l,l'} (q_l\cdot q_{l'})\Theta(\pi(l)-\pi(l'))  \\
&= -\sum_{\substack{l,l'=1 \\ l<l'}}^{n} \Theta(\Delta \tau_{l,l'}) 2 i \Delta \tau_{l,l'} (q_l\cdot q_{l'}).
\end{align}

We see that every term in the sum corresponds to a defined ordering of $l$ and $l'$, which extends to a partial ordering of the set of $\{1,...,n\}$. 

\textbf{General formula} For $m<n$, we can write 
\begin{align}
\ell^{(m)}(V,\{\},p)&=\frac{(-i)^m}{m!}\sum_{\pi\in\mathfrak{S}_n}\left[\prod_{j=0}^{n}\Theta\big(-\Delta \tau_{\pi(j),\pi(j+1)}\big)\right] \left(\sum_{\substack{l,l'=1 \\ l>l'}}^{n} 2\Delta \tau_{l,l'} q_l\cdot q_{l'} \Theta(\pi(l)-\pi(l')) \right)^m \\
&=(-i)^m \sum_{k=0}^m\sum_{F\in \mathcal{P}_k(V)}\sum_{\vec{j}\in \{1,...,m\}^{|F|}}\delta_{j_1+...+j_k,m}\prod_{e=(v,v')\in F} \Theta(\Delta \tau_{e}) \frac{1}{j_e!}(2\Delta \tau_{e}  q_{v}\cdot q_{v'})^{j_e}. \label{eq:virtual_line}
\end{align}
Now each term of $\mathcal{L}^{(m)}(V,\{\},p)$ obtained by stripping away all the sums corresponds to a partial ordering of $\{1,...,n\}$ obtained by establishing the ordering of at most $m$ couplets. The sum over $k$ counts the amount of distinct pairs of vertices that are ordered. A summand with index $k$ will be feature a product of $k$ Heaviside theta functions imposing the particular ordering. When integrating back, this means that $n-k-1$ lines will be cut. 

\textbf{Integrating back, diagrammatics} More precisely, integrating $\ell^{(m)}(V,\{\},p)$ back from Schwinger parameter space will require to choose a different integration basis of proper time differences $\{\Delta\tau_e\}_{e\in F'}$ for each of the summands in a forest $F\in \mathcal{F}_k(V)$. Such integration basis are constrained by the request that $F\subseteq F'$, as we wish the integration variables to be aligned with the content of the Heaviside theta functions for ease of integration, which is then trivial. For $m<n$, this demand does not completely fix the change of variables, which will re-introduce a degree of arbitrariness in how the final result is written. In particular, by their very own nature, valid changes of variables can be mapped one-to-one to the potentially many spanning arborescences $F'\in\mathcal{F}_n(V)$ such that $F\subseteq F'$. Integrating these terms back undoes the canonicalisation that we achieved by going to proper time space. Consider the most general integral that can appear from the expansion of the quantum worldline, namely
\begin{align}
I=\int \left[\prod_{i=1}^{n-1}\mathrm{d}\Delta\tau_{i,i+1}e^{-i\Delta\tau_{i,i+1}D_i}\right] \prod_{e\in F}\Delta\tau_e^{n_e}\Theta(\Delta\tau_e^{y_e}).
\end{align}
with $y_e\in\{0,1\}$, $n_e\in\mathbb{Z}$, $e\in \mathcal{K}(V)$, where $F\subset\mathcal{K}(V)$, and $\mathcal{K}(V)$ is the set of edges of the complete graph on $V$. We choose a spanning arborescence $F'$ such that $F\subseteq F'$, and write
\begin{align}
I=\int \left[\prod_{e=(i,j)\in F'}\mathrm{d}\Delta\tau_{e}\text{exp}\left\{-i\Delta\tau_{e}\sum_{l=i}^{n-1}\sigma_{el} D_l\right\} (\chi^{F'\setminus F}_e+\chi^F_e\Delta\tau_e^{n_e}\Theta(\Delta\tau_e^{y_e}))\right],
\end{align}
where $\chi^F_e$ is the $e$-th component of the characteristic vector of the set $F$, and $\sigma_{el}$ are the components of the matrix expressing the change of variables from the basis $\{(1,2),...,(n-1,n)\}$ to the basis $F'$. The integral is now in factorised form. 

\begin{figure}
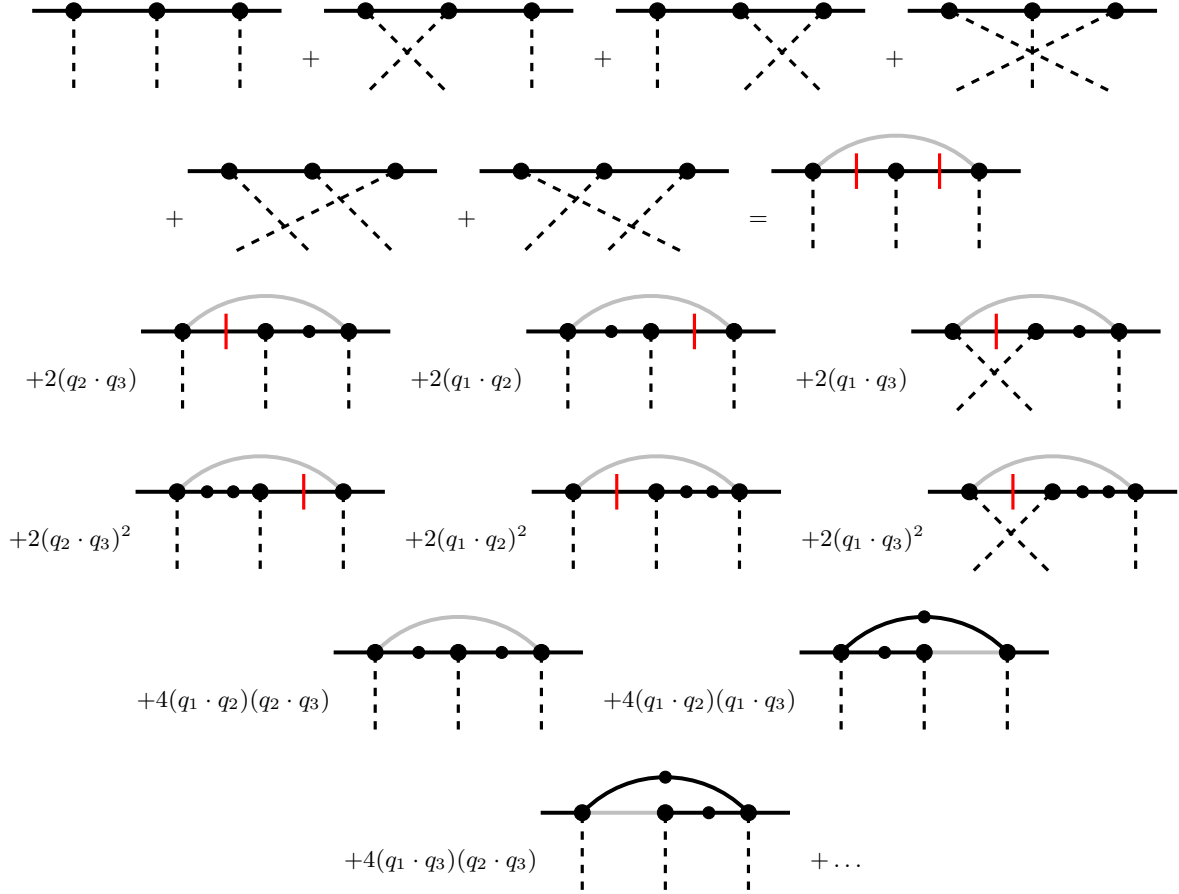

\centering
\raisebox{-0.4cm}{\input{3att_1}}$+$\raisebox{-0.4cm}{\input{3att_2}}$+$\raisebox{-0.4cm}{\input{3att_3}}$+$\raisebox{-0.4cm}{\input{3att_4}} \\[1em]
$+$\raisebox{-0.4cm}{\input{3att_5}}$+$\raisebox{-0.4cm}{\input{3att_6}}$=$\raisebox{-0.4cm}{\input{3att_wqft_0}}\\[1em]
$+2(q_2\cdot q_3)$\raisebox{-0.4cm}{\input{3att_wqft_11}}$+2(q_1\cdot q_2)$\raisebox{-0.4cm}{\input{3att_wqft_12}}$+2(q_1\cdot q_3)$\raisebox{-0.4cm}{\input{3att_wqft_13}}\\[1em]
$+2(q_2\cdot q_3)^2$\raisebox{-0.4cm}{\input{3att_wqft_22}}$+2(q_1\cdot q_2)^2$\raisebox{-0.4cm}{\input{3att_wqft_21}}$+2(q_1\cdot q_3)^2$\raisebox{-0.4cm}{\input{3att_wqft_23}}\\[1em]
$+4(q_1\cdot q_2)(q_2\cdot q_3)$\raisebox{-0.4cm}{\input{3att_wqft_24}}$+4(q_1\cdot q_2)(q_1\cdot q_3)$\raisebox{-0.4cm}{\input{3att_wqft_25}}\\[1em]
$+4(q_1\cdot q_3)(q_2\cdot q_3)$\raisebox{-0.4cm}{\input{3att_wqft_26}}$+\ldots$
\caption{The scalar quantum worldline with three attachments before and after the expansion of $\ell_3$. In accordance with the discussion of back-integration in sect.~\ref{sec:classical_expansion_quantum_worldline} we draw the complete graph on three vertices by using light-gray-coloured edges. In black, we overlay the edges corresponding to the chosen basis of proper times: for example, for the double cut contribution, we chose to integrate $\Delta\tau_{12}$ and $\Delta\tau_{23}$, leaving out $\Delta\tau_{13}$, which corresponds to the light-gray arc connecting the first and third vertex.}
\label{fig:ell3_exp}
\end{figure}

The implications of this freedom of choice are best explained diagrammatically. The forests live on the complete graph:
\begin{align}
V=\input{struct1} \qquad \Rightarrow \qquad \mathcal{K}(V)=\raisebox{-0.3cm}{\scalebox{1.1}{\input{theory_diags_tree_structures_complete_graph}}}.
\end{align}
We may remove the explicit orientation of the edges since it is anyway uniquely induced by the ordering of the vertices. The forest $F$ associated with the edges that receive a correction $\Delta\tau_e^{n_e}\Theta(-\Delta\tau_e^{y_e})$ can be drawn as solid black lines over the forest, say $F=\{e_1=(1,2),e_2=(1,4)\}$:
\begin{align}
\scalebox{1.1}{\input{complete_graph_forest_1}}.
\end{align}
Now we may decorate the black edges with $n_e$ (smaller) dots and a cut if $y_e=0$. For example, if $n_{e_1}=n_{e_2}=1$, $y_{e_1}=1$, $y_{e_2}=0$, we draw 
\begin{align}
\label{eq:freedom_of_choice}
\scalebox{1.1}{\input{complete_graph_decorated}}.
\end{align}
Since $F$ is not a spanning arborescence, we have the freedom to complete it to either $\{e_1,e_2,(3,4)\}$ or $\{e_1,e_2,(1,3)\}$. Choosing, say $\{e_1,e_2,(3,4)\}$, will yield the graph
\begin{align}
\scalebox{1.1}{\input{complete_graph_after_integration}}.
\end{align}
 The benefit of drawing diagrams as in eq.~\eqref{eq:freedom_of_choice} is that they only embed the constraint imposed by $F$ and they let the user choose the specific completion that they want, according to their needs (e.g. \emph{non-planarity issues}, see also ref.~\cite{Driesse:2024xad}). See fig.~\ref{fig:ell3_exp} for an example of the use of this diagrammatic convention.

\textbf{A remarkably compact expression} It also turns out that the expansion of eq.~\eqref{eq:virtual_line} can be rewritten in a much more compact way up to orders higher than $\lambda^{n}$. In particular, up to orders higher than $\lambda^{n}$, \emph{the corrections resum as follows}:
\begin{equation}
\boxed{
\ell(V,\{\},p)=\sum_{k=0}^n\sum_{F\in \mathcal{P}_k(V)} \prod_{e=(v,v')\in F}(e^{-2i\Delta\tau_e (q_v\cdot q_{v'})}-1) \Theta(\Delta \tau_{e})+\mathcal{O}(\lambda^{n}).}
\end{equation}
See sect.~\ref{sec:3att_scalar} for an explicit example of this resummation procedure.

\textbf{Expansion of the cut quantum worldline} We now discuss the cut quantum worldline. The result is the same apart from the product of Heaviside theta functions multiplying each term:
\begin{align}
\ell^{(m)}(S_1,S_2,p)&=i^m\sum_{k=1}^m\sum_{F\in \mathcal{P}_k(V)}\sum_{\vec{j}\in \{1,...,m\}^{|F|}}\delta_{j_1+...+j_k,m} \left[\prod_{e=(v,v')\in F} \frac{1}{j_e!}(-2 \Delta \tau_{e} q_{v}\cdot q_{v'})^{j_e}\right]\times\nonumber \\ &\times\left[ \prod_{e\in F\cap(S_1\times S_1)}\Theta(\Delta \tau_e)\right]\left[\prod_{e\in F\cap(S_2\times S_2)}\Theta(-\Delta \tau_e)\right].
\end{align}
For algorithmic purposes, it is more convenient to use the following form, where we include additional trivial factors of $1$:
\begin{align}
  \ell^{(m)}(S_1,S_2,p)&=i^m\sum_{k=1}^m\sum_{F\in \mathcal{P}_k(V)}\sum_{\vec{j}\in \{1,...,m\}^{|F|}}\delta_{j_1+...+j_k,m} \left[\prod_{e=(v,v')\in F} \frac{1}{j_e!}(-2 \Delta \tau_{e} q_{v}\cdot q_{v'})^{j_e}\right]\times \nonumber \\
  &\times\left[ \prod_{e\in F\cap(S_1\times S_1)}\Theta(-\Delta \tau_e)\right]\left[\prod_{e\in F\cap(S_2\times S_2)}\Theta(\Delta \tau_e)\right]\left[\prod_{\substack{e\in F\cap(S_1\times S_2)}}\big(\Theta(\Delta\tau_e)+\Theta(-\Delta\tau_e)\big)\right]. \label{eq:manifested_one}
\end{align}
Equivalently, one can write the cut quantum worldline as
\begin{equation}
\boxed{\ell(S_1,S_2,p)=\sum_{k=0}^n\sum_{F\in \mathcal{P}_k(\mathbb{Z}_n)} \prod_{e=(v,v')\in F}(e^{-2i\Delta\tau_e (q_v\cdot q_{v'})}-1) (\sigma_e^1\Theta(\Delta \tau_{e})+\sigma_e^2\Theta(-\Delta \tau_{e}))+\mathcal{O}(\lambda^{n}),}
\end{equation}
with 
\begin{equation}
\sigma_e^i =\begin{cases}
1 \quad \text{if }e\notin S_i\times S_i \\
0 \quad \text{otherwise}
\end{cases}.
\end{equation}
We discussed the case of a massive scalar whose interaction with the massless field is momentum independent. For a theory with non-trivial numerators associated to the massive lines, the expansion is more involved. The ordering structure of the expansion, encoded in $\mathcal{P}_k(V)$, remains the same, much like the Heaviside theta structure associated with a particular forest $F\in \mathcal{P}_k(V)$. However, the specific quantity that is associated to each expanded edge may vary. 

\textbf{Quantum worldline in scalar QED}  Let us take the example of scalar QED. We start by splitting the set of vertices into those of valence three, $V_3$, that couple to one photon only and those of valence four, $V_4$, that couple to two photons simultaneously. In order to write the numerator, let us number these vertices according to the following convention $V_3=\{1,...,l\}$, $V_4=\{l+1,...,n\}$, $V=V_3\cup V_4$ and consider a permutation $\pi$ of the full vertex set:
\begin{equation}
\bar{\mathcal{N}}_\pi^V=\prod_{v\in V_4} -2g^{\mu_v^1\mu_v^2}\prod_{v\in V_3} \left(q_{v}+2\left(p+\sum_{n=1}^{\pi(v)-1} q_{\pi(n)}\right)\right)^{\mu_{v}} ,
\end{equation}
where each three-point vertex $v$ is associated with one photon polarisation index $\mu_v$, each four-point vertex $v$ is associated with two photon polarisation indices $\mu_v^{1,2}$, and the momentum $q_v$ is the overall momentum entering the vertex $v$ coming from a single or double photon attachment. We canonicalise this numerator by choosing the same reference ordering we chose for the Schwinger parametrisation, namely $\pi=\mathds{1}$. This choice naturally leads to the rescaled numerator
\begin{equation}
\bar{\mathcal{N}}_\pi^V(\lambda)=\prod_{v\in V_4} -2g^{\mu_v^1\mu_v^2}\prod_{v\in V_3} \left(V_{v}+2\lambda \left(\sum_{n=1}^{\pi(v)-1} \bar{q}_{\pi(n)}-\sum_{j=1}^{v-1}\bar{q}_j\right)\right)^{\mu_{v}}, \quad V_{v}=
q_{v}+2\Bigg(p+\sum_{j=1}^{v-1}q_j\Bigg) .
\end{equation}
We have $\bar{\mathcal{N}}_\pi^V(0)=\bar{\mathcal{N}}_\mathds{1}^V$, left unexpanded. Expanding additional terms in the numerator in $\lambda$ under the soft limit rescaling, $q_k \to \lambda \bar q_k$, on top of the scalar products appearing in the exponent of the Schwinger parameterisation, we obtain a new expansion for the quantum worldline. In order to write it compactly, let us consider, for a given forest $F$, the forest $\tilde{F}$ obtained by flipping the orientation of all edges, and let $F_u$ be the underlying undirected graph of both $F$ and $\tilde{F}$. Let 
\begin{align}
\mathcal{T}(F)&=\{T\subset F\cup \tilde{F} \, | \, |\partial_T^-(v)|\le 1, \, \forall v\in V_3, \, |\partial_T^-(v)|=0, \, \forall v\in V_4\}, \label{eq:TF} \\
V(T)&=\{v\in V \, | \, |\partial_T^-(v)|= 0\},
\end{align}
where $\partial_T^-(v)$ is the set of edges that inject into $v$ in the graph $T$. An element in $\mathcal{T}(F)$ is associated to the part of the classical expansion due to the numerator. Then, we consider the object
\begin{equation}
\label{eq:WF}
W(F)=\{(F_1,F_2) \, | \, F_1\subseteq F, \ F_2\in \mathcal{T}(F), \ (F_1\cup F_2)_u=F_u\}.
\end{equation}
Observe that while $(F_1\cup F_2)$ may have multi-edges, its underlying undirected graph cannot, since multi-edges are mapped to the same edge. Then, we can write a sum over the partial orderings defined in eq.~\eqref{eq:partialOderingLEdges},
\begin{align}
\ell(V,\{\},p)=&\sum_{k=0}^n\sum_{F\in \mathcal{P}_k(V)}\left[\prod_{e=(v,v')\in F}\Theta(\Delta \tau_{e})\right] \sum_{(F_1,F_2)\in W(F)} \left[\prod_{e=(v,v')\in F_1}(e^{-2i\Delta\tau_e (q_v\cdot q_{v'})}-1) \right]\times \nonumber\\
\times&\left[\prod_{e=(v,v')\in F_2}-2 s_{e}^{F_2} q_v^{\mu_{v'}} \right]\prod_{\substack{v\in V(F_2)\cap V_3}} V_v^{\mu_v}\prod_{\substack{v\in V_4}} -2g^{\mu_v^1\mu_v^2}+\mathcal{O}(\lambda^{n}),\label{eq:simplified_QED}
\end{align}
where $s_e^F$ is equal to minus one if the orientation of $e$ has to be flipped in $F$ in order to match that of the corresponding edge in $F_1$, and is equal to one otherwise. This formula can be easily turned into a generation algorithm for quantum worldline expressions. We can see the first forest $F_1$ as describing the expansion terms of the exponentials, while the forest $F_2$ describes expansion terms of the numerator. If $(v_i,v_j)$ belongs to $F_2$, then $q_i^{\mu_j}$ appears in the summand corresponding to $(F_1,F_2)$. The constraint that $|\partial_T^-(v_i)|\le 1$ is related to the request that each Lorentz index should appear only once. Indeed, if we were to allow sets of the form $F_2=\{(1,2),(3,2)\}$, then they would correspond to a term of the form $q_1^{\mu_2}q_3^{\mu_2}$, which does not make sense.

For future purposes, let us also introduce the related quantity: given an arborescence $F$, rooted at any vertex (not necessarily at $v_1$ as established by the reference ordering)
\begin{align}
\ell^F(V,\{\},p)=&\left[\prod_{e=(v,v')\in F}\Theta(\Delta \tau_{e})\right]\sum_{(F_1,F_2)\in W(F)} \left[\prod_{e=(v,v')\in F_1}(e^{-2i\Delta\tau_e (q_v\cdot q_{v'})}-1) \right]\times \nonumber \\
&\hspace{4cm}\times\left[\prod_{e=(v,v')\in F_2}-2 s_{e}^{F_2} q_v^{\mu_{v'}} \right]\prod_{\substack{v\in V(F_2)\cap V_3}} V_v^{\mu_v}\prod_{\substack{v\in V_4}} -2g^{\mu_v^1\mu_v^2},\label{eq:flow-line}
\end{align}
where $s_e^{F_2}$ is defined as above with reference to the new canonical ordering given by $F$.

Let us generate the quantum worldline with two attachments of single photons using this machinery. We start by generating the elements of $\mathcal{P}_k$, which for the ground set $\{1,2\}$ give $\{\emptyset,\{(1,2)\}\}$. Define $F_1=\{(1,2)\}$, $F_2=\{(2,1)\}$ and $F_3=\{(1,2),(2,1)\}$. We have
\begin{align}
\mathcal{T}(\emptyset)=\{\emptyset\}, \quad \mathcal{T}(F_1)=\{\emptyset,F_1,F_2,F_3\},
\end{align}
and
\begin{align}
W(\emptyset)=\{(\emptyset,\emptyset)\}, \quad W(\{(1,2)\})=\{(F_1,\emptyset),(\emptyset,F_1),(\emptyset,F_2),(\emptyset,F_3),(F_1,F_2),(F_1,F_1),(F_1,F_3)\}.
\end{align}
Finally, we have
\begin{align}
V(\emptyset)=\{1,2\}, \quad  V(F_1)=\{1\}, \quad  V(F_2)=\{2\}, \quad V(F_3)=\{\emptyset\}.
\end{align}
To each element in $W(\emptyset)$ and $W(\{(1,2)\})$ we associate now a term in the quantum worldline with two attachments:
\begin{align}
(\emptyset,\emptyset)&\rightarrow V^{\mu_1}V^{\mu_2} \\
(F_1,\emptyset)&\rightarrow \Theta(\Delta\tau_{12})(e^{-2i(q_1\cdot q_2)\Delta\tau_{12}}-1)V_1^{\mu_1}V_2^{\mu_2}=-2i\Theta(\Delta\tau_{12})(q_1\cdot q_2)\Delta\tau_{12}V_1^{\mu_1}V_2^{\mu_2}+\mathcal{O}(\lambda^2), \\
(\emptyset,F_1)&\rightarrow\Theta(\Delta\tau_{12})(-2q_1^{\mu_2})V_1^{\mu_1}, \\
(\emptyset,F_2)&\rightarrow\Theta(\Delta\tau_{12})(2q_2^{\mu_1})V_2^{\mu_2}, \\
(\emptyset,F_3)&\rightarrow\Theta(\Delta\tau_{12})(-2q_1^{\mu_2})(2q_2^{\mu_1})=\mathcal{O}(\lambda^2), \\
(F_1,F_2)&\rightarrow -2i\Theta(\Delta\tau_{12})(q_1\cdot q_2)\Delta\tau_{12}(2q_2^{\mu_1})V_2^{\mu_2}+\mathcal{O}(\lambda^3)=\mathcal{O}(\lambda^2), \\
(F_1,F_1)&\rightarrow -2i\Theta(\Delta\tau_{12})(q_1\cdot q_2)\Delta\tau_{12}(-2q_1^{\mu_2})V_1^{\mu_1}+\mathcal{O}(\lambda^3)=\mathcal{O}(\lambda^2), \\
(F_1,F_3)&\rightarrow -2i\Theta(\Delta\tau_{12})(q_1\cdot q_2)\Delta\tau_{12}(-2q_1^{\mu_2})(2q_2^{\mu_1})+\mathcal{O}(\lambda^4)=\mathcal{O}(\lambda^2).
\end{align}
The last four terms give quantum contributions, and can be discarded. The first four terms map one-to-one with the terms of the quantum worldline of eq.~\eqref{eq:scalar_worldline_sum_expanded}.

In general, the compact form of the quantum worldline, obtained after expanding and applying the triangulation identities, will depend on the theory under study. However, we see no conceptual obstacle that would forbid the application of the techniques showed for the scalar theory and scalar QED to be used for other theories of interest such as gravity.

\subsection{Examples}

In this section, we explicitly work out the expansion of the quantum worldline for different examples, taking advantage of the systematised method presented above to go beyond the simplest cases considered in Sections \ref{sec:scalar_worldline} and \ref{sec:one_loop_impulse}. In particular, the case of a quantum worldline with three attachments will be treated in both scalar theory and scalar QED. We will also discuss the issue of self-attachments.

\subsubsection{Three attachments in scalar theory}
\label{sec:3att_scalar}
\textbf{Schwinger parametrisation} For three attachments, we need to derive the leading, sub-leading and sub-sub-leading terms in the expansion. This is entirely consistent with the fact that the three-attachment quantum worldline appears in two-loop calculations. The quantum worldline with three attachments reads
\begin{align}
&\La(V,\{\},p)=\sum_{\pi\in\mathfrak{S}_3}\frac{i}{[(q_{\pi(1)}+p)^2-m^2]}\frac{i}{[(q_{\pi(1)}+q_{\pi(2)}+p)^2-m^2]}\tilde{\delta}((q_{\pi(1)}+q_{\pi(2)}+q_{\pi(3)}+p)^2-m^2) \nonumber\\
&=\int \mathrm{d}\Delta\tau_{12}\mathrm{d}\Delta\tau_{23} \mathrm{d}\tau_3 e^{-i\Delta\tau_{12}((q_1+p)^2-m^2)-i\Delta\tau_{23}((q_1+q_2+p)^2-m^2)-i\tau_3 ((q_1+q_2+q_3+p)^2-m^2)} \Theta(-\Delta\tau_{12}) \Theta(-\Delta\tau_{23}) \nonumber\\
&+\int \mathrm{d}\Delta\tau_{13}\mathrm{d}\Delta\tau_{32} \mathrm{d}\tau_2 e^{-i\Delta\tau_{13}((q_1+p)^2-m^2)-i\Delta\tau_{32}((q_1+q_3+p)^2-m^2)-i\tau_2 ((q_1+q_2+q_3+p)^2-m^2)} \Theta(-\Delta\tau_{13}) \Theta(-\Delta\tau_{32}) \nonumber\\
&+\int \mathrm{d}\Delta\tau_{21}\mathrm{d}\Delta\tau_{13} \mathrm{d}\tau_3 e^{-i\Delta\tau_{21}((q_2+p)^2-m^2)-i\Delta\tau_{13}((q_1+q_2+p)^2-m^2)-i\tau_3 ((q_1+q_2+q_3+p)^2-m^2)} \Theta(-\Delta\tau_{21}) \Theta(-\Delta\tau_{13}) \nonumber\\
&+\int \mathrm{d}\Delta\tau_{23}\mathrm{d}\Delta\tau_{31} \mathrm{d}\tau_1 e^{-i\Delta\tau_{23}((q_2+p)^2-m^2)-i\Delta\tau_{31}((q_2+q_3+p)^2-m^2)-i\tau_1 ((q_1+q_2+q_3+p)^2-m^2)} \Theta(-\Delta\tau_{23}) \Theta(-\Delta\tau_{31}) \nonumber\\
&+\int \mathrm{d}\Delta\tau_{32}\mathrm{d}\Delta\tau_{21} \mathrm{d}\tau_1 e^{-i\Delta\tau_{32}((q_3+p)^2-m^2)-i\Delta\tau_{21}((q_2+q_3+p)^2-m^2)-i\tau_1 ((q_1+q_2+q_3+p)^2-m^2)} \Theta(-\Delta\tau_{32}) \Theta(-\Delta\tau_{21}) \nonumber\\
&+\int \mathrm{d}\Delta\tau_{31}\mathrm{d}\Delta\tau_{12} \mathrm{d}\tau_2 e^{-i\Delta\tau_{31}((q_3+p)^2-m^2)-i\Delta\tau_{12}((q_1+q_3+p)^2-m^2)-i\tau_2 ((q_1+q_2+q_3+p)^2-m^2)} \Theta(-\Delta\tau_{31}) \Theta(-\Delta\tau_{12}) .
\end{align}
The required change of variables for each of the last five integrals follows from the linear relations that express all proper times in terms of $\Delta\tau_{12}$, $\Delta\tau_{23}$ and $\tau_3$:
\begin{align}
&\Delta\tau_{21}=-\Delta\tau_{12}, \quad \Delta\tau_{32}=-\Delta\tau_{23}, \quad \Delta\tau_{13}=\Delta \tau_{12}+\Delta \tau_{23}, \quad \Delta\tau_{31}=-\Delta \tau_{12}-\Delta \tau_{23} \\
&\tau_1=\Delta \tau_{12}+\Delta \tau_{23}+\tau_3, \quad \tau_2=\Delta\tau_{23}+\tau_3.
\end{align}
Consequently, we can use $\Delta\tau_{12}$, $\Delta\tau_{23}$ and $\tau_3$ as a common integration basis for all six contributions: 
\begin{align}
\La&(V,\{\},p)=\int \mathrm{d}\Delta\tau_{12}\mathrm{d}\Delta\tau_{23} \mathrm{d}\tau_3 e^{-i\Delta\tau_{12}((q_1+p)^2-m^2)-i\Delta\tau_{23}((q_1+q_2+p)^2-m^2)-i\tau_3 ((q_1+q_2+q_3+p)^2-m^2)} \times \nonumber\\
&\times (-\Theta(-\Delta\tau_{12}) \Theta(-\Delta\tau_{23})+\Theta(-\Delta\tau_{21})\Theta(-\Delta\tau_{13})e^{-2i q_1\cdot q_2 \Delta \tau_{12}}+\Theta(-\Delta\tau_{13})\Theta(-\Delta\tau_{32})e^{-2i q_2\cdot q_3 \Delta \tau_{23}} \nonumber\\
&+\Theta(-\Delta\tau_{23})\Theta(-\Delta\tau_{31})e^{-2i q_1\cdot q_2 \Delta \tau_{12}-2i q_1\cdot q_3 \Delta \tau_{13}} 
+\Theta(-\Delta\tau_{31})\Theta(-\Delta\tau_{12})e^{-2i q_2\cdot q_3 \Delta \tau_{23}-2i q_1\cdot q_3 \Delta \tau_{13}} \nonumber\\
&+\Theta(-\Delta\tau_{32})\Theta(-\Delta\tau_{21})e^{-2i q_1\cdot q_2 \Delta \tau_{12}-2i q_1\cdot q_3 \Delta \tau_{13}-2i q_2\cdot q_3 \Delta \tau_{23}}).\label{eq:3att_init}
\end{align}
Expanding up to second order in the soft expansion (again, corresponding to an expansion in the rescaling $q_i=\lambda \bar{q}_i$, $\Delta\tau_{ij}=\Delta \bar{\tau}_{ij}/\lambda$) and combining theta function together according to the following identities
\begin{align}
\Theta(\tau_i-\tau_j)\Theta(\tau_j-\tau_k)+\Theta(\tau_i-\tau_k)\Theta(\tau_k-\tau_j)+\Theta(\tau_k-\tau_i)\Theta(\tau_i-\tau_j)&=\Theta(\tau_i-\tau_j), \label{eq:theta_3att_1} \\
\Theta(\tau_i-\tau_j)\Theta(\tau_j-\tau_k)+\Theta(\tau_i-\tau_k)\Theta(\tau_k-\tau_j)&=\Theta(\tau_i-\tau_k)\Theta(\tau_i-\tau_j)\label{eq:theta_3att_2},
\end{align}
we obtain 
\begin{align}
\La&(V,\{\},p)=\int \mathrm{d}\Pi_3(p) \big[ 1 \label{eq:3att_forest1} \\
 &+ 2(-i (q_1\cdot q_2) \Delta \tau_{12}+(-i  (q_1\cdot q_2) \Delta \tau_{12})^2) \Theta(\Delta \tau_{12}) \label{eq:3att_forest21}\\
 &+ 2(-i (q_2\cdot q_3) \Delta \tau_{23}+(-i  (q_2\cdot q_3) \Delta \tau_{23})^2) \Theta(\Delta \tau_{23}) \label{eq:3att_forest22}\\
&+ 2(-i (q_1\cdot q_3) \Delta \tau_{13}+(-i  (q_1\cdot q_3) \Delta \tau_{13})^2) \Theta(\Delta \tau_{13}) \label{eq:3att_forest23}\\ 
&+4 i^2 (q_1\cdot q_2)(q_2\cdot q_3)\Delta \tau_{12}\Delta \tau_{23}\Theta(\Delta \tau_{12})\Theta(\Delta \tau_{23}) \label{eq:3att_forest31}\\
&+4 i^2 (q_1\cdot q_3)(q_2\cdot q_3)\Delta \tau_{13}\Delta \tau_{23}\Theta(\Delta \tau_{13})\Theta(\Delta \tau_{23}) \label{eq:3att_forest32}\\
&+4 i^2 (q_1\cdot q_3)(q_1\cdot q_2)\Delta \tau_{13}\Delta \tau_{12}\Theta(\Delta \tau_{13})\Theta(\Delta \tau_{12}) \big]+\mathcal{O}(\lambda^0)\label{eq:3att_forest33}, 
\end{align}
with
\begin{align}
\mathrm{d}\Pi_3(p)=\mathrm{d}\Delta\tau_{12}\mathrm{d}\Delta\tau_{23} \mathrm{d}\tau_3 e^{-i\Delta\tau_{12}((q_1+p)^2-m^2)-i\Delta\tau_{23}((q_1+q_2+p)^2-m^2)-i\tau_3 ((q_1+q_2+q_3+p)^2-m^2)}
\end{align}
\textbf{Resummation} The worldline can also be presented in resummed form. We consider the three sets of forests, organised according to cardinality:
\begin{align}
\mathcal{F}_0(\{1,2,3\})=\{\}, \quad \mathcal{F}_1(\{1,2,3\})=\{(1,2),(1,3),(2,3)\}, \label{eq:3forest_1}\\
\mathcal{F}_2(\{1,2,3\})=\{\{(1,2),(2,3)\},\{(1,3),(2,3)\},\{(1,3),(1,2)\}\},\label{eq:3forest_2}
\end{align}
and write
\begin{align}
\La&(V,\{\},p)=\int \mathrm{d}\Pi_3(p) \Bigg[\underbrace{1}_{\text{eq.}\eqref{eq:3att_forest1}}+\underbrace{\sum_{F\in\mathcal{F}_1(\{1,2,3\})}\prod_{e\in F}\left(e^{-2iq_v\cdot q_{v'}\Delta\tau_e}-1\right)\Theta(\Delta\tau_e)}_{\text{eq.}\eqref{eq:3att_forest21},\eqref{eq:3att_forest22},\eqref{eq:3att_forest23}} \nonumber\\
&+\underbrace{\sum_{F\in\mathcal{F}_2(\{1,2,3\})}\prod_{e\in F}\left(e^{-2iq_v\cdot q_{v'}\Delta\tau_e}-1\right)\Theta(\Delta\tau_e)}_{\text{eq.}\eqref{eq:3att_forest31},\eqref{eq:3att_forest32},\eqref{eq:3att_forest33}}\Bigg]+\mathcal{O}(\lambda^0).
\end{align}
The resummed form can also be obtained directly from eq~\eqref{eq:3att_init} by writing each exponential as
\begin{equation}
e^{-2i(q_i\cdot q_j)\Delta\tau_{ij}}=\underbrace{(e^{-2i(q_i\cdot q_j)\Delta\tau_{ij}}-1)}_{=E_{ij}}+\underbrace{1}_{=I},
\end{equation}
expanding in $E_{ij}$ and $I$ and combining Heaviside theta functions according to eq.~\eqref{eq:theta_3att_1} and eq.~\eqref{eq:theta_3att_2}. Finally, the term
\begin{equation}
E_{12}E_{13}E_{23}\Theta(\Delta\tau_{12})\Theta(\Delta\tau_{13})\Theta(\Delta\tau_{23})
\end{equation}
is purely quantum and can be dropped.

\textbf{Back-integration} Again, let us integrate this expression back for the purposes of highlighting the ambiguity in the choice of integration variables. Integrating the canonical expression back will require choosing integration variables for each expansion term. This choice re-introduces a level of arbitrariness. The result is expressed as follows
\begin{equation}
\ell(V,\{\},p)=\sum_{\substack{\vec{n}\in\mathbb{N}^3 \\ n_1+n_2+n_3\le 2}}\sum_{\substack{\vec{y}\in\{0,1\}^3}} c_{\vec{n},\vec{y}} \, \Delta \tau_{12}^{n_1}\Delta \tau_{23}^{n_2}\Delta \tau_{13}^{n_3}\Theta(\Delta \tau_{12}^{y_1})\Theta(\Delta \tau_{23}^{y_2})\Theta(\Delta \tau_{13}^{y_3})+\mathcal{O}(\lambda^0)
\end{equation}
For the orders that we are interested in, namely those such that $n_1+n_2+n_3\le 2$, we can impose that the integration variables that we choose for the term $n_1,n_2,n_3$ contain at least the $\Delta\tau_{ij}$ terms whose corresponding power $n_i\neq 0$. For example, for the a term $c\Delta\tau_{12}$, we can choose either the integration variables $\{\Delta\tau_{12},\Delta\tau_{23}\}$ or $\{\Delta\tau_{12},\Delta\tau_{13}\}$, while for the term $c\Delta\tau_{12}\Delta\tau_{23}$ we can only choose $\{\Delta\tau_{12},\Delta\tau_{23}\}$. Using this constraint, we may integrate the quantum worldline back and obtain
\begin{align}
\La(V,\{\},p)&=\tilde{\delta}((q_1+q_2+q_3+p)^2-m^2)\Bigg[\tilde{\delta}((q_1+p)^2-m^2)\tilde{\delta}((q_1+q_2+p)^2-m^2) \nonumber\\
&+2 i (q_1\cdot q_2)\frac{\tilde{\delta}((q_1+q_2+p)^2-m^2)}{((q_1+p)^2-m^2-i\varepsilon)^2}
+2 i(q_2\cdot q_3)\frac{\tilde{\delta}((q_1+p)^2-m^2)}{((q_1+q_2+p)^2-m^2-i\varepsilon)^2} \nonumber\\
&+2 i(q_1\cdot q_3)\frac{\tilde{\delta}((q_1+p)^2-(q_1+q_2+p)^2)}{((q_1+q_2+p)^2-m^2-i\varepsilon)^2}
+2 i(q_1\cdot q_2)^2\frac{\tilde{\delta}((q_1+q_2+p)^2-m^2)}{((q_1+p)^2-m^2-i\varepsilon)^3} \nonumber\\
&+2i(q_2\cdot q_3)^2\frac{\tilde{\delta}((q_1+p)^2-m^2)}{((q_1+q_2+p)^2-m^2-i\varepsilon)^3}
+2 i(q_1\cdot q_3)^2\frac{\tilde{\delta}((q_1+p)^2-(q_1+q_2+p)^2)}{((q_1+q_2+p)^2-m^2-i\varepsilon)^3} \nonumber\\
&+4 (q_1\cdot q_2)(q_2\cdot q_3)\frac{i^2}{((q_1+p)^2-m^2-i\varepsilon)^2((q_1+q_2+p)^2-m^2-i\varepsilon)^2} \nonumber\\
&+4 (q_1\cdot q_2)(q_1\cdot q_3)\frac{i^2}{((q_1+p)^2-i\varepsilon)^2((q_1+q_2+p)^2-(q_1+p)^2-i\varepsilon)^2} \nonumber\\
&+4 (q_1\cdot q_3)(q_2\cdot q_3)\frac{i^2}{((q_1+p)^2-(q_1+q_2+p)^2-i\varepsilon)^2((q_1+q_2+p)^2-m^2-i\varepsilon)^2}\Bigg] +\mathcal{O}(\lambda^0).
\end{align}
The result is reported diagrammatically in fig.~\ref{fig:ell3_exp}. The non-unique choice of basis for back-integrating from Schwinger parameter space fixes a forest (drawn in solid black lines) over the complete graph, whose edges outside of the forest are drawn in gray. We now proceed to discussing an example with non-trivial numerators.

\subsubsection{Three attachments in scalar QED}
\label{sec:3worldline_QED}

Scalar QED features a vertex that couples two photons to the quantum worldline at one vertex. More specifically, there is a contribution to the quantum worldline in the form
\begin{equation}
\label{eq:sQED_3att_nc}
\mathcal{L}(V,\{\},p)=\raisebox{-0.5cm}{\input{1p2att1}}+\raisebox{-0.5cm}{\input{1p2att2}},
\end{equation}
corresponding to a vertex set of the type $V=\{v_{12},v_3\}$. Although the worldline has three insertions, from a conceptual level the difficulty of the expansion is related to the number of vertices, which is the same as the two-insertion case treated in sect.~\ref{sec:worldline_QED}. eq.~\eqref{eq:sQED_3att_nc} equivalently reads
\begin{align}
\mathcal{L}(V,\{\},p)&=i\left[\frac{g^{\mu_1\mu_2}(q_3+2(q_1+q_2+p))^{\mu_3}}{(q_1+q_2+p)^2-m^2+i\varepsilon}+\frac{(q_3+p)^{\mu_3}g^{\mu_1\mu_2}}{(q_3+p)^2-m^2+i\varepsilon}\right]\tilde{\delta}\left((q_1+q_2+q_3+p)^2-m^2\right).
\end{align}
The vertex numerator associated to the four-valent vertex is already in canonical form, as it is momentum independent. We only need to canonicalise the vertex numerator associated to the cubic vertex; choosing the reference ordering $\{\{1,2\},\{3\}\}$:
\begin{align}
\mathcal{L}(V,\{\},p)&=\int \mathrm{d}\Pi_2(p)\Big(g^{\mu_1\mu_2}V_3^{\mu_3}+(-2i\Delta\tau_{12}(q_3\cdot (q_1+q_2)) g^{\mu_1\mu_2}V_3^{\mu_3}-2g^{\mu_1\mu_2}(q_1+q_2)^{\mu_3})\Theta(\Delta\tau_{12})\Big)+\mathcal{O}(\lambda^0),
\end{align}
with $V_3=q_3+2(q_1+q_2+p)$. Once integrated back, this term has exactly the same structure as eq.~\eqref{eq:sQED_2att}, the only difference being in the numerator factor for the last summand in parenthesis.

Let us discuss the expansion of the quantum worldline with three ``distinct'' attachments in scalar QED. Excluding the diagrams of eq.~\eqref{eq:sQED_3att_nc}, we remain with those featuring cubic vertices only. In particular, for a vertex set $V=\{v_1,v_2,v_3\}$:
\begin{align}
\mathcal{L}&(V,\{\},p)=\sum_{\pi\in\mathfrak{S}_3} \frac{i^2 (q_{\pi(1)}+2p)^{\mu_{\pi(1)}}(q_{\pi(2)}+2(q_{\pi(1)}+p))^{\mu_{\pi(2)}}(q_{\pi(3)}+2(q_{\pi(1)}+q_{\pi(2)}+p))^{\mu_{\pi(3)}}}{[(q_{\pi(1)}+p)^2-m^2+i\varepsilon][(q_{\pi(1)}+q_{\pi(2)}+p)^2-m^2+i\varepsilon]}\tilde{\delta}((q_{123}+p)^2-m^2)
\end{align}
We now Schwinger parametrise propagators and Dirac delta function, re-arrange the expression as usual, performing the appropriate change of variables and extracting a common integration measure, corresponding to the reference ordering $(1,2,3)$:
\begin{align}
&\mathcal{L}(V,\{\},p)=\int \mathrm{d}\Delta\tau_{12}\mathrm{d}\Delta\tau_{23} \mathrm{d}\tau_3 e^{-i\Delta\tau_{12}((q_1+p)^2-m^2)-i\Delta\tau_{23}((q_1+q_2+p)^2-m^2)-i\tau_3 ((q_1+q_2+q_3+p)^2-m^2)} \times \nonumber\\
\times &(\Theta(-\Delta\tau_{12}) \Theta(-\Delta\tau_{23}) V_1^{\mu_1}V_2^{\mu_2}V_3^{\mu_3} \nonumber\\
+&\Theta(-\Delta\tau_{21})\Theta(-\Delta\tau_{13})e^{-2i q_1\cdot q_2 \Delta \tau_{12}} (V_1+2q_2)^{\mu_1}(V_2-2q_1)^{\mu_2}V_3^{\mu_3}\nonumber\\
+&\Theta(-\Delta\tau_{13})\Theta(-\Delta\tau_{32})e^{-2i q_2\cdot q_3 \Delta \tau_{23}} V_1^{\mu_1}(V_2+2q_3)^{\mu_2}(V_3-2q_2)^{\mu_3}\nonumber\\
+&\Theta(-\Delta\tau_{23})\Theta(-\Delta\tau_{31})e^{-2i q_1\cdot q_2 \Delta \tau_{12}-2i q_1\cdot q_3 \Delta \tau_{13}} (V_1+2q_2+2q_3)^{\mu_1}(V_2-2q_1)^{\mu_2}(V_3-2q_1)^{\mu_3}\nonumber\\
+&\Theta(-\Delta\tau_{31})\Theta(-\Delta\tau_{12})e^{-2i q_2\cdot q_3 \Delta \tau_{23}-2i q_1\cdot q_3 \Delta \tau_{13}} (V_1+2q_3)^{\mu_1}(V_2+2q_3)^{\mu_2}(V_3-2q_1-2q_2)^{\mu_3}\nonumber\\
+&\Theta(-\Delta\tau_{32})\Theta(-\Delta\tau_{21})e^{-2i q_1\cdot q_2 \Delta \tau_{12}-2i q_1\cdot q_3 \Delta \tau_{13}-2i q_2\cdot q_3 \Delta \tau_{23}} (V_1+2q_2+2q_3)^{\mu_1}(V_2-2q_1)^{\mu_2}(V_3-2q_1-2q_2)^{\mu_3})
\end{align}
with $V_1=q_1+2p$, $V_2=q_2+2(q_1+p)$, $V_3=q_3+2(q_1+q_2+p)$. We may now expand and combine Heaviside theta functions. We obtain: 
\begin{align}
\mathcal{L}&(V,\{\},p)=\int \mathrm{d}\Pi_3 (V_1^{\mu_1}V_2^{\mu_2}V_3^{\mu_3} \nonumber\\
 +& [2q_2^{\mu_1}V_2^{\mu_2}-2q_1^{\mu_2}V_1^{\mu_1}-4 q_2^{\mu_1}q_1^{\mu_2}+ 2(-i \ (q_1\cdot q_2) \Delta \tau_{12}+(-i  (q_1\cdot q_2) \Delta \tau_{12})^2) \nonumber\\
 &-4 i q_2^{\mu_1}V_2^{\mu_2} (q_1\cdot q_2)\Delta \tau_{12}+4 i q_1^{\mu_2}V_1^{\mu_1} (q_1\cdot q_2)\Delta \tau_{12}]V_3^{\mu_3} \Theta(\Delta \tau_{12}) \nonumber\\
 +& [2q_3^{\mu_2}V_3^{\mu_3}-2q_2^{\mu_3}V_2^{\mu_2}-4q_2^{\mu_3}q_3^{\mu_2}+2(-i  (q_2\cdot q_3) \Delta \tau_{23}+(-i  (q_2\cdot q_3) \Delta \tau_{23})^2) \nonumber\\
 &-4 i q_3^{\mu_2}V_3^{\mu_3} (q_2\cdot q_3)\Delta \tau_{23}+4 i q_2^{\mu_3}V_2^{\mu_2} (q_2\cdot q_3)\Delta \tau_{23}]V_1^{\mu_1} \Theta(\Delta \tau_{23}) \nonumber\\
+& [2q_3^{\mu_1}V_3^{\mu_3}-2q_1^{\mu_3}V_1^{\mu_1}-4q_1^{\mu_3}q_3^{\mu_1}+2(-i  (q_1\cdot q_3) \Delta \tau_{13}+(-i  (q_1\cdot q_3) \Delta \tau_{13})^2) \nonumber\\
 &-4 i q_3^{\mu_1}V_3^{\mu_3} (q_1\cdot q_3)\Delta \tau_{13}+4 i q_1^{\mu_3}V_1^{\mu_1} (q_1\cdot q_3)\Delta \tau_{13}]V_2^{\mu_2} \Theta(\Delta \tau_{13}) \nonumber\\ 
+&[4 (-i)^2  (q_1\cdot q_2)(q_2\cdot q_3)\Delta \tau_{12}\Delta \tau_{23}V_1^{\mu_1}V_2^{\mu_2}V_3^{\mu_3}+4 q_1^{\mu_2}q_2^{\mu_3} V_1^{\mu_1}-4 q_2^{\mu_1}q_2^{\mu_3}V_2^{\mu_2}+4 q_2^{\mu_1}q_3^{\mu_2}V_3^{\mu_3} \nonumber\\
&+4i q_1^{\mu_2}V_1^{\mu_1}V_3^{\mu_3}(q_2\cdot q_3)\Delta \tau_{23}-4i q_2^{\mu_1}V_2^{\mu_2}V_3^{\mu_3}(q_2\cdot q_3)\Delta \tau_{23} \nonumber\\
&+4i q_2^{\mu_3}V_1^{\mu_1}V_2^{\mu_2}(q_1\cdot q_2)\Delta \tau_{12}-4i q_3^{\mu_2}V_1^{\mu_1}V_3^{\mu_3}(q_1\cdot q_2)\Delta \tau_{12} ]\Theta(\Delta \tau_{12})\Theta(\Delta \tau_{23}) \nonumber\\
+&[4 (-i)^2  (q_1\cdot q_3)(q_2\cdot q_3)\Delta \tau_{13}\Delta \tau_{23}V_1^{\mu_1}V_2^{\mu_2}V_3^{\mu_3}-4 q_1^{\mu_3}q_3^{\mu_2} V_1^{\mu_1}+4 q_3^{\mu_1}q_3^{\mu_2}V_3^{\mu_3}-4 q_3^{\mu_1}q_2^{\mu_3}V_2^{\mu_2} \nonumber\\
&+4i q_1^{\mu_3}V_2^{\mu_2}V_1^{\mu_1}(q_2\cdot q_3)\Delta \tau_{23}-4i q_3^{\mu_1}V_2^{\mu_2}V_3^{\mu_3}(q_2\cdot q_3)\Delta \tau_{23} \nonumber\\
&+4i q_2^{\mu_3}V_1^{\mu_1}V_2^{\mu_2}(q_1\cdot q_3)\Delta \tau_{13}-4i q_3^{\mu_2}V_1^{\mu_1}V_3^{\mu_3}(q_1\cdot q_3)\Delta \tau_{13} ]\Theta(\Delta \tau_{13})\Theta(\Delta \tau_{23}) \nonumber\\
+&[4 (-i)^2  (q_1\cdot q_2)(q_1\cdot q_3)\Delta \tau_{12}\Delta \tau_{13}V_1^{\mu_1}V_2^{\mu_2}V_3^{\mu_3}+4 q_1^{\mu_2}q_1^{\mu_3} V_1^{\mu_1}-4 q_2^{\mu_1}q_1^{\mu_3}V_2^{\mu_2}-4 q_1^{\mu_2}q_3^{\mu_1}V_3^{\mu_3} \nonumber\\
&+4i q_1^{\mu_2}V_1^{\mu_1}V_3^{\mu_3}(q_1\cdot q_3)\Delta \tau_{13}-4i q_2^{\mu_1}V_2^{\mu_2}V_3^{\mu_3}(q_1\cdot q_3)\Delta \tau_{13} \nonumber\\
&+4i q_1^{\mu_3}V_1^{\mu_1}V_2^{\mu_2}(q_1\cdot q_2)\Delta \tau_{12}-4i q_3^{\mu_1}V_2^{\mu_2}V_3^{\mu_3}(q_1\cdot q_2)\Delta \tau_{12} ]\Theta(\Delta \tau_{12})\Theta(\Delta \tau_{13})) +\mathcal{O}(\lambda^0). \label{eq:expanded_QED_worldline}
\end{align}
This concludes the derivation of the quantum worldline with three attachments in scalar QED. These results are consistent with the compact form of the worldline that we have provided in eq.~\eqref{eq:simplified_QED}. Let us see that explicitly by generating the expression of eq.~\eqref{eq:flow-line} for the forest $F=\{(1,2),(1,3)\}$. In this case $\tilde{F}=\{(2,1),(3,1)\}$. Then
\begin{align}
\mathcal{T}(F)=\{&F_1=\{(1,2)\},F_2=\{(2,1)\},F_3=\{(1,3)\},F_4=\{(3,1)\}, \\
&F_5=\{(1,2),(2,1)\},F_6=\{(1,3),(3,1)\},F_7=\{(1,3),(1,2)\},F_8=\{(1,3),(2,1)\},F_9=\{(3,1),(1,2)\}, \\
&F_{10}=\{(1,2),(2,1),(1,3)\},F_{11}=\{(1,2),(1,3),(3,1)\}\}.
\end{align}
$\mathcal{T}(F)$ contains subsets of $F\cup\tilde{F}$ that satisfy the following property: if we take such subsets, for example $\{(1,2),(2,1),(1,3)\}$, and we extract the last entry of each of the edges in it, giving $(2,1,3)$, there should not be repetition. In terms of graph orientation, this property is rephrased into the requirement that each vertex must have at most one edge injecting into it. For any element of $T\in\mathcal{T}(F)$ the function $V(T)$ precisely extract those vertices that have no edges injecting into them. For example $V(F_2)=\{2,3\}$ and $V(F_6)=\{2\}$.

We are now able to construct $W(F)$, which is obtained by gluing together subforests of $F$ (in the first entry below) with elements in $\mathcal{T}(F)$ in such a way that covers the original forest $F$
\begin{align}
W(F)=\{(F_1,F_3),(F_1,F_4),(F_3,F_1),(F_3,F_2),(F_7,\emptyset),(\emptyset,F_7),(\emptyset,F_8),(\emptyset,F_9),...\}.
\end{align}
where $...$ indicates all tuples $(F',F'')$ such that $|F'|+|F''|>2$, which will give quantum terms upon application of eq.~\eqref{eq:expanded_QED_worldline}. Leaving these terms out, $W(F)$ contains eight elements, exactly the number of terms that multiply $\Theta(\Delta\tau_{12})\Theta(\Delta\tau_{13})$ in eq.~\eqref{eq:expanded_QED_worldline}. We can use eq.~\eqref{eq:flow-line} to translate one into the other. For example, for the element $(F_7,\emptyset)$, we have $V(\emptyset)=\{1,2,3\}$ and thus:
\begin{align}
(F_7,\emptyset)&\rightarrow \Theta(\Delta\tau_{12})\Theta(\Delta\tau_{13})V_1^{\mu_1}V_2^{\mu_2}V_3^{\mu_3}(e^{-2i(q_1\cdot q_2)\Delta\tau_{12}}-1)(e^{-2i(q_1\cdot q_3)\Delta\tau_{13}}-1)\\
&=(2i)^2(q_1\cdot q_2)\Delta\tau_{12}(q_1\cdot q_3)\Delta\tau_{13}V_1^{\mu_1}V_2^{\mu_2}V_3^{\mu_3}\Theta(\Delta\tau_{12})\Theta(\Delta\tau_{13})+\mathcal{O}(\lambda^3).
\end{align}
Instead, for the element $(F_3,F_1)$, we have
\begin{align}
(F_3,F_1)&\rightarrow \Theta(\Delta\tau_{12})\Theta(\Delta\tau_{13})V_1^{\mu_1}(-2q_{1}^{\mu_2})V_3^{\mu_3}(e^{-2i(q_1\cdot q_3)\Delta\tau_{13}}-1)\\
&=4i\Theta(\Delta\tau_{12})\Theta(\Delta\tau_{13})V_1^{\mu_1}q_{1}^{\mu_2}V_3^{\mu_3}(q_1\cdot q_3)\Delta\tau_{13}+\mathcal{O}(\lambda^3).
\end{align}
As a final example, let us consider the term $(\emptyset,F_8)$, which gives the term
\begin{align}
(\emptyset,F_8)&\rightarrow \Theta(\Delta\tau_{12})\Theta(\Delta\tau_{13})(-2q_1^{\mu_3})V_2^{\mu_2}(2q_{2}^{\mu_1}).
\end{align}

\subsubsection{Cut quantum worldline with three attachments}

\begin{figure}
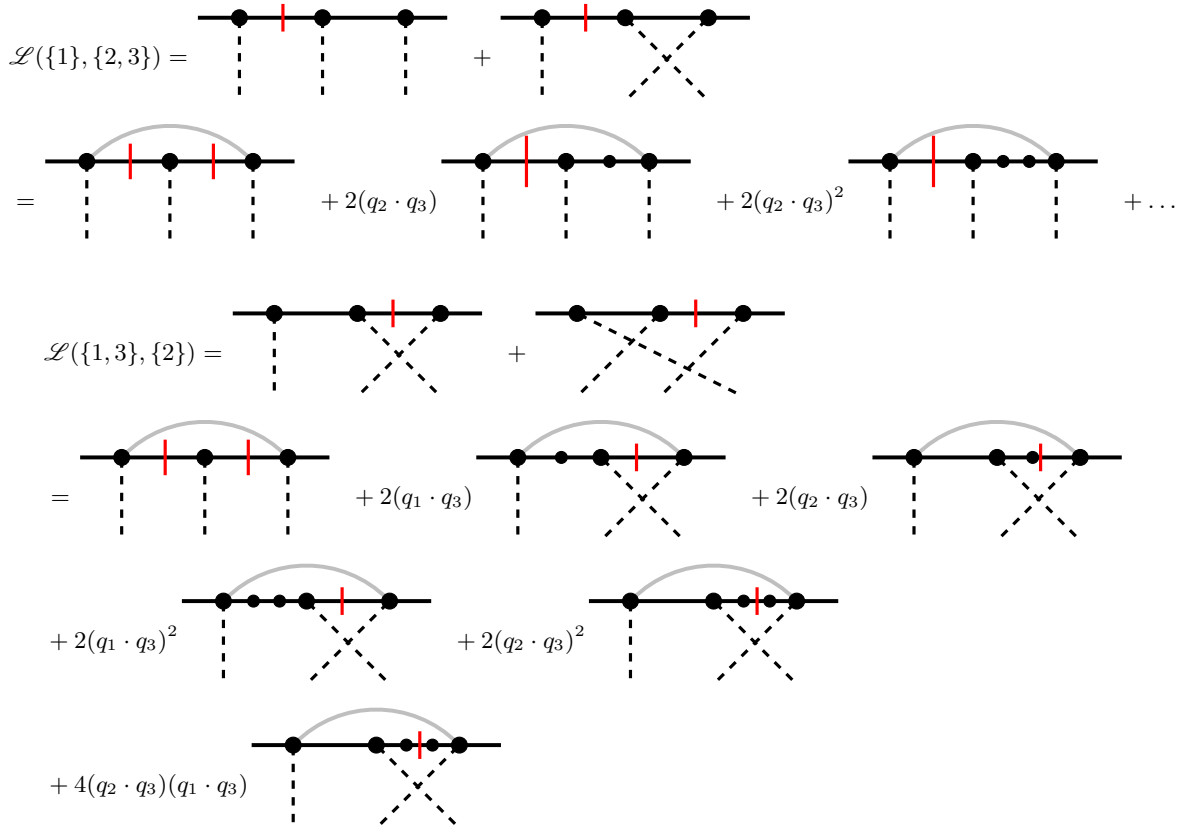

\centering
\begin{align*}&\mathcal{L}(\{1\},\{2,3\})=\raisebox{-0.5cm}{\input{3att_cut1}}+\raisebox{-0.5cm}{\input{3att_cut2}} \\
&=\raisebox{-0.5cm}{\input{3att_wqft_0}}+2(q_2\cdot q_3)\raisebox{-0.5cm}{\input{3att_cut_1_wqft1}}+2(q_2\cdot q_3)^2\raisebox{-0.5cm}{\input{3att_cut1_wqft2}}+\ldots\end{align*} \\
\begin{align*}&\mathcal{L}(\{1,3\},\{2\})=\raisebox{-0.5cm}{\input{3att_cut3}}+\raisebox{-0.5cm}{\input{3att_cut4}} \\
&=\raisebox{-0.5cm}{\input{3att_wqft_0}}+2(q_1\cdot q_3)\raisebox{-0.5cm}{\input{3att_cut3_wqft11}}+2(q_2\cdot q_3)\raisebox{-0.5cm}{\input{3att_cut3_wqft12}} \\
&+2(q_1\cdot q_3)^2\raisebox{-0.5cm}{\input{3att_cut3_wqft21}}+2(q_2\cdot q_3)^2\raisebox{-0.5cm}{\input{3att_cut3_wqft22}} \\
&+4(q_2\cdot q_3)(q_1\cdot q_3)\raisebox{-0.5cm}{\input{3att_cut3_wqft22}}
\end{align*}
\caption{Cut scalar quantum worldlines corresponding to two specific choices of $S_1$ and $S_2$.}
\label{fig:cut_quantum}
\end{figure}

We saw that the definition of the cut quantum worldline depends on a partition of the set of vertices. The partitions of $\{1,2,3\}$ are
\begin{align}
&S_1^1=\{1\}, \ S_2^1=\{2,3\}, \quad S_1^2=\{2\}, \ S_2^2=\{1,3\}, \quad S_1^3=\{3\}, \ S_2^3=\{1,2\}, \nonumber\\
&S_1^4=\{1,2\}, \ S_2^4=\{3\}, \quad S_1^5=\{1,3\}, \ S_2^5=\{2\}, \quad S_1^6=\{2,3\}, \ S_2^6=\{1\}.
\end{align}
Let us show how to write the cut quantum worldline for the partitions $S_1^1,S_2^1$ and $S_1^5,S_2^5$. We have: 
\begin{align}
\mathcal{L}(S_1^1,S_2^1,p)&=\int \mathrm{d}\Pi_3(p) \left(\Theta(-\Delta\tau_{23})+e^{-2i(q_2\cdot q_3)\Delta\tau_{23}}\Theta(\Delta\tau_{23})\right) \label{eq:cut_qw_1}\\
&=\int \mathrm{d}\Pi_3(p) \left(1-2i(q_2\cdot q_3)\Delta\tau_{23}(1-i(q_2\cdot q_3)\Delta\tau_{23})\Theta(\Delta\tau_{23})\right)+\mathcal{O}(\lambda^0). 
\end{align}
In resummed form:
\begin{align}
\mathcal{L}(S_1^1,S_2^1,p)
&=\int \mathrm{d}\Pi_3(p) \left(1+(e^{-2i(q_2\cdot q_3)\Delta\tau_{23}}-1)\Theta(\Delta\tau_{23})\right)+\mathcal{O}(\lambda^0). 
\end{align}
$\mathcal{L}(S_1^1,S_2^1,p)$ is represented in fig.~\ref{fig:cut_quantum} after integration of the Schwinger parameters. It gives
\begin{align}
\mathcal{L}(S_1^1,S_2^1,p)&=\tilde{\delta}((q_1+p)^2-m^2)\tilde{\delta}((q_1+q_2+p)^2-m^2)\tilde{\delta}((q_1+q_2+q_3+p)^2-m^2) \nonumber\\
&+2i(q_2\cdot q_3)\frac{\tilde{\delta}((q_1+p)^2-m^2)\tilde{\delta}((q_1+q_2+q_3+p)^2-m^2)}{((q_1+q_2+p)^2-m^2-i\varepsilon)^2} \nonumber\\
&+2i(q_2\cdot q_3)^2\frac{\tilde{\delta}((q_1+p)^2-m^2)\delta((q_1+q_2+q_3+p)^2-m^2)}{((q_1+q_2+p)^2-m^2-i\varepsilon)^3}+\ldots,
\end{align}
In fig.~\ref{fig:cut_quantum} we also represent $\mathcal{L}(S_1^5,S_2^5,p)$. In this case, after introducing Schwinger parameters, the quantum worldline reads: 
\begin{align}
\mathcal{L}&(S_1^5,S_2^5,p)=\int \mathrm{d}\Pi_3(p) \left(e^{-2i(q_2\cdot q_3)\Delta\tau_{23}}\Theta(-\Delta\tau_{13})+e^{-2i(q_2\cdot q_3)\Delta\tau_{23}-2i(q_1\cdot q_3)\Delta\tau_{13}}\Theta(\Delta\tau_{13})\right) \nonumber\\
&=\int \mathrm{d}\Pi_3(p) \big(1-2i(q_2\cdot q_3)\Delta\tau_{23}(1-i(q_2\cdot q_3)\Delta\tau_{23})-2i(q_1\cdot q_3)\Delta\tau_{13}(1-i(q_1\cdot q_3)\Delta\tau_{13})\Theta(\Delta\tau_{13}) \nonumber\\
&+(-2i)^2(q_1\cdot q_3)\Delta\tau_{13}(q_2\cdot q_3)\Delta\tau_{23}\Theta(\Delta\tau_{13})\big). 
\end{align}
After the integration of Schwinger parameters, we obtain
\begin{align}
\mathcal{L}(S_1^5,S_2^5,p)&=\tilde{\delta}((q_1+p)^2-m^2)\tilde{\delta}((q_1+q_2+p)^2-m^2)\tilde{\delta}((q_1+q_2+q_3+p)^2-m^2) \nonumber\\
&+2(q_2\cdot q_3)\tilde{\delta}((q_1+p)^2-m^2)\tilde{\delta}'((q_1+q_2+p)^2-m^2)\tilde{\delta}((q_1+q_2+q_3+p)^2-m^2)\nonumber\\
&+2i(q_1\cdot q_3)\frac{\tilde{\delta}((q_1+q_2+p)^2-(q_1+p)^2)\tilde{\delta}((q_1+q_2+q_3+p)^2-m^2)}{((q_1+p)^2-m^2-i\varepsilon)^2}\nonumber\\
&+2(q_2\cdot q_3)^2\tilde{\delta}((q_1+p)^2-m^2)\tilde{\delta}''((q_1+q_2+p)^2-m^2)\tilde{\delta}((q_1+q_2+q_3+p)^2-m^2)\nonumber\\
&+2i(q_1\cdot q_3)^2\frac{\tilde{\delta}((q_1+q_2+p)^2-(q_1+p)^2)\tilde{\delta}((q_1+q_2+q_3+p)^2-m^2)}{((q_1+p)^2-m^2-i\varepsilon)^3}\nonumber\\
&+2i(q_1\cdot q_3)(q_2\cdot q_3)\frac{\tilde{\delta}'((q_1+q_2+p)^2-(q_1+p)^2)\tilde{\delta}((q_1+q_2+q_3+p)^2-m^2)}{((q_1+p)^2-m^2-i\varepsilon)^2}+\ldots.
\end{align}
The difference in amount of expansion terms in $\mathcal{L}(S_1^1,S_2^1,p)$ and $\mathcal{L}(S_1^5,S_2^5,p)$ is explained by how ``close'' the partitions entering the cut worldline are to the reference ordering $(1,2,3)$. 

Introducing numerators does not change significantly the derivation shown here for the scalar case which, analogously to the case of the non-cut quantum worldline, has to be canonicalised. 

The cut quantum worldline features terms including derivatives of Dirac delta functions. For all the examples treated in this paper, these terms will however drop in the expression of the classical impulse. Furthermore, cancellations between virtual and real contributions are manifested more easily when the expression is written in the form eq.~\eqref{eq:manifested_one}. For $\mathcal{L}(S_1^2,S_2^2,p)$, for example, this prescription amounts to the following representation for its expansion:
\begin{align}
\mathcal{L}&(S_1^2,S_2^2,p)=\int \mathrm{d}\Pi_3(p) \left(e^{-2i(q_1\cdot q_2)\Delta\tau_{12}}\big(\Theta(\Delta\tau_{13})+e^{-2i(q_1\cdot q_3)\Delta\tau_{13}}\Theta(-\Delta\tau_{13})\big)\right) \\
&=\int \mathrm{d}\Pi_3(p) \big(1-2i(q_1\cdot q_2)\Delta\tau_{12}(1-i(q_1\cdot q_2)\Delta\tau_{12})\underbrace{(\Theta(-\Delta\tau_{12})+\Theta(\Delta\tau_{12}))}_{=1} \nonumber\\
&-2i(q_1\cdot q_3)\Delta\tau_{13}(1-i(q_1\cdot q_3)\Delta\tau_{13}))\Theta(-\Delta\tau_{13}) \nonumber\\
&+(-2i)^2(q_1\cdot q_3)\Delta\tau_{13}(q_1\cdot q_2)\Delta\tau_{12} \Theta(-\Delta\tau_{13})\underbrace{(\Theta(\Delta\tau_{12})+\Theta(-\Delta\tau_{12}))}_{=1}\big)+\mathcal{O}(\lambda^2),
\end{align}
where we simply used the partition of the identity $1=\Theta(\Delta\tau_{12})+\Theta(-\Delta\tau_{12})$. In other words, we rewrote the expansion of the cut quantum worldline in the following form
\begin{align}
\mathcal{L}&(S_1,S_2,p)=\sum_{n_1,n_2,n_3=1}^n\sum_{\sigma_1,\sigma_2,\sigma_3\in\{\pm 1\}} c_{n_1,n_2,n_3}^{\sigma_1,\sigma_2,\sigma_3}\Delta \tau_{12}^{n_1}\Delta \tau_{23}^{n_2}\Delta \tau_{13}^{n_3} \Theta(\sigma_1 \Delta \tau_{12}) \Theta(\sigma_1 \Delta \tau_{23})\Theta(\sigma_1 \Delta \tau_{13}).
\end{align}
This organisation of the result in a polynomial of $\Delta\tau_{ij}$ and $\Theta(\pm \Delta\tau_{ij})$ will lead to an algebraic realisation of super-leading cancellation, as well as of propagators raised to third power or higher and derivatives of Dirac delta functions. It will also help in manifesting the worldline form.

\section{Expansion of the KMOC integrand}
\label{sec:method}

The soft expansion for the integrand of an equivalence class of diagrams, $F_{[G]}$ defined in eq.~\eqref{eq:integrandEquivClass}, can be organised in steps: first, one expresses $F_{[G]}$ in terms of the quantum worldline in Schwinger space, and expands the worldlines $\ell(V,\{\},p_i)$ and $\ell(S_1,S_2,p_i)$ appearing in the virtual contribution and real contribution respectively. This is sufficient to realising the cancellation of super-leading contributions, up to some subtleties in the treatment of massless propagators in the radiation region. Then, one integrates such expansion using the integration measure $\mathrm{d}\Pi_n(p_1)\mathrm{d}\Pi_m(-p_2)$, and finally the result is re-expanded in the classical limit one last time. 

At each step, we will obtain expressions that are equivalent up to higher order in the classical expansion. We divide the expansion in two distinct stages as it will be beneficial to use properties of the intermediate expansion terms to derive general considerations on the shape of the final expansion. In particular, the two-step procedure is meant to eliminate certain ambiguities that would arise by expanding diagrams such as mushroom diagrams and to reduce considerably the amount of intermediate terms.

\subsection{Zero-measured cuts}
\label{sec:zero_measured}

In order to both get cancellations and re-write the classical result in terms of retarded propagators, one needs to frequently invoke the vanishing of certain collections of cuts. In the context of this paper, a zero-measured cut is any collection of signed cuts (namely selecting a specific energy sign for the cut particle) that upon deletion leaves out a graph that has at most one incoming particle with positive energy or at most one outgoing particle with positive energy. In other words, they correspond to vacuum decays or decays of the particles in the theory (and their time-reversed counter-parts) which are either kinematically degenerate or kinematically forbidden. We already presented an instance of such cuts in the context of the example of sect~\ref{sec:one_loop_impulse}. In that instance, the kinematically forbidden process is identified by the $v_1$ vertex, and we can schematically write:
\begin{equation}
\raisebox{-0.6cm}{\begin{tikzpicture}

    \node[inner sep=0pt] (1) {};
    \node[inner sep=0pt] (F1) [right = 1cm of 1] {};
    \node[inner sep=0pt] (F2) [below = 0.2cm of F1] {$v_1$};
    \node[inner sep=0pt] (2) [above right = 0.65cm and 0.65cm of F1] {};
    \node[inner sep=0pt] (3) [below right = 0.65cm and 0.65cm of F1] {};
    
        \begin{feynman}

     \draw[-, thick, black, line width=0.4mm] (1) to (F1);

     \draw[-, thick, black, line width=0.4mm] (F1) to (2);
     \draw[photon, thick, black, line width=0.4mm] (F1) to (3);
    	
    	\end{feynman}

	\path[draw=black, fill=black] (F1) circle[radius=0.1];

\end{tikzpicture}}
\subset \raisebox{-1.3cm}{\input{zero_measured_box}} 
\end{equation}

\begin{figure}
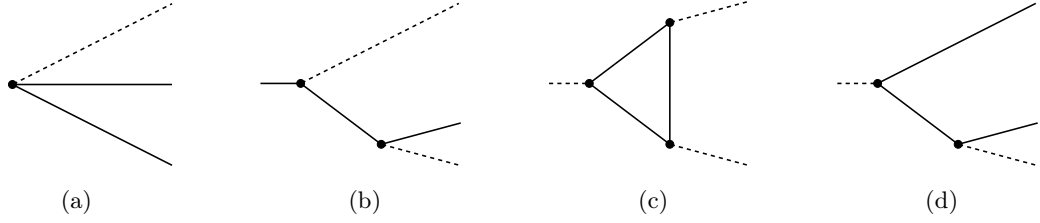

\centering
\begin{subfigure}{0.2\textwidth}
\scalebox{0.5}{\input{vacuum_decay}}
\caption{}
\label{fig:zero_a}
\end{subfigure}
\begin{subfigure}{0.2\textwidth}
\scalebox{0.5}{\input{zero_measured_decay}}
\caption{}
\label{fig:zero_b}
\end{subfigure}
\begin{subfigure}{0.2\textwidth}
\scalebox{0.5}{\input{forbidden_decay}}
\caption{}
\label{fig:zero_c}
\end{subfigure}
\begin{subfigure}{0.2\textwidth}
\scalebox{0.5}{\input{forbidden_decay_2}}
\caption{}
\label{fig:zero_d}
\end{subfigure}
\caption{Processes that are kinematically forbidden or degenerate. Cuts of Feynman diagrams that isolate such processes are set to zero. Dashed lines correspond to massless particles and solid lines correspond to massive particles with equal mass.}
\label{fig:zero_measured_ex}
\end{figure}

Fig.~\ref{fig:zero_measured_ex} provides some more examples of processes that are kinematically forbidden: if the cut graph contains such processes, then we say it is a zero-measured cut. Indeed, in all cases, the phase-space for these processes is either empty or degenerate.

These cuts are ill-defined since the surfaces identifies by these cuts, when they are non-empty, are in fact pinched. This implies that the coordinate transformation rule for the Dirac delta function does not work anymore, since the derivative of the argument of the Dirac delta function is either zero or discontinuous. Take, as an example, the following integration of a test function $g$:
\begin{equation}
I=\int \mathrm{d}^4 k \mathrm{d}^4p \, \tilde{\delta}^-(k^2)\tilde{\delta}^+((k+p)^2-m^2) \tilde{\delta}^+(p^2-m^2) g(k^0,\vec{k};p^0,\vec{p})
\end{equation}
Solving two out of the three Dirac delta functions

\begin{align}
I&=\pi^3\int \mathrm{d}^3 \vec{k} \mathrm{d}^3\vec{p}\frac{\delta\left(\sqrt{|\vec{k}+\vec{p}|^2+m^2}+|\vec{k}|-\sqrt{|\vec{p}|^2+m^2}\right)}{|\vec{k}|\sqrt{|\vec{k}+\vec{p}|^2+m^2}\sqrt{|\vec{p}|^2+m^2}}g(-|\vec{k}|,\vec{k};|\vec{p}|,\vec{p}) \\
&=\pi^3\int \mathrm{d}^3 \vec{k} \mathrm{d}^3\vec{p} \frac{1}{1+\frac{|\vec{k}|+\hat{k}\cdot \vec{p}}{\sqrt{|\vec{k}+\vec{p}|^2+m^2}}}\frac{\delta\big(|\vec{k}|\big)}{|\vec{k}|\sqrt{|\vec{k}+\vec{p}|^2+m^2}\sqrt{|\vec{p}|^2+m^2}}g(-|\vec{k}|,\vec{k};|\vec{p}|,\vec{p}).
\end{align}
$\hat{k}=\vec{k}/|\vec{k}|$ does not have a well-defined limit when $|\vec{k}|$ vanishes. In other words, the surface defined by setting the argument of the Dirac delta function to zero does not allow for a well-defined normal vector field. The situation is rescued if the test function $g$ or the integral measure provide enough suppression so that the integrand vanishes, but in general this may not be the case. For certain other observables, the ill behaviour of zero-measured cuts is lifted (for example, due to the offshellness granted to one of the cut particles) and give rise to subtle effects associated with long-range Coulomb like potentials (see e.g.\ recent work in a KMOC context in refs.~\cite{Caron-Huot:2023vxl, Elkhidir:2024izo}). A systematic incorporation of such effects into our framework is beyond the scope of the current work.

More in general, any time the argument of a Dirac delta function has ill-defined or vanishing gradient at the points where the argument itself is zero we can conclude that the Dirac delta function itself is ill-defined (as the change-of-variable rule is not allowed). The surfaces identified by setting such arguments to zero are said to be ``pinched''~\cite{Landau:1959fi,Coleman1965SingularitiesIT}.

The reason these ill-defined cuts arise in the first place can be traced back to the (mis-)handling of $i\varepsilon$ prescription that is inherent to working in the Minkowski representation. In three-dimensional representations, obtained by integrating out from the covariant representation the energy component of loop integrals~\cite{Catani_2008,Bierenbaum_2010,Runkel_2019,Capatti:2019ypt,Capatti:2022mly,Aguilera_Verdugo_2021,Sborlini_2021}, Cutkosky cuts appear as denominators of the integrand, and their $i\varepsilon$-prescriptions can be handled separately: we can set the $i\varepsilon$-prescription to zero in those denominators that identify pinched singularities and leave it in the rest of the denominators. It follows that zero-measured cuts simply would never appear if we were to take discontinuities or compute imaginary parts of the three-dimensional representation. As a simple example, consider the following identities
\begin{align}
\frac{1}{(|\vec{k}|+i\varepsilon)(|\vec{k}|-Q^0+i\varepsilon)}-\frac{1}{(|\vec{k}|-i\varepsilon)(|\vec{k}|-Q^0-i\varepsilon)}&=2\pi i \left[\frac{\delta(|\vec{k}|)}{|\vec{k}|-Q^0}+\frac{\delta(|\vec{k}|-Q^0)}{|\vec{k}|} \right]\\
\frac{1}{|\vec{k}|(|\vec{k}|-Q^0+i\varepsilon)}-\frac{1}{|\vec{k}|(|\vec{k}|-Q^0-i\varepsilon)}&=2\pi i\frac{\delta(|\vec{k}|-Q^0)}{|\vec{k}|}.
\end{align}
In the first case, a zero-measured cut $\delta(|\vec{k}|)$ appears, while in the second, where we dropped the $i\varepsilon$-prescription in virtue of $|\vec{k}|=0$ being a pinched singularity, it doesn't. In general, carefully dropping the $i\varepsilon$-prescription from those denominators of the three-dimensional representation that identify pinched singularities will allow to avoid the appearance of zero-measured cuts. However, since for the scope of this paper we would like to work in four dimensions, we will simply set these contributions to zero by hand whenever they appear. This approach is somewhat different to that proposed in the appendix of ref.~\cite{Bourjaily:2020wvq} in that the zero-measured cuts, in our prescription, can never contribute, consistently with the idea that pinched surfaces do not allow for any finite $i\varepsilon$-prescription. 

\subsection{Classical expansion of the quantum integrand}

We start by writing virtual contributions in terms of the quantum worldline. After symmetrisation, in accordance with the picture of eq.~\eqref{eq:virtual_worldline}, we can rewrite $F_{[G]}^V$ as: 
\begin{align}
F_{[G]}^V&=(-i)^{N_G}\mathcal{L}(V_{\text{up}},\{\},p_1)\mathcal{L}(V_{\text{bot}},\{\},-p_2)\bar{\mathcal{N}}_{E_{\text{m}}(V)}\prod_{e\in E_\text{m}(V)}\frac{1}{q_e^2+i\varepsilon} \\
&=(-i)^{N_G}\int \mathrm{d}\Pi_{|V_{\text{up}}|}(p_1)\mathrm{d}\Pi_{|V_{\text{bot}}|}(-p_2) \frac{\bar{\mathcal{N}}_{E_{\text{m}}(V)}\ell(V_{\text{up}},\{\},p_1)\ell(V_{\text{bot}},\{\},-p_2)}{\prod_{e\in E_\text{m}(V)}[q_e^2+i\varepsilon]}.
\end{align}
where $\bar{\mathcal{N}}_{E_{\text{m}}(V)}$ is the numerator associated to the massless blobs (which have no quantum loops), stripped of $-i$ factors for the vertices and $i$ factors for the propagators. This is compensated by the $(-i)^{N_G}$ overall factor. In order to write the cut contributions in terms of the quantum worldline, let us introduce an equivalence relation between cut diagrams. Starting from the equivalence class $[G]$, we let $[G]_c$ be the set of all graphs in $[G]$ that have a certain cut $c\in \mathcal{C}_{[G]}$ (a consistent labelling based on the map $\phi_{\pi,\pi'}$ is assumed in order to identify the same cut $c$ across different graphs in the same equivalence class $[G]$), where
\begin{equation}
\mathcal{C}_{[G]}=\bigcup_{G'\in [G]}\mathcal{C}_{G'}.
\end{equation}
Then, for any $G\in [G]_c$, let $G_L^c=(V_L^c,E_L^c)$ and $G_R^c=(V_R^c,E_R^c)$ be the two graphs separated by the cut $c$, laying on the left and right of the cut, respectively. Let $N_{G_R^c}$ and $N_{G_L^c}$ be the number of connected massless components that are strictly on the right and left of the Cutkosky cut respectively (strictly as in: they are not cut through by the Cutkosky cut itself). Then, we can write the real integrand $F_{[G]}^R$ in Schwinger-parameter space as 
\begin{align}
\label{eq:FG_schwinger_parameters}
&F_{[G]}^R=(-i)^{N_{G}}\int \mathrm{d}\Pi_{|V_{\text{up}}|}(p_1)\mathrm{d}\Pi_{|V_{\text{bot}}|}(-p_2) \bar{\mathcal{N}}_{E_{\text{m}}(V)}\sum_{c\in\mathcal{C}_{[G]}} (-1)^{N_{G_R^c}} \ell(V_L^c\cap V_{\text{up}},V_R^c\cap V_{\text{up}},p_1)\times \nonumber\\
&\times \ell(V_L^c\cap V_{\text{bot}},V_R^c\cap V_{\text{bot}},-p_2)\prod_{e\in E_\text{m}(V)\cap E_L^c}\frac{1}{q_e^2+i\varepsilon} \prod_{e\in E_\text{m}(V)\cap E_R^c}\left[\frac{1}{q_e^2+i\varepsilon}+i\tilde{\delta}^+(q_e)+i\tilde{\delta}^-(q_e)\right] \prod_{e\in c} \tilde{\delta}^{+}(\sigma_e(V_L^c)q_e),
\end{align}
with
\begin{equation}
\sigma_e(V')=\begin{cases}
1 \quad &\text{if }e=(v,v'), v\in V', \, v'\in V\setminus V' \\
-1 \quad &\text{if }e=(v,v'), v\in V\setminus V', \, v'\in V', 
\end{cases} 
\end{equation}
Observe that the integration measure for real and virtual contributions is exactly the same. In fact, we could promote the virtual contribution to a real one by simply adding to $\mathcal{C}_{[G]}$ the cut crossing the two final-state massive particles. In particular, we may write
\begin{align}
F_{[G]}=F_{[G]}^V+F_{[G]}^R=\int \mathrm{d}\Pi_{|V_{\text{up}}|}(p_1)\mathrm{d}\Pi_{|V_{\text{bot}}|}(-p_2) f_{[G]}.
\end{align}
We may proceed to discussing the classical expansion of the quantum worldline and the associated classical expansion of the quantum observable.
The classical expansion of $\ell$ induces an expansion of the integrand $f_{[G]}$. Defining $D_{\tau}^G=D_{\text{soft}}^G-4(L+1)+|V_{\text{up}}|+|V_{\text{bot}}|$, which is obtained by subtracting the scaling of the loop integration measure and proper-time integration measure to the overall contribution of the graph to the observable, we have
\begin{align}
f_{[G]}&=\sum_{m=0}^{N_G-1}\lambda^{D_{\tau}^G+m} f_{[G]}^{(m)}+\mathcal{O}(\lambda^{D_{\tau}^G+N_G}).
\end{align}

\begin{figure}
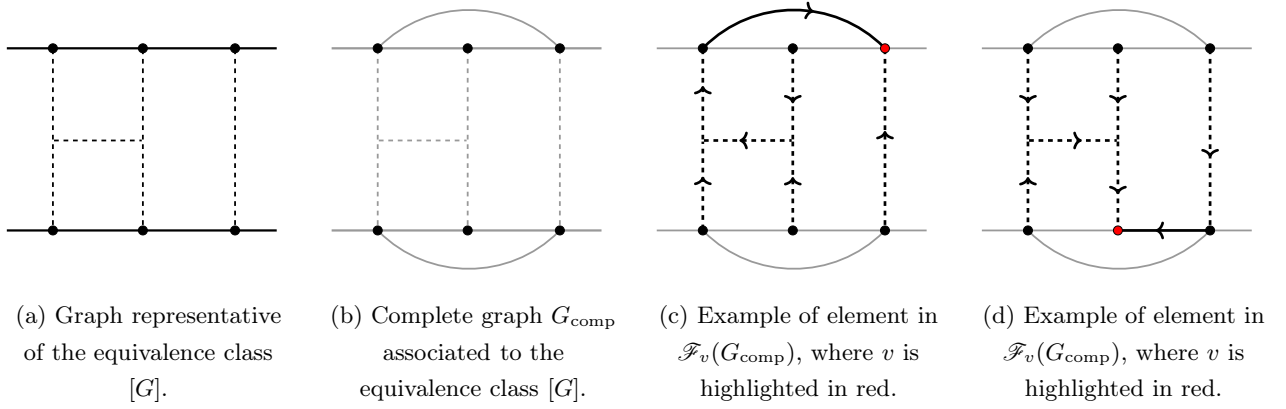

\centering
\begin{subfigure}{3.8cm}
\centering
\scalebox{0.75}{\input{example_graph}}
\caption{Graph representative of the equivalence class $[G]$.}
\end{subfigure}%
\hspace{0.5cm}\begin{subfigure}{3.8cm}
\centering
\scalebox{0.75}{\input{method_examples_diags_complete_graph}}
\caption{Complete graph $G_{\text{comp}}$ associated to the equivalence class $[G]$.}
\end{subfigure}%
\hspace{0.5cm}\begin{subfigure}{3.8cm}
\centering
\scalebox{0.75}{\input{example_tree_1}}\caption{Example of element in $\mathcal{F}_v(G_{\text{comp}})$, where $v$ is highlighted in red.}
\end{subfigure}%
\hspace{0.5cm}\begin{subfigure}{3.8cm}
\centering
\scalebox{0.75}{\input{example_tree_2}}
\caption{Example of element in $\mathcal{F}_v(G_{\text{comp}})$, where $v$ is highlighted in red.}
\end{subfigure}
\caption{An equivalence class $[G]$ (as usual, equivalence under permutation of vertices on the upper or lower massive line) defines a unique graph $G_{\text{comp}}$. A choice of a vertex $v$ further defines a set of spanning arborescences $\mathcal{F}_v(G_{\text{comp}})$ rooted at $v$ whose constraint is that they contain all massless (dashed) edges. In the language of WQFT, at the red vertex one has the observable insertion, which we will later simply denote by a half-edge, following the convention from the existing literature.}
\end{figure}

\textbf{Cancellation of super-leading contributions} The statement of integrand-level cancellation of super-leading contributions is summarised by the request that $f_{[G]}^{(m)}=0$ for $0\le m<N_G-1$. In fact, as mentioned, if we think of $f_{[G]}^{(m)}$ as a multivariate polynomial in the expansion parameters of the worldline and the Feynman propagators and Dirac delta functions associated with massless propagators
\begin{equation}
f_{[G]}^{(m)}=P^{(m)}(\{\Delta\tau_{e}\}_{e\in\mathcal{K}(V_{\text{up}})},\{\Delta\tau_{e}'\}_{e\in\mathcal{K}(V_{\text{bot}})},\{G_F(q_e)\}_{e\in E_{\text{m}(V)}},\{\tilde{\delta}^{\pm}(q_e)\}_{e\in E_{\text{m}(V)}}, \{q_e\}_{e\in E_{\text{m}}(V)},p_1,p_2),
\end{equation}
with coefficients given by sums and differences of products of Heaviside theta functions, then requiring cancellation of super-leading contributions implies that either \textbf{(a)} the coefficients of this polynomial are all zero or \textbf{(b)} the non-vanishing terms integrate back to a zero-measured cut. These two conditions can be checked algorithmically. Let us stress that if we use the form of the quantum worldline given in eq.~\eqref{eq:virtual_line} and eq.~\eqref{eq:manifested_one}, the vanishing of the coefficients (implication (a)) is algebraic, \emph{meaning that no additional theta identity needs to be used}. For what concerns implication (b), it should also be checked without need of simplifying the result using further identities. In other words, once eq.~\eqref{eq:virtual_line} and eq.~\eqref{eq:manifested_one} have been used to write down $P^{(m)}_{[G]}$, one only needs to collect the non-zero terms contributing to it and check that they integrate back to zero-measured cuts.

\textbf{Worldline form} The second fact that we would like to check is the extent to which our integrand reproduces that which can be derived using the worldline formalisms. Take $f_{[G]}^{(N_G-1)}$ and, again, eliminate all terms that integrate to zero measured cuts. Taking $V_{\text{up}}=\{v_1,...,v_n\}$ and $V_{\text{bot}}=\{w_1,...,w_m\}$, let $G_{\text{comp}}=(V,E_{\text{comp}})$ be the graph with the same set of vertices $V$ as that of the graph $G$ and the edge set
\begin{equation}
E_{\text{comp}}=\mathcal{K}(V_{\text{up}})\cup\mathcal{K}(V_{\text{bot}}) \cup E_{\text{m}}(V),
\end{equation}
recalling that $\mathcal{K}(V')$ is the set of edges of the complete graph on the ground set $V'$. For any fixed vertex on the upper line, $v\in V_{\text{up}}$, we define the following set
\begin{equation}
\mathcal{F}_v(G_{\text{comp}})=\{T \, | \, T \text{ is a spanning arborescence in }G_{\text{comp}}\text{ rooted at the sink }v, \ T\cap E_{\text{m}}(V)=E_{\text{m}}(V)\}.
\end{equation}
Then, we hypothesise that, up to pieces that integrate to purely imaginary quantities, one has
\begin{align}
\label{eq:worldline_hyp}
f_{[G]}^{(N_G-1)}&=(-i)^{N_G}\sum_{v\in V_{\text{up}}}\left(\sum_{e\in E_{\text{m}}(v)}q_e^\mu\right)\sum_{T\in\mathcal{F}_v(G_{\text{comp}})} \ell^{T\cap \mathcal{K}(V_{\text{up}})}(V_{\text{up}},\{\},p) \, \ell^{T\cap \mathcal{K}(V_{\text{bot}})}(V_{\text{bot}},\{\},p)\prod_{e\in E_{\text{m}}(V)} G_R( s_e^T q_e).
\end{align} 
where $\ell^F$ was defined in eq.~\eqref{eq:flow-line}. In words, the polynomial can be written as a sum of terms, each corresponding to a specific vertex on the upper line, $v$, where the impulse is measured. For each $v$ the integrand is a sum of spanning arborescences on the graph $G_{\text{comp}}$ with an orientation defined by the choice of the vertex $v$. In particular, for fixed spanning arborescence $T$ and vertex $v'$, there is a unique directed path from $v'$ to $v$.

\section{Applications and examples}
\label{sec:applications_examples}

\subsection{Classical limit for diagrams with one connected massless component}

\begin{figure}
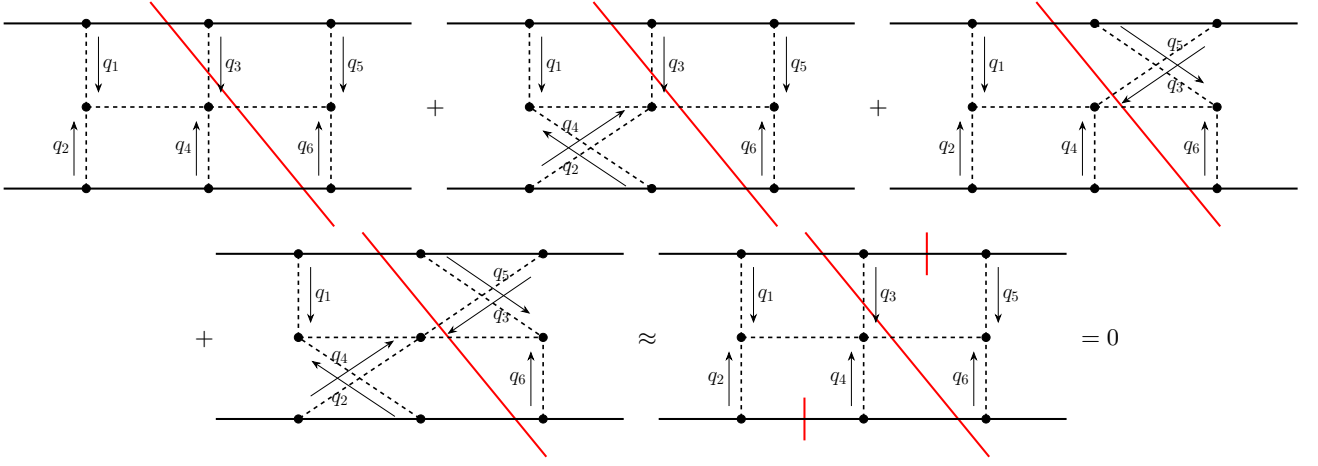

\centering
\raisebox{-1.5cm}{\input{cut_zero_m_pre_1}}$+$\raisebox{-1.5cm}{\input{cut_zero_m_pre_2}}$+$\raisebox{-1.5cm}{\input{cut_zero_m_pre_3}} \\
$+$\raisebox{-1.5cm}{\input{cut_zero_m_pre_4}}$\approx$\raisebox{-1.5cm}{\input{cut_zero_m_post}}$=0$
\caption{An example of sum of cut diagrams that gives a classical contribution containing a zero-measured cut, which directly vanishes.}
\label{fig:zero_measured}
\end{figure}

For diagrams with one connected massless component, the classical limit coincides with the leading term in the soft expansion. As we saw, for theories with a non-renormalisable self-interaction between massless particles only, power-counting alone allows to conclude that the connected massless component must be a tree diagram. 

\textbf{Example} Let us start with an example. Consider the sum of graphs in the equivalence class with the following representative:
\begin{equation}
\input{one_connected}
\end{equation}
with all the diagrams obtained by permuting vertices on its upper or lower massive line. The original diagram is then identified with the couplet of identity permutations $\pi=\mathds{1}$, $\pi'=\mathds{1}$. A loop momentum basis of this graph is given, for example, by $q_1,...,q_5$, and the classical limit corresponds to taking their soft limit. The leading term in the classical expansion of $F_G$, excluding all the zero measured cuts (namely all those in which any of the massless propagators attaching to the massive lines are cut, since all massive propagators are cut), reads
\begin{align}
F_G=\frac{1}{3!}\mathcal{N}&\tilde{\delta}(2q_1\cdot p_1)\tilde{\delta}(2q_3\cdot p_1)\tilde{\delta}(2q_5\cdot p_1)\tilde{\delta}(2q_2\cdot p_2)\tilde{\delta}(2q_4\cdot p_2)\tilde{\delta}(2q_6\cdot p_2)\times \nonumber \\ \times&\Bigg[q_1^\mu G_F(q_1)G_F(q_2)\tilde{\delta}^+(q_{12})G_F(q_3)^\star G_F(q_{1234})^\star G_F(q_4)^\star G_F(q_5)^\star G_F(q_6)^\star \nonumber\\ 
+&(q_1+q_3)^\mu G_F(q_1)G_F(q_2)G_F(q_{12})G_F(q_3) G_F(q_4)\tilde{\delta}^+(q_{1234})  G_F(q_5)^\star G_F(q_6)^\star\nonumber\\
+&i q_3^\mu G_F(q_3)G_F(q_4)\tilde{\delta}^-(q_{12})\tilde{\delta}^+(q_{1234})G_F(q_1)^\star G_F(q_2)^\star G_F(q_5)^\star G_F(q_6)^\star \nonumber\\
+&q_5^\mu G_F(q_5)G_F(q_6)\tilde{\delta}^-(q_{1234})G_F(q_{12})^\star G_F(q_1)^\star G_F(q_2)^\star G_F(q_3)^\star G_F(q_4)^\star\nonumber\\
-i&(q_1+q_5)^\mu G_F(q_5)G_F(q_6)G_F(q_1) G_F(q_2)\tilde{\delta}^-(q_{1234})\tilde{\delta}^+(q_{12})  G_F(q_3)^\star G_F(q_4)^\star
\nonumber\\
+&(q_3+q_5)^\mu G_F(q_5)G_F(q_6)G_F(q_3) G_F(q_4)G_F(q_{1234})\tilde{\delta}^-(q_{12})  G_F(q_1)^\star G_F(q_2)^\star \nonumber\\
-i&(q_1+q_3+q_5)^\mu G_F(q_1)G_F(q_2)G_F(q_{12})G_F(q_3) G_F(q_{1234}) G_F(q_4) G_F(q_5) G_F(q_6)
\Bigg],\label{eq:one_con}
\end{align} 
with $G_F(q_e)=(q_e^2+i\varepsilon)^{-1}$. $\mathcal{N}$ is the numerator factor including all of the contribution from the massless as well as massive propagators. It factorises all contributions in the leading classical expansion since any canonicalisation step requires $\mathcal{N}_\pi=\mathcal{N}_\mathds{1}$ at leading order in the classical expansion.

For completeness, let us draw the diagrammatic contributions to $F_G$ that give terms proportional to $q_1^\mu$, 
\begin{equation}
\raisebox{-1.1cm}{\scalebox{1.1}{\input{fishnet_q1_1}}}+\raisebox{-1.1cm}{\scalebox{1.1}{\input{fishnet_q1_2}}}+\raisebox{-1.1cm}{\scalebox{1.1}{\input{fishnet_q1_3}}}+\raisebox{-1.1cm}{\scalebox{1.1}{\input{fishnet_q1_4}}},
\end{equation}
where we highlighted in blue the KMOC cut and in red the additional cuts arising from taking the classical limit. We see that the particle with momentum $q_1$ always lies on the left-hand side of the KMOC cut, as expected. 

Now we may turn conjugate propagators associated with the lines $q_1,...,q_6$ into ordinary Feynman propagators, using the identity $G_F(q)^\star=G_F(q)+i\tilde{\delta}(q^2)$. The cut contributions arising from this procedure vanish, because together with the cut massive lines they identity zero-measured cuts (see fig.~\ref{fig:zero_measured}). In other words, the $i\varepsilon$-prescription can be fully dropped for these propagators, and we can write, identifying terms that factorise the same $q_i^\mu$ observable insertion 
\begin{align}
F_G=&\frac{1}{3!}\mathcal{N}\tilde{\delta}(2q_1\cdot p_1)\tilde{\delta}(2q_3\cdot p_1)\tilde{\delta}(2q_5\cdot p_1)\tilde{\delta}(2q_2\cdot p_2)\tilde{\delta}(2q_4\cdot p_2)\tilde{\delta}(2q_6\cdot p_2)\times \nonumber\\
\times&(G_F(q_1)G_F(q_2)G_F(q_3)G_F(q_4)G_F(q_5)G_F(q_6))|_{\varepsilon=0}\times  \nonumber\\ \times&\Big[q_1^\mu(-iG_F(q_{12})G_F(q_{1234})+G_F(q_{12})\tilde{\delta}^+(q_{1234})+G_F(q_{1234})^\star\tilde{\delta}^+(q_{12})-i\tilde{\delta}^-(q_{1234})\tilde{\delta}^+(q_{12}))  \nonumber\\
&+q_3^\mu(G_F(q_{12})\tilde{\delta}^+(q_{1234})-iG_F(q_{12})G_F(q_{1234})+\tilde{\delta}^-(q_{12})G_F(q_{1234})+i\tilde{\delta}^-(q_{1234})\tilde{\delta}^+(q_{12}) ) \nonumber\\
&+q_5^\mu(G_F(q_{12})^\star\tilde{\delta}^-(q_{1234})-iG_F(q_{12})G_F(q_{1234})+\tilde{\delta}^-(q_{12})G_F(q_{1234})-i\tilde{\delta}^-(q_{1234})\tilde{\delta}^-(q_{12}) )
\Big].
\end{align}

Using the following distributional identities
\begin{align}
i\tilde{\delta}^\pm(q_e)=G_R(\mp q_e)-G_F(q_e), \quad -i\tilde{\delta}^\pm(q_e)=G_R(\pm q_e)-G_F(q_e)^\star,
\end{align}
we obtain 
\begin{align}
F_G=&-\frac{i}{3!}\mathcal{N}\tilde{\delta}(2q_1\cdot p_1)\tilde{\delta}(2q_3\cdot p_1)\tilde{\delta}(2q_5\cdot p_1)\tilde{\delta}(2q_2\cdot p_2)\tilde{\delta}(2q_4\cdot p_2)\tilde{\delta}(2q_6\cdot p_2) \nonumber\\
\times &(G_F(q_1)G_F(q_2)G_F(q_3)G_F(q_4)G_F(q_5)G_F(q_6))|_{\varepsilon=0}\times  \nonumber\\ \times&\Big[q_1^\mu G_R(-q_{12})G_R(-q_{1234})+q_3^\mu G_R(q_{12})G_R(-q_{1234})+q_5^\mu G_R(q_{12})G_R(q_{1234})
\Big].
\end{align}
Upon relabelling
\begin{align}
F_G=&-\frac{i}{2}q_5^\mu \mathcal{N}\tilde{\delta}(2q_1\cdot p_1)\tilde{\delta}(2q_3\cdot p_1)\tilde{\delta}(2q_5\cdot p_1)\tilde{\delta}(2q_2\cdot p_2)\tilde{\delta}(2q_4\cdot p_2)\tilde{\delta}(2q_6\cdot p_2) \times\nonumber
 \\ &\times(G_F(q_1)G_F(q_2)G_F(q_3)G_F(q_4)G_F(q_5)G_F(q_6))|_{\varepsilon=0}  G_R(q_{12})G_R(q_{1234})
\Big],
\end{align}
which is fully consistent with the result directly obtained by applying WQFT Feynman rules, and in particular reproduces the WQFT diagram \\
\begin{center}
\input{one_connected_wqft}.
\end{center}

\textbf{General ladder} At this point, a general strategy for reproducing the correct retarded propagator prescription for any diagram is already clear, as no particular complication is expected in extending the result for the example described above: for each impulse numerator $q_i^\mu$, collect all the diagrams contributing to it, and use distributional identities to express the result in terms of retarded propagators only.  However, for completeness, and to highlight how precisely achieve this last step, let us sketch the procedure at all orders for a specific class of diagrams, namely those of the following type:
\begin{center}
\input{one_connected_general}
\end{center}
Stripping away from the graph all the edges that correspond to massive propagators as well as those that attach to the massive lines, we remain with an oriented tree, and in this particular case an open directed path $T$ (whose orientation should be identified with momentum flow). In the example above, it would correspond to the two massless edges that receive a retarded propagator prescription. Let us consider its vertices $V_T=\{v_1,...,v_m\}$ and edges $E_T=\{e_1,...,e_{m-1}\}$. The KMOC integrand, once we collect all contributions that factorise the same impulse numerator, equals
\begin{align}
&F_G=(-i)^{|V_T|} \prod_{v\in V_{\text{up}}}\left[\tilde{\delta}\Bigg(\sum_{e\in E_{\text{m}}(v)}2q_e\cdot p_1\Bigg)\prod_{e\in E_{\text{m}}(v)} G_F(q_e)|_{\varepsilon=0}\right]\prod_{v'\in V_{\text{bot}}}\left[\tilde{\delta}\Bigg(\sum_{e\in E_{\text{m}}(v')}2q_e\cdot p_2\Bigg)\prod_{e\in E_{\text{m}}(v')} G_F(q_e)|_{\varepsilon=0}\right]\times \nonumber\\
&\times\sum_{v\in V_{\text{up}}} \Bigg(\sum_{e\in E_{\text{m}}(v)}q_e^{\mu}\Bigg)\Bigg[\sum_{\substack{P_1\cup P_2=V_T \\ v\in E_{\text{m}}(P_1)}}
(-1)^{|P_2|}\prod_{e\in T \cap P_1^2 } [iG_F(q_e)] \prod_{e\in T \cap (P_1\times P_2) } \tilde{\delta}^{+}(\sigma_e(P_1) q_e) \prod_{e\in T \cap P_2^2 } [-iG_F(q_e)^\star]\Bigg],
\end{align}
We are now ready to manifest the retarded propagator prescription. Let us define
\begin{equation}
\Delta_v=\sum_{\substack{P_1\cup P_2=V_T \\ v\in P_1}}
(-1)^{|P_2|}\prod_{e\in T \cap P_1^2 } [iG_F(q_e)] \prod_{e\in T \cap (P_1\times P_2) } \tilde{\delta}^{+}(\sigma_e(P_1) q_e) \prod_{e\in T \cap P_2^2 } [-iG_F(q_e)^\star]
\end{equation}
We will use the following identities
\begin{equation}
\label{eq:distr_identities}
i\tilde{\delta}^+(\sigma q_e)=G_R(-\sigma q_e)-G_F(q_e), \quad G_F(q_e)^\star=-G_F(q_e)+G_R(q_e)+G_R(-q_e).
\end{equation}
After substitution, we expect the result to only be expressed in terms of Feynman and retarded propagators. In fact, define $A(v,T)$, for any $v\in V_T$, to be the only arborescence with same underlying undirected tree graph as $T$, rooted at the only vertex $v'\in V_T$ such that $v'\in E_{\text{m}}(v)$. Recall the definition of the vector $s^A$ such that $s_e^A=1$ if $e$ has the same orientation in $A(v,T)$ and $T$, and $-1$ otherwise. Then:
\begin{align}
\Delta_v= -i\prod_{e\in A(v,T)} G_R(s_e^A q_e).
\end{align}
Let us check this identity for a tree $\{(1,2),(2,3),(2,4)\}$, whose edges correspond to the momenta $q_1$, $q_2$ and $q_3$. Fix a vertex, say $1$. The partitions relevant to the sum are 
\begin{align}
&\{\{1\},\{2,3,4\}\}, \ \{\{1,2\},\{3,4\}\}, \ \{\{1,3\},\{2,4\}\}, \nonumber \\  &\{\{1,4\},\{3,2\}\}, \ \{\{1,2,3\},\{4\}\}, \ \{\{1,2,4\},\{3\}\}, \ \{\{1,4,3\},\{2\}\}, 
\end{align}
Correspondingly, $\Delta_1$ reads
\begin{align}
\Delta_1&=-i \tilde{\delta}^-(q_3) \tilde{\delta}^+(q_2) G_F(q_1)-i
   \tilde{\delta}^-(q_2) \tilde{\delta}^+(q_1) G_F(q_3)-i
   \tilde{\delta}^-(q_3) \tilde{\delta}^+(q_1)
   G_F(q_2)^\star \nonumber\\
   &+\tilde{\delta}^-(q_2) \tilde{\delta}^+(q_1)
   \tilde{\delta}^+(q_3)+\tilde{\delta}^+(q_2) G_F(q_1)
   G_F(q_3)^\star+\tilde{\delta}^+(q_3) G_F(q_1)
   G_F(q_2)+\tilde{\delta}^+(q_1) G_F(q_2)^\star G_F(q_3)^\star \nonumber\\
   &-i
   G_F(q_1) G_F(q_2) G_F(q_3)
\end{align}
Inserting the identities for the conjugate propagators and Dirac delta functions of eq.~\eqref{eq:distr_identities}, we get
\begin{equation}
\Delta_1=-iG_R(-q_1)G_R(-q_2)G_R(-q_3).
\end{equation}
For one connected component, we were able to derive the classical limit and the retarded propagator prescription. In fact, the result we found is, up to zero measured cuts operated on massless edges connecting to the massive line, equivalent to that stated in eq.~\eqref{eq:worldline_hyp}. For diagrams with more than one connected component the situation is subtler.

\subsection{Classical limit for iterations}

While for diagrams with a single connected massless component we have been able to easily retrieve the classical case, for iterations this is highly non-trivial, due to the complicated combinatorics involved in the calculation and the non-canonical nature of the integrand. In this section, we will show how the general procedure highlighted before unfolds in specific examples in scalar theory and scalar QED, and we will arrange the result so as to reproduce the classical worldline integrand, as derived through WQFT.

\subsubsection{Two-loop ladder in scalar theory}

Let us now show a two loop example with two iterations. We report here the result of the manipulations highlighted above, and provide an explicit computation in appendix~\ref{sec:explicit_two_loop}. We write the symmetrised integrand as
\begin{equation}
F_{\scalebox{0.8}{\rotatebox[origin=c]{90}{$\boxminus$}}}=\frac{g^6}{3!}\left(F_{\scalebox{0.8}{\rotatebox[origin=c]{90}{$\boxminus$}}}^V+F_{\scalebox{0.8}{\rotatebox[origin=c]{-90}{$\slashed{\boxminus}$}}}^R\right)
\end{equation}
The virtual contributions for the ladder diagrams read
\begin{align}
F_{\scalebox{0.8}{\rotatebox[origin=c]{90}{$\boxminus$}}}^V=i(q_1+q_2+q_3)^\mu&\left[\sum_{\pi\in\mathfrak{S}_3}\frac{i^2\delta((q_{\pi(1)}+q_{\pi(2)}+q_{\pi(3)}+p_1)^2-m_1^2)}{[(q_{\pi(1)}+p_1)^2-m_1^2][(q_{\pi(1)}+q_{\pi(2)}+p_1)^2-m_1^2]}\right] \nonumber\\
\times & \left[\sum_{\pi'\in\mathfrak{S}_3}\frac{i^2\delta((q_{\pi'(1)}+q_{\pi'(2)}+q_{\pi'(3)}-p_2)^2-m_2^2)}{[(q_{\pi'(1)}-p_2)^2-m_2^2][(q_{\pi'(1)}+q_{\pi'(2)}-p_2)^2-m_2^2]}\right]  G_F(q_1)G_F(q_2) G_F(q_3) \nonumber\\
=(q_1+q_2+q_3)^\mu &\mathcal{L}(\{1,2,3\},\{\},p_1)\mathcal{L}(\{1,2,3\},\{\},-p_2)G_F(q_1)G_F(q_2) G_F(q_3).
\end{align}
For what concerns the cut contributions, we divide them into two sets
\begin{equation}
F_{\scalebox{0.8}{\rotatebox[origin=c]{-90}{$\slashed{\boxminus}$}}}^R=F_{\scalebox{0.8}{\rotatebox[origin=c]{-90}{$\slashed{\boxminus}$}},2}^R+F_{\scalebox{0.8}{\rotatebox[origin=c]{-90}{$\slashed{\boxminus}$}},3}^R.
\end{equation}
We have that the sum of contributions with cuts of valence two is:
\begin{align}
iF_{\scalebox{0.8}{\rotatebox[origin=c]{-90}{$\slashed{\boxminus}$}},2}^R
=&-q_1^\mu \mathcal{L}(\{1\},\{2,3\};p_1)\mathcal{L}(\{1\},\{2,3\};-p_2)G_F(q_1)^\star G_F(q_2)G_F(q_3) \nonumber\\
&-q_2^\mu \mathcal{L}(\{2\},\{1,3\};p_1)\mathcal{L}(\{2\},\{1,3\};-p_2)G_F(q_2)^\star G_F(q_1)G_F(q_3) \nonumber\\
&-q_3^\mu \mathcal{L}(\{3\},\{1,2\};p_1)\mathcal{L}(\{3\},\{1,2\};-p_2)G_F(q_3)^\star G_F(q_1)G_F(q_2) \nonumber\\
&+(q_1+q_2)^\mu \mathcal{L}(\{1,2\},\{3\};p_1)\mathcal{L}(\{1,2\},\{3\};-p_2)G_F(q_1)^\star G_F(q_2)^\star G_F(q_3) \nonumber\\
&+(q_1+q_3)^\mu \mathcal{L}(\{1,3\},\{2\};p_1)\mathcal{L}(\{1,3\},\{2\};-p_2)G_F(q_1)^\star G_F(q_3)^\star G_F(q_2) \nonumber\\
&+(q_2+q_3)^\mu \mathcal{L}(\{2,3\},\{1\};p_1)\mathcal{L}(\{2,3\},\{1\};-p_2)G_F(q_2)^\star G_F(q_3)^\star G_F(q_1). 
\end{align}
Similarly, the sum of contributions with cuts of valence three reads
\begin{align}
-iF_{\scalebox{0.8}{\rotatebox[origin=c]{-90}{$\slashed{\boxminus}$}},3}^R=& q_1^\mu \mathcal{L}(\{1\},\{2,3\};p_1)\mathcal{L}(\{1,2\},\{3\};-p_2)G_F(q_1)^\star \delta^+(q_2)G_F(q_3) \nonumber\\
+& q_1^\mu \mathcal{L}(\{1\},\{2,3\};p_1)\mathcal{L}(\{1,3\},\{2\};-p_2)G_F(q_1)^\star \delta^+(q_3)G_F(q_2) \nonumber\\
+& q_2^\mu \mathcal{L}(\{2\},\{1,3\};p_1)\mathcal{L}(\{1,2\},\{3\};-p_2)G_F(q_2)^\star \delta^+(q_1)G_F(q_3) \nonumber\\
+& q_2^\mu \mathcal{L}(\{2\},\{1,3\};p_1)\mathcal{L}(\{3,2\},\{1\};-p_2)G_F(q_2)^\star \delta^+(q_3)G_F(q_1) \nonumber\\
+& q_3^\mu \mathcal{L}(\{3\},\{1,2\};p_1)\mathcal{L}(\{1,3\},\{2\};-p_2)G_F(q_3)^\star \delta^+(q_1)G_F(q_2) \nonumber\\
+& q_3^\mu \mathcal{L}(\{3\},\{1,2\};p_1)\mathcal{L}(\{3,2\},\{1\};-p_2)G_F(q_3)^\star \delta^+(q_2)G_F(q_1) \nonumber\\
+& (q_1+q_2)^\mu \mathcal{L}(\{1,2\},\{3\};p_1)\mathcal{L}(\{1\},\{2,3\};-p_2)G_F(q_1)^\star \delta^-(q_2)G_F(q_3) \nonumber\\
+& (q_1+q_2)^\mu \mathcal{L}(\{1,2\},\{3\};p_1)\mathcal{L}(\{2\},\{1,3\};-p_2)G_F(q_2)^\star \delta^-(q_1)G_F(q_3) \nonumber\\
+& (q_1+q_3)^\mu \mathcal{L}(\{1,3\},\{2\};p_1)\mathcal{L}(\{1\},\{2,3\};-p_2)G_F(q_1)^\star \delta^-(q_3)G_F(q_2) \nonumber\\
+& (q_1+q_3)^\mu \mathcal{L}(\{1,3\},\{2\};p_1)\mathcal{L}(\{3\},\{1,2\};-p_2)G_F(q_3)^\star \delta^-(q_1)G_F(q_2) \nonumber\\
+& (q_2+q_3)^\mu \mathcal{L}(\{2,3\},\{1\};p_1)\mathcal{L}(\{2\},\{1,3\};-p_2)G_F(q_3)^\star \delta^-(q_2)G_F(q_1) \nonumber\\
+& (q_2+q_3)^\mu \mathcal{L}(\{2,3\},\{1\};p_1)\mathcal{L}(\{3\},\{1,2\};-p_2)G_F(q_2)^\star \delta^-(q_3)G_F(q_1) 
\end{align}
The same expressions hold for the sum of the ladder diagrams in scalar QED, with the scalar quantum worldline substituted with its analogue in scalar QED, and with additional contractions due to the massless propagators. In appendix~\ref{sec:explicit_two_loop} we provide an explicit derivation of the classical integrand. Here we only report the result. Using the fact that 
\begin{align}
\int \left[\prod_{j=1}^3 \frac{\mathrm{d}^4 q_j}{(2\pi)^4} e^{iq_j\cdot b}\right] F_{\scalebox{0.8}{\rotatebox[origin=c]{90}{$\boxminus$}}}&=\int \left[\prod_{j=1}^3 \frac{\mathrm{d}^4 q_j}{(2\pi)^4} e^{iq_j\cdot b}\right]\frac{1}{3!}\left(q_1^\mu F_{\scalebox{0.8}{\rotatebox[origin=c]{90}{$\boxminus$}}}^1+q_2^\mu F_{\scalebox{0.8}{\rotatebox[origin=c]{90}{$\boxminus$}}}^2+q_3^\mu F_{\scalebox{0.8}{\rotatebox[origin=c]{90}{$\boxminus$}}}^3\right)=\int \left[\prod_{j=1}^3 \frac{\mathrm{d}^4 q_j}{(2\pi)^4} e^{iq_j\cdot b}\right]\frac{q_3^\mu}{2}F_{\scalebox{0.8}{\rotatebox[origin=c]{90}{$\boxminus$}}}^{3},
\label{eq:F_triple_ladder_q3_anchor}
\end{align}
we only need to determine the part of the classical integrand that is proportional to $q_3^\mu$. Splitting $F_{\scalebox{0.8}{\rotatebox[origin=c]{90}{$\boxminus$}}}^{3}=F_{\scalebox{0.8}{\rotatebox[origin=c]{90}{$\boxminus$}}}^{3,\text{Re}}+F_{\scalebox{0.8}{\rotatebox[origin=c]{90}{$\boxminus$}}}^{3,\text{Im}}$ for reasons that will soon be clear, we have
\begin{align}
F_{\scalebox{0.8}{\rotatebox[origin=c]{90}{$\boxminus$}}}^{3,\text{Re}}=&-i\Bigg[4\frac{(q_1\cdot q_3)(q_2\cdot q_3)\delta(2 q_1\cdot p_1)\delta(2(q_1+q_2)\cdot p_1)}{(2q_1\cdot p_2+i\varepsilon)^2(2q_2\cdot p_2+i\varepsilon)^2}G_R(-q_1)G_R(-q_2)G_R(q_3) \nonumber\\
+&4\frac{(q_1\cdot q_3)(q_2\cdot q_3)\delta(2 q_1\cdot p_2)\delta(2(q_1+q_2)\cdot p_2)}{(2q_1\cdot p_1+i\varepsilon)^2(2q_2\cdot p_1+i\varepsilon)^2}G_R(q_1)G_R(q_2)G_R(q_3)\nonumber\\
+&4\frac{(q_1\cdot q_2)(q_2\cdot q_3)\delta(2 q_1\cdot p_1)\delta(2(q_1+q_2)\cdot p_1)}{(2q_1\cdot p_2+i\varepsilon)^2(2(q_1+q_2)\cdot p_2+i\varepsilon)^2} G_R(-q_1)G_R(-q_2)G_R(-q_3)\nonumber\\
+&4\frac{(q_1\cdot q_2)(q_1\cdot q_3)\delta(2 q_2\cdot p_2)\delta(2(q_1+q_2)\cdot p_1)}{(2q_1\cdot p_1-i\varepsilon)^2(2(q_1+q_2)\cdot p_2+i\varepsilon)^2} G_R(-q_1)G_R(q_2)G_R(q_3)\nonumber\\
+&4\frac{(q_1\cdot q_2)(q_2\cdot q_3)\delta(2 q_1\cdot p_2)\delta(2(q_1+q_2)\cdot p_1)}{(2q_1\cdot p_1+i\varepsilon)^2(2(q_1+q_2)\cdot p_2+i\varepsilon)^2}G_R(q_1)G_R(-q_2)G_R(q_3)\nonumber\\
+&4\frac{(q_1\cdot q_2)(q_1\cdot q_3)\delta(2 q_1\cdot p_1)\delta(2(q_1+q_2)\cdot p_1)}{(2q_2\cdot p_2+i\varepsilon)^2(2(q_1+q_2)\cdot p_2+i\varepsilon)^2}G_R(-q_1)G_R(-q_2)G_R(q_3)\nonumber\\
+&4\frac{(q_1\cdot q_2)(q_1\cdot q_3)\delta(2 q_2\cdot p_1)\delta(2(q_1+q_2)\cdot p_2)}{(2q_1\cdot p_2-i\varepsilon)^2(2(q_1+q_2)\cdot p_1+i\varepsilon)^2}G_R(q_1)G_R(-q_2)G_R(q_3)\nonumber\\
+&4\frac{(q_1\cdot q_2)(q_1\cdot q_3)\delta(2 q_1\cdot p_1)\delta(2(q_1+q_2)\cdot p_2)}{(2q_1\cdot p_2+i\varepsilon)^2(2(q_1+q_2)\cdot p_1+i\varepsilon)^2}G_R(-q_1)G_R(q_2)G_R(q_3)\nonumber\\
+&4\frac{(q_1\cdot q_2)(q_2\cdot q_3)\delta(2 q_1\cdot p_2)\delta(2(q_1+q_2)\cdot p_2)}{(2q_1\cdot p_1+i\varepsilon)^2(2(q_1+q_2)\cdot p_1+i\varepsilon)^2}G_R(q_1)G_R(q_2)G_R(q_3)\nonumber\\
+&4\frac{(q_1\cdot q_2)(q_1\cdot q_3)\delta(2 q_1\cdot p_2)\delta(2(q_1+q_2)\cdot p_2)}{(2q_2\cdot p_1+i\varepsilon)^2(2(q_1+q_2)\cdot p_1+i\varepsilon)^2}G_R(q_1)G_R(q_2)G_R(q_3)\nonumber\\
+&4\frac{(q_1\cdot q_3)(q_2\cdot q_3)\delta(2 q_1\cdot p_2)\delta(2 q_2\cdot p_1)}{(2(q_1+q_2)\cdot p_2+i\varepsilon)^2(2(q_1+q_2)\cdot p_1+i\varepsilon)^2}G_R(q_1)G_R(-q_2)G_R(q_3)\nonumber\\
+&4\frac{(q_1\cdot q_3)(q_2\cdot q_3)\delta(2 q_1\cdot p_1)\delta(2 q_2\cdot p_2)}{(2(q_1+q_2)\cdot p_2+i\varepsilon)^2(2(q_1+q_2)\cdot p_1+i\varepsilon)^2}G_R(-q_1)G_R(-q_2)G_R(q_3)\Bigg]\times \nonumber\\
\times&\delta(2(q_1+q_2+q_3)\cdot p_1)\delta(2(q_1+q_2+q_3)\cdot p_2), \label{eq:classical_limit_scalar}
\end{align}
and
\begin{align}
\label{eq:cycle_contr}
F_{\scalebox{0.8}{\rotatebox[origin=c]{90}{$\boxminus$}}}^{3,\text{Im}}=&-4iG_F(q_3)\left[\frac{(q_1\cdot q_2)^2 G_R(-q_2)G_R(q_1)}{(2q_2\cdot p_1-i\varepsilon)^2(2q_2\cdot p_2-i\varepsilon)^2}+\frac{(q_1\cdot q_2)^2G_R(q_2)G_R(-q_1)}{(2q_2\cdot p_1+i\varepsilon)^2(2q_2\cdot p_2+i\varepsilon)^2}\right]\times \nonumber\\
\times&\delta(2q_3\cdot p_1)\delta(2q_3\cdot p_2)\delta(2q_{12}\cdot p_1)\delta(2q_{12}\cdot p_2) 
\end{align}
Notice that the diagrams contributing to $F_{\scalebox{0.8}{\rotatebox[origin=c]{90}{$\boxminus$}}}^{3,\text{Im}}$ have one uncut loop, and thus do not match with the expected worldline form. The time flow associated to these loops makes it so that there is a directed cycle:
\begin{align}
F^{3,\text{Im}}_{\scalebox{0.8}{\rotatebox[origin=c]{90}{$\boxminus$}}}=
\raisebox{-1.2cm}{\scalebox{0.7}{\input{diagloop1}}}
+\raisebox{-1.2cm}{\scalebox{0.7}{\input{diagloop2}}}
\end{align}
These terms give imaginary contribution upon being Fourier integrated. We will set them to zero. We note that now each of the surviving terms can be written as a worldline integrand in the sense of eq.~\eqref{eq:worldline_hyp}: 
\begin{align}
F^{3,\text{Re}}_{\scalebox{0.8}{\rotatebox[origin=c]{90}{$\boxminus$}}}=
&\raisebox{-1.2cm}{\scalebox{0.7}{\input{diag4}}}
+\raisebox{-1.2cm}{\scalebox{0.7}{\input{diag2}}}
+\raisebox{-1.2cm}{\scalebox{0.7}{\input{diag3}}}
+\raisebox{-1.2cm}{\scalebox{0.7}{\input{diag12}}}\nonumber \\
+&\raisebox{-1.2cm}{\scalebox{0.7}{\input{diag5}}}
+\raisebox{-1.2cm}{\scalebox{0.7}{\input{diag10}}}
+\raisebox{-1.2cm}{\scalebox{0.7}{\input{diag11}}}
+\raisebox{-1.2cm}{\scalebox{0.7}{\input{diag7}}}\nonumber \\
+&\raisebox{-1.2cm}{\scalebox{0.7}{\input{diag1}}}
+\raisebox{-1.2cm}{\scalebox{0.7}{\input{diag9}}}
+\raisebox{-1.2cm}{\scalebox{0.7}{\input{diag6}}}
+\raisebox{-1.2cm}{\scalebox{0.7}{\input{diag8}}},
\label{eq:F_triple_ladder_Re}
\end{align}
where each solid black line should be understood as a quadratic eikonal propagator. Note that massless propagators have the correct retarded propagator prescription, as predicted by the time-flow arguments. Comparison with appendix~\ref{sec:wqft} shows that these are exactly the diagrams that arise from applying Feynman rules in the WQFT formalism, including the correct symmetry factors. To elaborate, each WQFT diagrams with a unit symmetry factor appears twice, in isomorphic versions, in Eq.~\eqref{eq:F_triple_ladder_Re}, and the factor of $2$ cancels with the factor of $1/2$ multiplying $q_3^\mu$ in Eq.~\eqref{eq:F_triple_ladder_q3_anchor}. Meanwhile, EFT WQFT diagram with a symmetry factor of $1/2$ appears once in Eq.~\eqref{eq:F_triple_ladder_Re}, again as is appropriate.

\subsubsection{Two-loop mushroom in scalar theory}

At two loops in a scalar theory, we also encounter diagrams of the form
\begin{equation}
\raisebox{-1cm}{\scalebox{0.5}{\input{mushroom2L}}}.
\end{equation}
As opposed to the one-loop mushroom diagrams, which after taking the classical limit individually give homogeneous vanishing contributions, for two-loop mushroom diagrams homogeneous integrals only appear upon combining different diagrams. The virtual integrand, in terms of the quantum worldline, is given by
\begin{equation}
F_{\text{mush}}^V=ig^6\frac{(q_2+q_3)^\mu}{4} \mathcal{L}(\{1,2,3,4\},\{\},p_1)|_{q_4=-q_1}\mathcal{L}(\{2,3\},\{\},p_2)G_F(q_1)G_F(q_2)G_F(q_3),
\end{equation}
where the self-attaching massless particle stretches between vertex $1$ and $4$. Because the graph has three massless connected components, and one is self-attaching to the upper massive line, we need three orders of expansion for the ``upper'' quantum worldline and two for the "lower" one. As we have argued in sect.~\ref{sec:classical_expansion_quantum_worldline}, each term of the quantum worldline maps to an acyclic graph on the respective set of vertices, and if we are interested in the first three orders in the classical expansion, we can focus on arborescences (directed forests) only. For the "upper" quantum worldline, since it has a self-attachment, we will also draw the corresponding dashed massless edge further connecting vertices in this forest. The expansion of the quantum worldline then takes the following diagrammatic form
\begin{align}
\mathcal{L}(\{1,2,3,4\},\{\},p_1)|_{q_4=-q_1}&=\input{forest1}+\lambda(\input{forest2}+...) \nonumber\\
&+\lambda^2(\input{forest3}+\input{forest4}+...)+\mathcal{O}(\lambda^0).
\end{align}
It is trivial to check that up to order $\lambda^2$, the only non-vanishing contributions are given by
\begin{align}
\label{eq:surviving_self_att}
\mathcal{L}(\{1,2,3,4\},\{\},p_1)|_{q_4=-q_1}&=\input{forest4}+\raisebox{-0.3cm}{\input{forest7}}+\mathcal{O}(\lambda^0),
\end{align}
while everything else, once integrated back from Schwinger parameter space, factorises a homogeneous integral. As an example, consider the following contribution:
\begin{align}
&\input{forest3}=\frac{1}{q_1^2}\int \mathrm{d}\Pi_4(p_1)[2i(q_1\cdot q_2)\Delta\tau_{12}] \Theta(\Delta\tau_{12}) [2i(q_2\cdot q_3)\Delta\tau_{23}] \Theta(\Delta\tau_{23}) \nonumber\\
&=\frac{1}{q_1^2}\frac{2i(q_1\cdot q_2)}{[(q_1+p_1)^2-m^2-i\varepsilon]^{2}}\frac{2i(q_2\cdot q_3)}{[(q_1+q_2+p_1)^2-m^2-i\varepsilon]^{2}}\tilde{\delta}((q_{13}+p_1)^2-m^2)\tilde{\delta}((q_{23}+p_1)^2-m^2) \nonumber\\
&=\frac{1}{q_1^2}\frac{2i(q_1\cdot q_2)}{[2q_1\cdot (q_2+q_3)+i\varepsilon]^{2}}\frac{2i(q_2\cdot q_3)}{[(q_1+q_2+p_1)^2-m^2-i\varepsilon]^{2}}\tilde{\delta}((q_1+p_1)^2-m^2+2q_1\cdot(q_2+q_3))\tilde{\delta}((q_{23}+p_1)^2-m^2) \nonumber \\
&=\frac{1}{q_1^2}\frac{2i(q_1\cdot q_2)}{[2q_1\cdot (q_2+q_3)+i\varepsilon]^{2}}\frac{2i(q_2\cdot q_3)}{[2q_2\cdot p_1-i\varepsilon]^{2}}\tilde{\delta}(2q_1\cdot p_1)\tilde{\delta}(2(q_2+q_3)\cdot p_1)+\ldots.
\end{align}
which is homogeneous in $q_1$, and thus vanishes in dimensional regularisation upon integration. Let us look at a second example:
\begin{align}
&\raisebox{-0.33cm}{\input{forest8}}=\frac{1}{q_1^2}\int \mathrm{d}\Pi_4(p_1)[2i(q_1\cdot q_3)\Delta\tau_{13}] \Theta(\Delta\tau_{13}) [2i(q_3\cdot q_4)\Delta\tau_{34}] \Theta(\Delta\tau_{34}) \Bigg|_{q_4=-q_1} \nonumber\\
&=\frac{1}{q_1^2}\frac{2i(q_1\cdot q_3)}{[(q_1+q_2+p_1)^2-m^2-i\varepsilon]^{2}}\frac{-2i(q_3\cdot q_1)}{[(q_{13}+p_1)^2-m^2-i\varepsilon]^{2}}\tilde{\delta}((q_{1}+p_1)^2-(q_{12}+p_1)^2)\tilde{\delta}((q_{23}+p_1)^2-m^2) \nonumber\\
&=\frac{1}{q_1^2}\frac{2i(q_1\cdot q_3)}{[2q_1\cdot p_1-i\varepsilon]^{2}}\frac{-2i(q_3\cdot q_1)}{[2q_1\cdot p_1-i\varepsilon]^{2}}\tilde{\delta}(2q_2\cdot p_1)\tilde{\delta}(2q_3\cdot p_1)+\ldots,
\end{align}
which, again, is homogeneous. Since subleading orders, which we collected in the $\ldots$, differ from the leading order only by higher powers of the denominators, derivatives of the Dirac delta functions, and polynomials in the numerator, the homogeneity argument also holds for them, and we can conclude that, in the soft region,
\begin{align}
\input{forest3}\hspace{-0.2cm}|_{\text{soft region}}=\raisebox{-0.35cm}{\input{forest8}}\hspace{-0.2cm}|_{\text{soft region}}=0. 
\end{align}
The same argument we presented for these two examples can be applied to all order $\lambda^{-n}$, $n\ge 0$  contributions to the quantum worldline with a self-attachment, except the two given in eq.~\eqref{eq:surviving_self_att}. Indeed, we have that
\begin{align}
\input{forest4}\hspace{-0.2cm}&=\frac{1}{q_1^2}\frac{2i(q_1\cdot q_2)}{[2q_1\cdot p_1-i\varepsilon]^2}\frac{-2i(q_3\cdot q_1)}{[2(q_2+q_3)\cdot p_1-i\varepsilon]^2}\tilde{\delta}(2(q_1+q_2)\cdot p_1)\tilde{\delta}(2(q_1+q_2+q_3)\cdot p_1)+\ldots, \\
\raisebox{-0.35cm}{\input{forest7}}\hspace{-0.2cm}&=\frac{1}{q_1^2}\frac{2i(q_1\cdot q_3)}{[2q_1\cdot p_1-i\varepsilon]^2}\frac{-2i(q_2\cdot q_1)}{[2(q_1+q_2+q_3)\cdot p_1-i\varepsilon]^2}\tilde{\delta}(2(q_1+q_3)\cdot p_1)\tilde{\delta}(2(q_2+q_3)\cdot p_1)+\ldots,
\end{align}
are not homogeneous in $q_1$. In summary, we can immediately conclude that there is only one worldline diagram (up to $q_2\leftrightarrow q_3$ transformation) that survives the classical limit for the virtual contributions, and in particular
\begin{align}
F_{\text{mush}}^V=&ig^6\frac{q_2^\mu}{2}\frac{4(q_1\cdot q_2)(q_3\cdot q_1)\tilde{\delta}(2(q_1+q_2)\cdot p_1)\tilde{\delta}(2(q_2+q_3)\cdot p_1)\tilde{\delta}(2q_2\cdot p_2)\tilde{\delta}(2q_3\cdot p_2)}{(2q_2\cdot p_1+i\varepsilon)^4}G_F(q_1)G_F(q_2)G_F(q_3) \nonumber\\
&ig^6\frac{q_3^\mu}{2}\frac{4(q_1\cdot q_3)(q_2\cdot q_1)\tilde{\delta}(2(q_1+q_2)\cdot p_1)\tilde{\delta}(2(q_2+q_3)\cdot p_1)\tilde{\delta}(2q_2\cdot p_2)\tilde{\delta}(2q_3\cdot p_2)}{(2q_3\cdot p_1-i\varepsilon)^4}G_F(q_1)G_F(q_2)G_F(q_3)
\end{align}
Continuing to treat the real contributions, we focus on the only existing equivalence class of real diagrams, with representative:
\begin{equation}
\label{eq:right_mushroom}
\raisebox{-1cm}{\scalebox{0.5}{\input{mudhroom2L_cut}}},
\end{equation}
The other elements in the equivalence class are obtained by permuting the vertices of the quantum worldline on the left and right of the cut separately. All other equivalence classes (specifically, those in which the self-attachment corresponds to a vertex correction, or to an external self-energy correction) also give rise to vanishing homogeneous integrals in the classical limit. In summary, we may write
\begin{align}
F_{\text{mush}}^R&=g^6\frac{(q_1+q_2)^\mu}{4} \mathcal{L}(\{1,2\},\{3,4\},p_1)|_{q_4=-q_1}\mathcal{L}(\{2\},\{3\},p_2)\tilde{\delta}^-(q_1)G_F(q_2)G_F(q_3)^\star\nonumber \\
&+g^6\frac{(q_1+q_3)^\mu}{4} \mathcal{L}(\{1,3\},\{2,4\},p_1)|_{q_4=-q_1}\mathcal{L}(\{3\},\{2\},p_2)\tilde{\delta}^-(q_1)G_F(q_2)^\star G_F(q_3)\nonumber\\
&+g^6\frac{(-q_1+q_2)^\mu}{4} \mathcal{L}(\{4,2\},\{3,1\},p_1)|_{q_4=-q_1}\mathcal{L}(\{2\},\{3\},p_2)\tilde{\delta}^+(q_1)G_F(q_2)G_F(q_3)^\star \nonumber\\
&+g^6\frac{(-q_1+q_3)^\mu}{4} \mathcal{L}(\{4,3\},\{2,1\},p_1)|_{q_4=-q_1}\mathcal{L}(\{3\},\{2\},p_2)\tilde{\delta}^+(q_1)G_F(q_2)^\star G_F(q_3).
\end{align}
 Except for the matter propagator that is already initially in the KMOC cut of eq.~\eqref{eq:right_mushroom}, other ones on the upper massive lines can never be cut, for otherwise they would lead to zero-measured cuts. For the same reason, every cut of the non self-attaching massless propagators vanishes. It follows that the only non-vanishing contributions come from the terms in which the uncut matter propagators in the upper line are raised. In particular:
\begin{align}
F_{\text{mush}}^R=-g^6&\Bigg[\frac{(q_1+q_2)^\mu}{4}\frac{2(q_1\cdot q_2)}{[2q_1\cdot p_1+i\varepsilon]^2}\frac{2(q_1\cdot q_3)}{[2q_1\cdot p_1-i\varepsilon]^2}\delta(2(q_1+q_2)\cdot p_1)\tilde{\delta}^-(q_1)\nonumber \\
&+\frac{(q_1+q_3)^\mu}{4}\frac{2(q_1\cdot q_2)}{[2q_1\cdot p_1+i\varepsilon]^2}\frac{2(q_1\cdot q_3)}{[2q_1\cdot p_1-i\varepsilon]^2}\delta(2(q_1+q_3)\cdot p_1)\tilde{\delta}^-(q_1)\nonumber \\
&+\frac{(-q_1+q_2)^\mu}{4}\frac{2(q_1\cdot q_2)}{[2q_1\cdot p_1-i\varepsilon]^2}\frac{2(q_1\cdot q_3)}{[2q_1\cdot p_1+i\varepsilon]^2}\delta(2(-q_1+q_2)\cdot p_1)\tilde{\delta}^+(q_1)\nonumber \\
&+\frac{(-q_1+q_3)^\mu}{4}\frac{2(q_1\cdot q_2)}{[2q_1\cdot p_1-i\varepsilon]^2}\frac{2(q_1\cdot q_3)}{[2q_1\cdot p_1+i\varepsilon]^2}\delta(2(-q_1+q_3)\cdot p_1)\tilde{\delta}^+(q_1)\Bigg]\times\nonumber \\
&\times \delta(2q_{23}\cdot p_1)\delta(2q_2\cdot p_2)\delta(2q_3\cdot p_2) G_F(q_2)G_F(q_3)
\end{align}
Upon undoing the $q_1\leftrightarrow q_4=-q_1$ symmetrisation, we get
\begin{align}
F_{\text{mush}}^R=-g^6&\Bigg[\frac{(q_1+q_2)^\mu}{2}\frac{2(q_1\cdot q_2)}{[2q_1\cdot p_1+i\varepsilon]^2}\frac{2(q_1\cdot q_3)}{[2q_1\cdot p_1-i\varepsilon]^2}\delta(2(q_1+q_2)\cdot p_1)\tilde{\delta}^-(q_1)\nonumber \\
&+\frac{(q_1+q_3)^\mu}{2}\frac{2(q_1\cdot q_2)}{[2q_1\cdot p_1+i\varepsilon]^2}\frac{2(q_1\cdot q_3)}{[2q_1\cdot p_1-i\varepsilon]^2}\delta(2(q_1+q_3)\cdot p_1)\tilde{\delta}^-(q_1)\Bigg]\times \nonumber \\
&\times \delta(2q_{23}\cdot p_1)\delta(2q_2\cdot p_2)\delta(2q_3\cdot p_2) G_F(q_2)G_F(q_3)
\end{align}
Note that the $i\varepsilon$ prescription of the matter propagators can be changed, since the difference amounts, again, to zero-measured cuts. Indeed, they are ``tree'' propagators of the cut diagram. We observe that if we decompose $F_{\text{mush}}^R=q_1^\mu F_{\text{mush}}^{R,1}+q_2^\mu F_{\text{mush}}^{R,2}+q_3^\mu F_{\text{mush}}^{R,3}$, we have that $q_1^\mu F_{\text{mush}}^{R,1}$ vanishes upon being integrated, and can thus be dropped. Indeed $q_1^\mu F_{\text{mush}}^{R,1}$ is real and even upon the simultaneous reflection of $q_2\rightarrow -q_2$ and $q_3\rightarrow -q_3$, and thus gives an imaginary contribution to the observable.

Finally, combining the surviving contributions with the virtual contributions:
\begin{align}
F_{\text{mush}}=F_{\text{mush}}^V+F_{\text{mush}}^R=ig^6&\Bigg[\frac{q_2^\mu}{2}\frac{4(q_1\cdot q_2)(q_1\cdot q_3)\delta(2(q_1+q_2)\cdot p_1)}{(2q_1\cdot p_1+i\varepsilon)^4}G_R(-q_1) \nonumber\\
+&\frac{q_3^\mu}{2}\frac{4(q_1\cdot q_2)(q_1\cdot q_3)\delta(2(q_1+q_2)\cdot p_1)}{(2q_1\cdot p_1-i\varepsilon)^4}G_R(q_1)\Bigg]\times\nonumber \\
&\times \delta(2(q_2+q_3)\cdot p_1)\delta(2q_2\cdot p_2)\delta(2q_3\cdot p_2) G_F(q_2)G_F(q_3).
\end{align}
We are free to turn the remaining Feynman propagators $G_F(q_2)$ and $G_F(q_3)$ into retarded propagators, as the difference between the two leads to zero-measured cuts. In particular their prescription can be chosen to match that of the worldline form. Again, these two diagrams are in the worldline form:
\begin{equation}
\scalebox{0.5}{\input{mushroom_classical}}\quad \scalebox{0.5}{\input{mushroom_classical_2}}.
\end{equation}
Only one-diagram is obtained after mapping $q_2\rightarrow q_3$ and $q_1\rightarrow -q_1$. This concludes the derivation of the classical limit for the mushroom diagram.

\subsubsection{Term-by-term comparison of Worldline and KMOC integrands for the two-loop impulse in scalar QED}

 We make a direct comparison between our integrand and the worldline integrand for the two-loop impulse in scalar QED and in the potential region. The KMOC amplitude receives contributions from the following four equivalence classes of diagrams:
\begin{align}
\text{III}=\raisebox{-0.9cm}{\scalebox{0.65}{\input{double_box}}}, \quad \text{IV}=\raisebox{-0.9cm}{\scalebox{0.65}{\input{IV}}}, \quad \rotatebox[origin=c]{180}{\text{IV}}=\raisebox{-0.9cm}{\scalebox{0.65}{\input{IV_flipped}}}, \\
\text{VV}=\raisebox{-0.9cm}{\scalebox{0.65}{\input{VV}}}, \quad \rotatebox[origin=c]{180}{\text{VV}}=\raisebox{-0.9cm}{\scalebox{0.65}{\input{VV_flipped}}}
\end{align}
Cut contributions arise from Cutkosky cuts of these diagrams and their permutations. In the potential region, any cut photon propagator vanishes. After symmetrisation, the KMOC-derived integrand reads
\begin{align}
F^\text{KMOC}=-\frac{i e_1^3e_2^3}{3!}&\Big[ \\
\text{III}&=\begin{cases}
&+(q_1+q_2+q_3)^\mu \mathcal{L}(\{v_1,v_2,v_3\},\emptyset;p_1)\mathcal{L}(\{w_1,w_2,w_3\},\emptyset;-p_2) \\
&-q_1^\mu \mathcal{L}(\{v_1\},\{v_2,v_3\};p_1)\mathcal{L}(\{w_1\},\{w_2,w_3\};-p_2) \\
&-q_2^\mu \mathcal{L}(\{v_2\},\{v_1,v_3\};p_1)\mathcal{L}(\{v_2\},\{v_1,v_3\};-p_2) \\
&-q_3^\mu \mathcal{L}(\{v_3\},\{v_1,v_2\};p_1)\mathcal{L}(\{w_3\},\{w_1,w_2\};-p_2) \\
&+(q_1+q_2)^\mu \mathcal{L}(\{v_1,v_2\},\{v_3\};p_1)\mathcal{L}(\{w_1,w_2\},\{w_3\};-p_2) \\
&+(q_1+q_3)^\mu \mathcal{L}(\{v_1,v_3\},\{v_2\};p_1)\mathcal{L}(\{w_1,w_3\},\{w_2\};-p_2) \\
&+(q_2+q_3)^\mu \mathcal{L}(\{v_2,v_3\},\{v_1\};p_1)\mathcal{L}(\{w_2,w_3\},\{w_1\};-p_2)
\end{cases}\\
\scalebox{1}{\rotatebox[origin=c]{180}{\text{IV}}}&=\begin{cases}&-(q_1+q_2+q_3)^\mu \mathcal{L}(\{v_{12},v_3\},\emptyset;p_1)\mathcal{L}(\{w_1,w_2,w_3\},\emptyset;-p_2)\\
&+(q_1+q_2)^\mu \mathcal{L}(\{v_{12}\},\{v_3\};p_1)\mathcal{L}(\{w_1,w_2\},\{w_3\};-p_2)\\
&+q_3^\mu \mathcal{L}(\{v_{3}\},\{v_{12}\};p_1)\mathcal{L}(\{w_3\},\{w_{12}\};-p_2)\end{cases}\\
\text{IV}&=\begin{cases}&-(q_1+q_2+q_3)^\mu \mathcal{L}(\{v_{1},v_{2},v_3\},\emptyset;p_1)\mathcal{L}(\{w_{12},w_3\},\emptyset;-p_2) \\
&-(q_1+q_2)^\mu \mathcal{L}(\{v_{12}\},\{v_3\};p_1)\mathcal{L}(\{w_{12}\},\{w_3\};-p_2) \\
&-q_3^\mu \mathcal{L}(\{v_{3}\},\{v_{12}\};p_1)\mathcal{L}(\{w_{3}\},\{w_{12}\};-p_2)\Big]G_F(q_2) G_F(q_3) G_F(q_1) \end{cases} \\
\scalebox{1}{\rotatebox[origin=c]{180}{\text{VV}}}&=\begin{cases}&+(q_1+q_2+q_3)^\mu \mathcal{L}(\{v_{12},v_3\},\emptyset;p_1)\mathcal{L}(\{w_{1},w_{23}\},\emptyset;-p_2) \end{cases} \\
\text{VV}&=\begin{cases}&+(q_1+q_2+q_3)^\mu \mathcal{L}(\{v_{1},v_{23}\},\emptyset;p_1)\mathcal{L}(\{w_{12},w_{3}\},\emptyset;-p_2)
\end{cases}
\Big]G_F(q_2) G_F(q_3) G_F(q_1)
\end{align}
where $\mathcal{L}$ is, of course, the quantum worldline for scalar QED, and vertices with two indices denote the vertices with valence two. Contraction of all relevant Lorentz indexes is assumed. Note that every conjugate photon propagator is turned into an ordinary Feynman propagator since we are working in the potential region. In summary, we can write
\begin{equation}
F^\text{KMOC}=\frac{1}{3!}(q_1^\mu F_1^\text{KMOC}+q_2^\mu F_2^\text{KMOC}+q_3^\mu F_3^\text{KMOC}).
\end{equation}
Much like in the two-loop scalar case, we rewrite the integrand in terms of one impulse numerator, by performing the corresponding change of variables for each of the $F_i$ integrand above:
\begin{equation}
\int \left[\prod_{j=1}^3 \mathrm{d}^4q_j e^{iq_j\cdot b}\right] F^\text{KMOC}=\int \left[\prod_{j=1}^3 \mathrm{d}^4q_j e^{iq_j\cdot b}\right]\frac{q_1^\mu}{2} F_1^\text{KMOC}.
\end{equation}
Finally, $F_1^\text{KMOC}$ is rewritten in Schwinger parameter space, using the corresponding definition of the quantum worldlines and the reference ordering $1,2,3$ for each of the two worldlines,
\begin{align}
F_1^\text{KMOC}&=\int \mathrm{d}\Pi_3(p_1)\mathrm{d}\Pi_3'(-p_2) F_{\text{III}}+\int \mathrm{d}\Pi_2(p_1)\mathrm{d}\Pi_3'(-p_2) F_{\scalebox{0.75}{\rotatebox[origin=c]{180}{\text{IV}}}}+\int \mathrm{d}\Pi_3(p_1)\mathrm{d}\Pi_2'(-p_2) F_{\text{IV}} \nonumber\\
&+\int \mathrm{d}\Pi_2(p_1)\mathrm{d}\Pi_2'(-p_2) F_{\text{VV}}+\int \mathrm{d}\Pi_2(p_1)\mathrm{d}\Pi_2'(-p_2) F_{\scalebox{0.75}{\rotatebox[origin=c]{180}{\text{VV}}}}.
\end{align}
Each of the five integrands on the left-hand-side is then expanded in the classical limit, simplified resulting in a forest-like expansion as in eq.~\eqref{eq:simplified_QED}, and turns out, as expected, to be free of super-leading contributions. Once truncated to the classical order, the expression is integrated back, giving the classical integrand arising from the KMOC formalism.

Let us now turn to the worldline integrand. The KMOC integrand $F_1$ derived above still retains residual permutational symmetry in $q_2\leftrightarrow q_3$. In order to set up the comparison, we thus derive the worldline integrand from the WQFT Feynman rules, obtaining a function $F^{\text{WL}}$. We impose that the photon adjacent to the vertex at which the measurement is performed is labelled as $q_1$, and then symmetrise it over $q_2\leftrightarrow q_3$ to set up the local, term-by-term comparison. In other words, we will show that
\begin{equation}
\frac{q_1^\mu}{2} F_1^\text{KMOC}=F^{\text{WL}}+F^{\text{WL}}|_{q_2\leftrightarrow q_3}.
\end{equation}
Let us see how the procedure unfolds with the probe contributions. In the following, we will use the short-hand notation of $\sigma=u_1\cdot u_2$ with $u_i=p_i/m_i$.

\textbf{Probe contributions} The worldline diagrams for the probe contribution are 
\begin{align}
\frac{1}{2}\left[\raisebox{-1cm}{\scalebox{0.65}{\input{worldline_diag_probe1}}}+\raisebox{-1cm}{\scalebox{0.65}{\input{worldline_diag_probe2}}}+\raisebox{-1cm}{\scalebox{0.65}{\input{worldline_diag_probe3}}}\right],
\end{align}
where we include explicitly the correct symmetry factor of $1/2$. The corresponding worldline integrand reads
\begin{align}
\lim_{m_2\rightarrow 0}F^{\text{WL}}/m_2=\frac{ie_1^3e_2^3q_1^\mu}{2} \Bigg[&\frac{2\sigma^3(q_1\cdot q_2)(q_2\cdot q_3)}{(q_1\cdot u_1-i\varepsilon)^2((q_1+q_2)\cdot u_1-i\varepsilon)^2}+\frac{\sigma(q_2\cdot q_3)}{((q_1+q_2)\cdot u_1-i\varepsilon)^2}+\frac{\sigma(q_1\cdot q_2)}{(q_1\cdot u_1-i\varepsilon)^2} \nonumber\\
+&\frac{\sigma^3(q_1\cdot q_2)(q_1\cdot q_3)}{(q_2\cdot u_1+i\varepsilon)^2((q_1+q_2)\cdot u_1-i\varepsilon)^2}+\frac{\sigma(q_1\cdot q_3)}{((q_1+q_2)\cdot u_1-i\varepsilon)^2}+\frac{\sigma(q_1\cdot q_2)}{(q_2\cdot u_1+i\varepsilon)^2} \nonumber\\
+&\frac{\sigma(q_1\cdot q_2)}{((q_1+q_2)\cdot u_1-i\varepsilon)^2}+\frac{\sigma(q_2\cdot q_3)}{(q_1\cdot u_1-i\varepsilon)^2}\Bigg]\times \nonumber\\
\times&\delta(q_1\cdot u_2)\delta(q_2\cdot u_2)\delta(q_3\cdot u_2)\delta((q_1+q_2+q_3)\cdot u_1).
\end{align}
This integrand is not symmetric under exchange of $q_2$ and $q_3$. We thus symmetrise this expression with respect to $q_2$ and $q_3$. The result is
\begin{align}
\lim_{m_2\rightarrow 0}\frac{1}{2m_2}&(F^{\text{WL}}+F^{\text{WL}}|_{q_2\leftrightarrow q_3})=\frac{ie_1^3e_2^3q_1^\mu}{2} \Bigg[\frac{\sigma^3(q_1\cdot q_2)(q_2\cdot q_3)}{(q_1\cdot u_1-i\varepsilon)^2((q_1+q_2)\cdot u_1-i\varepsilon)^2}+\frac{\sigma(q_2\cdot q_3)}{((q_1+q_2)\cdot u_1-i\varepsilon)^2} \nonumber\\
+&\frac{\sigma(q_1\cdot q_2)}{(q_1\cdot u_1-i\varepsilon)^2} 
+\frac{\sigma^3(q_1\cdot q_2)(q_1\cdot q_3)}{(q_2\cdot u_1+i\varepsilon)^2((q_1+q_2)\cdot u_1-i\varepsilon)^2}+\frac{\sigma(q_1\cdot q_3)}{((q_1+q_2)\cdot u_1-i\varepsilon)^2} \nonumber\\
+&\frac{\sigma(q_1\cdot q_2)}{(q_2\cdot u_1+i\varepsilon)^2} 
+\frac{\sigma^3(q_1\cdot q_3)(q_2\cdot q_3)}{(q_1\cdot u_1-i\varepsilon)^2(q_1\cdot u_1-i\varepsilon)^2}+\frac{\sigma(q_1\cdot q_3)}{(q_2\cdot u_1+i\varepsilon)^2}+\frac{\sigma(q_2\cdot q_3)}{(q_2\cdot u_1+i\varepsilon)^2}\Bigg]\times \nonumber\\
\times&\delta(q_1\cdot u_2)\delta(q_2\cdot u_2)\delta(q_3\cdot u_2)\delta((q_1+q_2+q_3)\cdot u_1).
\end{align}
This expression matches term-by-term with the probe contribution that is derived within the KMOC formalism, namely
\begin{equation}
\lim_{m_2\rightarrow 0}\frac{1}{2m_2}(F^{\text{WL}}+F^{\text{WL}}|_{q_2\leftrightarrow q_3})=\lim_{m_2\rightarrow 0}\frac{q_1^\mu}{2m_2}F^{\text{KMOC}}_1
\end{equation}

\textbf{Beyond probe contributions} The local comparison can be extended to the full integrand in exactly the same way as done for the probe contributions. In particular, we find term-by-term agreement between the classical worldline integrand (derived through WQFT) and KMOC integrand. We have
\begin{equation}
\frac{1}{2}(F^{\text{WL}}+F^{\text{WL}}|_{q_2\leftrightarrow q_3})=\frac{q_1^\mu}{2}F^{\text{KMOC}}_1=\frac{ie_1^3e_2^3}{2}(c_0+\sigma c_1+\sigma^2c_2+\sigma^3c_3).
\end{equation}
$c_3$ is proportional to the contribution we found for the two-loop ladder in the scalar field theory case:
\begin{align}
c_3=\Bigg[&\frac{(q_1\cdot q_3)(q_2\cdot q_3)\delta(q_1\cdot u_1)\delta((q_1+q_2)\cdot u_1)}{(q_1\cdot u_2+i\varepsilon)^2(q_2\cdot u_2+i\varepsilon)^2}+\frac{(q_1\cdot q_3)(q_2\cdot q_3)\delta(q_1\cdot u_2)\delta((q_1+q_2)\cdot u_2)}{(q_1\cdot u_1+i\varepsilon)^2(q_2\cdot u_1+i\varepsilon)^2}\nonumber\\
+&\frac{(q_1\cdot q_2)(q_2\cdot q_3)\delta(q_1\cdot u_1)\delta((q_1+q_2)\cdot u_1)}{(q_1\cdot u_2+i\varepsilon)^2((q_1+q_2)\cdot u_2+i\varepsilon)^2}+\frac{(q_1\cdot q_2)(q_1\cdot q_3)\delta(q_2\cdot u_2)\delta((q_1+q_2)\cdot u_1)}{(q_1\cdot u_1-i\varepsilon)^2((q_1+q_2)\cdot u_2+i\varepsilon)^2}\nonumber\\
+&\frac{(q_1\cdot q_2)(q_2\cdot q_3)\delta(q_1\cdot u_2)\delta((q_1+q_2)\cdot u_1)}{(q_1\cdot u_1+i\varepsilon)^2((q_1+q_2)\cdot u_2+i\varepsilon)^2}+\frac{(q_1\cdot q_2)(q_1\cdot q_3)\delta(q_1\cdot u_1)\delta((q_1+q_2)\cdot u_1)}{(q_2\cdot u_2+i\varepsilon)^2((q_1+q_2)\cdot u_2+i\varepsilon)^2}\nonumber\\
+&\frac{(q_1\cdot q_2)(q_1\cdot q_3)\delta(q_2\cdot u_1)\delta((q_1+q_2)\cdot u_2)}{(q_1\cdot u_2-i\varepsilon)^2((q_1+q_2)\cdot u_1+i\varepsilon)^2}+\frac{(q_1\cdot q_2)(q_1\cdot q_3)\delta(q_1\cdot u_1)\delta((q_1+q_2)\cdot u_2)}{(q_1\cdot u_2+i\varepsilon)^2((q_1+q_2)\cdot u_1+i\varepsilon)^2}\nonumber\\
+&\frac{(q_1\cdot q_2)(q_2\cdot q_3)\delta(q_1\cdot u_2)\delta((q_1+q_2)\cdot u_2)}{(q_1\cdot u_1+i\varepsilon)^2((q_1+q_2)\cdot u_1+i\varepsilon)^2}+\frac{(q_1\cdot q_2)(q_1\cdot q_3)\delta(q_1\cdot u_2)\delta((q_1+q_2)\cdot u_2)}{(q_2\cdot u_1+i\varepsilon)^2((q_1+q_2)\cdot u_1+i\varepsilon)^2}\nonumber\\
+&\frac{(q_1\cdot q_3)(q_2\cdot q_3)\delta( q_1\cdot u_2)\delta( q_2\cdot u_1)}{((q_1+q_2)\cdot u_2+i\varepsilon)^2((q_1+q_2)\cdot u_1+i\varepsilon)^2}+\frac{(q_1\cdot q_3)(q_2\cdot q_3)\delta(q_1\cdot u_1)\delta(q_2\cdot u_2)}{((q_1+q_2)\cdot u_2+i\varepsilon)^2((q_1+q_2)\cdot u_1+i\varepsilon)^2}\Bigg]\times \nonumber\\
\times&\delta((q_1+q_2+q_3)\cdot u_1)\delta((q_1+q_2+q_3)\cdot u_2).
\end{align}
Much like in the scalar case, the agreement of $c_3$ in WQFT and KMOC holds up to terms that should vanish after Fourier transformation (the agreement will instead be exact for $c_2$, $c_1$, $c_0$). In other words, we excluded from the comparison the following two terms:
\begin{equation}
c_3^{\text{Im}}=(q_2\cdot q_3)^2\left[\frac{1}{(q_2\cdot u_1-i\varepsilon)^2(q_2\cdot u_2-i\varepsilon)^2}+\frac{1}{(q_2\cdot u_1+i\varepsilon)^2(q_2\cdot u_2+i\varepsilon)^2}\right]\delta(q_1\cdot u_1)\delta(q_1\cdot u_2).
\end{equation}
Conversely $c_2$, $c_1$ and $c_0$ also receive contributions from contact terms and the expansion of the numerator. For example, the coefficient $c_2$ receives contributions entirely from the expansion of the numerator of the ladder diagrams:
\begin{align}
c_2=&-\frac{(q_1\cdot q_2+q_1\cdot q_3+q_1\cdot q_3)}{(q_1\cdot u_1-i\varepsilon)(q_1\cdot u_2+i\varepsilon)}\Bigg[\delta((q_1+q_2)\cdot u_1)\delta(q_2\cdot u_2)+\delta((q_1+q_2)\cdot u_2)\delta(q_2\cdot u_1)\Bigg] \nonumber\\
&+\frac{(q_1\cdot q_2+q_1\cdot q_3+q_1\cdot q_3)}{(q_1\cdot u_2+i\varepsilon)(q_2\cdot u_1+i\varepsilon)}\Bigg[\delta(q_1\cdot u_1)\delta((q_1+q_2)\cdot u_2)-\delta(q_1\cdot u_1)\delta(q_2\cdot u_2)\Bigg] \nonumber\\
&+\frac{(q_1\cdot q_2+q_1\cdot q_3+q_1\cdot q_3)}{(q_1\cdot u_1-i\varepsilon)(q_2\cdot u_2+i\varepsilon)}\Bigg[\delta(q_1\cdot u_2)\delta((q_1+q_2)\cdot u_1)-\delta(q_1\cdot u_2)\delta(q_2\cdot u_1)\Bigg].
\end{align}
while $c_1$ also receive contribution from the IV and VV diagrams:
\begin{align}
c_1=(q_1\cdot q_3)&\Bigg[\frac{\delta(q_1\cdot u_2)[\delta(q_2\cdot u_1)+\delta((q_1+q_2)\cdot u_2)]+\delta(q_2\cdot u_1)\delta((q_1+q_2)\cdot u_2)}{(q_1\cdot u_1-i\varepsilon)^2}  \nonumber\\
&+\frac{\delta(q_1\cdot u_1)[\delta(q_2\cdot u_2)+\delta((q_1+q_2)\cdot u_1)]+\delta(q_2\cdot u_2)\delta((q_1+q_2)\cdot u_1)}{(q_1\cdot u_2+i\varepsilon)^2} \nonumber\\
&+\frac{\delta(q_1\cdot u_1)\delta((q_1+q_2)\cdot u_1)}{((q_1+q_2)\cdot u_2+i\varepsilon)^2}+\frac{\delta(q_1\cdot u_2)\delta((q_1+q_2)\cdot u_2)}{((q_1+q_2)\cdot u_1-i\varepsilon)^2}\Bigg] \nonumber\\
+(q_2\cdot q_3)&\Bigg[\frac{\delta(q_1\cdot u_1)\delta((q_1+q_2)\cdot u_2)}{(q_2\cdot u_1-i\varepsilon)^2}+\frac{\delta(q_1\cdot u_1)\delta(q_2\cdot u_2)+\delta(q_1\cdot u_2)\delta((q_1+q_2)\cdot u_2)}{(q_2\cdot u_1+i\varepsilon)^2} \nonumber\\
&+\frac{\delta(q_1\cdot u_2)\delta(q_2\cdot u_1)+\delta(q_1\cdot u_1)\delta((q_1+q_2)\cdot u_1)}{(q_2\cdot u_2-i\varepsilon)^2}+\frac{\delta(q_1\cdot u_2)\delta((q_1+q_2)\cdot u_1)}{(q_2\cdot u_2+i\varepsilon)^2} \nonumber\\
&+\frac{\delta(q_1\cdot u_2)\delta((q_1+q_2)\cdot u_2)}{((q_1+q_2)\cdot u_1-i\varepsilon)^2}+\frac{\delta(q_1\cdot u_1)\delta((q_1+q_2)\cdot u_1)}{((q_1+q_2)\cdot u_2+i\varepsilon)^2}\Bigg] \nonumber\\
+(q_2\cdot q_1)&\Bigg[\frac{\delta(q_1\cdot u_2)[\delta((q_1+q_2)\cdot u_1)+\delta((q_1+q_2)\cdot u_2)]+\delta(q_2\cdot u_2)\delta((q_1+q_2)\cdot u_1)}{(q_1\cdot u_1-i\varepsilon)^2}  \nonumber\\
&+\frac{\delta(q_1\cdot u_1)[\delta((q_1+q_2)\cdot u_1)+\delta(q_2\cdot u_1)]+\delta(q_2\cdot u_1)\delta((q_1+q_2)\cdot u_2)}{(q_1\cdot u_2+i\varepsilon)^2} \nonumber\\
&+\frac{\delta(q_1\cdot u_2)\delta((q_1+q_2)\cdot u_2)}{(q_2\cdot u_1+i\varepsilon)^2}+\frac{\delta(q_1\cdot u_1)\delta((q_1+q_2)\cdot u_1)}{(q_2\cdot u_2-i\varepsilon)^2}\Bigg],
\end{align}
and $c_0$ only receives contributions from the IV diagrams and their numerator expansion:
\begin{align}
c_0&=\frac{(q_2\cdot u_2)\delta(q_1\cdot u_2)[\delta((q_1+q_2)\cdot u_1)-\delta(q_2\cdot u_1)]}{(q_1\cdot u_1-i\varepsilon)}+\frac{(q_1\cdot u_2)\delta(q_1\cdot u_1)\delta((q_1+q_2)\cdot u_2)}{(q_2\cdot u_1-i\varepsilon)} \nonumber\\
&-\frac{(q_1\cdot u_2)\delta(q_1\cdot u_1)\delta(q_2\cdot u_2)}{(q_2\cdot u_1+i\varepsilon)}-\frac{(q_1\cdot u_2)[\delta((q_1+q_2)\cdot u_1)\delta(q_2\cdot u_2)+\delta(q_2\cdot u_1)\delta((q_1+q_2)\cdot u_2)]}{(q_1\cdot u_1-i\varepsilon)} \nonumber\\
&-\frac{(q_1\cdot u_1)[\delta((q_1+q_2)\cdot u_1)\delta(q_2\cdot u_2)+\delta(q_2\cdot u_1)\delta((q_1+q_2)\cdot u_2)]}{(q_1\cdot u_2+i\varepsilon)}-\frac{(q_1\cdot u_1)\delta(q_1\cdot u_2)\delta(q_2\cdot u_1)}{(q_2\cdot u_2-i\varepsilon)} \nonumber\\
&+\frac{(q_1\cdot u_1)\delta(q_1\cdot u_2)\delta((q_1+q_2)\cdot u_1)}{(q_2\cdot u_2+i\varepsilon)}+\frac{(q_2\cdot u_1)\delta(q_1\cdot u_1)[\delta((q_1+q_2)\cdot u_2)-\delta(q_2\cdot u_2)]}{(q_1\cdot u_2+i\varepsilon)}.
\end{align}
Having checked that all this coefficients match when generated with the KMOC and classical worldline formalisms, we conclude that the KMOC and classical worldline formalisms generate the same integrand for the full two-loop impulse in scalar QED in the potential region.

\section{Conclusion}

We presented a novel and general method to demonstrate the cancellation of super-leading divergences (namely, divergences in the limit $\hbar\rightarrow 0$) within the KMOC formalism at a \emph{local} level, meaning point-by-point in loop momentum space for the integrand, extending beyond the one-loop level previously explored in the original paper \cite{Kosower:2018adc}. This is achieved by recasting the open massive lines that appear in Feynman diagrams contributing to quantum observables into worldline-like expressions. For brevity, we refer to these expressions as ``quantum worldlines'' to distinguish them from manifestly classical worldline formulations.

We gave a representation of quantum worldlines in Schwinger-parameter space, and obtained compact expressions for the subsequent classical expansions of the quantum worldlines, facilitated by the notion of equivalence classes of graphs, which generalise the symmetrisation procedure of \cite{Kosower:2018adc} as well as directed forests, which embed the notion of a causality flow. The use of a Schwinger parametrisation also allows to democratically combine real and virtual diagrams appearing in the KMOC integrand. After combining all the KMOC diagrams, super-leading divergences cancel, leaving a finite classical integrand that exhibits proper-time orderings consistent with classical causality flow, an emergent feature of our formalism. 

The finite classical integrand takes the same form as its counterpart in worldline approaches, e.g.\ the worldline quantum field theory (WQFT) formalism for concreteness. We have shown this explicitly for the impulse observable and for all contributions in a model of long-range scalar interactions at two loops, as well as for the conservative contributions in the scattering of charged massive bodies in electrodynamics, also at two loops. This result is reminiscent of ref.~\cite{Bern:2024vqs}, which demonstrated that different representations of the loop integrand for the same scattering amplitude—e.g., as obtained from different gauge-fixing conditions—can be mapped to identical expressions when one chooses an appropriate basis of irreducible numerators for each diagram topology and resolves all symmetry relations identifying trivially equivalent terms.

However, our work shows that it is possible to go even further than ref.~\cite{Bern:2024vqs}, beyond the amplitude level, and obtain identical loop integrands for specific observables (such as the impulse) starting from very different formalisms: one based on scattering amplitudes and the other on worldlines. Unlike Refs.~\cite{Caron-Huot:2023vxl, Biswas:2024ept} that reformulated field theory correlators with retarded propagators in the Schwinger-Keldysh formalism, our work starts from the original KMOC formalism, involving in-out amplitudes and their complex conjugates, and proceeds with purely algebraic manipulations to recover the same integrand from worldline theory. Classical causality is an emergent feature in our work, as ``quantum'' orderings of worldline vertices appear in individual terms before they are summed up. While our formalism is presented through electrodynamics and a scalar model where amplitudes are easily computed from Feynman diagrams, the essential physics of super-leading divergences in scattering amplitudes is general, so we do not expect obstruction in applying our formalism to gravitational theories and in using loop integrands constructed from modern amplitude methods.In the future, we plan to provide a fully automated code that converts KMOC observables into a form with manifestly finite classical limits using the general ideas presented in this paper.

Our work unifies scattering amplitudes and worldline approaches to computing classical gravity observables and will open a new venue in cross-fertilisation between theoretical advances developed in these two approaches. For example, advanced techniques for loop integrand construction developed for scattering amplitudes, such as generalised unitarity \cite{Bern:1994zx, Bern:1994cg, Bern:1995db, Bern:1997sc, Britto:2004nc} and colour-kinematic duality \cite{Kawai:1985xq, Bern:2008qj, Bern:2010ue, Bern:2019prr}, can be adapted for our framework, while we can also exploit desirable features of worldline calculations, such as the manifest absence of unphysical classically divergent integrals as well as techniques for reducing the number of nonplanar diagrams ~\cite{Driesse:2024xad}.

Finally, the techniques presented here might be of broader use in studying soft physics for collider observables, such as extending the LBK theorem (see ref.~\cite{Engel:2023ifn} and references therein) to sub-subleading power, the search for a convenient representation for the computation of soft functions~\cite{Liu:2024hfa}, generalised Wilson lines (see e.g.\ refs.~\cite{Laenen:2008gt, White:2011yy, Bonocore:2020xuj, Bonocore:2021qxh}) and worldline formulations of QFTs~\cite{Schubert:2001he, Edwards:2019eby, Feal:2022iyn}.

\begin{acknowledgments}
  We thank Thomas Becher, Zvi Bern, Hofie Hannesdottir, Aidan Herderschee, Enrico Herrmann, Valentin Hirschi, Lucien Huber, Sebastian Jaskiewicz, Gustav Mogull, Donal O'Connell, Radu Roiban, and Michael Ruf for valuable discussions related to this work. We also thank Zvi Bern and Radu Roiban for their comments on a draft of this paper. Z.C.'s work is supported by the Swiss National Science Foundation (SNSF) under grant number PCEFP2\_203335.
  M.Z.’s work is supported in part by the U.K.\ Royal Society through
  Grant URF\textbackslash R1\textbackslash 20109.
  For the purpose of open access, the authors have applied a Creative
  Commons Attribution (CC BY) license to any Author Accepted
  Manuscript version arising from this submission.
\end{acknowledgments}

\appendix

\section{Explicit computation for the scalar two-loop ladder}
\label{sec:explicit_two_loop}

We may write the integrand in $\tau$ space, taking advantage of the fact that we have already studied the expansion of the quantum worldline. Let us focus on the terms that are proportional to $q_1^\mu$:
\begin{align}
&F_{\scalebox{0.8}{\rotatebox[origin=c]{90}{$\boxminus$}}}=q_1^\mu F_{\scalebox{0.8}{\rotatebox[origin=c]{90}{$\boxminus$}}}^1+q_2^\mu F_{\scalebox{0.8}{\rotatebox[origin=c]{90}{$\boxminus$}}}^2+q_3^\mu F_{\scalebox{0.8}{\rotatebox[origin=c]{90}{$\boxminus$}}}^3, \quad F^V_{\scalebox{0.8}{\rotatebox[origin=c]{90}{$\boxminus$}}}=q_1^\mu F_{\scalebox{0.8}{\rotatebox[origin=c]{90}{$\boxminus$}}}^{V,1}+q_2^\mu F_{\scalebox{0.8}{\rotatebox[origin=c]{90}{$\boxminus$}}}^{V,2}+q_3^\mu F_{\scalebox{0.8}{\rotatebox[origin=c]{90}{$\boxminus$}}}^{V,3}, \\
&F_{\scalebox{0.8}{\rotatebox[origin=c]{90}{$\slashed{\boxminus}$}},2}=q_1^\mu F_{\scalebox{0.8}{\rotatebox[origin=c]{90}{$\slashed{\boxminus}$}},2}^{R,1}+q_2^\mu F_{\scalebox{0.8}{\rotatebox[origin=c]{90}{$\slashed{\boxminus}$}},2}^{R,2}+q_3^\mu F_{\scalebox{0.8}{\rotatebox[origin=c]{90}{$\slashed{\boxminus}$}},2}^{R,3}, \quad F_{\scalebox{0.8}{\rotatebox[origin=c]{90}{$\slashed{\boxminus}$}},3}=q_1^\mu F_{\scalebox{0.8}{\rotatebox[origin=c]{90}{$\slashed{\boxminus}$}},3}^{R,1}+q_2^\mu F_{\scalebox{0.8}{\rotatebox[origin=c]{90}{$\slashed{\boxminus}$}},3}^{R,2}+q_3^\mu F_{\scalebox{0.8}{\rotatebox[origin=c]{90}{$\slashed{\boxminus}$}},3}^{R,3}
\end{align}
We use the resummed version of the quantum worldline to write the virtual contributions proportional to $q_1^\mu$
\begin{align}
F_{\scalebox{0.8}{\rotatebox[origin=c]{90}{$\boxminus$}}}^{V,1}=&
i\int \mathrm{d}\Pi_3 \mathrm{d}\Pi_3' \Big[1+E_{12}\Theta(\Delta\tau_{12})+E_{13}\Theta(\Delta\tau_{13})+E_{23}\Theta(\Delta\tau_{23})+E_{12}'\Theta(\Delta\tau_{12}')+E_{13}'\Theta(\Delta\tau_{13}')  \nonumber\\
+&E_{23}'\Theta(\Delta\tau_{23}')+E_{12}E_{12}'\Theta(\Delta\tau_{12})\Theta(\Delta\tau_{12}') +E_{23}E_{23}'\Theta(\Delta\tau_{23})\Theta(\Delta\tau_{23}')  \nonumber\\
+&E_{13}E_{13}'\Theta(\Delta\tau_{13})\Theta(\Delta\tau_{13}')+E_{12}E_{23}'\Theta(\Delta\tau_{12})\Theta(\Delta\tau_{23}')+E_{12}E_{13}'\Theta(\Delta\tau_{12})\Theta(\Delta\tau_{13}')  \nonumber\\
+&E_{23}E_{13}'\Theta(\Delta\tau_{23})\Theta(\Delta\tau_{13}')+E_{23}E_{12}'\Theta(\Delta\tau_{23})\Theta(\Delta\tau_{12}')+E_{13}E_{12}'\Theta(\Delta\tau_{13})\Theta(\Delta\tau_{12}')  \nonumber\\
+&E_{13}E_{23}'\Theta(\Delta\tau_{13})\Theta(\Delta\tau_{23}')\Big]G_F(q_1)G_F(q_2)G_F(q_3),
\end{align}
with
\begin{equation}
E_{ij}=e^{-2i(q_i\cdot q_j)\Delta\tau_{ij}}-1.
\end{equation}
The factor of $i$ combines the $-i$ contribution from each of the six vertices as well as the $i$ contribution from each of the three photon propagators. For what concerns the cut contributions, we have
\begin{align}
F_{\scalebox{0.8}{\rotatebox[origin=c]{-90}{$\slashed{\boxminus}$}},2}^{R,1}=i\int \mathrm{d}\Pi_3 \mathrm{d}\Pi_3' \big[&G_F(q_1)G_F(q_2)^\star G_F(q_3)^\star  \big(1+E_{23}\Theta(-\Delta\tau_{23}) +E_{23}'\Theta(-\Delta\tau_{23}')+E_{23}E_{23}'\Theta(-\Delta\tau_{23})\Theta(-\Delta\tau_{23}')\big)  \nonumber\\
-G_F(q_1)G_F(q_2) G_F(q_3)^\star & \big(1+E_{12}\Theta(\Delta\tau_{12}) +E_{12}'\Theta(\Delta\tau_{12}')+E_{12}E_{12}'\Theta(\Delta\tau_{12})\Theta(\Delta\tau_{12}')\big)  \nonumber\\
-G_F(q_1)G_F(q_3) G_F(q_2)^\star & \big(1+E_{23}+E_{23}'+E_{13}\Theta(\Delta\tau_{13})+E_{13}'\Theta(\Delta\tau_{13}')  \nonumber\\
+E_{23}E_{23}'+&E_{23}E_{13}\Theta(\Delta\tau_{13})+E_{23}'E_{13}'\Theta(\Delta\tau_{13}')+E_{13}E_{13}'\Theta(\Delta\tau_{13})\Theta(\Delta\tau_{13}')\big) \big].
\end{align}
Using extensively the vanishing of zero-measured cuts, we also have
\begin{align}
F_{\scalebox{0.8}{\rotatebox[origin=c]{-90}{$\boxminus$}},3}^{R,1}=\int \mathrm{d}\Pi_3 \mathrm{d}\Pi_3'&\big(E_{12}E_{23}'\Theta(\Delta\tau_{12})\Theta(-\Delta\tau_{23}')G_F(q_1)\tilde{\delta}^-(q_2)G_F(q_3) \nonumber\\
+&E_{12}'E_{23}\Theta(-\Delta \tau_{12}')\Theta(\Delta \tau_{23})G_F(q_1)\tilde{\delta}^+(q_2)G_F(q_3)  \nonumber\\
+&E_{13}E_{23}'\Theta(\Delta \tau_{13})\Theta(-\Delta \tau_{23}')G_F(q_1)G_F(q_2)\tilde{\delta}^+(q_3)  \nonumber\\
+&E_{13}'E_{23}\Theta(-\Delta \tau_{13}')\Theta(\Delta \tau_{23})G_F(q_1)G_F(q_2)\tilde{\delta}^-(q_3)  \nonumber\\
+&E_{13}E_{12}'\Theta(\Delta \tau_{13})\Theta(-\Delta \tau_{12}')\tilde{\delta}^+(q_1)G_F(q_2)G_F(q_3)  \nonumber\\
+&E_{13}'E_{12}\Theta(-\Delta \tau_{13}')\Theta(\Delta \tau_{12})\tilde{\delta}^-(q_1)G_F(q_2)G_F(q_3).
\big)
\end{align}
Note that in each of the terms of the previous expression, we can freely turn $\Theta(-\Delta \tau)$ into $\Theta(\Delta \tau)$, as any extra term is also a vanishing cut. In other words, for the cut contributions with cuts of valence three, the $i\varepsilon$-prescription does not matter, in line with the intuition that the $i\varepsilon$-prescription of tree propagators can be dropped. In particular, we can choose to write $F_{\scalebox{0.8}{\rotatebox[origin=c]{-90}{$\boxminus$}},3}^{R,1}$ as
\begin{align}
F_{\scalebox{0.8}{\rotatebox[origin=c]{-90}{$\boxminus$}},3}^{R,1}=\int \mathrm{d}\Pi_3 \mathrm{d}\Pi_3'&\big(-E_{12}E_{23}'\Theta(\Delta\tau_{12})\Theta(\Delta\tau_{23}')G_F(q_1)\tilde{\delta}^-(q_2)G_F(q_3) \nonumber\\
-&E_{12}'E_{23}\Theta(\Delta \tau_{12}')\Theta(\Delta \tau_{23})G_F(q_1)\tilde{\delta}^+(q_2)G_F(q_3)  \nonumber\\
+&E_{13}E_{23}'\Theta(\Delta \tau_{13})\Theta(-\Delta \tau_{23}')G_F(q_1)G_F(q_2)\tilde{\delta}^+(q_3)  \nonumber\\
+&E_{13}'E_{23}\Theta(\Delta \tau_{13}')\Theta(-\Delta \tau_{23})G_F(q_1)G_F(q_2)\tilde{\delta}^-(q_3)  \nonumber\\
-&E_{13}E_{12}'\Theta(\Delta \tau_{13})\Theta(\Delta \tau_{12}')\tilde{\delta}^+(q_1)G_F(q_2)G_F(q_3)  \nonumber\\
-&E_{13}'E_{12}\Theta(\Delta \tau_{13}')\Theta(\Delta \tau_{12})\tilde{\delta}^-(q_1)G_F(q_2)G_F(q_3)
\big)
\end{align}
We are now ready to combine contributions. Summing $F_{\scalebox{0.8}{\rotatebox[origin=c]{90}{$\boxminus$}}}^{R,1}$ and $F_{\scalebox{0.8}{\rotatebox[origin=c]{-90}{$\slashed{\boxminus}$}},2}^{R,1}$ and turning conjugate Feynman propagators into ordinary Feynman propagators according to the identity $G_F(q)^\star=G_F(q)+i\tilde{\delta}(q^2)$, and dropping all zero-measured cuts, we obtain
\begin{align}
F_{\scalebox{0.8}{\rotatebox[origin=c]{90}{$\boxminus$}}}^{R,1}+&F_{\scalebox{0.8}{\rotatebox[origin=c]{-90}{$\slashed{\boxminus}$}},2}^{R,1}=\int \mathrm{d}\Pi_3 \mathrm{d}\Pi_3' G_F(q_1)G_F(q_2)G_F(q_3)i\Big[E_{12}E_{13}\Theta(\Delta\tau_{13})\Theta(\Delta\tau_{12})-E_{23}E_{13}\Theta(-\Delta\tau_{23})\Theta(\Delta\tau_{13}) \nonumber\\
+&E_{12}E_{23}\Theta(\Delta\tau_{23})\Theta(\Delta\tau_{12})+E_{12}'E_{13}\Theta(\Delta\tau_{12}')\Theta(\Delta\tau_{13})+E_{12}'E_{23}\Theta(\Delta\tau_{12}')\Theta(\Delta\tau_{23})  \nonumber\\
-&E_{13}E_{23}'\Theta(\Delta\tau_{13})\Theta(-\Delta\tau_{23}') -E_{23}E_{23}'\Theta(\Delta\tau_{23})\Theta(-\Delta\tau_{23}') +E_{12}E_{13}'\Theta(\Delta\tau_{12})\Theta(\Delta\tau_{13}') \nonumber\\
-&E_{23}E_{13}'\Theta(-\Delta\tau_{23})\Theta(\Delta\tau_{13}') +E_{12}'E_{13}'\Theta(\Delta\tau_{12}')\Theta(\Delta\tau_{13}') -E_{23}'E_{13}'\Theta(-\Delta\tau_{23}')\Theta(\Delta\tau_{13}')  \nonumber\\
+&E_{12}E_{23}'\Theta(\Delta\tau_{23}')\Theta(\Delta\tau_{12}) -E_{23}E_{23}'\Theta(-\Delta\tau_{23})\Theta(\Delta\tau_{23}')+E_{12}'E_{23}'\Theta(\Delta\tau_{23}')\Theta(\Delta\tau_{12}')\Big]  \nonumber\\
-&iE_{23}E_{23}'G_F(q_1)G_F(q_2)G_F(q_3)[\Theta(\Delta\tau_{23})\Theta(-\Delta\tau_{23}')+\Theta(-\Delta\tau_{23})\Theta(\Delta\tau_{23}')] \\
+&iE_{23}E_{23}G_F(q_1)(iG_F(q_2)\tilde{\delta}(q_1^2)+iG_F(q_1)\tilde{\delta}(q_2^2)-\tilde{\delta}(q_1^2)\tilde{\delta}(q_2^2))\Theta(-\Delta\tau_{23})\Theta(-\Delta\tau_{23}')
\end{align}
where we used the identity $G_R(\pm q)=G_F(q)-\delta^\pm (q)$ to obtain a compact form for the contributions in the last line. This quantity scales classically. Adding the valence-three cuts and using the identity $G_R(\pm q)=G_F(q)-\delta^\pm (q)$, we write
\begin{align}
F^1_{\scalebox{0.8}{\rotatebox[origin=c]{90}{$\boxminus$}}}=&i\int \mathrm{d}\Pi_3 \mathrm{d}\Pi_3'\Big[E_{12}E_{23}'\Theta(\Delta \tau_{12})\Theta(\Delta \tau_{23}')G_F(q_1)G_R(-q_2)G_F(q_3)  \nonumber\\
+&E_{12}'E_{23}\Theta(\Delta \tau_{12}')\Theta(\Delta \tau_{23})G_F(q_1)G_R(q_2)G_F(q_3)  \nonumber\\
-&E_{13}E_{23}'\Theta(\Delta \tau_{13})\Theta(-\Delta \tau_{23}')G_F(q_1)G_F(q_2)G_R(q_3)  \nonumber\\
-&E_{13}'E_{23}\Theta(-\Delta \tau_{13}')\Theta(\Delta \tau_{23})G_F(q_1)G_F(q_2)G_R(-q_3) \nonumber \\
+&E_{13}E_{12}'\Theta(\Delta \tau_{13})\Theta(\Delta \tau_{12}')G_R(q_1)G_F(q_2)G_F(q_3)  \nonumber\\
+&E_{13}'E_{12}\Theta(\Delta \tau_{13}')\Theta(\Delta \tau_{12})G_R(-q_1)G_F(q_2)G_F(q_3)  \nonumber\\
+&\big[E_{12}E_{13}\Theta(\Delta\tau_{13})\Theta(\Delta\tau_{12})-E_{23}E_{13}\Theta(-\Delta\tau_{23})\Theta(\Delta\tau_{13})+E_{12}E_{23}\Theta(\Delta\tau_{23})\Theta(\Delta\tau_{12}) \nonumber\\
+&E_{12}'E_{13}'\Theta(\Delta\tau_{12}')\Theta(\Delta\tau_{13}')-E_{23}'E_{13}'\Theta(-\Delta\tau_{23}')\Theta(\Delta\tau_{13}')+E_{12}'E_{23}'\Theta(\Delta\tau_{23}')\Theta(\Delta\tau_{12}')\big]\times  \nonumber\\
\times& G_F(q_1)G_F(q_2)G_F(q_3)\Big]  \nonumber\\
-&iE_{23}E_{23}'[-(G_F(q_1)G_R(q_2)G_R(-q_3)\Theta(\Delta\tau_{23})\Theta(-\Delta\tau_{23}')+G_F(q_1)G_R(-q_2)G_R(q_3)\Theta(-\Delta\tau_{23})\Theta(\Delta\tau_{23}'))]. 
\end{align}
Mapping $q_1$ to $q_3$ and integrating back gives the twelve contributions listed in eq.~\eqref{eq:classical_limit_scalar} plus the two contributions listed in eq.~\eqref{eq:cycle_contr} (We are free to turn the remaining Feynman propagators of the expression above into retarded propagators with whichever prescription, as the difference gives zero-measured cuts).

\section{Tests at higher loops}

To test the robustness of the approach, we generated automatically the Schwinger-parametrised classical integrand for the three and four-loop ladders in scalar QED in the potential region, and checked that the cancellation of super-leading contributions takes place. Specifically, defining the set $J=\{1,...,n\}$ the integrand for the ladder equivalence class at $n-1$ loops can be conveniently written in the following way
\begin{equation}
F^{\text{KMOC}}=\frac{\prod_{j=1}^n G_F(q_j)}{n!} \sum_{\substack{I\subseteq \{1,...,n\} \\
I\neq\emptyset}} \left(\sum_{i\in I}q_i^\mu\right)\mathcal{L}(\{v_i\}_{i\in I}, \{v_i\}_{i\in J\setminus I}, p_1)\mathcal{L}(\{w_i\}_{i\in I}, \{w_i\}_{i\in J\setminus I}, -p_2).
\end{equation}
where the planar ladder is defined by the following vertex and (internal) edge sets
\begin{equation}
V=\{v_i\}_{i\in J} \cup \{w_i\}_{i\in J}, \quad E=\{\{v_i,v_{i+1}\}\}_{i=1}^{n-1} \cup \{\{w_i,w_{i+1}\}\}_{i=1}^{n-1}\cup \{\{v_i,w_i\}\}_{i=1}^n,
\end{equation}
and satisfies that $L=n-1$. The KMOC integrand is directly written in Schwinger-parameter space by using the expanded form of the worldline (eq.~\eqref{eq:simplified_QED}) for each of the two massive lines:
\begin{align}
F^{\text{KMOC}}=\mathcal{O}(\lambda^{-3n+1})+\frac{\prod_{j=1}^n G_F(q_j)}{n!} \int \mathrm{d}\Pi_n(p_1) &\mathrm{d}\Pi_n(-p_2)\sum_{\substack{I\subseteq \{1,...,n\}  \nonumber\\
I\neq\emptyset}}\sum_{\substack{m,m'=0 \\ m+m'\le n}}^n \left(\sum_{i\in I}q_i^\mu\right)\times \\ &\times\ell^{(m)}(\{v_i\}_{i\in I}, \{v_i\}_{i\in J\setminus I}, p_1)\ell^{(m')}(\{w_i\}_{i\in I}, \{w_i\}_{i\in J\setminus I}, -p_2).
\end{align}
At this point, one may check that the cancellation of super-leading contributions has taken place locally. In other words, one ensures that
\begin{equation}
\sum_{\substack{I\subseteq \{1,...,n\} \\
I\neq\emptyset}}\sum_{\substack{m,m'=0 \\ m+m'\le n}}^n \left(\sum_{i\in I}q_i^\mu\right)\ell^{(m)}(\{v_i\}_{i\in I}, \{v_i\}_{i\in J\setminus I}, p_1)\ell^{(m')}(\{w_i\}_{i\in I}, \{w_i\}_{i\in J\setminus I}, -p_2)=\mathcal{O}(\lambda^{-{n}}).
\end{equation}
We checked this fact for the three and four-loop ladders in scalar QED.

\section{WQFT integrand for scalar model and electrodynamics}
\label{sec:wqft}
\subsection{Feynman rules}
We refer readers to the original paper \cite{Mogull:2020sak} for the
basics of WQFT. Here we summarise the Feynman rules in the models we
use.

We consider a scalar particle of mass $m$ coupled to a
massless scalar field $\phi$ which mediates an attractive interaction
between the massive scalar particles. The Lagrangian is, with a
coupling constant $\lambda$ and spacetime dimension $d$,
\begin{equation}
  \label{eq:scalar_wqft_lag}
  \mathcal L = \int d \tau \left( -\frac m 2 \dot X^\mu(\tau) \dot
    X_\mu (\tau) - \frac 1 2 \lambda \phi(X(\tau)) \right)
  + \int d^d x \left(\frac 1 2 \partial^\mu \phi (x) \partial_\mu \phi
    (x)  \right) \, ,
\end{equation}
The position coordinates of the massive particle are $\dot X^\mu$,
dependent on the proper time, also the worldline parameter,
$\tau$. The worldline position is expanded around the free-field limit
of a straight line trajectory,
\begin{equation}
  X^\mu = b^\mu + v^\mu \cdot \tau + z^\mu \, ,
\end{equation}
where $z^\mu$, treated as a quantum field, is the small perturbation,
while the straight line trajectory $b^\mu + v^\mu \cdot \tau$ is
treated as a background source. $\phi(x(\tau))$ in the interaction
term in eq.~\eqref{eq:scalar_wqft_lag} can be Taylor expanded in terms
of $\phi(b + v \cdot \tau)$ and its spacetime derivatives. The
resulting momentum-space Feynman rules are, with the normalised Dirac
delta function notation $\tilde \delta (x) = 2 \pi \delta(x)$:
\begin{itemize}
\item Propagator between $z^\mu$ and $z^\nu$ with worldline momentum
  $\omega$:
  $$-i \frac {\eta^{\mu \nu}} {m (\omega \pm i\varepsilon)^2} \, ,$$
  where the $+$ sign is taken if $\omega$ flows in the same direction
  of the classical causality flow, and the $-$ sign is taken otherwise
  \cite{Jakobsen:2022psy}.
\item Background source (straight line trajectory) coupling to a
  massless scalar line of outgoing momentum $q$:
  $$ - \frac 1 2 i \lambda \, e^{-i q \cdot b} \tilde \delta (q \cdot v) \, .$$
\item Coupling between a massless scalar line of outgoing momentum $q$
  and $n$ worldline deviation fields $z^{\rho_i}$ of outgoing worldline
  energies $\omega_i$, where $i = 1, 2, \dots , n$:
  $$- \frac 1 2 i^{n+1} \lambda \, e^{-i q \cdot b} \tilde \delta (q \cdot v + \omega_1 +
  \omega_2 + \dots + \omega_n) q^{\rho_1} q^{\rho_2} \dots q^{\rho_n}
  \, .$$
\end{itemize}
We omit the well-known propagator for the massless scalar field
$\phi$. Note that the retarded needs to be used when one takes both
conservative and dissipative effects into account \cite{Jakobsen:2022psy}.

The Lagrangian for a charged scalar particle of mass $m$ coupled to the
electromagnetic field is, with spacetime dimension $d$,
\begin{equation}
  \label{eq:qed_wqft_lag}
  \mathcal L = \int d \tau \left( -\frac m 2 \dot X^\mu(\tau) \dot
    X_\mu (\tau) - e \dot  X_\mu (\tau) A^\mu (X(\tau)) \right)
  + \int d^d x \left( - \frac 1 4 F^{\mu \nu} (x) F_{\mu \nu} (x)
  \right) \, ,
\end{equation}
with $F^{\mu \nu} = \partial^\mu A^\nu - \partial^\nu A^\mu$.
The worldline propagator is unchanged from the previous model, and we
give the photon couplings to background sources and worldline
deviation fields:
\begin{itemize}
\item Background source (straight line trajectory) coupling to a
  photon of outgoing momentum $q$ and polarisation index $\mu$:
  $$ -i \lambda \, e^{-i q \cdot b} \tilde \delta (q \cdot v) v^\mu \, .$$
\item Coupling between a photon, with outgoing momentum $q$ and
  polarisation index $\mu$, and $n$ worldline deviation fields
  $z^{\rho_i}$ of outgoing worldline energies $\omega_i$, where
  $i = 1, 2, \dots , n$:
  \begin{align}
    & - i^{n+1}\lambda \, e^{-i q \cdot b} \tilde \delta (q \cdot v + \omega_1 +
    \omega_2 + \dots + \omega_n) \nonumber \\
    & \quad \times \left[
      v^\mu \prod_{i=1}^n q^{\rho_i} + \sum_{j=1}^n
      \omega_j \, \eta^{\rho_j \mu} \prod_{1 \leq i \leq n, i \neq j}
      q^{\rho_i}\right] \, .
    \label{eq:zeroEnergyFactorization}
  \end{align}
\end{itemize}

\subsection{Impulse observables and factorisation of zero-energy vertices}

Impulse observables in WQFT are one-point functions with an amputated
zero-energy external leg for the $z^\mu$ worldline deflection field
multiplied by an extra factor of $i$ \cite{Mogull:2020sak}.

\begin{figure}[h]
  \centering
  \includegraphics[width=0.5\textwidth]{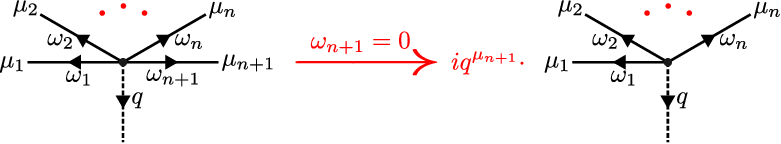}
  \caption{Zero-energy factorisation of WQFT vertices for the scalar
    model and electrodynamics. Note that the solid lines (worldline
    deflection fields are unordered, so the removed line can as well
    be a line other than the $(n+1)$-th line.}
  \label{fig:zeroEnergyFactorization}
\end{figure}

For both the scalar model and electrodynamics,
the $(n+1)$-point vertex, coupling $n$ worldline deflection field with
one bulk field (massless scalar or photon), factorise when one of the
external worldline deflection fields have zero energy, shown in
fig.~\ref{fig:zeroEnergyFactorization}.

In fact, this factorisation holds for any theory in the WQFT framework,
as shown in ref.~\cite{Jakobsen:2021zvh}.

\subsection{One-loop integrands}

\subsubsection{Scalar model}
We present the integrand for the impulse received by particle 1 of
mass $m_1$ and initial normalised velocity $u_1$, with an impact
parameter, i.e.\ the relative transverse displacement of particle 1
w.r.t.\ particle 2 in the far past, being $b$.

\begin{figure}[h]
  \centering
  \includegraphics[width=0.22\textwidth]{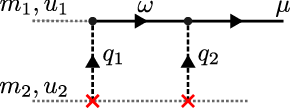}
  \caption{One of the two WQFT Feynman diagrams, at one loop, for
    the impulse of particle 1 in the scalar model
    eq.~\eqref{eq:scalar_wqft_lag}. This diagram dominates in the
    limit $m_2 \gg m_1$.}
  \label{fig:oneLoopDiag1}
\end{figure}

\begin{figure}[h]
  \centering
  \includegraphics[width=0.22\textwidth]{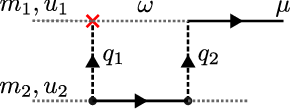}
  \caption{One of the two WQFT Feynman diagrams, at one loop, for
    the impulse of particle 1 in the scalar model
    eq.~\eqref{eq:scalar_wqft_lag}. This diagram dominates in the
    limit $m_1 \gg m_2$.}
  \label{fig:oneLoopDiag2}
\end{figure}

There are two one-loop diagrams. The one that dominates in the probe
limit $m_2 \gg m_1$ is fig.~\ref{fig:oneLoopDiag1}. The integrand
from the diagram is
\begin{align}
  & \qquad e^{i (q_1+q_2) \cdot b} \frac{i}{q_1^2} \frac{i}{q_2^2}
    \frac{-i}{(\omega + i\varepsilon)^2} \left( \frac 1 2 \right)^4
           \nonumber \\
  & \quad \times (-i \lambda) \tilde \delta(q_1 \cdot u_2) (-i
  \lambda) \tilde \delta(q_2 \cdot u_2)  \nonumber \\
  & \quad \times (-\lambda) \tilde \delta(-q_1.u_1 + \omega) (-q_1^\nu)  (-\lambda)
           \tilde \delta(-q_2.u_1 - \omega) (-q_{2 \nu}) \nonumber \\
  & \quad \times i (-i q_2^\mu) \label{eq:oneLoopIntegrand1a} \\
  & = -i \left( \frac \lambda 2 \right)^4 e^{i (q_1+q_2) \cdot b} \tilde \delta((q_1+q_2) \cdot u_1)
    \tilde \delta((q_1+q_2) \cdot u_2) \frac {1}{q_1^2 q_2^2 (q_1 \cdot u_1
    + i\varepsilon)^2} \nonumber \\
  & \quad \times (q_1 \cdot q_2) \tilde \delta(q_1 \cdot u_2)
    q_2^\mu \, .
  \label{eq:oneLoopIntegrand1}
\end{align}
where $\nu$ is the Lorentz index of the internal worldline deviation
line, and we exploited the factorisation of the zero-energy external
leg due to eq.~\eqref{eq:zeroEnergyFactorization} to write down the
integrand as a product of the diagram without the external leg on the
top right and the extra factor $(-i q_2^\mu)$. Another factor of $i$
in the line eq.~\eqref{eq:oneLoopIntegrand1a} is the necessary extra
factor for the impulse observable in WQFT.
Similarly, the diagram fig.~\ref{fig:oneLoopDiag2} which dominates in
the opposite probe limit $m_1 \gg m_2$ gives the integrand
\begin{align}
  \label{eq:oneLoopIntegrand2}
  & \qquad e^{i (q_1+q_2) \cdot b} \frac{i}{q_1^2} \frac{i}{q_2^2}
    \frac{-i}{(\omega + i\varepsilon)^2} \left( \frac 1 2 \right)^4
    \nonumber \\
  & \quad \times (-i \lambda) \tilde \delta(-q_1 \cdot u_1) (-i
           \lambda) \tilde \delta(-q_2 \cdot u_1) \nonumber \\
  & \quad \times (-\lambda) \tilde \delta(-q_1.u_2 - \omega) q_1^\nu  (-\lambda)
           \tilde \delta(-q_2.u_2 + \omega) q_{2 \nu} \nonumber \\
  & \quad \times i (-i q_2^\mu) \nonumber \\
  & = - i \left( \frac \lambda 2 \right)^4 e^{i (q_1+q_2) \cdot b} \tilde \delta((q_1+q_2) \cdot u_1)
    \tilde \delta((q_1+q_2) \cdot u_2) \frac {1}{q_1^2 q_2^2 (-q_1 \cdot u_2
    + i\varepsilon)^2} \nonumber \\
  & \quad \times (q_1 \cdot q_2) \tilde \delta(q_1 \cdot u_1) q_2^\mu \, .
\end{align}

\subsubsection{Electrodynamics}
\label{sec:worldline_QED_1loop}

For scalar QED, the diagram analogous to fig.~\ref{fig:oneLoopDiag1},
but with massless scalar lines replaced by massless photons and with
charges $e_1$ and $e_2$ for the two matter lines, gives the integrand
\begin{align}
  \label{eq:oneLoopQEDIntegrand1}
  & \quad -i e_1^2 e_2^2  e^{i (q_1+q_2) \cdot b} \tilde \delta((q_1+q_2) \cdot u_1)
    \tilde \delta((q_1+q_2) \cdot u_2) \frac {1}{q_1^2 q_2^2 (q_1 \cdot u_1
    + i\varepsilon)^2} \nonumber \\
  & \times \left[ (u_1 \cdot u_2)^2 q_1 \cdot q_2 - (q_1 \cdot u_1)^2 \right] \tilde \delta(q_1 \cdot u_2) q_2^\mu \, .
\end{align}
Similarly, the QED diagram analogous to fig.~\ref{fig:oneLoopDiag2}
gives the integrand
\begin{align}
  \label{eq:oneLoopQEDIntegrand2}
  &\quad -i e_1^2 e_2^2  e^{i (q_1+q_2) \cdot b} \tilde \delta((q_1+q_2) \cdot u_1)
    \tilde \delta((q_1+q_2) \cdot u_2) \frac {1}{q_1^2 q_2^2 (-q_1 \cdot u_2
    + i\varepsilon)^2} \nonumber \\
  & \times \left[ (u_1 \cdot u_2)^2 q_1 \cdot q_2 - (-q_1 \cdot u_2)^2 \right] \tilde \delta(q_1 \cdot u_1) q_2^\mu \, .
\end{align}

\subsection{Two-loop integrands}
\subsubsection{Scalar model}
We only give the integrands for some sample WQFT Feynman diagrams at
two loops for the purpose of illustration.

\begin{figure}[h]
  \centering
  \includegraphics[width=0.3\textwidth]{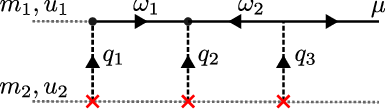}
  \caption{One of the WQFT Feynman diagrams, at two loops, for
    the impulse of particle 1 in the scalar model
    eq.~\eqref{eq:scalar_wqft_lag}.}
  \label{fig:twoLoopDiag1}
\end{figure}

\begin{figure}[h]
  \centering
  \includegraphics[width=0.25\textwidth]{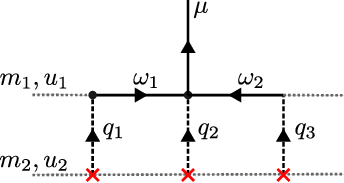}
  \caption{One of the WQFT Feynman diagrams, at two loops, for
    the impulse of particle 1 in the scalar model
    eq.~\eqref{eq:scalar_wqft_lag}.}
  \label{fig:twoLoopDiag2}
\end{figure}

\begin{figure}[h]
  \centering
  \includegraphics[width=0.29\textwidth]{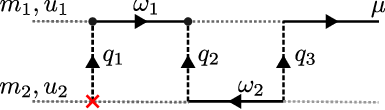}
  \caption{One of the WQFT Feynman diagrams, at two loops, for
    the impulse of particle 1 in the scalar model
    eq.~\eqref{eq:scalar_wqft_lag}.}
  \label{fig:twoLoopDiag3}
\end{figure}

\begin{figure}[h]
  \centering
  \includegraphics[width=0.3\textwidth]{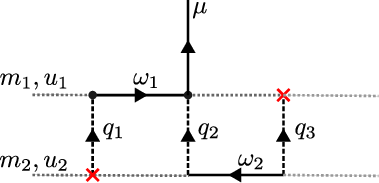}
  \caption{One of the WQFT Feynman diagrams, at two loops, for
    the impulse of particle 1 in the scalar model
    eq.~\eqref{eq:scalar_wqft_lag}.}
  \label{fig:twoLoopDiag4}
\end{figure}

For the scalar model, the probe-limit diagram
fig.~\ref{fig:twoLoopDiag1} gives the integrand
\begin{align}
  \label{eq:twoLoopIntegrand1}
  & \quad -i \left( \frac \lambda 2 \right)^6  e^{i (q_1+q_2+q_3) \cdot b} \tilde \delta((q_1+q_2+q_3) \cdot u_1)
    \tilde \delta((q_1+q_2+q_3) \cdot u_2) \frac {1}{q_1^2 q_2^2 q_3^2 (q_1 \cdot u_1
    + i\varepsilon)^2 (q_3 \cdot u_1 - i\varepsilon)^2} \nonumber \\
  & \times \left[ (q_1 \cdot q_2) (q_2 \cdot q_3) \right]
    \tilde \delta(q_1 \cdot u_2) \tilde \delta(q_3 \cdot u_2) q_3^\mu \, .
\end{align}
Another probe limit diagram, fig.~\ref{fig:twoLoopDiag2}, with the
external worldline field attached to the $q_2$ line, gives an
integrand that is similar but slightly different in several respects:
(1) there is a symmetry factor of $1/2$, (2) $i q_3^\mu$ is replaced by
$i q_2^\mu$, and (3) an opposite $i\varepsilon$ prescription is used for a
worldline propagator due to a change in the causality flow. The
expression is
\begin{align}
  \label{eq:twoLoopIntegrand2}
  & \quad -\frac i 2 \left( \frac \lambda 2 \right)^6  e^{i (q_1+q_2+q_3) \cdot b} \tilde \delta((q_1+q_2+q_3) \cdot u_1)
    \tilde \delta((q_1+q_2+q_3) \cdot u_2) \frac {1}{q_1^2 q_2^2 q_3^2 (q_1 \cdot u_1
    + i\varepsilon)^2 (q_3 \cdot u_1 + i\varepsilon)^2} \nonumber \\
  & \times \left[ (q_1 \cdot q_2) (q_2 \cdot q_3) \right]
    \tilde \delta(q_1 \cdot u_2) \tilde \delta(q_3 \cdot u_2) q_2^\mu \, .
\end{align}
A diagram that does not contribute to the probe limit but to the
``first-self-force'' correction is fig.~\ref{fig:twoLoopDiag3}. The
resulting integrand is
\begin{align}
  \label{eq:twoLoopIntegrand3}
  & \quad -i \frac i 2 \left( \frac \lambda 2 \right)^6  e^{i (q_1+q_2+q_3) \cdot b} \tilde \delta((q_1+q_2+q_3) \cdot u_1)
    \tilde \delta((q_1+q_2+q_3) \cdot u_2) \frac {1}{q_1^2 q_2^2 q_3^2 (q_1 \cdot u_1
    + i\varepsilon)^2 (q_3 \cdot u_2 - i\varepsilon)^2} \nonumber \\
  & \times \left[ (q_1 \cdot q_2) (q_2 \cdot q_3) \right]
    \tilde \delta(q_1 \cdot u_2) \tilde \delta(q_3 \cdot u_1) q_3^\mu \, .
\end{align}
If the external line is attached to $q_2$ instead, the diagram is
fig.~\ref{fig:twoLoopDiag4}, giving a similar integrand except for
changing $i q_3^\mu$ to $i q_2^\mu$ and flipping the $i\varepsilon$ prescription
of a worldline propagator to follow the altered causality flow,
\begin{align}
  \label{eq:twoLoopIntegrand4}
  & \quad -i \frac i 2 \left( \frac \lambda 2 \right)^6  e^{i (q_1+q_2+q_3) \cdot b} \tilde \delta((q_1+q_2+q_3) \cdot u_1)
    \tilde \delta((q_1+q_2+q_3) \cdot u_2) \frac {1}{q_1^2 q_2^2 q_3^2 (q_1 \cdot u_1
    + i\varepsilon)^2 (q_3 \cdot u_2 + i\varepsilon)^2} \nonumber \\
  & \times \left[ (q_1 \cdot q_2) (q_2 \cdot q_3) \right]
    \tilde \delta(q_1 \cdot u_2) \tilde \delta(q_3 \cdot u_1) q_2^\mu \, .
\end{align}
The external line can also be attached $q_1$, and the integrand is not
written here but can be easily obtained from the Feynman rules. Other
two-loop diagrams, such as ``mushroom diagrams'' which contribute to
dissipative physics, are also omitted.

\subsubsection{Electrodynamics}
The diagram analogous to fig.~\ref{fig:twoLoopDiag1}, with massless
scalars replaced by photons, gives the integrand
\begin{align}
  \label{eq:twoLoopQEDIntegrand1}
  & \quad i e_1^3 e_2^3  e^{i (q_1+q_2+q_3) \cdot b} \tilde \delta((q_1+q_2+q_3) \cdot u_1)
    \tilde \delta((q_1+q_2+q_3) \cdot u_2) \frac {1}{q_1^2 q_2^2 q_3^2 (q_1 \cdot u_1
    + i\varepsilon)^2 (q_3 \cdot u_1 - i\varepsilon)^2} \nonumber \\
  & \times \left[ (u_1 \cdot u_2)^3 (q_1 \cdot q_2) (q_2 \cdot q_3)
    - (u_1 \cdot u_2) (q_1 \cdot u_1)^2 q_2 \cdot q_3
    - (u_1 \cdot u_2) (q_3 \cdot u_1)^2 q_1 \cdot q_2 \right]
    \tilde \delta(q_1 \cdot u_2) \tilde \delta(q_3 \cdot u_2) q_3^\mu \, .
\end{align}
As in the scalar model, the diagram analogous to
fig.~\ref{fig:twoLoopDiag2} gives a similar integrand with a few
changes: (1) there is a symmetry factor of $1/2$, (2) $i q_3^\mu$ is
replaced by $i q_2^\mu$, and (3) an opposite $i\varepsilon$ prescription is used
for a worldline propagator due to a change in the causality flow. We
do not write down the expression here.

The diagram analogous to fig.~\ref{fig:twoLoopDiag3} gives the integrand
\begin{align}
  \label{eq:twoLoopQEDIntegrand3}
  & i e_1^3 e_2^3  e^{i (q_1+q_2+q_3) \cdot b} \tilde \delta((q_1+q_2+q_3) \cdot u_1)
    \tilde \delta((q_1+q_2+q_3) \cdot u_2) \frac {1}{q_1^2 q_2^2 q_3^2 (q_1 \cdot u_1
    + i\varepsilon)^2 (q_3 \cdot u_2 - i\varepsilon)^2} \nonumber \\
  & \times \big [ (u_1 \cdot u_2)^3 (q_1 \cdot q_2) (q_2 \cdot q_3)
    - (u_1 \cdot u_2) (q_1 \cdot u_1)^2 q_2 \cdot q_3
    - (u_1 \cdot u_2) (q_3 \cdot u_2)^2 q_1 \cdot q_2
    \nonumber \\
  & \quad + (u_1 \cdot u_2)^2 (q_1 \cdot u_1) (q_3 \cdot u_2) (q_1
    \cdot q_2 + q_2 \cdot q_3 + q_3 \cdot q_1) \nonumber \\
  & \quad -  (q_1 \cdot u_1)^3 (q_3 \cdot u_2) -  (q_1 \cdot u_1) (q_3 \cdot u_2)^3
    \big] \nonumber \\
  & \times \tilde \delta(q_1 \cdot u_2) \tilde \delta(q_3 \cdot u_1) q_3^\mu \, .
\end{align}
As in the scalar model, the diagram analogous to
fig.~\ref{fig:twoLoopDiag4} gives a similar integrand except for
changing $i q_3^\mu$ to $i q_2^\mu$ and flipping the $i\varepsilon$ prescription
of a worldline propagator to follow the altered causality flow. We
omit the expression here.

\flushbottom

\bibliography{biblio}

\begin{thebibliography}{102}%
\makeatletter
\providecommand \@ifxundefined [1]{%
 \@ifx{#1\undefined}
}%
\providecommand \@ifnum [1]{%
 \ifnum #1\expandafter \@firstoftwo
 \else \expandafter \@secondoftwo
 \fi
}%
\providecommand \@ifx [1]{%
 \ifx #1\expandafter \@firstoftwo
 \else \expandafter \@secondoftwo
 \fi
}%
\providecommand \natexlab [1]{#1}%
\providecommand \enquote  [1]{``#1''}%
\providecommand \bibnamefont  [1]{#1}%
\providecommand \bibfnamefont [1]{#1}%
\providecommand \citenamefont [1]{#1}%
\providecommand \href@noop [0]{\@secondoftwo}%
\providecommand \href [0]{\begingroup \@sanitize@url \@href}%
\providecommand \@href[1]{\@@startlink{#1}\@@href}%
\providecommand \@@href[1]{\endgroup#1\@@endlink}%
\providecommand \@sanitize@url [0]{\catcode `\\12\catcode `\$12\catcode `\&12\catcode `\#12\catcode `\^12\catcode `\_12\catcode `\%12\relax}%
\providecommand \@@startlink[1]{}%
\providecommand \@@endlink[0]{}%
\providecommand \url  [0]{\begingroup\@sanitize@url \@url }%
\providecommand \@url [1]{\endgroup\@href {#1}{\urlprefix }}%
\providecommand \urlprefix  [0]{URL }%
\providecommand \Eprint [0]{\href }%
\providecommand \doibase [0]{https://doi.org/}%
\providecommand \selectlanguage [0]{\@gobble}%
\providecommand \bibinfo  [0]{\@secondoftwo}%
\providecommand \bibfield  [0]{\@secondoftwo}%
\providecommand \translation [1]{[#1]}%
\providecommand \BibitemOpen [0]{}%
\providecommand \bibitemStop [0]{}%
\providecommand \bibitemNoStop [0]{.\EOS\space}%
\providecommand \EOS [0]{\spacefactor3000\relax}%
\providecommand \BibitemShut  [1]{\csname bibitem#1\endcsname}%
\let\auto@bib@innerbib\@empty
\bibitem [{\citenamefont {Abbott}\ \emph {et~al.}(2016)\citenamefont {Abbott} \emph {et~al.}}]{LIGOScientific:2016aoc}%
  \BibitemOpen
  \bibfield  {author} {\bibinfo {author} {\bibfnamefont {B.~P.}\ \bibnamefont {Abbott}} \emph {et~al.} (\bibinfo {collaboration} {LIGO Scientific, Virgo}),\ }\bibfield  {title} {\bibinfo {title} {{Observation of Gravitational Waves from a Binary Black Hole Merger}},\ }\href {https://doi.org/10.1103/PhysRevLett.116.061102} {\bibfield  {journal} {\bibinfo  {journal} {Phys. Rev. Lett.}\ }\textbf {\bibinfo {volume} {116}},\ \bibinfo {pages} {061102} (\bibinfo {year} {2016})},\ \Eprint {https://arxiv.org/abs/1602.03837} {arXiv:1602.03837 [gr-qc]} \BibitemShut {NoStop}%
\bibitem [{\citenamefont {Punturo}\ \emph {et~al.}(2010)\citenamefont {Punturo} \emph {et~al.}}]{Punturo_2010}%
  \BibitemOpen
  \bibfield  {author} {\bibinfo {author} {\bibfnamefont {M.}~\bibnamefont {Punturo}} \emph {et~al.},\ }\bibfield  {title} {\bibinfo {title} {The einstein telescope: a third-generation gravitational wave observatory},\ }\href {https://doi.org/10.1088/0264-9381/27/19/194002} {\bibfield  {journal} {\bibinfo  {journal} {Classical and Quantum Gravity}\ }\textbf {\bibinfo {volume} {27}},\ \bibinfo {pages} {194002} (\bibinfo {year} {2010})}\BibitemShut {NoStop}%
\bibitem [{\citenamefont {Amaro-Seoane}\ \emph {et~al.}(2017)\citenamefont {Amaro-Seoane} \emph {et~al.}}]{LISA:2017pwj}%
  \BibitemOpen
  \bibfield  {author} {\bibinfo {author} {\bibfnamefont {P.}~\bibnamefont {Amaro-Seoane}} \emph {et~al.} (\bibinfo {collaboration} {LISA}),\ }\bibfield  {title} {\bibinfo {title} {{Laser Interferometer Space Antenna}},\ }\href@noop {} {\  (\bibinfo {year} {2017})},\ \Eprint {https://arxiv.org/abs/1702.00786} {arXiv:1702.00786 [astro-ph.IM]} \BibitemShut {NoStop}%
\bibitem [{\citenamefont {Reitze}\ \emph {et~al.}(2019)\citenamefont {Reitze} \emph {et~al.}}]{Reitze:2019iox}%
  \BibitemOpen
  \bibfield  {author} {\bibinfo {author} {\bibfnamefont {D.}~\bibnamefont {Reitze}} \emph {et~al.},\ }\bibfield  {title} {\bibinfo {title} {{Cosmic Explorer: The U.S. Contribution to Gravitational-Wave Astronomy beyond LIGO}},\ }\href@noop {} {\bibfield  {journal} {\bibinfo  {journal} {Bull. Am. Astron. Soc.}\ }\textbf {\bibinfo {volume} {51}},\ \bibinfo {pages} {035} (\bibinfo {year} {2019})},\ \Eprint {https://arxiv.org/abs/1907.04833} {arXiv:1907.04833 [astro-ph.IM]} \BibitemShut {NoStop}%
\bibitem [{\citenamefont {P\"urrer}\ and\ \citenamefont {Haster}(2020)}]{Purrer:2019jcp}%
  \BibitemOpen
  \bibfield  {author} {\bibinfo {author} {\bibfnamefont {M.}~\bibnamefont {P\"urrer}}\ and\ \bibinfo {author} {\bibfnamefont {C.-J.}\ \bibnamefont {Haster}},\ }\bibfield  {title} {\bibinfo {title} {{Gravitational waveform accuracy requirements for future ground-based detectors}},\ }\href {https://doi.org/10.1103/PhysRevResearch.2.023151} {\bibfield  {journal} {\bibinfo  {journal} {Phys. Rev. Res.}\ }\textbf {\bibinfo {volume} {2}},\ \bibinfo {pages} {023151} (\bibinfo {year} {2020})},\ \Eprint {https://arxiv.org/abs/1912.10055} {arXiv:1912.10055 [gr-qc]} \BibitemShut {NoStop}%
\bibitem [{\citenamefont {Buonanno}\ \emph {et~al.}(2022)\citenamefont {Buonanno}, \citenamefont {Khalil}, \citenamefont {O'Connell}, \citenamefont {Roiban}, \citenamefont {Solon},\ and\ \citenamefont {Zeng}}]{Buonanno:2022pgc}%
  \BibitemOpen
  \bibfield  {author} {\bibinfo {author} {\bibfnamefont {A.}~\bibnamefont {Buonanno}}, \bibinfo {author} {\bibfnamefont {M.}~\bibnamefont {Khalil}}, \bibinfo {author} {\bibfnamefont {D.}~\bibnamefont {O'Connell}}, \bibinfo {author} {\bibfnamefont {R.}~\bibnamefont {Roiban}}, \bibinfo {author} {\bibfnamefont {M.~P.}\ \bibnamefont {Solon}},\ and\ \bibinfo {author} {\bibfnamefont {M.}~\bibnamefont {Zeng}},\ }\bibfield  {title} {\bibinfo {title} {{Snowmass White Paper: Gravitational Waves and Scattering Amplitudes}},\ }in\ \href@noop {} {\emph {\bibinfo {booktitle} {{Snowmass 2021}}}}\ (\bibinfo {year} {2022})\ \Eprint {https://arxiv.org/abs/2204.05194} {arXiv:2204.05194 [hep-th]} \BibitemShut {NoStop}%
\bibitem [{\citenamefont {Bertotti}(1956)}]{Bertotti:1956pxu}%
  \BibitemOpen
  \bibfield  {author} {\bibinfo {author} {\bibfnamefont {B.}~\bibnamefont {Bertotti}},\ }\bibfield  {title} {\bibinfo {title} {{On gravitational motion}},\ }\href {https://doi.org/10.1007/bf02746175} {\bibfield  {journal} {\bibinfo  {journal} {Nuovo Cim.}\ }\textbf {\bibinfo {volume} {4}},\ \bibinfo {pages} {898} (\bibinfo {year} {1956})}\BibitemShut {NoStop}%
\bibitem [{\citenamefont {Kerr}(1959)}]{Kerr:1959zlt}%
  \BibitemOpen
  \bibfield  {author} {\bibinfo {author} {\bibfnamefont {R.~P.}\ \bibnamefont {Kerr}},\ }\bibfield  {title} {\bibinfo {title} {{The Lorentz-covariant approximation method in general relativity I}},\ }\href {https://doi.org/10.1007/bf02732767} {\bibfield  {journal} {\bibinfo  {journal} {Nuovo Cim.}\ }\textbf {\bibinfo {volume} {13}},\ \bibinfo {pages} {469} (\bibinfo {year} {1959})}\BibitemShut {NoStop}%
\bibitem [{\citenamefont {Bertotti}\ and\ \citenamefont {Plebanski}(1960)}]{Bertotti:1960wuq}%
  \BibitemOpen
  \bibfield  {author} {\bibinfo {author} {\bibfnamefont {B.}~\bibnamefont {Bertotti}}\ and\ \bibinfo {author} {\bibfnamefont {J.}~\bibnamefont {Plebanski}},\ }\bibfield  {title} {\bibinfo {title} {{Theory of gravitational perturbations in the fast motion approximation}},\ }\href {https://doi.org/10.1016/0003-4916(60)90132-9} {\bibfield  {journal} {\bibinfo  {journal} {Annals Phys.}\ }\textbf {\bibinfo {volume} {11}},\ \bibinfo {pages} {169} (\bibinfo {year} {1960})}\BibitemShut {NoStop}%
\bibitem [{\citenamefont {Westpfahl}\ and\ \citenamefont {Goller}(1979)}]{Westpfahl:1979gu}%
  \BibitemOpen
  \bibfield  {author} {\bibinfo {author} {\bibfnamefont {K.}~\bibnamefont {Westpfahl}}\ and\ \bibinfo {author} {\bibfnamefont {M.}~\bibnamefont {Goller}},\ }\bibfield  {title} {\bibinfo {title} {{Gravitational scattering of two relativistic particles in the postlinear approximation}},\ }\href {https://doi.org/10.1007/BF02817047} {\bibfield  {journal} {\bibinfo  {journal} {Lett. Nuovo Cim.}\ }\textbf {\bibinfo {volume} {26}},\ \bibinfo {pages} {573} (\bibinfo {year} {1979})}\BibitemShut {NoStop}%
\bibitem [{\citenamefont {Portilla}(1980)}]{Portilla:1980uz}%
  \BibitemOpen
  \bibfield  {author} {\bibinfo {author} {\bibfnamefont {M.}~\bibnamefont {Portilla}},\ }\bibfield  {title} {\bibinfo {title} {{Scattering of two gravitating particles: classical approach}},\ }\href {https://doi.org/10.1088/0305-4470/13/12/017} {\bibfield  {journal} {\bibinfo  {journal} {J. Phys. A}\ }\textbf {\bibinfo {volume} {13}},\ \bibinfo {pages} {3677} (\bibinfo {year} {1980})}\BibitemShut {NoStop}%
\bibitem [{\citenamefont {Bel}\ \emph {et~al.}(1981)\citenamefont {Bel}, \citenamefont {Damour}, \citenamefont {Deruelle}, \citenamefont {Ibanez},\ and\ \citenamefont {Martin}}]{Bel:1981be}%
  \BibitemOpen
  \bibfield  {author} {\bibinfo {author} {\bibfnamefont {L.}~\bibnamefont {Bel}}, \bibinfo {author} {\bibfnamefont {T.}~\bibnamefont {Damour}}, \bibinfo {author} {\bibfnamefont {N.}~\bibnamefont {Deruelle}}, \bibinfo {author} {\bibfnamefont {J.}~\bibnamefont {Ibanez}},\ and\ \bibinfo {author} {\bibfnamefont {J.}~\bibnamefont {Martin}},\ }\bibfield  {title} {\bibinfo {title} {{Poincar\'e-invariant gravitational field and equations of motion of two pointlike objects: The postlinear approximation of general relativity}},\ }\href {https://doi.org/10.1007/BF00756073} {\bibfield  {journal} {\bibinfo  {journal} {Gen. Rel. Grav.}\ }\textbf {\bibinfo {volume} {13}},\ \bibinfo {pages} {963} (\bibinfo {year} {1981})}\BibitemShut {NoStop}%
\bibitem [{\citenamefont {Cheung}\ \emph {et~al.}(2018)\citenamefont {Cheung}, \citenamefont {Rothstein},\ and\ \citenamefont {Solon}}]{Cheung:2018wkq}%
  \BibitemOpen
  \bibfield  {author} {\bibinfo {author} {\bibfnamefont {C.}~\bibnamefont {Cheung}}, \bibinfo {author} {\bibfnamefont {I.~Z.}\ \bibnamefont {Rothstein}},\ and\ \bibinfo {author} {\bibfnamefont {M.~P.}\ \bibnamefont {Solon}},\ }\bibfield  {title} {\bibinfo {title} {{From Scattering Amplitudes to Classical Potentials in the Post-Minkowskian Expansion}},\ }\href {https://doi.org/10.1103/PhysRevLett.121.251101} {\bibfield  {journal} {\bibinfo  {journal} {Phys. Rev. Lett.}\ }\textbf {\bibinfo {volume} {121}},\ \bibinfo {pages} {251101} (\bibinfo {year} {2018})},\ \Eprint {https://arxiv.org/abs/1808.02489} {arXiv:1808.02489 [hep-th]} \BibitemShut {NoStop}%
\bibitem [{\citenamefont {Kosower}\ \emph {et~al.}(2019)\citenamefont {Kosower}, \citenamefont {Maybee},\ and\ \citenamefont {O'Connell}}]{Kosower:2018adc}%
  \BibitemOpen
  \bibfield  {author} {\bibinfo {author} {\bibfnamefont {D.~A.}\ \bibnamefont {Kosower}}, \bibinfo {author} {\bibfnamefont {B.}~\bibnamefont {Maybee}},\ and\ \bibinfo {author} {\bibfnamefont {D.}~\bibnamefont {O'Connell}},\ }\bibfield  {title} {\bibinfo {title} {{Amplitudes, Observables, and Classical Scattering}},\ }\href {https://doi.org/10.1007/JHEP02(2019)137} {\bibfield  {journal} {\bibinfo  {journal} {JHEP}\ }\textbf {\bibinfo {volume} {02}},\ \bibinfo {pages} {137}},\ \Eprint {https://arxiv.org/abs/1811.10950} {arXiv:1811.10950 [hep-th]} \BibitemShut {NoStop}%
\bibitem [{\citenamefont {Bern}\ \emph {et~al.}(2019{\natexlab{a}})\citenamefont {Bern}, \citenamefont {Cheung}, \citenamefont {Roiban}, \citenamefont {Shen}, \citenamefont {Solon},\ and\ \citenamefont {Zeng}}]{Bern:2019nnu}%
  \BibitemOpen
  \bibfield  {author} {\bibinfo {author} {\bibfnamefont {Z.}~\bibnamefont {Bern}}, \bibinfo {author} {\bibfnamefont {C.}~\bibnamefont {Cheung}}, \bibinfo {author} {\bibfnamefont {R.}~\bibnamefont {Roiban}}, \bibinfo {author} {\bibfnamefont {C.-H.}\ \bibnamefont {Shen}}, \bibinfo {author} {\bibfnamefont {M.~P.}\ \bibnamefont {Solon}},\ and\ \bibinfo {author} {\bibfnamefont {M.}~\bibnamefont {Zeng}},\ }\bibfield  {title} {\bibinfo {title} {{Scattering Amplitudes and the Conservative Hamiltonian for Binary Systems at Third Post-Minkowskian Order}},\ }\href {https://doi.org/10.1103/PhysRevLett.122.201603} {\bibfield  {journal} {\bibinfo  {journal} {Phys. Rev. Lett.}\ }\textbf {\bibinfo {volume} {122}},\ \bibinfo {pages} {201603} (\bibinfo {year} {2019}{\natexlab{a}})},\ \Eprint {https://arxiv.org/abs/1901.04424} {arXiv:1901.04424 [hep-th]} \BibitemShut {NoStop}%
\bibitem [{\citenamefont {Bern}\ \emph {et~al.}(2019{\natexlab{b}})\citenamefont {Bern}, \citenamefont {Cheung}, \citenamefont {Roiban}, \citenamefont {Shen}, \citenamefont {Solon},\ and\ \citenamefont {Zeng}}]{Bern:2019crd}%
  \BibitemOpen
  \bibfield  {author} {\bibinfo {author} {\bibfnamefont {Z.}~\bibnamefont {Bern}}, \bibinfo {author} {\bibfnamefont {C.}~\bibnamefont {Cheung}}, \bibinfo {author} {\bibfnamefont {R.}~\bibnamefont {Roiban}}, \bibinfo {author} {\bibfnamefont {C.-H.}\ \bibnamefont {Shen}}, \bibinfo {author} {\bibfnamefont {M.~P.}\ \bibnamefont {Solon}},\ and\ \bibinfo {author} {\bibfnamefont {M.}~\bibnamefont {Zeng}},\ }\bibfield  {title} {\bibinfo {title} {{Black Hole Binary Dynamics from the Double Copy and Effective Theory}},\ }\href {https://doi.org/10.1007/JHEP10(2019)206} {\bibfield  {journal} {\bibinfo  {journal} {JHEP}\ }\textbf {\bibinfo {volume} {10}},\ \bibinfo {pages} {206}},\ \Eprint {https://arxiv.org/abs/1908.01493} {arXiv:1908.01493 [hep-th]} \BibitemShut {NoStop}%
\bibitem [{\citenamefont {Cristofoli}\ \emph {et~al.}(2019)\citenamefont {Cristofoli}, \citenamefont {Bjerrum-Bohr}, \citenamefont {Damgaard},\ and\ \citenamefont {Vanhove}}]{Cristofoli:2019neg}%
  \BibitemOpen
  \bibfield  {author} {\bibinfo {author} {\bibfnamefont {A.}~\bibnamefont {Cristofoli}}, \bibinfo {author} {\bibfnamefont {N.~E.~J.}\ \bibnamefont {Bjerrum-Bohr}}, \bibinfo {author} {\bibfnamefont {P.~H.}\ \bibnamefont {Damgaard}},\ and\ \bibinfo {author} {\bibfnamefont {P.}~\bibnamefont {Vanhove}},\ }\bibfield  {title} {\bibinfo {title} {{Post-Minkowskian Hamiltonians in general relativity}},\ }\href {https://doi.org/10.1103/PhysRevD.100.084040} {\bibfield  {journal} {\bibinfo  {journal} {Phys. Rev. D}\ }\textbf {\bibinfo {volume} {100}},\ \bibinfo {pages} {084040} (\bibinfo {year} {2019})},\ \Eprint {https://arxiv.org/abs/1906.01579} {arXiv:1906.01579 [hep-th]} \BibitemShut {NoStop}%
\bibitem [{\citenamefont {Bjerrum-Bohr}\ \emph {et~al.}(2020)\citenamefont {Bjerrum-Bohr}, \citenamefont {Cristofoli},\ and\ \citenamefont {Damgaard}}]{Bjerrum-Bohr:2019kec}%
  \BibitemOpen
  \bibfield  {author} {\bibinfo {author} {\bibfnamefont {N.~E.~J.}\ \bibnamefont {Bjerrum-Bohr}}, \bibinfo {author} {\bibfnamefont {A.}~\bibnamefont {Cristofoli}},\ and\ \bibinfo {author} {\bibfnamefont {P.~H.}\ \bibnamefont {Damgaard}},\ }\bibfield  {title} {\bibinfo {title} {{Post-Minkowskian Scattering Angle in Einstein Gravity}},\ }\href {https://doi.org/10.1007/JHEP08(2020)038} {\bibfield  {journal} {\bibinfo  {journal} {JHEP}\ }\textbf {\bibinfo {volume} {08}},\ \bibinfo {pages} {038}},\ \Eprint {https://arxiv.org/abs/1910.09366} {arXiv:1910.09366 [hep-th]} \BibitemShut {NoStop}%
\bibitem [{\citenamefont {Brandhuber}\ \emph {et~al.}(2021)\citenamefont {Brandhuber}, \citenamefont {Chen}, \citenamefont {Travaglini},\ and\ \citenamefont {Wen}}]{Brandhuber:2021eyq}%
  \BibitemOpen
  \bibfield  {author} {\bibinfo {author} {\bibfnamefont {A.}~\bibnamefont {Brandhuber}}, \bibinfo {author} {\bibfnamefont {G.}~\bibnamefont {Chen}}, \bibinfo {author} {\bibfnamefont {G.}~\bibnamefont {Travaglini}},\ and\ \bibinfo {author} {\bibfnamefont {C.}~\bibnamefont {Wen}},\ }\bibfield  {title} {\bibinfo {title} {{Classical gravitational scattering from a gauge-invariant double copy}},\ }\href {https://doi.org/10.1007/JHEP10(2021)118} {\bibfield  {journal} {\bibinfo  {journal} {JHEP}\ }\textbf {\bibinfo {volume} {10}},\ \bibinfo {pages} {118}},\ \Eprint {https://arxiv.org/abs/2108.04216} {arXiv:2108.04216 [hep-th]} \BibitemShut {NoStop}%
\bibitem [{\citenamefont {Bern}\ \emph {et~al.}(2021)\citenamefont {Bern}, \citenamefont {Parra-Martinez}, \citenamefont {Roiban}, \citenamefont {Ruf}, \citenamefont {Shen}, \citenamefont {Solon},\ and\ \citenamefont {Zeng}}]{Bern:2021dqo}%
  \BibitemOpen
  \bibfield  {author} {\bibinfo {author} {\bibfnamefont {Z.}~\bibnamefont {Bern}}, \bibinfo {author} {\bibfnamefont {J.}~\bibnamefont {Parra-Martinez}}, \bibinfo {author} {\bibfnamefont {R.}~\bibnamefont {Roiban}}, \bibinfo {author} {\bibfnamefont {M.~S.}\ \bibnamefont {Ruf}}, \bibinfo {author} {\bibfnamefont {C.-H.}\ \bibnamefont {Shen}}, \bibinfo {author} {\bibfnamefont {M.~P.}\ \bibnamefont {Solon}},\ and\ \bibinfo {author} {\bibfnamefont {M.}~\bibnamefont {Zeng}},\ }\bibfield  {title} {\bibinfo {title} {{Scattering Amplitudes and Conservative Binary Dynamics at ${\cal O}(G^4)$}},\ }\href {https://doi.org/10.1103/PhysRevLett.126.171601} {\bibfield  {journal} {\bibinfo  {journal} {Phys. Rev. Lett.}\ }\textbf {\bibinfo {volume} {126}},\ \bibinfo {pages} {171601} (\bibinfo {year} {2021})},\ \Eprint {https://arxiv.org/abs/2101.07254} {arXiv:2101.07254 [hep-th]} \BibitemShut {NoStop}%
\bibitem [{\citenamefont {Bern}\ \emph {et~al.}(2022{\natexlab{a}})\citenamefont {Bern}, \citenamefont {Parra-Martinez}, \citenamefont {Roiban}, \citenamefont {Ruf}, \citenamefont {Shen}, \citenamefont {Solon},\ and\ \citenamefont {Zeng}}]{Bern:2021yeh}%
  \BibitemOpen
  \bibfield  {author} {\bibinfo {author} {\bibfnamefont {Z.}~\bibnamefont {Bern}}, \bibinfo {author} {\bibfnamefont {J.}~\bibnamefont {Parra-Martinez}}, \bibinfo {author} {\bibfnamefont {R.}~\bibnamefont {Roiban}}, \bibinfo {author} {\bibfnamefont {M.~S.}\ \bibnamefont {Ruf}}, \bibinfo {author} {\bibfnamefont {C.-H.}\ \bibnamefont {Shen}}, \bibinfo {author} {\bibfnamefont {M.~P.}\ \bibnamefont {Solon}},\ and\ \bibinfo {author} {\bibfnamefont {M.}~\bibnamefont {Zeng}},\ }\bibfield  {title} {\bibinfo {title} {{Scattering Amplitudes, the Tail Effect, and Conservative Binary Dynamics at $\mathcal{O}(G^4)$}},\ }\href {https://doi.org/10.1103/PhysRevLett.128.161103} {\bibfield  {journal} {\bibinfo  {journal} {Phys. Rev. Lett.}\ }\textbf {\bibinfo {volume} {128}},\ \bibinfo {pages} {161103} (\bibinfo {year} {2022}{\natexlab{a}})},\ \Eprint {https://arxiv.org/abs/2112.10750} {arXiv:2112.10750 [hep-th]} \BibitemShut {NoStop}%
\bibitem [{\citenamefont {Damgaard}\ \emph {et~al.}(2023{\natexlab{a}})\citenamefont {Damgaard}, \citenamefont {Hansen}, \citenamefont {Plant\'e},\ and\ \citenamefont {Vanhove}}]{Damgaard:2023ttc}%
  \BibitemOpen
  \bibfield  {author} {\bibinfo {author} {\bibfnamefont {P.~H.}\ \bibnamefont {Damgaard}}, \bibinfo {author} {\bibfnamefont {E.~R.}\ \bibnamefont {Hansen}}, \bibinfo {author} {\bibfnamefont {L.}~\bibnamefont {Plant\'e}},\ and\ \bibinfo {author} {\bibfnamefont {P.}~\bibnamefont {Vanhove}},\ }\bibfield  {title} {\bibinfo {title} {{Classical observables from the exponential representation of the gravitational S-matrix}},\ }\href {https://doi.org/10.1007/JHEP09(2023)183} {\bibfield  {journal} {\bibinfo  {journal} {JHEP}\ }\textbf {\bibinfo {volume} {09}},\ \bibinfo {pages} {183}},\ \Eprint {https://arxiv.org/abs/2307.04746} {arXiv:2307.04746 [hep-th]} \BibitemShut {NoStop}%
\bibitem [{\citenamefont {K\"alin}\ and\ \citenamefont {Porto}(2020)}]{Kalin:2020mvi}%
  \BibitemOpen
  \bibfield  {author} {\bibinfo {author} {\bibfnamefont {G.}~\bibnamefont {K\"alin}}\ and\ \bibinfo {author} {\bibfnamefont {R.~A.}\ \bibnamefont {Porto}},\ }\bibfield  {title} {\bibinfo {title} {{Post-Minkowskian Effective Field Theory for Conservative Binary Dynamics}},\ }\href {https://doi.org/10.1007/JHEP11(2020)106} {\bibfield  {journal} {\bibinfo  {journal} {JHEP}\ }\textbf {\bibinfo {volume} {11}},\ \bibinfo {pages} {106}},\ \Eprint {https://arxiv.org/abs/2006.01184} {arXiv:2006.01184 [hep-th]} \BibitemShut {NoStop}%
\bibitem [{\citenamefont {K\"alin}\ \emph {et~al.}(2020{\natexlab{a}})\citenamefont {K\"alin}, \citenamefont {Liu},\ and\ \citenamefont {Porto}}]{Kalin:2020fhe}%
  \BibitemOpen
  \bibfield  {author} {\bibinfo {author} {\bibfnamefont {G.}~\bibnamefont {K\"alin}}, \bibinfo {author} {\bibfnamefont {Z.}~\bibnamefont {Liu}},\ and\ \bibinfo {author} {\bibfnamefont {R.~A.}\ \bibnamefont {Porto}},\ }\bibfield  {title} {\bibinfo {title} {{Conservative Dynamics of Binary Systems to Third Post-Minkowskian Order from the Effective Field Theory Approach}},\ }\href {https://doi.org/10.1103/PhysRevLett.125.261103} {\bibfield  {journal} {\bibinfo  {journal} {Phys. Rev. Lett.}\ }\textbf {\bibinfo {volume} {125}},\ \bibinfo {pages} {261103} (\bibinfo {year} {2020}{\natexlab{a}})},\ \Eprint {https://arxiv.org/abs/2007.04977} {arXiv:2007.04977 [hep-th]} \BibitemShut {NoStop}%
\bibitem [{\citenamefont {K\"alin}\ \emph {et~al.}(2020{\natexlab{b}})\citenamefont {K\"alin}, \citenamefont {Liu},\ and\ \citenamefont {Porto}}]{Kalin:2020lmz}%
  \BibitemOpen
  \bibfield  {author} {\bibinfo {author} {\bibfnamefont {G.}~\bibnamefont {K\"alin}}, \bibinfo {author} {\bibfnamefont {Z.}~\bibnamefont {Liu}},\ and\ \bibinfo {author} {\bibfnamefont {R.~A.}\ \bibnamefont {Porto}},\ }\bibfield  {title} {\bibinfo {title} {{Conservative Tidal Effects in Compact Binary Systems to Next-to-Leading Post-Minkowskian Order}},\ }\href {https://doi.org/10.1103/PhysRevD.102.124025} {\bibfield  {journal} {\bibinfo  {journal} {Phys. Rev. D}\ }\textbf {\bibinfo {volume} {102}},\ \bibinfo {pages} {124025} (\bibinfo {year} {2020}{\natexlab{b}})},\ \Eprint {https://arxiv.org/abs/2008.06047} {arXiv:2008.06047 [hep-th]} \BibitemShut {NoStop}%
\bibitem [{\citenamefont {Mogull}\ \emph {et~al.}(2021)\citenamefont {Mogull}, \citenamefont {Plefka},\ and\ \citenamefont {Steinhoff}}]{Mogull:2020sak}%
  \BibitemOpen
  \bibfield  {author} {\bibinfo {author} {\bibfnamefont {G.}~\bibnamefont {Mogull}}, \bibinfo {author} {\bibfnamefont {J.}~\bibnamefont {Plefka}},\ and\ \bibinfo {author} {\bibfnamefont {J.}~\bibnamefont {Steinhoff}},\ }\bibfield  {title} {\bibinfo {title} {{Classical black hole scattering from a worldline quantum field theory}},\ }\href {https://doi.org/10.1007/JHEP02(2021)048} {\bibfield  {journal} {\bibinfo  {journal} {JHEP}\ }\textbf {\bibinfo {volume} {02}},\ \bibinfo {pages} {048}},\ \Eprint {https://arxiv.org/abs/2010.02865} {arXiv:2010.02865 [hep-th]} \BibitemShut {NoStop}%
\bibitem [{\citenamefont {Jakobsen}\ \emph {et~al.}(2021)\citenamefont {Jakobsen}, \citenamefont {Mogull}, \citenamefont {Plefka},\ and\ \citenamefont {Steinhoff}}]{Jakobsen:2021smu}%
  \BibitemOpen
  \bibfield  {author} {\bibinfo {author} {\bibfnamefont {G.~U.}\ \bibnamefont {Jakobsen}}, \bibinfo {author} {\bibfnamefont {G.}~\bibnamefont {Mogull}}, \bibinfo {author} {\bibfnamefont {J.}~\bibnamefont {Plefka}},\ and\ \bibinfo {author} {\bibfnamefont {J.}~\bibnamefont {Steinhoff}},\ }\bibfield  {title} {\bibinfo {title} {{Classical Gravitational Bremsstrahlung from a Worldline Quantum Field Theory}},\ }\href {https://doi.org/10.1103/PhysRevLett.126.201103} {\bibfield  {journal} {\bibinfo  {journal} {Phys. Rev. Lett.}\ }\textbf {\bibinfo {volume} {126}},\ \bibinfo {pages} {201103} (\bibinfo {year} {2021})},\ \Eprint {https://arxiv.org/abs/2101.12688} {arXiv:2101.12688 [gr-qc]} \BibitemShut {NoStop}%
\bibitem [{\citenamefont {Dlapa}\ \emph {et~al.}(2022{\natexlab{a}})\citenamefont {Dlapa}, \citenamefont {K\"alin}, \citenamefont {Liu},\ and\ \citenamefont {Porto}}]{Dlapa:2021npj}%
  \BibitemOpen
  \bibfield  {author} {\bibinfo {author} {\bibfnamefont {C.}~\bibnamefont {Dlapa}}, \bibinfo {author} {\bibfnamefont {G.}~\bibnamefont {K\"alin}}, \bibinfo {author} {\bibfnamefont {Z.}~\bibnamefont {Liu}},\ and\ \bibinfo {author} {\bibfnamefont {R.~A.}\ \bibnamefont {Porto}},\ }\bibfield  {title} {\bibinfo {title} {{Dynamics of binary systems to fourth Post-Minkowskian order from the effective field theory approach}},\ }\href {https://doi.org/10.1016/j.physletb.2022.137203} {\bibfield  {journal} {\bibinfo  {journal} {Phys. Lett. B}\ }\textbf {\bibinfo {volume} {831}},\ \bibinfo {pages} {137203} (\bibinfo {year} {2022}{\natexlab{a}})},\ \Eprint {https://arxiv.org/abs/2106.08276} {arXiv:2106.08276 [hep-th]} \BibitemShut {NoStop}%
\bibitem [{\citenamefont {Dlapa}\ \emph {et~al.}(2022{\natexlab{b}})\citenamefont {Dlapa}, \citenamefont {K\"alin}, \citenamefont {Liu},\ and\ \citenamefont {Porto}}]{Dlapa:2021vgp}%
  \BibitemOpen
  \bibfield  {author} {\bibinfo {author} {\bibfnamefont {C.}~\bibnamefont {Dlapa}}, \bibinfo {author} {\bibfnamefont {G.}~\bibnamefont {K\"alin}}, \bibinfo {author} {\bibfnamefont {Z.}~\bibnamefont {Liu}},\ and\ \bibinfo {author} {\bibfnamefont {R.~A.}\ \bibnamefont {Porto}},\ }\bibfield  {title} {\bibinfo {title} {{Conservative Dynamics of Binary Systems at Fourth Post-Minkowskian Order in the Large-Eccentricity Expansion}},\ }\href {https://doi.org/10.1103/PhysRevLett.128.161104} {\bibfield  {journal} {\bibinfo  {journal} {Phys. Rev. Lett.}\ }\textbf {\bibinfo {volume} {128}},\ \bibinfo {pages} {161104} (\bibinfo {year} {2022}{\natexlab{b}})},\ \Eprint {https://arxiv.org/abs/2112.11296} {arXiv:2112.11296 [hep-th]} \BibitemShut {NoStop}%
\bibitem [{\citenamefont {Mougiakakos}\ \emph {et~al.}(2021)\citenamefont {Mougiakakos}, \citenamefont {Riva},\ and\ \citenamefont {Vernizzi}}]{Mougiakakos:2021ckm}%
  \BibitemOpen
  \bibfield  {author} {\bibinfo {author} {\bibfnamefont {S.}~\bibnamefont {Mougiakakos}}, \bibinfo {author} {\bibfnamefont {M.~M.}\ \bibnamefont {Riva}},\ and\ \bibinfo {author} {\bibfnamefont {F.}~\bibnamefont {Vernizzi}},\ }\bibfield  {title} {\bibinfo {title} {{Gravitational Bremsstrahlung in the post-Minkowskian effective field theory}},\ }\href {https://doi.org/10.1103/PhysRevD.104.024041} {\bibfield  {journal} {\bibinfo  {journal} {Phys. Rev. D}\ }\textbf {\bibinfo {volume} {104}},\ \bibinfo {pages} {024041} (\bibinfo {year} {2021})},\ \Eprint {https://arxiv.org/abs/2102.08339} {arXiv:2102.08339 [gr-qc]} \BibitemShut {NoStop}%
\bibitem [{\citenamefont {Jakobsen}\ \emph {et~al.}(2022{\natexlab{a}})\citenamefont {Jakobsen}, \citenamefont {Mogull}, \citenamefont {Plefka},\ and\ \citenamefont {Sauer}}]{Jakobsen:2022psy}%
  \BibitemOpen
  \bibfield  {author} {\bibinfo {author} {\bibfnamefont {G.~U.}\ \bibnamefont {Jakobsen}}, \bibinfo {author} {\bibfnamefont {G.}~\bibnamefont {Mogull}}, \bibinfo {author} {\bibfnamefont {J.}~\bibnamefont {Plefka}},\ and\ \bibinfo {author} {\bibfnamefont {B.}~\bibnamefont {Sauer}},\ }\bibfield  {title} {\bibinfo {title} {{All things retarded: radiation-reaction in worldline quantum field theory}},\ }\href {https://doi.org/10.1007/JHEP10(2022)128} {\bibfield  {journal} {\bibinfo  {journal} {JHEP}\ }\textbf {\bibinfo {volume} {10}},\ \bibinfo {pages} {128}},\ \Eprint {https://arxiv.org/abs/2207.00569} {arXiv:2207.00569 [hep-th]} \BibitemShut {NoStop}%
\bibitem [{\citenamefont {Riva}\ and\ \citenamefont {Vernizzi}(2021)}]{Riva:2021vnj}%
  \BibitemOpen
  \bibfield  {author} {\bibinfo {author} {\bibfnamefont {M.~M.}\ \bibnamefont {Riva}}\ and\ \bibinfo {author} {\bibfnamefont {F.}~\bibnamefont {Vernizzi}},\ }\bibfield  {title} {\bibinfo {title} {{Radiated momentum in the post-Minkowskian worldline approach via reverse unitarity}},\ }\href {https://doi.org/10.1007/JHEP11(2021)228} {\bibfield  {journal} {\bibinfo  {journal} {JHEP}\ }\textbf {\bibinfo {volume} {11}},\ \bibinfo {pages} {228}},\ \Eprint {https://arxiv.org/abs/2110.10140} {arXiv:2110.10140 [hep-th]} \BibitemShut {NoStop}%
\bibitem [{\citenamefont {K\"alin}\ \emph {et~al.}(2023)\citenamefont {K\"alin}, \citenamefont {Neef},\ and\ \citenamefont {Porto}}]{Kalin:2022hph}%
  \BibitemOpen
  \bibfield  {author} {\bibinfo {author} {\bibfnamefont {G.}~\bibnamefont {K\"alin}}, \bibinfo {author} {\bibfnamefont {J.}~\bibnamefont {Neef}},\ and\ \bibinfo {author} {\bibfnamefont {R.~A.}\ \bibnamefont {Porto}},\ }\bibfield  {title} {\bibinfo {title} {{Radiation-reaction in the Effective Field Theory approach to Post-Minkowskian dynamics}},\ }\href {https://doi.org/10.1007/JHEP01(2023)140} {\bibfield  {journal} {\bibinfo  {journal} {JHEP}\ }\textbf {\bibinfo {volume} {01}},\ \bibinfo {pages} {140}},\ \Eprint {https://arxiv.org/abs/2207.00580} {arXiv:2207.00580 [hep-th]} \BibitemShut {NoStop}%
\bibitem [{\citenamefont {Dlapa}\ \emph {et~al.}(2023{\natexlab{a}})\citenamefont {Dlapa}, \citenamefont {K\"alin}, \citenamefont {Liu}, \citenamefont {Neef},\ and\ \citenamefont {Porto}}]{Dlapa:2022lmu}%
  \BibitemOpen
  \bibfield  {author} {\bibinfo {author} {\bibfnamefont {C.}~\bibnamefont {Dlapa}}, \bibinfo {author} {\bibfnamefont {G.}~\bibnamefont {K\"alin}}, \bibinfo {author} {\bibfnamefont {Z.}~\bibnamefont {Liu}}, \bibinfo {author} {\bibfnamefont {J.}~\bibnamefont {Neef}},\ and\ \bibinfo {author} {\bibfnamefont {R.~A.}\ \bibnamefont {Porto}},\ }\bibfield  {title} {\bibinfo {title} {{Radiation Reaction and Gravitational Waves at Fourth Post-Minkowskian Order}},\ }\href {https://doi.org/10.1103/PhysRevLett.130.101401} {\bibfield  {journal} {\bibinfo  {journal} {Phys. Rev. Lett.}\ }\textbf {\bibinfo {volume} {130}},\ \bibinfo {pages} {101401} (\bibinfo {year} {2023}{\natexlab{a}})},\ \Eprint {https://arxiv.org/abs/2210.05541} {arXiv:2210.05541 [hep-th]} \BibitemShut {NoStop}%
\bibitem [{\citenamefont {Dlapa}\ \emph {et~al.}(2023{\natexlab{b}})\citenamefont {Dlapa}, \citenamefont {K\"alin}, \citenamefont {Liu},\ and\ \citenamefont {Porto}}]{Dlapa:2023hsl}%
  \BibitemOpen
  \bibfield  {author} {\bibinfo {author} {\bibfnamefont {C.}~\bibnamefont {Dlapa}}, \bibinfo {author} {\bibfnamefont {G.}~\bibnamefont {K\"alin}}, \bibinfo {author} {\bibfnamefont {Z.}~\bibnamefont {Liu}},\ and\ \bibinfo {author} {\bibfnamefont {R.~A.}\ \bibnamefont {Porto}},\ }\bibfield  {title} {\bibinfo {title} {{Bootstrapping the relativistic two-body problem}},\ }\href {https://doi.org/10.1007/JHEP08(2023)109} {\bibfield  {journal} {\bibinfo  {journal} {JHEP}\ }\textbf {\bibinfo {volume} {08}},\ \bibinfo {pages} {109}},\ \Eprint {https://arxiv.org/abs/2304.01275} {arXiv:2304.01275 [hep-th]} \BibitemShut {NoStop}%
\bibitem [{\citenamefont {Driesse}\ \emph {et~al.}(2024{\natexlab{a}})\citenamefont {Driesse}, \citenamefont {Jakobsen}, \citenamefont {Klemm}, \citenamefont {Mogull}, \citenamefont {Nega}, \citenamefont {Plefka}, \citenamefont {Sauer},\ and\ \citenamefont {Usovitsch}}]{Driesse:2024feo}%
  \BibitemOpen
  \bibfield  {author} {\bibinfo {author} {\bibfnamefont {M.}~\bibnamefont {Driesse}}, \bibinfo {author} {\bibfnamefont {G.~U.}\ \bibnamefont {Jakobsen}}, \bibinfo {author} {\bibfnamefont {A.}~\bibnamefont {Klemm}}, \bibinfo {author} {\bibfnamefont {G.}~\bibnamefont {Mogull}}, \bibinfo {author} {\bibfnamefont {C.}~\bibnamefont {Nega}}, \bibinfo {author} {\bibfnamefont {J.}~\bibnamefont {Plefka}}, \bibinfo {author} {\bibfnamefont {B.}~\bibnamefont {Sauer}},\ and\ \bibinfo {author} {\bibfnamefont {J.}~\bibnamefont {Usovitsch}},\ }\bibfield  {title} {\bibinfo {title} {{High-precision black hole scattering with Calabi-Yau manifolds}},\ }\href@noop {} {\  (\bibinfo {year} {2024}{\natexlab{a}})},\ \Eprint {https://arxiv.org/abs/2411.11846} {arXiv:2411.11846 [hep-th]} \BibitemShut {NoStop}%
\bibitem [{\citenamefont {Bern}\ \emph {et~al.}(1994)\citenamefont {Bern}, \citenamefont {Dixon}, \citenamefont {Dunbar},\ and\ \citenamefont {Kosower}}]{Bern:1994zx}%
  \BibitemOpen
  \bibfield  {author} {\bibinfo {author} {\bibfnamefont {Z.}~\bibnamefont {Bern}}, \bibinfo {author} {\bibfnamefont {L.~J.}\ \bibnamefont {Dixon}}, \bibinfo {author} {\bibfnamefont {D.~C.}\ \bibnamefont {Dunbar}},\ and\ \bibinfo {author} {\bibfnamefont {D.~A.}\ \bibnamefont {Kosower}},\ }\bibfield  {title} {\bibinfo {title} {{One loop n point gauge theory amplitudes, unitarity and collinear limits}},\ }\href {https://doi.org/10.1016/0550-3213(94)90179-1} {\bibfield  {journal} {\bibinfo  {journal} {Nucl. Phys. B}\ }\textbf {\bibinfo {volume} {425}},\ \bibinfo {pages} {217} (\bibinfo {year} {1994})},\ \Eprint {https://arxiv.org/abs/hep-ph/9403226} {arXiv:hep-ph/9403226} \BibitemShut {NoStop}%
\bibitem [{\citenamefont {Bern}\ \emph {et~al.}(1995)\citenamefont {Bern}, \citenamefont {Dixon}, \citenamefont {Dunbar},\ and\ \citenamefont {Kosower}}]{Bern:1994cg}%
  \BibitemOpen
  \bibfield  {author} {\bibinfo {author} {\bibfnamefont {Z.}~\bibnamefont {Bern}}, \bibinfo {author} {\bibfnamefont {L.~J.}\ \bibnamefont {Dixon}}, \bibinfo {author} {\bibfnamefont {D.~C.}\ \bibnamefont {Dunbar}},\ and\ \bibinfo {author} {\bibfnamefont {D.~A.}\ \bibnamefont {Kosower}},\ }\bibfield  {title} {\bibinfo {title} {{Fusing gauge theory tree amplitudes into loop amplitudes}},\ }\href {https://doi.org/10.1016/0550-3213(94)00488-Z} {\bibfield  {journal} {\bibinfo  {journal} {Nucl. Phys. B}\ }\textbf {\bibinfo {volume} {435}},\ \bibinfo {pages} {59} (\bibinfo {year} {1995})},\ \Eprint {https://arxiv.org/abs/hep-ph/9409265} {arXiv:hep-ph/9409265} \BibitemShut {NoStop}%
\bibitem [{\citenamefont {Bern}\ and\ \citenamefont {Morgan}(1996)}]{Bern:1995db}%
  \BibitemOpen
  \bibfield  {author} {\bibinfo {author} {\bibfnamefont {Z.}~\bibnamefont {Bern}}\ and\ \bibinfo {author} {\bibfnamefont {A.~G.}\ \bibnamefont {Morgan}},\ }\bibfield  {title} {\bibinfo {title} {{Massive loop amplitudes from unitarity}},\ }\href {https://doi.org/10.1016/0550-3213(96)00078-8} {\bibfield  {journal} {\bibinfo  {journal} {Nucl. Phys. B}\ }\textbf {\bibinfo {volume} {467}},\ \bibinfo {pages} {479} (\bibinfo {year} {1996})},\ \Eprint {https://arxiv.org/abs/hep-ph/9511336} {arXiv:hep-ph/9511336} \BibitemShut {NoStop}%
\bibitem [{\citenamefont {Bern}\ \emph {et~al.}(1998)\citenamefont {Bern}, \citenamefont {Dixon},\ and\ \citenamefont {Kosower}}]{Bern:1997sc}%
  \BibitemOpen
  \bibfield  {author} {\bibinfo {author} {\bibfnamefont {Z.}~\bibnamefont {Bern}}, \bibinfo {author} {\bibfnamefont {L.~J.}\ \bibnamefont {Dixon}},\ and\ \bibinfo {author} {\bibfnamefont {D.~A.}\ \bibnamefont {Kosower}},\ }\bibfield  {title} {\bibinfo {title} {{One loop amplitudes for e+ e- to four partons}},\ }\href {https://doi.org/10.1016/S0550-3213(97)00703-7} {\bibfield  {journal} {\bibinfo  {journal} {Nucl. Phys. B}\ }\textbf {\bibinfo {volume} {513}},\ \bibinfo {pages} {3} (\bibinfo {year} {1998})},\ \Eprint {https://arxiv.org/abs/hep-ph/9708239} {arXiv:hep-ph/9708239} \BibitemShut {NoStop}%
\bibitem [{\citenamefont {Britto}\ \emph {et~al.}(2005)\citenamefont {Britto}, \citenamefont {Cachazo},\ and\ \citenamefont {Feng}}]{Britto:2004nc}%
  \BibitemOpen
  \bibfield  {author} {\bibinfo {author} {\bibfnamefont {R.}~\bibnamefont {Britto}}, \bibinfo {author} {\bibfnamefont {F.}~\bibnamefont {Cachazo}},\ and\ \bibinfo {author} {\bibfnamefont {B.}~\bibnamefont {Feng}},\ }\bibfield  {title} {\bibinfo {title} {{Generalized unitarity and one-loop amplitudes in N=4 super-Yang-Mills}},\ }\href {https://doi.org/10.1016/j.nuclphysb.2005.07.014} {\bibfield  {journal} {\bibinfo  {journal} {Nucl. Phys. B}\ }\textbf {\bibinfo {volume} {725}},\ \bibinfo {pages} {275} (\bibinfo {year} {2005})},\ \Eprint {https://arxiv.org/abs/hep-th/0412103} {arXiv:hep-th/0412103} \BibitemShut {NoStop}%
\bibitem [{\citenamefont {Bern}\ \emph {et~al.}(2008)\citenamefont {Bern}, \citenamefont {Carrasco},\ and\ \citenamefont {Johansson}}]{Bern:2008qj}%
  \BibitemOpen
  \bibfield  {author} {\bibinfo {author} {\bibfnamefont {Z.}~\bibnamefont {Bern}}, \bibinfo {author} {\bibfnamefont {J.~J.~M.}\ \bibnamefont {Carrasco}},\ and\ \bibinfo {author} {\bibfnamefont {H.}~\bibnamefont {Johansson}},\ }\bibfield  {title} {\bibinfo {title} {{New Relations for Gauge-Theory Amplitudes}},\ }\href {https://doi.org/10.1103/PhysRevD.78.085011} {\bibfield  {journal} {\bibinfo  {journal} {Phys. Rev. D}\ }\textbf {\bibinfo {volume} {78}},\ \bibinfo {pages} {085011} (\bibinfo {year} {2008})},\ \Eprint {https://arxiv.org/abs/0805.3993} {arXiv:0805.3993 [hep-ph]} \BibitemShut {NoStop}%
\bibitem [{\citenamefont {Bern}\ \emph {et~al.}(2010)\citenamefont {Bern}, \citenamefont {Carrasco},\ and\ \citenamefont {Johansson}}]{Bern:2010ue}%
  \BibitemOpen
  \bibfield  {author} {\bibinfo {author} {\bibfnamefont {Z.}~\bibnamefont {Bern}}, \bibinfo {author} {\bibfnamefont {J.~J.~M.}\ \bibnamefont {Carrasco}},\ and\ \bibinfo {author} {\bibfnamefont {H.}~\bibnamefont {Johansson}},\ }\bibfield  {title} {\bibinfo {title} {{Perturbative Quantum Gravity as a Double Copy of Gauge Theory}},\ }\href {https://doi.org/10.1103/PhysRevLett.105.061602} {\bibfield  {journal} {\bibinfo  {journal} {Phys. Rev. Lett.}\ }\textbf {\bibinfo {volume} {105}},\ \bibinfo {pages} {061602} (\bibinfo {year} {2010})},\ \Eprint {https://arxiv.org/abs/1004.0476} {arXiv:1004.0476 [hep-th]} \BibitemShut {NoStop}%
\bibitem [{\citenamefont {Parra-Martinez}\ \emph {et~al.}(2020)\citenamefont {Parra-Martinez}, \citenamefont {Ruf},\ and\ \citenamefont {Zeng}}]{Parra-Martinez:2020dzs}%
  \BibitemOpen
  \bibfield  {author} {\bibinfo {author} {\bibfnamefont {J.}~\bibnamefont {Parra-Martinez}}, \bibinfo {author} {\bibfnamefont {M.~S.}\ \bibnamefont {Ruf}},\ and\ \bibinfo {author} {\bibfnamefont {M.}~\bibnamefont {Zeng}},\ }\bibfield  {title} {\bibinfo {title} {{Extremal black hole scattering at $\mathcal{O}(G^3)$: graviton dominance, eikonal exponentiation, and differential equations}},\ }\href {https://doi.org/10.1007/JHEP11(2020)023} {\bibfield  {journal} {\bibinfo  {journal} {JHEP}\ }\textbf {\bibinfo {volume} {11}},\ \bibinfo {pages} {023}},\ \Eprint {https://arxiv.org/abs/2005.04236} {arXiv:2005.04236 [hep-th]} \BibitemShut {NoStop}%
\bibitem [{\citenamefont {Beneke}\ and\ \citenamefont {Smirnov}(1998)}]{Beneke:1997zp}%
  \BibitemOpen
  \bibfield  {author} {\bibinfo {author} {\bibfnamefont {M.}~\bibnamefont {Beneke}}\ and\ \bibinfo {author} {\bibfnamefont {V.~A.}\ \bibnamefont {Smirnov}},\ }\bibfield  {title} {\bibinfo {title} {{Asymptotic expansion of Feynman integrals near threshold}},\ }\href {https://doi.org/10.1016/S0550-3213(98)00138-2} {\bibfield  {journal} {\bibinfo  {journal} {Nucl. Phys. B}\ }\textbf {\bibinfo {volume} {522}},\ \bibinfo {pages} {321} (\bibinfo {year} {1998})},\ \Eprint {https://arxiv.org/abs/hep-ph/9711391} {arXiv:hep-ph/9711391} \BibitemShut {NoStop}%
\bibitem [{\citenamefont {Keldysh}(1964)}]{Keldysh:1964ud}%
  \BibitemOpen
  \bibfield  {author} {\bibinfo {author} {\bibfnamefont {L.~V.}\ \bibnamefont {Keldysh}},\ }\bibfield  {title} {\bibinfo {title} {{Diagram technique for nonequilibrium processes}},\ }\href {https://doi.org/10.1142/9789811279461_0007} {\bibfield  {journal} {\bibinfo  {journal} {Zh. Eksp. Teor. Fiz.}\ }\textbf {\bibinfo {volume} {47}},\ \bibinfo {pages} {1515} (\bibinfo {year} {1964})}\BibitemShut {NoStop}%
\bibitem [{\citenamefont {Schwinger}(1961)}]{Schwinger:1960qe}%
  \BibitemOpen
  \bibfield  {author} {\bibinfo {author} {\bibfnamefont {J.~S.}\ \bibnamefont {Schwinger}},\ }\bibfield  {title} {\bibinfo {title} {{Brownian motion of a quantum oscillator}},\ }\href {https://doi.org/10.1063/1.1703727} {\bibfield  {journal} {\bibinfo  {journal} {J. Math. Phys.}\ }\textbf {\bibinfo {volume} {2}},\ \bibinfo {pages} {407} (\bibinfo {year} {1961})}\BibitemShut {NoStop}%
\bibitem [{\citenamefont {Goldberger}\ and\ \citenamefont {Rothstein}(2006)}]{Goldberger:2004jt}%
  \BibitemOpen
  \bibfield  {author} {\bibinfo {author} {\bibfnamefont {W.~D.}\ \bibnamefont {Goldberger}}\ and\ \bibinfo {author} {\bibfnamefont {I.~Z.}\ \bibnamefont {Rothstein}},\ }\bibfield  {title} {\bibinfo {title} {{An Effective field theory of gravity for extended objects}},\ }\href {https://doi.org/10.1103/PhysRevD.73.104029} {\bibfield  {journal} {\bibinfo  {journal} {Phys. Rev. D}\ }\textbf {\bibinfo {volume} {73}},\ \bibinfo {pages} {104029} (\bibinfo {year} {2006})},\ \Eprint {https://arxiv.org/abs/hep-th/0409156} {arXiv:hep-th/0409156} \BibitemShut {NoStop}%
\bibitem [{\citenamefont {Damgaard}\ \emph {et~al.}(2021)\citenamefont {Damgaard}, \citenamefont {Plante},\ and\ \citenamefont {Vanhove}}]{Damgaard:2021ipf}%
  \BibitemOpen
  \bibfield  {author} {\bibinfo {author} {\bibfnamefont {P.~H.}\ \bibnamefont {Damgaard}}, \bibinfo {author} {\bibfnamefont {L.}~\bibnamefont {Plante}},\ and\ \bibinfo {author} {\bibfnamefont {P.}~\bibnamefont {Vanhove}},\ }\bibfield  {title} {\bibinfo {title} {{On an exponential representation of the gravitational S-matrix}},\ }\href {https://doi.org/10.1007/JHEP11(2021)213} {\bibfield  {journal} {\bibinfo  {journal} {JHEP}\ }\textbf {\bibinfo {volume} {11}},\ \bibinfo {pages} {213}},\ \Eprint {https://arxiv.org/abs/2107.12891} {arXiv:2107.12891 [hep-th]} \BibitemShut {NoStop}%
\bibitem [{\citenamefont {Barack}\ and\ \citenamefont {Long}(2022)}]{Barack:2022pde}%
  \BibitemOpen
  \bibfield  {author} {\bibinfo {author} {\bibfnamefont {L.}~\bibnamefont {Barack}}\ and\ \bibinfo {author} {\bibfnamefont {O.}~\bibnamefont {Long}},\ }\bibfield  {title} {\bibinfo {title} {{Self-force correction to the deflection angle in black-hole scattering: A scalar charge toy model}},\ }\href {https://doi.org/10.1103/PhysRevD.106.104031} {\bibfield  {journal} {\bibinfo  {journal} {Phys. Rev. D}\ }\textbf {\bibinfo {volume} {106}},\ \bibinfo {pages} {104031} (\bibinfo {year} {2022})},\ \Eprint {https://arxiv.org/abs/2209.03740} {arXiv:2209.03740 [gr-qc]} \BibitemShut {NoStop}%
\bibitem [{\citenamefont {Barack}\ \emph {et~al.}(2023)\citenamefont {Barack} \emph {et~al.}}]{Barack:2023oqp}%
  \BibitemOpen
  \bibfield  {author} {\bibinfo {author} {\bibfnamefont {L.}~\bibnamefont {Barack}} \emph {et~al.},\ }\bibfield  {title} {\bibinfo {title} {{Comparison of post-Minkowskian and self-force expansions: Scattering in a scalar charge toy model}},\ }\href {https://doi.org/10.1103/PhysRevD.108.024025} {\bibfield  {journal} {\bibinfo  {journal} {Phys. Rev. D}\ }\textbf {\bibinfo {volume} {108}},\ \bibinfo {pages} {024025} (\bibinfo {year} {2023})},\ \Eprint {https://arxiv.org/abs/2304.09200} {arXiv:2304.09200 [hep-th]} \BibitemShut {NoStop}%
\bibitem [{\citenamefont {Saketh}\ \emph {et~al.}(2022)\citenamefont {Saketh}, \citenamefont {Vines}, \citenamefont {Steinhoff},\ and\ \citenamefont {Buonanno}}]{Saketh:2021sri}%
  \BibitemOpen
  \bibfield  {author} {\bibinfo {author} {\bibfnamefont {M.~V.~S.}\ \bibnamefont {Saketh}}, \bibinfo {author} {\bibfnamefont {J.}~\bibnamefont {Vines}}, \bibinfo {author} {\bibfnamefont {J.}~\bibnamefont {Steinhoff}},\ and\ \bibinfo {author} {\bibfnamefont {A.}~\bibnamefont {Buonanno}},\ }\bibfield  {title} {\bibinfo {title} {{Conservative and radiative dynamics in classical relativistic scattering and bound systems}},\ }\href {https://doi.org/10.1103/PhysRevResearch.4.013127} {\bibfield  {journal} {\bibinfo  {journal} {Phys. Rev. Res.}\ }\textbf {\bibinfo {volume} {4}},\ \bibinfo {pages} {013127} (\bibinfo {year} {2022})},\ \Eprint {https://arxiv.org/abs/2109.05994} {arXiv:2109.05994 [gr-qc]} \BibitemShut {NoStop}%
\bibitem [{\citenamefont {Bern}\ \emph {et~al.}(2022{\natexlab{b}})\citenamefont {Bern}, \citenamefont {Gatica}, \citenamefont {Herrmann}, \citenamefont {Luna},\ and\ \citenamefont {Zeng}}]{Bern:2021xze}%
  \BibitemOpen
  \bibfield  {author} {\bibinfo {author} {\bibfnamefont {Z.}~\bibnamefont {Bern}}, \bibinfo {author} {\bibfnamefont {J.~P.}\ \bibnamefont {Gatica}}, \bibinfo {author} {\bibfnamefont {E.}~\bibnamefont {Herrmann}}, \bibinfo {author} {\bibfnamefont {A.}~\bibnamefont {Luna}},\ and\ \bibinfo {author} {\bibfnamefont {M.}~\bibnamefont {Zeng}},\ }\bibfield  {title} {\bibinfo {title} {{Scalar QED as a toy model for higher-order effects in classical gravitational scattering}},\ }\href {https://doi.org/10.1007/JHEP08(2022)131} {\bibfield  {journal} {\bibinfo  {journal} {JHEP}\ }\textbf {\bibinfo {volume} {08}},\ \bibinfo {pages} {131}},\ \Eprint {https://arxiv.org/abs/2112.12243} {arXiv:2112.12243 [hep-th]} \BibitemShut {NoStop}%
\bibitem [{\citenamefont {Bern}\ \emph {et~al.}(2024{\natexlab{a}})\citenamefont {Bern}, \citenamefont {Herrmann}, \citenamefont {Roiban}, \citenamefont {Ruf}, \citenamefont {Smirnov}, \citenamefont {Smirnov},\ and\ \citenamefont {Zeng}}]{Bern:2023ccb}%
  \BibitemOpen
  \bibfield  {author} {\bibinfo {author} {\bibfnamefont {Z.}~\bibnamefont {Bern}}, \bibinfo {author} {\bibfnamefont {E.}~\bibnamefont {Herrmann}}, \bibinfo {author} {\bibfnamefont {R.}~\bibnamefont {Roiban}}, \bibinfo {author} {\bibfnamefont {M.~S.}\ \bibnamefont {Ruf}}, \bibinfo {author} {\bibfnamefont {A.~V.}\ \bibnamefont {Smirnov}}, \bibinfo {author} {\bibfnamefont {V.~A.}\ \bibnamefont {Smirnov}},\ and\ \bibinfo {author} {\bibfnamefont {M.}~\bibnamefont {Zeng}},\ }\bibfield  {title} {\bibinfo {title} {{Conservative Binary Dynamics at Order \ensuremath{\alpha}5 in Electrodynamics}},\ }\href {https://doi.org/10.1103/PhysRevLett.132.251601} {\bibfield  {journal} {\bibinfo  {journal} {Phys. Rev. Lett.}\ }\textbf {\bibinfo {volume} {132}},\ \bibinfo {pages} {251601} (\bibinfo {year} {2024}{\natexlab{a}})},\ \Eprint {https://arxiv.org/abs/2305.08981} {arXiv:2305.08981 [hep-th]} \BibitemShut {NoStop}%
\bibitem [{\citenamefont {Jakobsen}(2024)}]{Jakobsen:2023tvm}%
  \BibitemOpen
  \bibfield  {author} {\bibinfo {author} {\bibfnamefont {G.~U.}\ \bibnamefont {Jakobsen}},\ }\bibfield  {title} {\bibinfo {title} {{Spin and Susceptibility Effects of Electromagnetic Self-Force in Effective Field Theory}},\ }\href {https://doi.org/10.1103/PhysRevLett.132.151601} {\bibfield  {journal} {\bibinfo  {journal} {Phys. Rev. Lett.}\ }\textbf {\bibinfo {volume} {132}},\ \bibinfo {pages} {151601} (\bibinfo {year} {2024})},\ \Eprint {https://arxiv.org/abs/2311.04151} {arXiv:2311.04151 [hep-th]} \BibitemShut {NoStop}%
\bibitem [{\citenamefont {Akpinar}\ \emph {et~al.}(2024)\citenamefont {Akpinar}, \citenamefont {Febres~Cordero}, \citenamefont {Kraus}, \citenamefont {Ruf},\ and\ \citenamefont {Zeng}}]{Akpinar:2024meg}%
  \BibitemOpen
  \bibfield  {author} {\bibinfo {author} {\bibfnamefont {D.}~\bibnamefont {Akpinar}}, \bibinfo {author} {\bibfnamefont {F.}~\bibnamefont {Febres~Cordero}}, \bibinfo {author} {\bibfnamefont {M.}~\bibnamefont {Kraus}}, \bibinfo {author} {\bibfnamefont {M.~S.}\ \bibnamefont {Ruf}},\ and\ \bibinfo {author} {\bibfnamefont {M.}~\bibnamefont {Zeng}},\ }\bibfield  {title} {\bibinfo {title} {{Spinning Black Hole Scattering at $\mathcal{O}(G^3 S^2)$: Casimir Terms, Radial Action and Hidden Symmetry}},\ }\href@noop {} {\  (\bibinfo {year} {2024})},\ \Eprint {https://arxiv.org/abs/2407.19005} {arXiv:2407.19005 [hep-th]} \BibitemShut {NoStop}%
\bibitem [{\citenamefont {Damgaard}\ \emph {et~al.}(2023{\natexlab{b}})\citenamefont {Damgaard}, \citenamefont {Hansen}, \citenamefont {Plant\'e},\ and\ \citenamefont {Vanhove}}]{Damgaard:2023vnx}%
  \BibitemOpen
  \bibfield  {author} {\bibinfo {author} {\bibfnamefont {P.~H.}\ \bibnamefont {Damgaard}}, \bibinfo {author} {\bibfnamefont {E.~R.}\ \bibnamefont {Hansen}}, \bibinfo {author} {\bibfnamefont {L.}~\bibnamefont {Plant\'e}},\ and\ \bibinfo {author} {\bibfnamefont {P.}~\bibnamefont {Vanhove}},\ }\bibfield  {title} {\bibinfo {title} {{The relation between KMOC and worldline formalisms for classical gravity}},\ }\href {https://doi.org/10.1007/JHEP09(2023)059} {\bibfield  {journal} {\bibinfo  {journal} {JHEP}\ }\textbf {\bibinfo {volume} {09}},\ \bibinfo {pages} {059}},\ \Eprint {https://arxiv.org/abs/2306.11454} {arXiv:2306.11454 [hep-th]} \BibitemShut {NoStop}%
\bibitem [{\citenamefont {Comberiati}\ and\ \citenamefont {de~la Cruz}(2023)}]{Comberiati:2022ldk}%
  \BibitemOpen
  \bibfield  {author} {\bibinfo {author} {\bibfnamefont {F.}~\bibnamefont {Comberiati}}\ and\ \bibinfo {author} {\bibfnamefont {L.}~\bibnamefont {de~la Cruz}},\ }\bibfield  {title} {\bibinfo {title} {{Classical off-shell currents}},\ }\href {https://doi.org/10.1007/JHEP03(2023)068} {\bibfield  {journal} {\bibinfo  {journal} {JHEP}\ }\textbf {\bibinfo {volume} {03}},\ \bibinfo {pages} {068}},\ \Eprint {https://arxiv.org/abs/2212.09259} {arXiv:2212.09259 [hep-th]} \BibitemShut {NoStop}%
\bibitem [{\citenamefont {Du}\ \emph {et~al.}(2024)\citenamefont {Du}, \citenamefont {Ajith}, \citenamefont {Rajagopal},\ and\ \citenamefont {Vaman}}]{Du:2024rkf}%
  \BibitemOpen
  \bibfield  {author} {\bibinfo {author} {\bibfnamefont {Y.}~\bibnamefont {Du}}, \bibinfo {author} {\bibfnamefont {S.}~\bibnamefont {Ajith}}, \bibinfo {author} {\bibfnamefont {R.}~\bibnamefont {Rajagopal}},\ and\ \bibinfo {author} {\bibfnamefont {D.}~\bibnamefont {Vaman}},\ }\bibfield  {title} {\bibinfo {title} {{Worldline Proof of Eikonal Exponentiation}},\ }\href@noop {} {\  (\bibinfo {year} {2024})},\ \Eprint {https://arxiv.org/abs/2409.12895} {arXiv:2409.12895 [hep-th]} \BibitemShut {NoStop}%
\bibitem [{\citenamefont {Ajith}\ \emph {et~al.}(2024)\citenamefont {Ajith}, \citenamefont {Du}, \citenamefont {Rajagopal},\ and\ \citenamefont {Vaman}}]{Ajith:2024fna}%
  \BibitemOpen
  \bibfield  {author} {\bibinfo {author} {\bibfnamefont {S.}~\bibnamefont {Ajith}}, \bibinfo {author} {\bibfnamefont {Y.}~\bibnamefont {Du}}, \bibinfo {author} {\bibfnamefont {R.}~\bibnamefont {Rajagopal}},\ and\ \bibinfo {author} {\bibfnamefont {D.}~\bibnamefont {Vaman}},\ }\bibfield  {title} {\bibinfo {title} {{Worldline Formalism, Eikonal Expansion and the Classical Limit of Scattering Amplitudes}},\ }\href@noop {} {\  (\bibinfo {year} {2024})},\ \Eprint {https://arxiv.org/abs/2409.17866} {arXiv:2409.17866 [hep-th]} \BibitemShut {NoStop}%
\bibitem [{\citenamefont {Kim}\ \emph {et~al.}(2024)\citenamefont {Kim}, \citenamefont {Kim}, \citenamefont {Kim},\ and\ \citenamefont {Lee}}]{Kim:2024svw}%
  \BibitemOpen
  \bibfield  {author} {\bibinfo {author} {\bibfnamefont {J.-H.}\ \bibnamefont {Kim}}, \bibinfo {author} {\bibfnamefont {J.-W.}\ \bibnamefont {Kim}}, \bibinfo {author} {\bibfnamefont {S.}~\bibnamefont {Kim}},\ and\ \bibinfo {author} {\bibfnamefont {S.}~\bibnamefont {Lee}},\ }\bibfield  {title} {\bibinfo {title} {{Classical eikonal from Magnus expansion}},\ }\href@noop {} {\  (\bibinfo {year} {2024})},\ \Eprint {https://arxiv.org/abs/2410.22988} {arXiv:2410.22988 [hep-th]} \BibitemShut {NoStop}%
\bibitem [{\citenamefont {Jes\'us Aguilera-Verdugo}\ \emph {et~al.}(2021)\citenamefont {Jes\'us Aguilera-Verdugo}, \citenamefont {Hern\'andez-Pinto}, \citenamefont {Rodrigo}, \citenamefont {Sborlini},\ and\ \citenamefont {Torres~Bobadilla}}]{Aguilera_Verdugo_2021}%
  \BibitemOpen
  \bibfield  {author} {\bibinfo {author} {\bibfnamefont {J.}~\bibnamefont {Jes\'us Aguilera-Verdugo}}, \bibinfo {author} {\bibfnamefont {R.~J.}\ \bibnamefont {Hern\'andez-Pinto}}, \bibinfo {author} {\bibfnamefont {G.}~\bibnamefont {Rodrigo}}, \bibinfo {author} {\bibfnamefont {G.~F.~R.}\ \bibnamefont {Sborlini}},\ and\ \bibinfo {author} {\bibfnamefont {W.~J.}\ \bibnamefont {Torres~Bobadilla}},\ }\bibfield  {title} {\bibinfo {title} {{Mathematical properties of nested residues and their application to multi-loop scattering amplitudes}},\ }\href {https://doi.org/10.1007/JHEP02(2021)112} {\bibfield  {journal} {\bibinfo  {journal} {JHEP}\ }\textbf {\bibinfo {volume} {02}},\ \bibinfo {pages} {112}},\ \Eprint {https://arxiv.org/abs/2010.12971} {arXiv:2010.12971 [hep-ph]} \BibitemShut {NoStop}%
\bibitem [{\citenamefont {Catani}\ \emph {et~al.}(2008)\citenamefont {Catani}, \citenamefont {Gleisberg}, \citenamefont {Krauss}, \citenamefont {Rodrigo},\ and\ \citenamefont {Winter}}]{Catani_2008}%
  \BibitemOpen
  \bibfield  {author} {\bibinfo {author} {\bibfnamefont {S.}~\bibnamefont {Catani}}, \bibinfo {author} {\bibfnamefont {T.}~\bibnamefont {Gleisberg}}, \bibinfo {author} {\bibfnamefont {F.}~\bibnamefont {Krauss}}, \bibinfo {author} {\bibfnamefont {G.}~\bibnamefont {Rodrigo}},\ and\ \bibinfo {author} {\bibfnamefont {J.-C.}\ \bibnamefont {Winter}},\ }\bibfield  {title} {\bibinfo {title} {{From loops to trees by-passing Feynman's theorem}},\ }\href {https://doi.org/10.1088/1126-6708/2008/09/065} {\bibfield  {journal} {\bibinfo  {journal} {JHEP}\ }\textbf {\bibinfo {volume} {09}},\ \bibinfo {pages} {065}},\ \Eprint {https://arxiv.org/abs/0804.3170} {arXiv:0804.3170 [hep-ph]} \BibitemShut {NoStop}%
\bibitem [{\citenamefont {Bierenbaum}\ \emph {et~al.}(2010)\citenamefont {Bierenbaum}, \citenamefont {Catani}, \citenamefont {Draggiotis},\ and\ \citenamefont {Rodrigo}}]{Bierenbaum_2010}%
  \BibitemOpen
  \bibfield  {author} {\bibinfo {author} {\bibfnamefont {I.}~\bibnamefont {Bierenbaum}}, \bibinfo {author} {\bibfnamefont {S.}~\bibnamefont {Catani}}, \bibinfo {author} {\bibfnamefont {P.}~\bibnamefont {Draggiotis}},\ and\ \bibinfo {author} {\bibfnamefont {G.}~\bibnamefont {Rodrigo}},\ }\bibfield  {title} {\bibinfo {title} {A tree-loop duality relation at two loops and beyond},\ }\href {https://doi.org/10.1007/JHEP10(2010)073} {\bibfield  {journal} {\bibinfo  {journal} {JHEP}\ }\textbf {\bibinfo {volume} {10}},\ \bibinfo {pages} {073}},\ \Eprint {https://arxiv.org/abs/1007.0194} {arXiv:1007.0194 [hep-ph]} \BibitemShut {NoStop}%
\bibitem [{\citenamefont {Runkel}\ \emph {et~al.}(2019)\citenamefont {Runkel}, \citenamefont {Sz\H{o}r}, \citenamefont {Vesga},\ and\ \citenamefont {Weinzierl}}]{Runkel_2019}%
  \BibitemOpen
  \bibfield  {author} {\bibinfo {author} {\bibfnamefont {R.}~\bibnamefont {Runkel}}, \bibinfo {author} {\bibfnamefont {Z.}~\bibnamefont {Sz\H{o}r}}, \bibinfo {author} {\bibfnamefont {J.~P.}\ \bibnamefont {Vesga}},\ and\ \bibinfo {author} {\bibfnamefont {S.}~\bibnamefont {Weinzierl}},\ }\bibfield  {title} {\bibinfo {title} {{Causality and loop-tree duality at higher loops}},\ }\href {https://doi.org/10.1103/PhysRevLett.122.111603} {\bibfield  {journal} {\bibinfo  {journal} {Phys. Rev. Lett.}\ }\textbf {\bibinfo {volume} {122}},\ \bibinfo {pages} {111603} (\bibinfo {year} {2019})},\ \bibinfo {note} {[Erratum: Phys.Rev.Lett. 123, 059902 (2019)]},\ \Eprint {https://arxiv.org/abs/1902.02135} {arXiv:1902.02135 [hep-ph]} \BibitemShut {NoStop}%
\bibitem [{\citenamefont {Sborlini}(2021)}]{Sborlini_2021}%
  \BibitemOpen
  \bibfield  {author} {\bibinfo {author} {\bibfnamefont {G.~F.~R.}\ \bibnamefont {Sborlini}},\ }\bibfield  {title} {\bibinfo {title} {Geometrical approach to causality in multiloop amplitudes},\ }\bibfield  {journal} {\bibinfo  {journal} {Physical Review D}\ }\textbf {\bibinfo {volume} {104}},\ \href {https://doi.org/10.1103/physrevd.104.036014} {10.1103/physrevd.104.036014} (\bibinfo {year} {2021})\BibitemShut {NoStop}%
\bibitem [{\citenamefont {Capatti}\ \emph {et~al.}(2019)\citenamefont {Capatti}, \citenamefont {Hirschi}, \citenamefont {Kermanschah},\ and\ \citenamefont {Ruijl}}]{Capatti:2019ypt}%
  \BibitemOpen
  \bibfield  {author} {\bibinfo {author} {\bibfnamefont {Z.}~\bibnamefont {Capatti}}, \bibinfo {author} {\bibfnamefont {V.}~\bibnamefont {Hirschi}}, \bibinfo {author} {\bibfnamefont {D.}~\bibnamefont {Kermanschah}},\ and\ \bibinfo {author} {\bibfnamefont {B.}~\bibnamefont {Ruijl}},\ }\bibfield  {title} {\bibinfo {title} {{Loop-Tree Duality for Multiloop Numerical Integration}},\ }\href {https://doi.org/10.1103/PhysRevLett.123.151602} {\bibfield  {journal} {\bibinfo  {journal} {Phys. Rev. Lett.}\ }\textbf {\bibinfo {volume} {123}},\ \bibinfo {pages} {151602} (\bibinfo {year} {2019})},\ \Eprint {https://arxiv.org/abs/1906.06138} {arXiv:1906.06138 [hep-ph]} \BibitemShut {NoStop}%
\bibitem [{\citenamefont {Capatti}(2023{\natexlab{a}})}]{Capatti:2022mly}%
  \BibitemOpen
  \bibfield  {author} {\bibinfo {author} {\bibfnamefont {Z.}~\bibnamefont {Capatti}},\ }\bibfield  {title} {\bibinfo {title} {{Exposing the threshold structure of loop integrals}},\ }\href {https://doi.org/10.1103/PhysRevD.107.L051902} {\bibfield  {journal} {\bibinfo  {journal} {Phys. Rev. D}\ }\textbf {\bibinfo {volume} {107}},\ \bibinfo {pages} {L051902} (\bibinfo {year} {2023}{\natexlab{a}})},\ \Eprint {https://arxiv.org/abs/2211.09653} {arXiv:2211.09653 [hep-th]} \BibitemShut {NoStop}%
\bibitem [{\citenamefont {Capatti}(2024)}]{Capatti:2023shz}%
  \BibitemOpen
  \bibfield  {author} {\bibinfo {author} {\bibfnamefont {Z.}~\bibnamefont {Capatti}},\ }\bibfield  {title} {\bibinfo {title} {{Derivation of the Cross-Free Family representation for the box diagram}},\ }\href {https://doi.org/10.22323/1.432.0027} {\bibfield  {journal} {\bibinfo  {journal} {PoS}\ }\textbf {\bibinfo {volume} {RADCOR2023}},\ \bibinfo {pages} {027} (\bibinfo {year} {2024})},\ \Eprint {https://arxiv.org/abs/2311.14374} {arXiv:2311.14374 [hep-ph]} \BibitemShut {NoStop}%
\bibitem [{\citenamefont {Sterman}(1993)}]{Sterman:1993hfp}%
  \BibitemOpen
  \bibfield  {author} {\bibinfo {author} {\bibfnamefont {G.~F.}\ \bibnamefont {Sterman}},\ }\href@noop {} {\emph {\bibinfo {title} {{An Introduction to quantum field theory}}}}\ (\bibinfo  {publisher} {Cambridge University Press},\ \bibinfo {year} {1993})\BibitemShut {NoStop}%
\bibitem [{\citenamefont {Sterman}(1995)}]{tasi_sterman}%
  \BibitemOpen
  \bibfield  {author} {\bibinfo {author} {\bibfnamefont {G.~F.}\ \bibnamefont {Sterman}},\ }\bibfield  {title} {\bibinfo {title} {{Partons, factorization and resummation, TASI 95}},\ }in\ \href@noop {} {\emph {\bibinfo {booktitle} {{Theoretical Advanced Study Institute in Elementary Particle Physics (TASI 95): QCD and Beyond}}}}\ (\bibinfo {year} {1995})\ pp.\ \bibinfo {pages} {327--408},\ \Eprint {https://arxiv.org/abs/hep-ph/9606312} {arXiv:hep-ph/9606312} \BibitemShut {NoStop}%
\bibitem [{\citenamefont {Sterman}\ and\ \citenamefont {Venkata}(2024)}]{Sterman:2023xdj}%
  \BibitemOpen
  \bibfield  {author} {\bibinfo {author} {\bibfnamefont {G.}~\bibnamefont {Sterman}}\ and\ \bibinfo {author} {\bibfnamefont {A.}~\bibnamefont {Venkata}},\ }\bibfield  {title} {\bibinfo {title} {{Local infrared safety in time-ordered perturbation theory}},\ }\href {https://doi.org/10.1007/JHEP02(2024)101} {\bibfield  {journal} {\bibinfo  {journal} {JHEP}\ }\textbf {\bibinfo {volume} {02}},\ \bibinfo {pages} {101}},\ \Eprint {https://arxiv.org/abs/2309.13023} {arXiv:2309.13023 [hep-ph]} \BibitemShut {NoStop}%
\bibitem [{\citenamefont {Borinsky}\ \emph {et~al.}(2023)\citenamefont {Borinsky}, \citenamefont {Capatti}, \citenamefont {Laenen},\ and\ \citenamefont {Salas-Bern\'ardez}}]{Borinsky:2022msp}%
  \BibitemOpen
  \bibfield  {author} {\bibinfo {author} {\bibfnamefont {M.}~\bibnamefont {Borinsky}}, \bibinfo {author} {\bibfnamefont {Z.}~\bibnamefont {Capatti}}, \bibinfo {author} {\bibfnamefont {E.}~\bibnamefont {Laenen}},\ and\ \bibinfo {author} {\bibfnamefont {A.}~\bibnamefont {Salas-Bern\'ardez}},\ }\bibfield  {title} {\bibinfo {title} {{Flow-oriented perturbation theory}},\ }\href {https://doi.org/10.1007/JHEP01(2023)172} {\bibfield  {journal} {\bibinfo  {journal} {JHEP}\ }\textbf {\bibinfo {volume} {01}},\ \bibinfo {pages} {172}},\ \Eprint {https://arxiv.org/abs/2210.05532} {arXiv:2210.05532 [hep-th]} \BibitemShut {NoStop}%
\bibitem [{\citenamefont {Capatti}\ \emph {et~al.}(2021)\citenamefont {Capatti}, \citenamefont {Hirschi}, \citenamefont {Pelloni},\ and\ \citenamefont {Ruijl}}]{Capatti:2020xjc}%
  \BibitemOpen
  \bibfield  {author} {\bibinfo {author} {\bibfnamefont {Z.}~\bibnamefont {Capatti}}, \bibinfo {author} {\bibfnamefont {V.}~\bibnamefont {Hirschi}}, \bibinfo {author} {\bibfnamefont {A.}~\bibnamefont {Pelloni}},\ and\ \bibinfo {author} {\bibfnamefont {B.}~\bibnamefont {Ruijl}},\ }\bibfield  {title} {\bibinfo {title} {{Local Unitarity: a representation of differential cross-sections that is locally free of infrared singularities at any order}},\ }\href {https://doi.org/10.1007/JHEP04(2021)104} {\bibfield  {journal} {\bibinfo  {journal} {JHEP}\ }\textbf {\bibinfo {volume} {04}},\ \bibinfo {pages} {104}},\ \Eprint {https://arxiv.org/abs/2010.01068} {arXiv:2010.01068 [hep-ph]} \BibitemShut {NoStop}%
\bibitem [{\citenamefont {Capatti}(2022)}]{Capatti:2021bsm}%
  \BibitemOpen
  \bibfield  {author} {\bibinfo {author} {\bibfnamefont {Z.}~\bibnamefont {Capatti}},\ }\bibfield  {title} {\bibinfo {title} {{Local Unitarity}},\ }\href {https://doi.org/10.21468/SciPostPhysProc.7.024} {\bibfield  {journal} {\bibinfo  {journal} {SciPost Phys. Proc.}\ }\textbf {\bibinfo {volume} {7}},\ \bibinfo {pages} {024} (\bibinfo {year} {2022})},\ \Eprint {https://arxiv.org/abs/2110.15662} {arXiv:2110.15662 [hep-ph]} \BibitemShut {NoStop}%
\bibitem [{\citenamefont {Capatti}\ \emph {et~al.}(2022)\citenamefont {Capatti}, \citenamefont {Hirschi},\ and\ \citenamefont {Ruijl}}]{Capatti:2022tit}%
  \BibitemOpen
  \bibfield  {author} {\bibinfo {author} {\bibfnamefont {Z.}~\bibnamefont {Capatti}}, \bibinfo {author} {\bibfnamefont {V.}~\bibnamefont {Hirschi}},\ and\ \bibinfo {author} {\bibfnamefont {B.}~\bibnamefont {Ruijl}},\ }\bibfield  {title} {\bibinfo {title} {{Local unitarity: cutting raised propagators and localising renormalisation}},\ }\href {https://doi.org/10.1007/JHEP10(2022)120} {\bibfield  {journal} {\bibinfo  {journal} {JHEP}\ }\textbf {\bibinfo {volume} {10}},\ \bibinfo {pages} {120}},\ \Eprint {https://arxiv.org/abs/2203.11038} {arXiv:2203.11038 [hep-ph]} \BibitemShut {NoStop}%
\bibitem [{\citenamefont {Capatti}(2023{\natexlab{b}})}]{Capatti:2023omc}%
  \BibitemOpen
  \bibfield  {author} {\bibinfo {author} {\bibfnamefont {Z.}~\bibnamefont {Capatti}},\ }\emph {\bibinfo {title} {{Singularities of Feynman diagrams and their local cancellation in collider cross-sections}}},\ \href {https://doi.org/10.3929/ethz-b-000647466} {Ph.D. thesis},\ \bibinfo  {school} {Zurich, ETH} (\bibinfo {year} {2023}{\natexlab{b}})\BibitemShut {NoStop}%
\bibitem [{\citenamefont {Galley}\ \emph {et~al.}(2016)\citenamefont {Galley}, \citenamefont {Leibovich}, \citenamefont {Porto},\ and\ \citenamefont {Ross}}]{Galley:2015kus}%
  \BibitemOpen
  \bibfield  {author} {\bibinfo {author} {\bibfnamefont {C.~R.}\ \bibnamefont {Galley}}, \bibinfo {author} {\bibfnamefont {A.~K.}\ \bibnamefont {Leibovich}}, \bibinfo {author} {\bibfnamefont {R.~A.}\ \bibnamefont {Porto}},\ and\ \bibinfo {author} {\bibfnamefont {A.}~\bibnamefont {Ross}},\ }\bibfield  {title} {\bibinfo {title} {{Tail effect in gravitational radiation reaction: Time nonlocality and renormalization group evolution}},\ }\href {https://doi.org/10.1103/PhysRevD.93.124010} {\bibfield  {journal} {\bibinfo  {journal} {Phys. Rev. D}\ }\textbf {\bibinfo {volume} {93}},\ \bibinfo {pages} {124010} (\bibinfo {year} {2016})},\ \Eprint {https://arxiv.org/abs/1511.07379} {arXiv:1511.07379 [gr-qc]} \BibitemShut {NoStop}%
\bibitem [{\citenamefont {de~la Cruz}\ \emph {et~al.}(2020)\citenamefont {de~la Cruz}, \citenamefont {Maybee}, \citenamefont {O'Connell},\ and\ \citenamefont {Ross}}]{delaCruz:2020bbn}%
  \BibitemOpen
  \bibfield  {author} {\bibinfo {author} {\bibfnamefont {L.}~\bibnamefont {de~la Cruz}}, \bibinfo {author} {\bibfnamefont {B.}~\bibnamefont {Maybee}}, \bibinfo {author} {\bibfnamefont {D.}~\bibnamefont {O'Connell}},\ and\ \bibinfo {author} {\bibfnamefont {A.}~\bibnamefont {Ross}},\ }\bibfield  {title} {\bibinfo {title} {{Classical Yang-Mills observables from amplitudes}},\ }\href {https://doi.org/10.1007/JHEP12(2020)076} {\bibfield  {journal} {\bibinfo  {journal} {JHEP}\ }\textbf {\bibinfo {volume} {12}},\ \bibinfo {pages} {076}},\ \Eprint {https://arxiv.org/abs/2009.03842} {arXiv:2009.03842 [hep-th]} \BibitemShut {NoStop}%
\bibitem [{\citenamefont {de~la Cruz}\ \emph {et~al.}(2022)\citenamefont {de~la Cruz}, \citenamefont {Luna},\ and\ \citenamefont {Scheopner}}]{delaCruz:2021gjp}%
  \BibitemOpen
  \bibfield  {author} {\bibinfo {author} {\bibfnamefont {L.}~\bibnamefont {de~la Cruz}}, \bibinfo {author} {\bibfnamefont {A.}~\bibnamefont {Luna}},\ and\ \bibinfo {author} {\bibfnamefont {T.}~\bibnamefont {Scheopner}},\ }\bibfield  {title} {\bibinfo {title} {{Yang-Mills observables: from KMOC to eikonal through EFT}},\ }\href {https://doi.org/10.1007/JHEP01(2022)045} {\bibfield  {journal} {\bibinfo  {journal} {JHEP}\ }\textbf {\bibinfo {volume} {01}},\ \bibinfo {pages} {045}},\ \Eprint {https://arxiv.org/abs/2108.02178} {arXiv:2108.02178 [hep-th]} \BibitemShut {NoStop}%
\bibitem [{\citenamefont {Weinberg}(2005)}]{Weinberg:1995mt}%
  \BibitemOpen
  \bibfield  {author} {\bibinfo {author} {\bibfnamefont {S.}~\bibnamefont {Weinberg}},\ }\href {https://doi.org/10.1017/CBO9781139644167} {\emph {\bibinfo {title} {{The Quantum theory of fields. Vol. 1: Foundations}}}}\ (\bibinfo  {publisher} {Cambridge University Press},\ \bibinfo {year} {2005})\BibitemShut {NoStop}%
\bibitem [{\citenamefont {Akhoury}\ \emph {et~al.}(2021)\citenamefont {Akhoury}, \citenamefont {Saotome},\ and\ \citenamefont {Sterman}}]{Akhoury:2013yua}%
  \BibitemOpen
  \bibfield  {author} {\bibinfo {author} {\bibfnamefont {R.}~\bibnamefont {Akhoury}}, \bibinfo {author} {\bibfnamefont {R.}~\bibnamefont {Saotome}},\ and\ \bibinfo {author} {\bibfnamefont {G.}~\bibnamefont {Sterman}},\ }\bibfield  {title} {\bibinfo {title} {{High Energy Scattering in Perturbative Quantum Gravity at Next to Leading Power}},\ }\href {https://doi.org/10.1103/PhysRevD.103.064036} {\bibfield  {journal} {\bibinfo  {journal} {Phys. Rev. D}\ }\textbf {\bibinfo {volume} {103}},\ \bibinfo {pages} {064036} (\bibinfo {year} {2021})},\ \Eprint {https://arxiv.org/abs/1308.5204} {arXiv:1308.5204 [hep-th]} \BibitemShut {NoStop}%
\bibitem [{\citenamefont {Driesse}\ \emph {et~al.}(2024{\natexlab{b}})\citenamefont {Driesse}, \citenamefont {Jakobsen}, \citenamefont {Mogull}, \citenamefont {Plefka}, \citenamefont {Sauer},\ and\ \citenamefont {Usovitsch}}]{Driesse:2024xad}%
  \BibitemOpen
  \bibfield  {author} {\bibinfo {author} {\bibfnamefont {M.}~\bibnamefont {Driesse}}, \bibinfo {author} {\bibfnamefont {G.~U.}\ \bibnamefont {Jakobsen}}, \bibinfo {author} {\bibfnamefont {G.}~\bibnamefont {Mogull}}, \bibinfo {author} {\bibfnamefont {J.}~\bibnamefont {Plefka}}, \bibinfo {author} {\bibfnamefont {B.}~\bibnamefont {Sauer}},\ and\ \bibinfo {author} {\bibfnamefont {J.}~\bibnamefont {Usovitsch}},\ }\bibfield  {title} {\bibinfo {title} {{Conservative Black Hole Scattering at Fifth Post-Minkowskian and First Self-Force Order}},\ }\href {https://doi.org/10.1103/PhysRevLett.132.241402} {\bibfield  {journal} {\bibinfo  {journal} {Phys. Rev. Lett.}\ }\textbf {\bibinfo {volume} {132}},\ \bibinfo {pages} {241402} (\bibinfo {year} {2024}{\natexlab{b}})},\ \Eprint {https://arxiv.org/abs/2403.07781} {arXiv:2403.07781 [hep-th]} \BibitemShut {NoStop}%
\bibitem [{\citenamefont {Caron-Huot}\ \emph {et~al.}(2024)\citenamefont {Caron-Huot}, \citenamefont {Giroux}, \citenamefont {Hannesdottir},\ and\ \citenamefont {Mizera}}]{Caron-Huot:2023vxl}%
  \BibitemOpen
  \bibfield  {author} {\bibinfo {author} {\bibfnamefont {S.}~\bibnamefont {Caron-Huot}}, \bibinfo {author} {\bibfnamefont {M.}~\bibnamefont {Giroux}}, \bibinfo {author} {\bibfnamefont {H.~S.}\ \bibnamefont {Hannesdottir}},\ and\ \bibinfo {author} {\bibfnamefont {S.}~\bibnamefont {Mizera}},\ }\bibfield  {title} {\bibinfo {title} {{What can be measured asymptotically?}},\ }\href {https://doi.org/10.1007/JHEP01(2024)139} {\bibfield  {journal} {\bibinfo  {journal} {JHEP}\ }\textbf {\bibinfo {volume} {01}},\ \bibinfo {pages} {139}},\ \Eprint {https://arxiv.org/abs/2308.02125} {arXiv:2308.02125 [hep-th]} \BibitemShut {NoStop}%
\bibitem [{\citenamefont {Elkhidir}\ \emph {et~al.}(2024)\citenamefont {Elkhidir}, \citenamefont {O'Connell},\ and\ \citenamefont {Roiban}}]{Elkhidir:2024izo}%
  \BibitemOpen
  \bibfield  {author} {\bibinfo {author} {\bibfnamefont {A.}~\bibnamefont {Elkhidir}}, \bibinfo {author} {\bibfnamefont {D.}~\bibnamefont {O'Connell}},\ and\ \bibinfo {author} {\bibfnamefont {R.}~\bibnamefont {Roiban}},\ }\bibfield  {title} {\bibinfo {title} {{Supertranslations from Scattering Amplitudes}},\ }\href@noop {} {\  (\bibinfo {year} {2024})},\ \Eprint {https://arxiv.org/abs/2408.15961} {arXiv:2408.15961 [hep-th]} \BibitemShut {NoStop}%
\bibitem [{\citenamefont {Landau}(1959)}]{Landau:1959fi}%
  \BibitemOpen
  \bibfield  {author} {\bibinfo {author} {\bibfnamefont {L.~D.}\ \bibnamefont {Landau}},\ }\bibfield  {title} {\bibinfo {title} {{On analytic properties of vertex parts in quantum field theory}},\ }\href {https://doi.org/10.1016/B978-0-08-010586-4.50103-6} {\bibfield  {journal} {\bibinfo  {journal} {Nucl. Phys.}\ }\textbf {\bibinfo {volume} {13}},\ \bibinfo {pages} {181} (\bibinfo {year} {1959})}\BibitemShut {NoStop}%
\bibitem [{\citenamefont {Coleman}\ and\ \citenamefont {Norton}(1965)}]{Coleman1965SingularitiesIT}%
  \BibitemOpen
  \bibfield  {author} {\bibinfo {author} {\bibfnamefont {S.~R.}\ \bibnamefont {Coleman}}\ and\ \bibinfo {author} {\bibfnamefont {R.~E.}\ \bibnamefont {Norton}},\ }\bibfield  {title} {\bibinfo {title} {Singularities in the physical region},\ }\href@noop {} {\bibfield  {journal} {\bibinfo  {journal} {Il Nuovo Cimento (1955-1965)}\ }\textbf {\bibinfo {volume} {38}},\ \bibinfo {pages} {438} (\bibinfo {year} {1965})}\BibitemShut {NoStop}%
\bibitem [{\citenamefont {Bourjaily}\ \emph {et~al.}(2021)\citenamefont {Bourjaily}, \citenamefont {Hannesdottir}, \citenamefont {McLeod}, \citenamefont {Schwartz},\ and\ \citenamefont {Vergu}}]{Bourjaily:2020wvq}%
  \BibitemOpen
  \bibfield  {author} {\bibinfo {author} {\bibfnamefont {J.~L.}\ \bibnamefont {Bourjaily}}, \bibinfo {author} {\bibfnamefont {H.}~\bibnamefont {Hannesdottir}}, \bibinfo {author} {\bibfnamefont {A.~J.}\ \bibnamefont {McLeod}}, \bibinfo {author} {\bibfnamefont {M.~D.}\ \bibnamefont {Schwartz}},\ and\ \bibinfo {author} {\bibfnamefont {C.}~\bibnamefont {Vergu}},\ }\bibfield  {title} {\bibinfo {title} {{Sequential Discontinuities of Feynman Integrals and the Monodromy Group}},\ }\href {https://doi.org/10.1007/JHEP01(2021)205} {\bibfield  {journal} {\bibinfo  {journal} {JHEP}\ }\textbf {\bibinfo {volume} {01}},\ \bibinfo {pages} {205}},\ \Eprint {https://arxiv.org/abs/2007.13747} {arXiv:2007.13747 [hep-th]} \BibitemShut {NoStop}%
\bibitem [{\citenamefont {Bern}\ \emph {et~al.}(2024{\natexlab{b}})\citenamefont {Bern}, \citenamefont {Herrmann}, \citenamefont {Roiban}, \citenamefont {Ruf},\ and\ \citenamefont {Zeng}}]{Bern:2024vqs}%
  \BibitemOpen
  \bibfield  {author} {\bibinfo {author} {\bibfnamefont {Z.}~\bibnamefont {Bern}}, \bibinfo {author} {\bibfnamefont {E.}~\bibnamefont {Herrmann}}, \bibinfo {author} {\bibfnamefont {R.}~\bibnamefont {Roiban}}, \bibinfo {author} {\bibfnamefont {M.~S.}\ \bibnamefont {Ruf}},\ and\ \bibinfo {author} {\bibfnamefont {M.}~\bibnamefont {Zeng}},\ }\bibfield  {title} {\bibinfo {title} {{Global Bases for Nonplanar Loop Integrands, Generalized Unitarity, and the Double Copy to All Loop Orders}},\ }\href@noop {} {\  (\bibinfo {year} {2024}{\natexlab{b}})},\ \Eprint {https://arxiv.org/abs/2408.06686} {arXiv:2408.06686 [hep-th]} \BibitemShut {NoStop}%
\bibitem [{\citenamefont {Biswas}\ and\ \citenamefont {Parra-Martinez}(2024)}]{Biswas:2024ept}%
  \BibitemOpen
  \bibfield  {author} {\bibinfo {author} {\bibfnamefont {S.}~\bibnamefont {Biswas}}\ and\ \bibinfo {author} {\bibfnamefont {J.}~\bibnamefont {Parra-Martinez}},\ }\bibfield  {title} {\bibinfo {title} {{Classical Observables from Causal Response Functions}},\ }\href@noop {} {\  (\bibinfo {year} {2024})},\ \Eprint {https://arxiv.org/abs/2411.09016} {arXiv:2411.09016 [hep-th]} \BibitemShut {NoStop}%
\bibitem [{\citenamefont {Kawai}\ \emph {et~al.}(1986)\citenamefont {Kawai}, \citenamefont {Lewellen},\ and\ \citenamefont {Tye}}]{Kawai:1985xq}%
  \BibitemOpen
  \bibfield  {author} {\bibinfo {author} {\bibfnamefont {H.}~\bibnamefont {Kawai}}, \bibinfo {author} {\bibfnamefont {D.~C.}\ \bibnamefont {Lewellen}},\ and\ \bibinfo {author} {\bibfnamefont {S.~H.~H.}\ \bibnamefont {Tye}},\ }\bibfield  {title} {\bibinfo {title} {{A Relation Between Tree Amplitudes of Closed and Open Strings}},\ }\href {https://doi.org/10.1016/0550-3213(86)90362-7} {\bibfield  {journal} {\bibinfo  {journal} {Nucl. Phys. B}\ }\textbf {\bibinfo {volume} {269}},\ \bibinfo {pages} {1} (\bibinfo {year} {1986})}\BibitemShut {NoStop}%
\bibitem [{\citenamefont {Bern}\ \emph {et~al.}(2024{\natexlab{c}})\citenamefont {Bern}, \citenamefont {Carrasco}, \citenamefont {Chiodaroli}, \citenamefont {Johansson},\ and\ \citenamefont {Roiban}}]{Bern:2019prr}%
  \BibitemOpen
  \bibfield  {author} {\bibinfo {author} {\bibfnamefont {Z.}~\bibnamefont {Bern}}, \bibinfo {author} {\bibfnamefont {J.~J.}\ \bibnamefont {Carrasco}}, \bibinfo {author} {\bibfnamefont {M.}~\bibnamefont {Chiodaroli}}, \bibinfo {author} {\bibfnamefont {H.}~\bibnamefont {Johansson}},\ and\ \bibinfo {author} {\bibfnamefont {R.}~\bibnamefont {Roiban}},\ }\bibfield  {title} {\bibinfo {title} {{The duality between color and kinematics and its applications}},\ }\href {https://doi.org/10.1088/1751-8121/ad5fd0} {\bibfield  {journal} {\bibinfo  {journal} {J. Phys. A}\ }\textbf {\bibinfo {volume} {57}},\ \bibinfo {pages} {333002} (\bibinfo {year} {2024}{\natexlab{c}})},\ \Eprint {https://arxiv.org/abs/1909.01358} {arXiv:1909.01358 [hep-th]} \BibitemShut {NoStop}%
\bibitem [{\citenamefont {Engel}(2023)}]{Engel:2023ifn}%
  \BibitemOpen
  \bibfield  {author} {\bibinfo {author} {\bibfnamefont {T.}~\bibnamefont {Engel}},\ }\bibfield  {title} {\bibinfo {title} {{The LBK theorem to all orders}},\ }\href {https://doi.org/10.1007/JHEP07(2023)177} {\bibfield  {journal} {\bibinfo  {journal} {JHEP}\ }\textbf {\bibinfo {volume} {07}},\ \bibinfo {pages} {177}},\ \Eprint {https://arxiv.org/abs/2304.11689} {arXiv:2304.11689 [hep-ph]} \BibitemShut {NoStop}%
\bibitem [{\citenamefont {Liu}\ and\ \citenamefont {Monni}(2024)}]{Liu:2024hfa}%
  \BibitemOpen
  \bibfield  {author} {\bibinfo {author} {\bibfnamefont {Z.~L.}\ \bibnamefont {Liu}}\ and\ \bibinfo {author} {\bibfnamefont {P.~F.}\ \bibnamefont {Monni}},\ }\bibfield  {title} {\bibinfo {title} {{The two-loop fully differential soft function for $Q\bar{Q}V$ production at lepton colliders}},\ }\href@noop {} {\  (\bibinfo {year} {2024})},\ \Eprint {https://arxiv.org/abs/2411.13466} {arXiv:2411.13466 [hep-ph]} \BibitemShut {NoStop}%
\bibitem [{\citenamefont {Laenen}\ \emph {et~al.}(2009)\citenamefont {Laenen}, \citenamefont {Stavenga},\ and\ \citenamefont {White}}]{Laenen:2008gt}%
  \BibitemOpen
  \bibfield  {author} {\bibinfo {author} {\bibfnamefont {E.}~\bibnamefont {Laenen}}, \bibinfo {author} {\bibfnamefont {G.}~\bibnamefont {Stavenga}},\ and\ \bibinfo {author} {\bibfnamefont {C.~D.}\ \bibnamefont {White}},\ }\bibfield  {title} {\bibinfo {title} {{Path integral approach to eikonal and next-to-eikonal exponentiation}},\ }\href {https://doi.org/10.1088/1126-6708/2009/03/054} {\bibfield  {journal} {\bibinfo  {journal} {JHEP}\ }\textbf {\bibinfo {volume} {03}},\ \bibinfo {pages} {054}},\ \Eprint {https://arxiv.org/abs/0811.2067} {arXiv:0811.2067 [hep-ph]} \BibitemShut {NoStop}%
\bibitem [{\citenamefont {White}(2011)}]{White:2011yy}%
  \BibitemOpen
  \bibfield  {author} {\bibinfo {author} {\bibfnamefont {C.~D.}\ \bibnamefont {White}},\ }\bibfield  {title} {\bibinfo {title} {{Factorization Properties of Soft Graviton Amplitudes}},\ }\href {https://doi.org/10.1007/JHEP05(2011)060} {\bibfield  {journal} {\bibinfo  {journal} {JHEP}\ }\textbf {\bibinfo {volume} {05}},\ \bibinfo {pages} {060}},\ \Eprint {https://arxiv.org/abs/1103.2981} {arXiv:1103.2981 [hep-th]} \BibitemShut {NoStop}%
\bibitem [{\citenamefont {Bonocore}(2021)}]{Bonocore:2020xuj}%
  \BibitemOpen
  \bibfield  {author} {\bibinfo {author} {\bibfnamefont {D.}~\bibnamefont {Bonocore}},\ }\bibfield  {title} {\bibinfo {title} {{Asymptotic dynamics on the worldline for spinning particles}},\ }\href {https://doi.org/10.1007/JHEP02(2021)007} {\bibfield  {journal} {\bibinfo  {journal} {JHEP}\ }\textbf {\bibinfo {volume} {02}},\ \bibinfo {pages} {007}},\ \Eprint {https://arxiv.org/abs/2009.07863} {arXiv:2009.07863 [hep-th]} \BibitemShut {NoStop}%
\bibitem [{\citenamefont {Bonocore}\ \emph {et~al.}(2022)\citenamefont {Bonocore}, \citenamefont {Kulesza},\ and\ \citenamefont {Pirsch}}]{Bonocore:2021qxh}%
  \BibitemOpen
  \bibfield  {author} {\bibinfo {author} {\bibfnamefont {D.}~\bibnamefont {Bonocore}}, \bibinfo {author} {\bibfnamefont {A.}~\bibnamefont {Kulesza}},\ and\ \bibinfo {author} {\bibfnamefont {J.}~\bibnamefont {Pirsch}},\ }\bibfield  {title} {\bibinfo {title} {{Classical and quantum gravitational scattering with Generalized Wilson Lines}},\ }\href {https://doi.org/10.1007/JHEP03(2022)147} {\bibfield  {journal} {\bibinfo  {journal} {JHEP}\ }\textbf {\bibinfo {volume} {03}},\ \bibinfo {pages} {147}},\ \Eprint {https://arxiv.org/abs/2112.02009} {arXiv:2112.02009 [hep-th]} \BibitemShut {NoStop}%
\bibitem [{\citenamefont {Schubert}(2001)}]{Schubert:2001he}%
  \BibitemOpen
  \bibfield  {author} {\bibinfo {author} {\bibfnamefont {C.}~\bibnamefont {Schubert}},\ }\bibfield  {title} {\bibinfo {title} {{Perturbative quantum field theory in the string inspired formalism}},\ }\href {https://doi.org/10.1016/S0370-1573(01)00013-8} {\bibfield  {journal} {\bibinfo  {journal} {Phys. Rept.}\ }\textbf {\bibinfo {volume} {355}},\ \bibinfo {pages} {73} (\bibinfo {year} {2001})},\ \Eprint {https://arxiv.org/abs/hep-th/0101036} {arXiv:hep-th/0101036} \BibitemShut {NoStop}%
\bibitem [{\citenamefont {Edwards}\ and\ \citenamefont {Schubert}(2019)}]{Edwards:2019eby}%
  \BibitemOpen
  \bibfield  {author} {\bibinfo {author} {\bibfnamefont {J.~P.}\ \bibnamefont {Edwards}}\ and\ \bibinfo {author} {\bibfnamefont {C.}~\bibnamefont {Schubert}},\ }\bibfield  {title} {\bibinfo {title} {{Quantum mechanical path integrals in the first quantised approach to quantum field theory}}\ }(\bibinfo {year} {2019})\ \Eprint {https://arxiv.org/abs/1912.10004} {arXiv:1912.10004 [hep-th]} \BibitemShut {NoStop}%
\bibitem [{\citenamefont {Feal}\ \emph {et~al.}(2022)\citenamefont {Feal}, \citenamefont {Tarasov},\ and\ \citenamefont {Venugopalan}}]{Feal:2022iyn}%
  \BibitemOpen
  \bibfield  {author} {\bibinfo {author} {\bibfnamefont {X.}~\bibnamefont {Feal}}, \bibinfo {author} {\bibfnamefont {A.}~\bibnamefont {Tarasov}},\ and\ \bibinfo {author} {\bibfnamefont {R.}~\bibnamefont {Venugopalan}},\ }\bibfield  {title} {\bibinfo {title} {{QED as a many-body theory of worldlines: General formalism and infrared structure}},\ }\href {https://doi.org/10.1103/PhysRevD.106.056009} {\bibfield  {journal} {\bibinfo  {journal} {Phys. Rev. D}\ }\textbf {\bibinfo {volume} {106}},\ \bibinfo {pages} {056009} (\bibinfo {year} {2022})},\ \Eprint {https://arxiv.org/abs/2206.04188} {arXiv:2206.04188 [hep-th]} \BibitemShut {NoStop}%
\bibitem [{\citenamefont {Jakobsen}\ \emph {et~al.}(2022{\natexlab{b}})\citenamefont {Jakobsen}, \citenamefont {Mogull}, \citenamefont {Plefka},\ and\ \citenamefont {Steinhoff}}]{Jakobsen:2021zvh}%
  \BibitemOpen
  \bibfield  {author} {\bibinfo {author} {\bibfnamefont {G.~U.}\ \bibnamefont {Jakobsen}}, \bibinfo {author} {\bibfnamefont {G.}~\bibnamefont {Mogull}}, \bibinfo {author} {\bibfnamefont {J.}~\bibnamefont {Plefka}},\ and\ \bibinfo {author} {\bibfnamefont {J.}~\bibnamefont {Steinhoff}},\ }\bibfield  {title} {\bibinfo {title} {{SUSY in the sky with gravitons}},\ }\href {https://doi.org/10.1007/JHEP01(2022)027} {\bibfield  {journal} {\bibinfo  {journal} {JHEP}\ }\textbf {\bibinfo {volume} {01}},\ \bibinfo {pages} {027}},\ \Eprint {https://arxiv.org/abs/2109.04465} {arXiv:2109.04465 [hep-th]} \BibitemShut {NoStop}%
\end{thebibliography}%

\end{document}